\documentclass[journal=jacsat,manuscript=article]{achemso}
\setkeys{acs}{articletitle = true}

\graphicspath{ {./images/} }

\usepackage[version=3]{mhchem} 

\usepackage{multicol}
\usepackage{graphicx} 
\usepackage{changepage}
\usepackage{amsmath}
\usepackage{verbatim}
\usepackage{color}
\usepackage{array}
\usepackage{soul}

\author{Fabian Berger}
\affiliation{Yusuf Hamied Department of Chemistry, University of Cambridge, CB2 1EW Cambridge, UK}
\alsoaffiliation{Lennard--Jones Centre, University of Cambridge, Trinity Ln, Cambridge, CB2 1TN, UK}
\alsoaffiliation{Max Planck Institute for Polymer Research, Ackermannweg 10, 55128 Mainz, Germany}
\email{fb593@cam.ac.uk}

\author{Yicheng Wang}
\affiliation{Department of Chemistry, Tufts University, Medford, MA, 02155, USA}

\author{E. Charles H. Sykes}
\affiliation{Department of Chemistry, Tufts University, Medford, MA, 02155, USA}
\alsoaffiliation{Department of Chemical and Biological Engineering, Tufts University, Medford, MA, 02155, USA}

\author{Angelos Michaelides}
\affiliation{Yusuf Hamied Department of Chemistry, University of Cambridge, CB2 1EW Cambridge, UK}
\alsoaffiliation{Lennard--Jones Centre, University of Cambridge, Trinity Ln, Cambridge, CB2 1TN, UK}
\email{am452@cam.ac.uk}

\title {An Atlas and Design Rules for Single- and Dual-Atom Alloys}


\begin{document}

\clearpage

\begin{abstract}

A long-standing goal across heterogeneous catalysis, materials science, and condensed matter physics is to design alloys with prescribed local atomic arrangements.
Recent experiments show that dilute trimetallic alloys unlock chemistries inaccessible to bimetallics, but realizing this potential requires knowing which dopant structures form across an enormous compositional space.
Using density functional theory screening, we construct an atlas spanning transition metal single- and dual-atom alloys in Cu and Ag surfaces, which we validate by scanning tunneling microscopy.
The stability follows an electron count: dopants pair most strongly when their combined \textit{d}-electron count approaches ten.
Host metal and surface facet can be used to tune the resulting active site motifs, while size mismatch and spin explain variations around this trend.
We further introduce a reactor--anchor concept, in which one dopant anchors a second, otherwise bulk-segregating dopant at the surface.
Together, these results establish design principles for engineering alloys with targeted active site structures.

\end{abstract}

\clearpage

\section{Introduction}
Catalysis underpins ~85\% of industrial chemical processes\cite{demand_catalysis} and contributes to roughly 35\% of global gross domestic product (GDP).\cite{ma2006heterogeneous}
Even modest improvements in catalyst performance therefore reduce material and energy demand, with substantial economic and environmental benefits.
Yet catalyst discovery has long relied on laborious empirical research, and it is only through the combined advances in surface and materials science, chemistry, and atomistic simulation that rational design is now becoming possible.\cite{greeley2004alloy, greeley2006computational, han2021single}

Single-atom alloys (SAAs) exemplify this shift.\cite{kyriakou2012isolated, yang2013single, darby2018lonely, cui2018bridging, hannagan2020single, reocreux2021one, spivey2021selective, lee2022dilute, chen2022electronic,schumann2024ten} 
In SAAs, isolated atoms of a reactive element are embedded in a more inert host metal. 
This architecture gives rise to distinct geometric and electronic properties of the active sites\cite{michaelides2007unhappy, inderwildi2007adding, greiner2018free, zhang2019catalytic, fung2020electronic, spivey2021selective, rosen2023free, he2024selective, berger2025dopant} and decouples activity and selectivity.\cite{sun2018breaking, perez2019strategies, monasterial2020more, nwaokorie2022alloy, stratton2023addressing}
The dopant provides an active site that selectively lowers activation barriers,\cite{hannagan2021first} thereby favoring specific reaction pathways.
The dopant can additionally bring reactants into proximity and stabilize their co-adsorbed configuration, promoting bond formation in coupling reactions.\cite{berger2024bringing, christiansen2024single, christiansen2025multiple, eren2018structure, wang2022observation}
In addition, reaction intermediates generally bind more weakly to the dopant sites in SAAs than on the corresponding pure dopant metal surface.\cite{hannagan2021first}
Upon spillover from the active site, the more inert host provides a less reactive environment that reduces undesired subsequent reactions, including those leading to coke formation and catalyst deactivation. 
Weaker binding on the host also facilitates product desorption.\cite{marcinkowski2013controlling, reocreux2022stick}
Together, these advantages allow SAAs to address the long-standing activity--selectivity trade-off in heterogeneous catalysis.\cite{sabatier1920catalyse, bronsted1928acid, bell1936theory, evans1937introduction, evans1938inertia, norskov2002universality, michaelides2003identification, che2013nobel} 
Accordingly, SAAs have demonstrated exceptional performance across a growing range of reactions, including hydrogenation,\cite{kyriakou2012isolated, lucci2015selective, jiang2020facet, pei2015ag} dehydrogenation,\cite{ngan2026early, hannagan2021first,zhou2020light,sun2018breaking} and selective oxidation.\cite{jalil2025nickel}

Despite these advantages, SAAs also face intrinsic limitations.
First, in multi-step reactions, a single dopant element may not efficiently catalyze every elementary step,\cite{laltrimetallic} limiting overall performance.
Second, in heteronuclear bond dissociation or cross-coupling reactions, only one reaction fragment may interact strongly with a given dopant element,\cite{kayode2021factors,kress2023priori,weaver2025self}
reducing the extent to which reaction barriers are lowered or intermediates are stabilized.
Third, many reactive dopants exhibit a thermodynamic preference for bulk segregation, limiting their accessibility as surface active sites at elevated temperatures.
Together, these challenges suggest that reactions may require active sites extending beyond isolated dopants, a concept that has been explored in other classes of atomically dispersed catalysts\cite{han2022single, he2022atomically,lin2024machine,ro2022bifunctional,chen2026dual}
but has received little attention in metal alloys.
These considerations motivate the exploration of trimetallic alloys, where a second dopant element can modify the electronic structure of the active site,\cite{zhang2022tuning}
introduce additional functionality, enable new classes of active sites, and potentially stabilize reactive dopants in the surface layer.
Notably, a few pioneering studies exist, which either examine the thermodynamics of small subsets of the compositional space\cite{zhang2022tuning, huang2025decoding} or focus on the effect of specific dopant combinations on a single reaction.\cite{kress2023priori,weaver2025self}
An illustrative example is provided by recent studies of PtCrAg.\cite{kress2023priori,weaver2025self}
In this trimetallic alloy, mixed Pt--Cr dopant pairs enable alcohol dehydrogenation chemistry that is inaccessible on either PtAg or CrAg bimetallics.

\begin{figure}[ht]
    \includegraphics[width=16cm,height=\textheight,keepaspectratio]{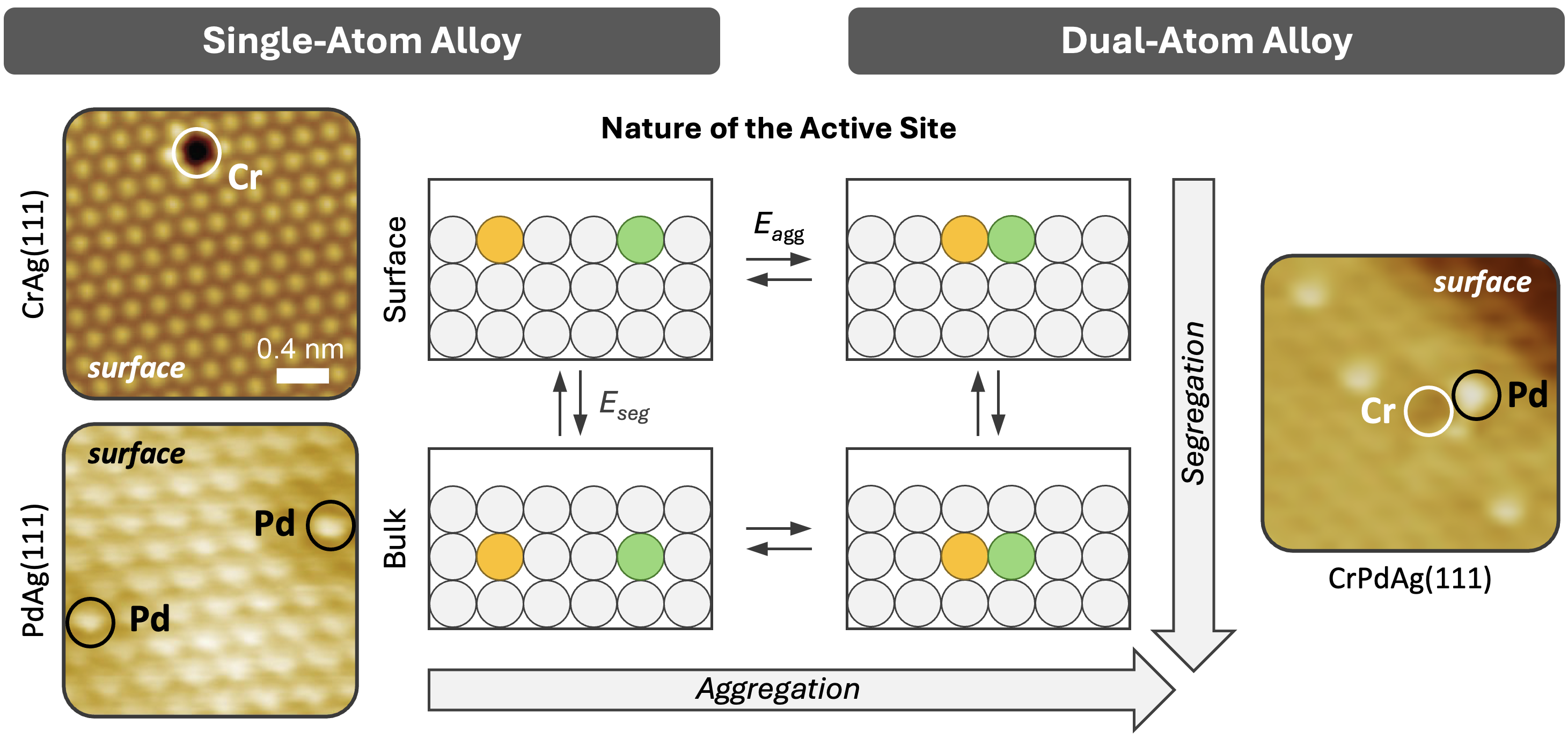}
    \caption{
    \textbf{Illustration of the active site types considered in this work and their experimental validation.}
    Center: Schematic of active site motifs and the associated process energies.
    The aggregation energy, $E_\text{agg}$ (Eq.~\ref{eq:aggregation_energy}), quantifies the conversion of two isolated single-atom sites into a dual-atom site, while the segregation energy, $E_\text{seg}$ (Eq.~\ref{eq:segregation_energy}), quantifies the preference for dopants to segregate from the surface into the bulk.
    Host atoms are shown in gray and dopant atoms in orange and green.
    Left and right: Scanning tunneling microscopy images of synthesized alloys.
    CrAg(111) and PdAg(111) exhibit single-atom sites, whereas hetero-dimers are observed for CrPdAg(111).
    Imaging conditions are:  CrAg(111), $-0.1$~V, 0.6~nA; PdAg(111), $-0.3$~V, 0.3~nA; and CrPdAg(111), $-0.02$~V, 40~nA.
    }
    \label{fig:scheme}
\end{figure}

The vastly expanded materials space of trimetallic alloys enables the construction of novel heteroatom pairs and interfaces, creating opportunities for materials discovery and design that go far beyond what is accessible with SAAs alone.
However, this added flexibility comes at the cost of increased structural complexity.\cite{rao2020extendable,he2025stability}
When two different dopants are introduced, active sites may remain atomically dispersed, resulting in SAA sites, or they may aggregate to form dimers, so-called dual-atom alloy (DAA) sites, as illustrated in Fig.~\ref{fig:scheme}.
Dopant dimers can consist of the same element (homo-dimers) or of two different elements (hetero-dimers), and their stability depends on the interplay between dopant elements, host metal, and surface facet.\cite{christensen1997phase}
This complexity calls for theoretical guidance to identify systems worthy of time-intensive experimental synthesis and testing.
Specifically, it raises the following key questions: 
\textit{Which dopant combinations thermodynamically favor SAAs and which favor DAAs? 
Which dopants are stabilized at the surface and thus remain accessible for reaction?
Can one dopant anchor a second dopant to the surface?
What fundamental principles govern these preferences, and can active site types be deliberately tuned?}

To address these points, we present an \textit{Atlas of Active Sites in Single- and Dual-Atom Alloys} that systematically maps the stability of combinations of transition metal (TM) dopants embedded in the catalytically important surfaces of Cu and Ag.
By surveying a broad compositional materials space, the Atlas identifies and rationalizes stability trends, establishes a predictive framework for anticipating active site types and their stability in the surface layer, and provides design handles for tuning active site type and location through the choice of host metal and surface facet.
Our DFT predictions are validated by scanning tunneling microscopy (STM) experiments on selected bi- and trimetallic alloys. 
The design principles established here also inform the design of other materials with atomically dispersed active sites, such as high-entropy alloys and other multimetallic systems.\cite{zhang2020short, chen2021direct, xie2021percolation, li2026synthesis,ding2026differentiable,mazitov2024surface}
Indeed, by establishing a bottom-up understanding of these interactions, this work provides a foundation for rationalizing, predicting, and ultimately designing increasingly complex local atomic arrangements across a broad range of functional materials.

\section{Results}
The Atlas spans nearly 3000 SAA and DAA configurations across all combinations of 3$d$, 4$d$, and 5$d$ transition metal dopants in Cu(100), Cu(111), and Ag(111).
Our computational results are based on periodic DFT calculations using the optB86b-vdW functional.\cite{optB86b-vdW}
Section~S1 in the Supporting Information (SI) demonstrates that our results are robust with respect to the choice of exchange-correlation functional, as validated using a set of functionals with distinct properties,\cite{schimka2010accurate} RPBE,\cite{hammer1999improved} PBE,\cite{PBE} and PBEsol,\cite{PBEsol, PBEsol-er} as well as the foundation machine learning potential MACE-MP-0.\cite{batatia2025foundation}

In the following, we first establish global stability trends for single- and dual-atom sites across the full space of transition metal combinations and distill these trends into simple physical principles.
We then show how host element, surface facet, and spin of the dopant atom provide design handles to tune active site type and location.
We focus on the thermodynamic stability of the different active site motifs because, once embedded into the metal surface through low-barrier incorporation pathways,\cite{karageorgiou2026mechanisms} thermodynamically stable surface motifs do not rely on kinetic trapping and can persist even at elevated temperatures where barriers to segregation into the bulk can be overcome.\cite{finzel2026metal}
Building on these trends, we discuss the role of kinetics in determining dopant accessibility and introduce a \textit{reactor--anchor design} concept that exploits dopant pairing to stabilize otherwise inaccessible reactive dopants at the surface.
Finally, we integrate these principles to identify the experimentally most promising hetero-dimer candidates.

\subsection{A ten-electron rule for active site formation.}

We begin by discussing the propensity for dimer formation, i.e., the aggregation energy, $E_\text{agg}$.\cite{kothakonda2025discovering, zhang2022tuning, he2025stability, rao2020extendable, ouyang2021directing}
This is defined as

\begin{align}
  E_\text{agg}^{ij} &= E^\text{DAA}_{ij} + E^\text{host} - E^\text{SAA}_{i} - E^\text{SAA}_{j}, \label{eq:aggregation_energy} 
\end{align}

\noindent where $E^\text{DAA}_{ij}$ is the total energy of the DAA containing a dimer of dopants $i$ and $j$, $E^\text{host}$ is the total energy of the pristine host surface, and $E^\text{SAA}_{i}$ and $E^\text{SAA}_{j}$ are the total energies of the corresponding SAAs with isolated dopants. 
Negative aggregation energies indicate that dopant pairing is energetically favored, while positive values indicate stabilization of single-atom sites.
In practical catalytic systems, where catalysts typically consist of supported nanoparticles, a preference for homo-dimer formation may ultimately drive larger-scale phase separation, giving rise to core--shell or Janus-type morphologies.\cite{van2021unlocking}
In contrast, favorable hetero-dimer formation suggests a tendency toward mixed-dopant arrangements.
As the smallest aggregation motif, dopant dimers therefore provide a reference configuration for understanding broader aggregation behavior in multimetallic systems.

\begin{figure}[htb]
    \includegraphics[width=16cm,height=\textheight,keepaspectratio]{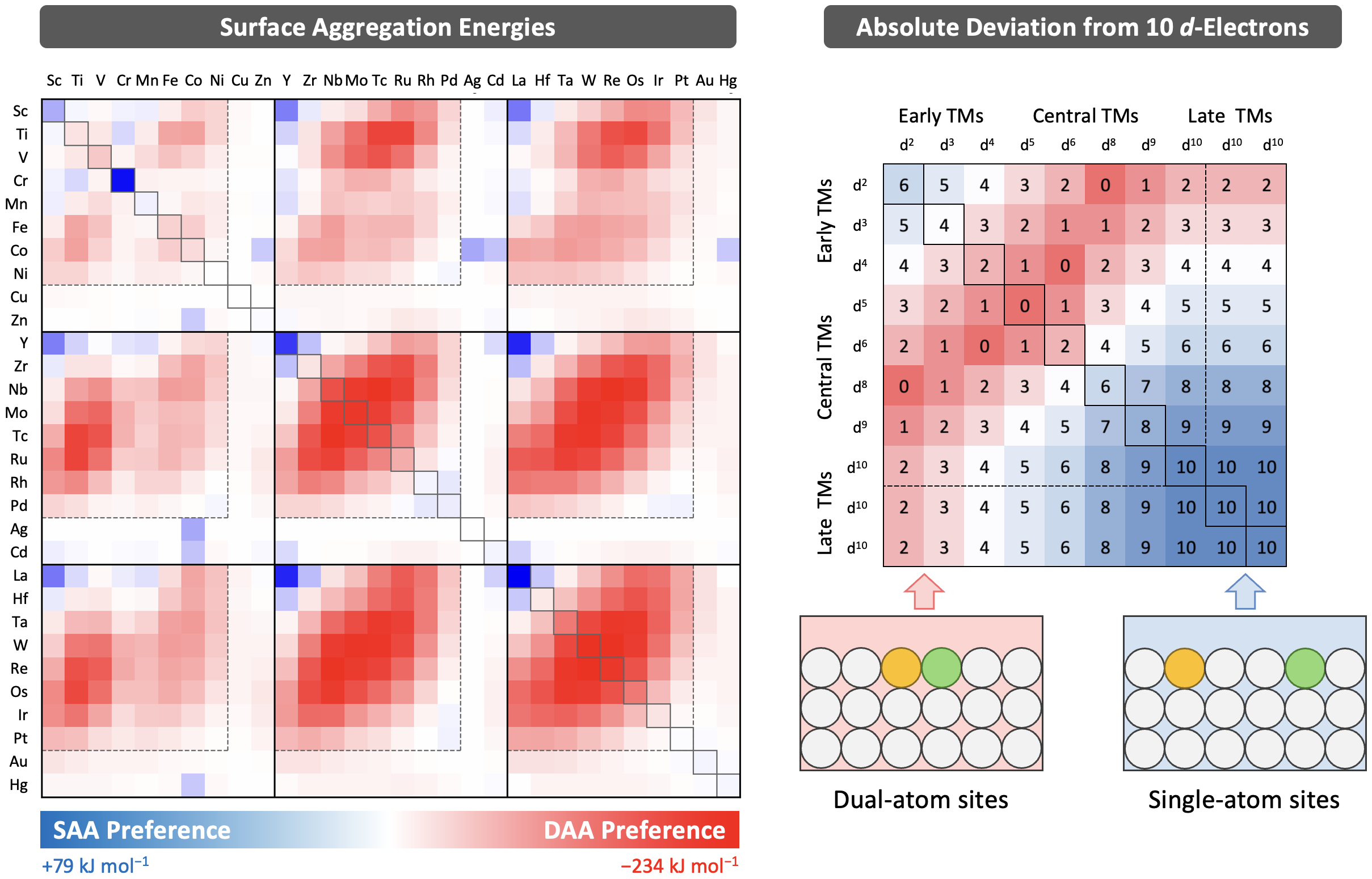}
    \caption{
    \textbf{Aggregation trends of dual-atom alloys and their rationalization by the ten-electron rule for dopant--dopant bond formation.}
    Left: Aggregation energies for the formation of all possible transition metal dopant dimers in the Ag(111) surface. 
    Preferences for single-atom sites are represented in blue, while preferences for dopant dimers are shown in red. 
    Illustrations in the bottom right depict single-atom sites and dopant dimers, with host atoms in gray and dopant atoms in orange and green.
    Right: Deviation of the sum of the \textit{d}-electrons of both transition metals forming a dopant dimer from a value of 10, using rounded integer \textit{d}-electron counts based on calculated \textit{d}-fillings from ref.~\citenum{schumann2024ten}.
    Small deviations are shown in red, while larger deviations are displayed in blue.
    The patterns observed on the left and right are similar, indicating a correlation. 
    }
    \label{fig:aggregation}
\end{figure}

Our screening reveals that the dopant element is the main factor determining whether atomically dispersed SAAs or DAAs form.
Specifically, very early and very late transition metals favor SAAs, whereas combinations of central TMs or early--late pairs favor DAAs.
These trends hold across all investigated host elements and surface facets, and are consistent with previous reports of homo-dimer formation for selected dopants.\cite{huang2025decoding, kothakonda2025discovering}
Figure~\ref{fig:aggregation} illustrates these relationships for Ag(111), and analogous plots for Cu(111) and Cu(100) are provided in Section~S2 of the SI.
Blue regions correspond to dopant combinations that remain as single-atom sites, while red regions indicate dimer formation. 
The right-hand panel shows the deviation of the total dopant \textit{d}-electron count from ten. 
The close match between the two panels demonstrates that aggregation is most favorable when the sum of the dopant \textit{d}-electrons is near ten, producing a characteristic anti-diagonal pattern. 

This behavior can be explained with a \textit{ten-electron rule: aggregation is strongest when the combined number of dopant \textit{d}-electrons approaches ten.}
A total of ten \textit{d}-electrons shared between two dopants is ideal for bond formation, since bonding molecular orbitals (MOs) derived from dopant \textit{d}-state hybridization are fully occupied while antibonding orbitals remain empty.
This electron configuration provides maximal stabilization of the dopant dimer, as it can formally support quintuple-bond character\cite{weinhold2007high,nguyen2005synthesis,brynda2006quantum} between two dopant atoms embedded in a metallic host.
In this sense, the rule explains the parabolic dependence of metal--metal bond strength on \textit{d}-band filling\cite{friedel1969physics, morse1986clusters} and generalizes it to the vast space of heteronuclear dopant pairs.
A related filling dependence also underlies descriptors recently proposed for the stability of homo-dimers.\cite{huang2025decoding}
Electron-counting rules are well established in chemistry, and a ten-electron rule has recently been shown to govern adsorbate binding at single-atom alloy sites.\cite{schumann2024ten}
Here the stability of the material itself follows an electron-counting rule, determining dopant--dopant interactions and the formation of stable dual-atom sites.

In general, combinations of TMs near the center of each TM row exhibit the strongest preference for dimer formation. 
These central TMs typically have half-filled \textit{d} bands, with about five electrons each, giving a total close to ten when paired.
A representative case is the Mo–Mo dimer, where both dopants formally contribute five \textit{d} electrons. 
This combination shows very strong aggregation, with an aggregation energy of $-214$ kJ mol$^{-1}$ in the Ag(111) surface.

Early–late combinations also tend to form dimers, although slightly less strongly.
In these cases, the late TM contributes nearly filled \textit{d} states, while the early metal contributes only a few \textit{d} electrons, together summing to about ten. 
An example is the Zr–Rh dimer, which has an aggregation energy of $-124$ kJ mol$^{-1}$ in Ag(111). 
This illustrates how pairing early and late TMs can stabilize dopant dimers through complementary electron counts.

In contrast, combinations of late TMs exhibit a slight preference for remaining as single-atom sites, with aggregation energies of about $+6$ kJ mol$^{-1}$ for a Pd homo dimer in Ag(111), as seen in Fig.~\ref{fig:scheme}. 
In such cases, isolated sites are further stabilized by the configurational entropy associated with the low dopant concentrations.
This near-thermoneutral behavior reflects the absence of significant bonding interactions between such dopants. 
When two nearly filled \textit{d} bands interact, both bonding and antibonding states become occupied, corresponding to close to 20 \textit{d} electrons and a net destabilization. 
As a result, dopant \textit{d} states avoid hybridization, and dimers among late TMs do not form readily. 
The same considerations also explain the weak mixing observed between late TM dopants and late coinage host metals in SAAs.\cite{berger2025dopant}

Combinations of very early TMs exhibit the strongest preference for single-atom sites.
A La–La pair in Ag(111), with La included here alongside the transition metals, has an aggregation energy of $+79$ kJ mol$^{-1}$, the most unfavorable value on this surface. 
This behavior cannot be explained solely by electron count, since similar deviations from ten electrons do not incur such large penalties for late TM dopants. 
Instead, the large atomic sizes of very early TMs introduce substantial lattice strain in the surrounding host surface, which is further amplified upon dimer formation.
This interpretation is supported by host dependence: a Zr–Zr dimer is 58 kJ mol$^{-1}$ more unfavorable in Cu(111) than in Ag(111), consistent with the larger lattice constant of Ag accommodating the pair more easily.
Although SAA research has so far focused primarily on late TM dopants, the strong thermodynamic preference of very early TMs for surface-stable isolated sites identifies early-TM SAAs as an attractive and largely unexplored material class. 
Their potential has recently been showcased for the selective dehydrogenation of propane to propylene.\cite{ngan2026early}

Taken together, these results show that the dopant element sets the baseline preference for whether single-atom or dual-atom sites are favored, and that the observed trends can be rationalized by a ten-electron rule combined with steric size mismatch for very early elements.

\subsection{Host element, surface facet, and dopant spin as design handles.}

\begin{figure}[htbp]
    \includegraphics[width=16cm, height=\textheight,keepaspectratio]{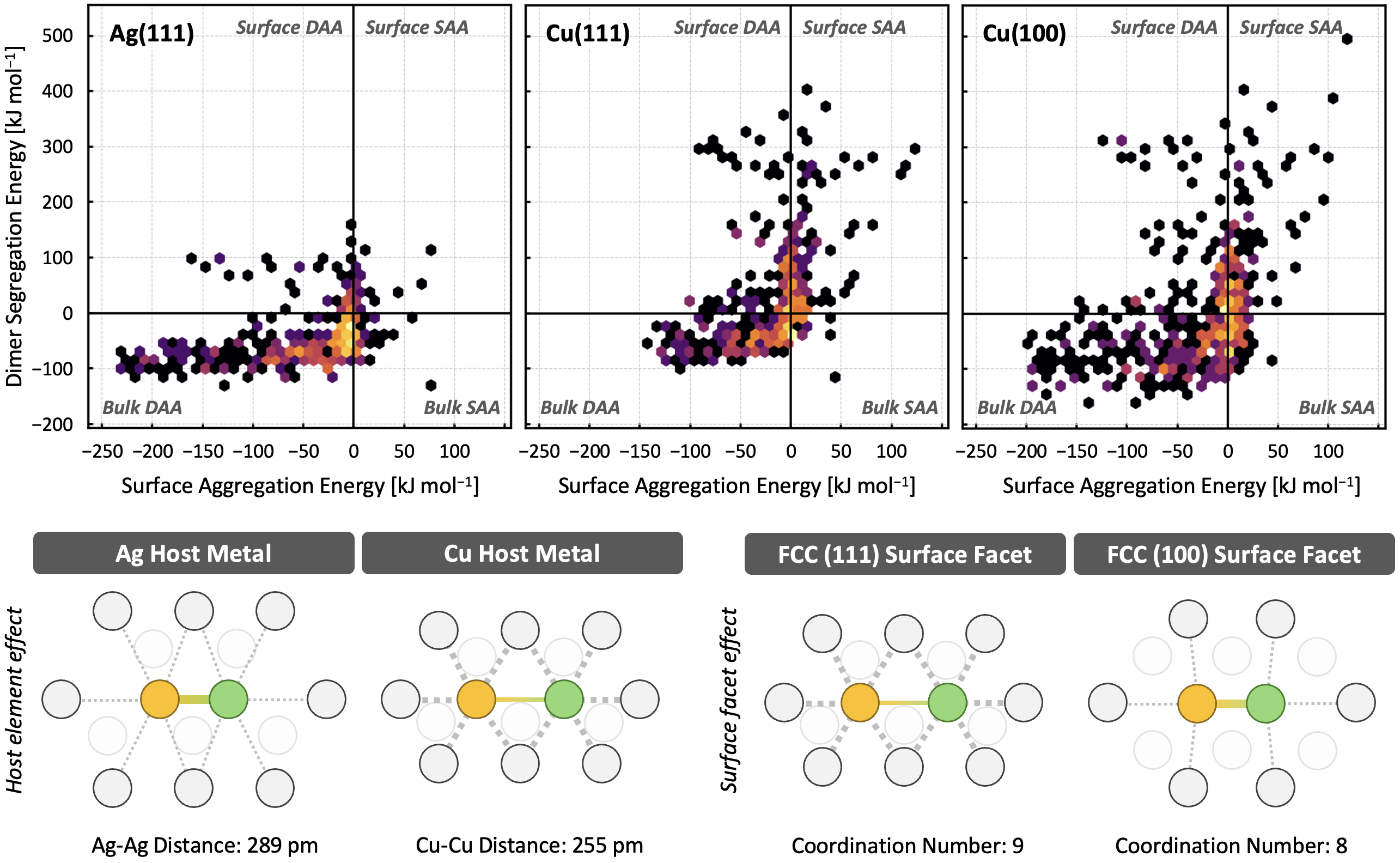}
    \caption{
    \textbf{Correlation between aggregation and segregation tendencies and the influence of surface facet and host metal.}
    Hexbin plots show that dopant combinations with a strong preference for dual-atom alloy (DAA) formation also tend to prefer bulk segregation, whereas combinations favoring single-atom alloy (SAA) sites tend to prefer the surface. 
    Thermodynamically stable surface DAAs are comparatively rare, and only very few dopant combinations favor bulk-incorporated SAAs. 
    Compared with Cu(111), Ag(111) generally exhibits stronger tendencies toward both dimer formation and bulk segregation whenever these processes are thermodynamically favored. 
    Similarly, trends are more pronounced in Cu(100) than in Cu(111) owing to the more open (100) surface facet. 
    Structural sketches illustrate the strengthening of dopant--host bonds and the decrease in interatomic distances when moving from Ag to Cu hosts, as well as the change in dopant coordination from the densely packed (111) to the more open (100) surface. 
    Host atoms are shown in gray and dopants in orange and green. 
    Aggregation and segregation energies are given in kJ mol$^{-1}$.
    A logarithmic color scale is used, with black indicating low and yellow indicating high data point density. 
    All aggregation and segregation energies are provided in the Supporting Information.
    }
    \label{fig:trends_aggregation}
\end{figure}

While aggregation energies identify many dopant combinations that favor the formation of DAAs, aggregation alone does not determine whether such sites are catalytically accessible.
Across the Atlas, a clear correlation emerges between preferred active site type and preferred location.
Surface stability can be quantified through the segregation energy,\cite{ruban1999surface, kothakonda2025discovering, he2025stability, rao2020extendable, zhang2020alloying, nilekar2009surface, mazitov2024surface}
\begin{align}
E_{\text{seg}} = E^\text{bulk-SAA/DAA} - E^\text{surface-SAA/DAA}, \label{eq:segregation_energy} 
\end{align}
where positive values indicate a thermodynamic preference for surface incorporation and negative values indicate a preference for bulk segregation.
Fig.~\ref{fig:trends_aggregation} shows that dopant combinations favoring DAAs generally exhibit a preference for bulk segregation, whereas combinations favoring SAAs tend to be stable in the surface.
As a result, only a small subset of dopant combinations remain atomically dispersed while preferring bulk segregation, and, more importantly, few form dimers while remaining stable in the surface layer.

Although the dopant elements are the main factor for the preference for active site types and location, these preferences are not immutable.
The surrounding environment, including the host element and surface facet, as well as the presence or absence of spin at a dopant, provide design levers that can bias dopant combinations toward SAAs or DAAs and toward surface or bulk incorporation.

\noindent\textbf{Host and facet effects.}
The overall aggregation and segregation patterns observed for Ag(111), Cu(111), and Cu(100) are qualitatively similar.
However, systematic shifts in these energies are observed that can be exploited as design levers.
For dopant combinations that favor dimer formation, aggregation is strongest in Ag(111), intermediate in Cu(100), and weakest in Cu(111).
For dopants that prefer to segregate into the bulk, the driving force for bulk segregation is largest in Cu(100), followed by Ag(111), and smallest in Cu(111).
Correspondingly, dopant combinations that favor isolated single-atom sites remain dispersed more readily in the Cu surfaces than in Ag(111), and dopants that are surface stable are held in the surface layer more strongly in Cu than in Ag(111).

These systematic shifts originate from differences in dopant--host bonding strength,\cite{berger2025dopant} local dopant coordination,\cite{bunting2023reactivity} and steric constraints.\cite{berger2025dopant}
Ag exhibits weaker host--host bonding, reflected in its lower cohesive and surface free energies relative to Cu.
Combined with its larger lattice constant and weaker dopant--host bonding, this imposes a smaller energetic penalty for dopant--dopant bond formation.
This allows dopant dimers to approach their preferred bond lengths, leading to stronger aggregation tendencies in Ag(111).
In contrast, stronger dopant--host bonding in Cu stabilizes isolated dopants in the surface and penalizes dimer formation.

Surface facet effects further modulate these trends.
In Cu(111), each surface dopant is coordinated by nine host atoms, compared to eight in Cu(100), increasing the energetic cost of weakening dopant--host bonds upon dimer formation.
As a result, aggregation preferences are reduced in Cu(111) relative to Cu(100).
However, the surface facet can also affect the kinetics of alloying and adatom diffusion, influencing the dopant motifs that form.\cite{wang2020surface}
The thermodynamic effects are illustrated in the structural sketches in Fig.~\ref{fig:trends_aggregation}.

Steric effects further contribute to differences between Cu and Ag hosts.
The smaller interatomic distances in Cu impose larger steric constraints on size-mismatched dopants.
In surface sites, this strain can be partially relieved by more flexible surface host atoms and dopant protrusion,\cite{berger2025dopant} leading to a comparatively stronger stabilization of large dopants in Cu surfaces than in Ag.
For sufficiently large dopants, this steric stabilization can overcompensate the stronger dopant--host bonding in Cu.

When segregation into the bulk is favored, the preference is most pronounced for the Cu(100) surface, where the coordination environment changes from eight neighbors in the surface layer to twelve in the bulk.
Gaining these additional dopant--host bonds upon bulk segregation results in the largest segregation driving forces among the surfaces considered.

Together, these results show that the open (100) facet and hosts that interact weakly with dopants (Ag) promote dimer formation, whereas the denser (111) facet and more strongly interacting hosts (Cu) stabilize single-atom sites. 
These trends directly guide experiments by significantly narrowing the number of candidates for experimental testing and providing practical handles to bias active site type and location for a given dopant combination.
For example, Co--Pd and Rh--Ir pairs slightly favor dimer formation in Ag(111), with aggregation energies of $-2$ and $-9$~kJ~mol$^{-1}$, respectively, whereas the same pairs prefer single-atom configurations in Cu(111), with aggregation energies of $+3$ and $+6$~kJ~mol$^{-1}$.
This demonstrates that changing the host alone can invert the preferred active site type.

\noindent\textbf{Spin effects.}
Spin polarization represents an additional factor that influences aggregation and surface stability, but unlike host element or surface facet, it cannot be directly tuned.
Instead, spin is an intrinsic property of the dopant element and its local confinement.\cite{berger2025dopant,zhangmagnetic}
Nevertheless, for dopants that exhibit pronounced spin polarization, particularly among central 3$d$ TMs, spin gives rise to distinct stability trends that can be exploited indirectly through dopant selection and environment engineering.

According to the ten-electron rule introduced here, combinations of central TMs are expected to strongly favor dimer formation.
This behavior is observed for 4$d$ and 5$d$ elements such as Mo and W, for which homo-dimers exhibit strongly negative aggregation energies, for example $-142~\text{kJ~mol}^{-1}$ for Mo--Mo in Cu(111).
The corresponding MoCu(111) SAA exhibits a pronounced preference for bulk incorporation, with a segregation energy of $-44~\text{kJ~mol}^{-1}$.
In contrast, some central 3$d$ elements, most notably Cr, remain stabilized as single-atom sites despite having nominal $d$-electron counts that would otherwise favor dimer formation.
For Cr--Cr in Cu(111), the aggregation energy becomes positive in the high-spin ground state, $+40~\text{kJ~mol}^{-1}$, while the CrCu(111) SAA exhibits an essentially thermoneutral segregation energy of $-1~\text{kJ~mol}^{-1}$.

This deviation arises from the pronounced spin polarization of certain 3$d$ dopants.
Upon dimerization, local magnetic moments may either be quenched, incurring a substantial spin-pairing penalty, or be retained. 
In the latter case, hybridization either leads to population of both bonding and antibonding MOs with unpaired electrons or is avoided altogether, thereby preventing bond formation.
In all scenarios, the energetic stabilization normally gained from dopant--dopant bonding is significantly reduced.
Similarly, local spin moments are typically larger in surface sites and decrease in the bulk, where stronger splitting of the dopant $d$-states occurs.\cite{berger2025dopant}
The associated spin-pairing energy can compensate the energetic gain from forming additional dopant--host bonds in the bulk.
As a result, spin polarization can offset or even reverse the aggregation tendencies predicted by the ten-electron rule. 
Central 3$d$ TMs can therefore remain stabilized as surface SAAs, whereas their 4$d$ and 5$d$ analogues aggregate strongly and preferentially segregate into the bulk.
Consequently, when multiple dopant candidates are available for a given catalytic function, for example TMs from the same group but different periods, consideration of spin provides an important, indirect design handle.

\subsection{Stabilizing surface active sites through reactor--anchor design.}

For dopants to act as catalytically active sites, they must reside in the surface layer.
Surface accessibility therefore represents an additional design constraint beyond aggregation tendencies.

\begin{figure}[htp]
\includegraphics[width=8cm,height=\textheight,keepaspectratio]{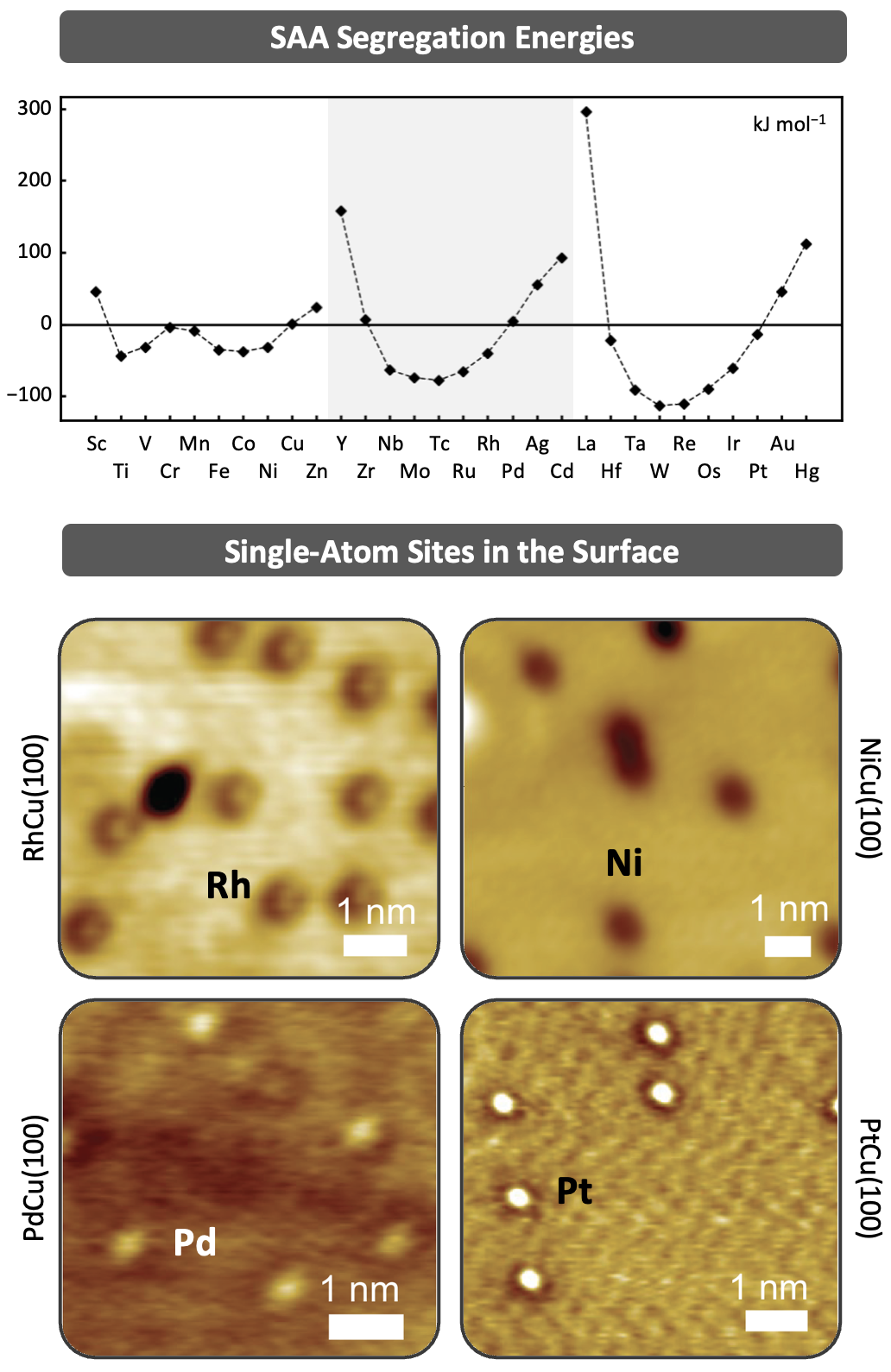}
    \caption{\textbf{Experimental STM images confirm that dopants remain embedded in the surface even when thermodynamics favors bulk segregation.}
    Top: Calculated segregation energies (kJ mol$^{-1}$) of transition metal (TM) dopants from the Cu(100) surface into the bulk. 
    Early and late TMs are stable in the surface, whereas most central 4$d$ and 5$d$ TMs exhibit a strong thermodynamic preference for bulk segregation. 
    In contrast, central 3$d$ TMs remain comparatively more surface stable. 
    Bottom: STM images of isolated Ni, Rh, Pd, and Pt dopants embedded in the Cu(100) surface.}
    \label{fig:segregation}
\end{figure}

Clear systematic trends emerge from the calculated bulk segregation energies, consistent with previous work.\cite{ruban1999surface, kothakonda2025discovering, he2025stability, rao2020extendable, zhang2020alloying, nilekar2009surface, mazitov2024surface}
For Cu(100)-based SAAs, these trends are shown in Fig.~\ref{fig:segregation}, with similar behavior observed for the other surfaces as shown in Section~S3 of the SI.
Very early TMs have larger atomic radii, and the resulting size mismatch is better accommodated at the surface than in the bulk, since surface host atoms are more flexible and the dopant can protrude.
Very late TMs can also favor surface sites, particularly when dopant--host bonds become weaker than host--host bonds, due to the lower coordination of dopants at the surface.
In contrast, most TMs without substantial size mismatch and with dopant--host bonds stronger than host--host bonds are thermodynamically predicted to prefer bulk segregation, reflecting the stabilization gained from higher coordination in bulk sites.
Additional deviations from the general trends arise from the aforementioned spin effects.
In particular, central 3$d$ TMs such as Cr and Mn can be energetically stabilized in the surface due to their local high-spin states.
Reduced spin polarization in bulk sites introduces a spin-pairing penalty that partially offsets the stabilization gained from higher coordination, leading to the characteristic W-shaped segregation trend for 3$d$ TMs.

However, calculated segregation tendencies do not directly translate into experimental observations.\cite{hannagan2021first}
For example, Ni, Rh, Pd, and Pt are experimentally observed in the surface layer of Cu(100), as shown in Fig.~\ref{fig:segregation}, despite calculated segregation energies that indicate a thermodynamic preference for bulk segregation.
This apparent discrepancy is kinetic in origin: the alloys are prepared at temperatures that allow dopants to overcome the low barriers for surface incorporation,\cite{karageorgiou2026mechanisms} but not the higher barriers for bulk segregation,\cite{lucci2014atomic} leaving them kinetically trapped in the surface. 
On prolonged annealing at sufficiently high temperatures, however, the dopants do gradually segregate into the bulk, consistent with the predicted thermodynamic preference.
This demonstrates that surface accessibility cannot be inferred from thermodynamic segregation energies alone, as kinetic effects can preserve dopants in metastable surface configurations.

Beyond kinetic trapping and spin effects, dopant pairing provides an additional lever to actively modulate the stability of dopants in the surface layer.
This forms the basis of the \textit{reactor--anchor design concept} introduced here.
So far, we have discussed dual-atom sites in which both dopants contribute to reactivity, creating catalytic synergy.\cite{kress2023priori}
If instead only one dopant, the \textit{reactor}, is required for activity, the second dopant can serve as an \textit{anchor} that stabilizes the reactor in the surface layer by modifying its segregation tendency through the formation of a stable hetero-dimer.
Such behavior has been observed for Sn and Pt dopants in Cu nanoparticles, where larger Sn atoms with a preference for surface sites pull Pt atoms from the particle interior to the surface.\cite{sun2025full}
To quantify this effect, we define the anchoring energy, $E_\text{anchor}^{ij}$, as the difference between the (per-atom) segregation energy of the hetero-dimer and that of the corresponding single-atom alloy:

\begin{align}
E_\text{anchor}^{ij} 
&= \frac{1}{2}E_\text{seg}^{\text{DAA},\,ij} - E_\text{seg}^{\text{SAA},\,j}.
\label{eq:anchoring_energy}
\end{align}

\begin{figure}[htp]
\includegraphics[width=8cm,height=\textheight,keepaspectratio]{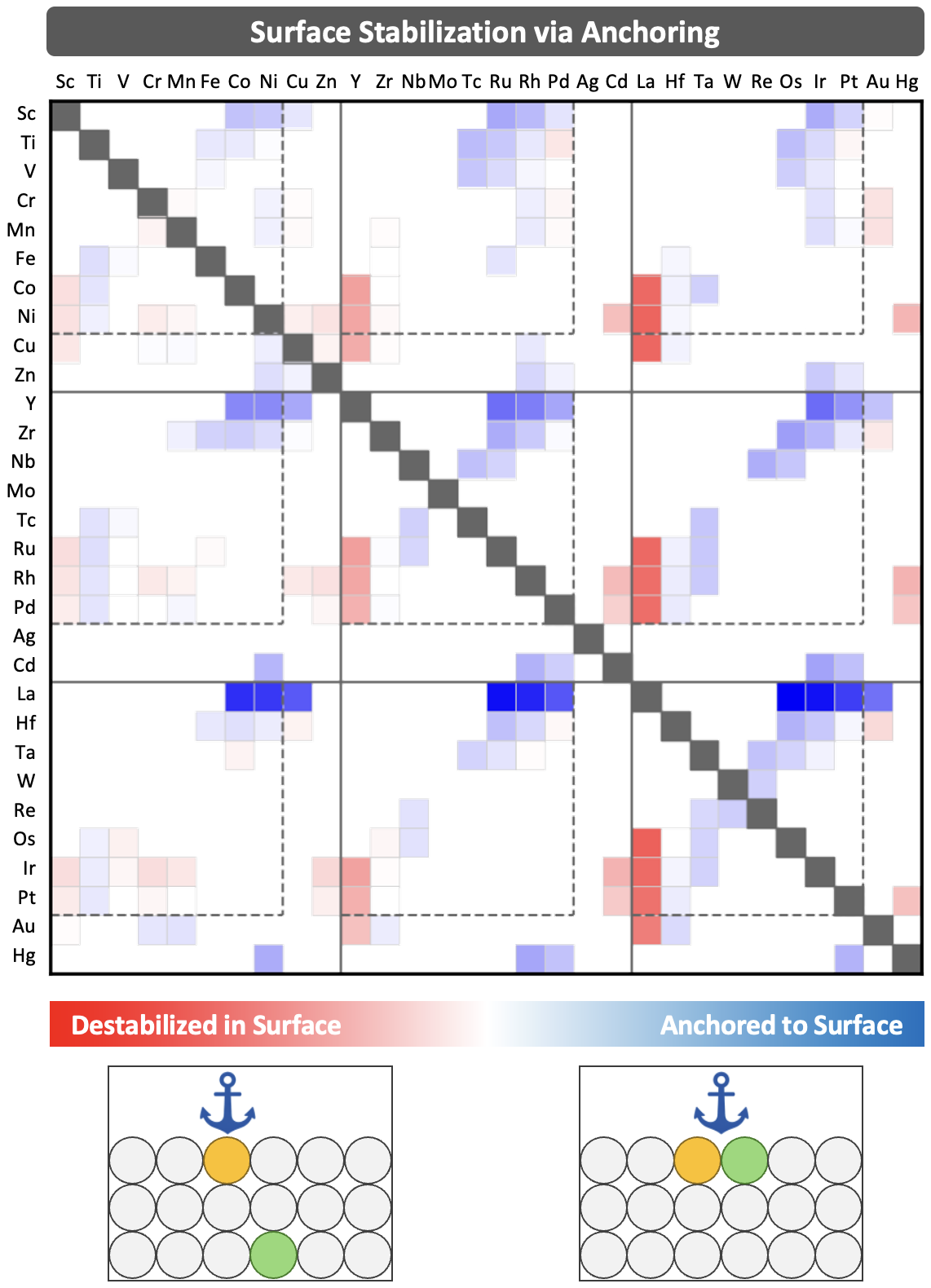}
    \caption{\textbf{The reactor--anchor concept enables thermodynamic stabilization of otherwise bulk-segregating dopants in the surface through hetero-dimer formation.}
    Anchoring energies ($E_\text{anchor}$), where blue indicates increased stabilization in the Cu(100) surface layer and red indicates increased stabilization in the bulk, relative to the reference SAA defined by the column.
    Only combinations that preferentially form hetero-dimers over both SAAs and homo-dimers are shown. 
    Illustrations depict dopants in the surface and bulk layers, with host atoms shown in gray and dopants in orange and green.}
    \label{fig:anchor}
\end{figure}

\noindent In Fig.~\ref{fig:anchor}, positive anchoring energies (blue) indicate when an element is stabilized at the surface,
whereas negative values (red) indicate reduced surface stability.
The plot serves as a lookup matrix, with each column showing how the surface stability of a given dopant is affected by pairing with the different anchoring elements indicated by the rows.

From this analysis, we find that while many late TMs exhibit a thermodynamic preference for bulk segregation as isolated SAAs, pairing them with early TMs can make the corresponding hetero-dimer thermodynamically surface-stable.
For example, Ni, Rh, and Ru exhibit negative segregation energies of –32, –40, and –65~kJ~mol$^{-1}$, respectively, as isolated SAAs in Cu(100), indicating a bulk preference.
When paired with the early TM Zr, however, the resulting hetero-dimers are relatively stable at the surface, with bulk segregation energies of 0, +8, and +10~kJ~mol$^{-1}$ per atom, respectively.
In such cases, the anchor not only enables hetero-dimer formation with the reactor element but can qualitatively invert the preferred location of the reactor from bulk to surface.
As a result, even at elevated temperatures where segregation barriers may be overcome, certain dopant combinations are predicted to remain thermodynamically stable at the surface and thus accessible to reactants as active sites.

While active site types and locations in real catalysts are influenced by dopant concentration, kinetic barriers, and adsorbate effects,\cite{zugic2017dynamic, papanikolaou2019co, darby2018carbon, papanikolaou2019engineering} the systematic energetic trends identified here define the underlying thermodynamic landscape governing site stability.
The reactor--anchor concept therefore represents not merely an alternative perspective on hetero-dimers, but a general design strategy.
By pairing late TMs with anchoring early or selected central 3$d$ TMs, dopants that would otherwise be thermodynamically unstable in the surface layer can be rendered stable.
This strategy complements host and facet selection as design handles and extends the accessible materials space.

\subsection{Guiding the discovery of hetero-dimers.}
The established trends and design handles for tuning active site motifs and their surface stability provide a framework for identifying experimentally viable hetero-dimer candidates.
Experimentally observed active site motifs in bimetallic and trimetallic alloys show excellent agreement with the predicted aggregation energetics.
In particular, our predictions are consistent with the experimental observation of SAAs in bimetallic alloys such as RhCu(100),\cite{wang2022observation} NiCu(100),\cite{meng2024coverage} PdCu(100), PtCu(100), CrAg(111),\cite{kress2023priori} PtAg(111),\cite{patel2019elucidating} and PdAg(111),\cite{muir2020adsorption} as well as with the formation of DAAs in the trimetallic alloy CrPtAg(111)\cite{kress2023priori} and, as shown here, CrPdAg(111).

To date, only two trimetallic highly dilute alloys containing hetero-dimer sites composed exclusively of transition metals have been synthesized, namely CrPtAg(111)\cite{kress2023priori} and the CrPdAg(111) system presented here.
In addition, two alloys containing hetero-dimers formed by a transition metal and a main group dopant, PdSnAg(111)\cite{mohrhusen2024modifying} and PtSnCu nanoparticles,\cite{sun2025full} have been reported.
Given the substantial work required to synthesize, characterize, and evaluate these new materials, computational guidance is essential for identifying promising targets and accelerating exploration, as the experimental development of even a single candidate system can be a several-year effort involving multiple laboratories.
Based on the calculated aggregation energies, a large number of dopant combinations are expected to favor hetero-dimers.
A hetero-dimer is a strong candidate when the aggregation energy for the mixed pair is negative and more favorable than for either of the two corresponding homo-dimers. 
These conditions yield a characteristic anti-diagonal pattern across the transition metal series, as illustrated in Figure~\ref{fig:anchor}.
All identified candidates together with their aggregation and segregation energies are also provided in Section~S4 of the SI.

As a representative example in Cu(111), W--Re illustrates a combination of central TMs with a mild relative preference for hetero-dimer formation.
The W--Re aggregation energy is $-122$~kJ~mol$^{-1}$, compared with $-115$~kJ~mol$^{-1}$ for W--W and $-113$~kJ~mol$^{-1}$ for Re--Re.
The strong but relatively nonselective aggregation tendency therefore leads to competition between homo- and hetero-dimer formation.
In contrast, Zr--Rh represents an early--late pair in which the hetero-dimer is strongly stabilized ($E_\text{agg} = -69$~kJ~mol$^{-1}$), while both homo-dimers are destabilized (+33 for Zr--Zr and +12 for Rh--Rh). 
Here, neither dopant aggregates individually, but together they form a strongly bound hetero-dimer. 

More generally, the most promising hetero-dimer candidates tend to combine early and late TMs.
Early TMs such as Sc, Ti, V, Y, Zr, Nb, Hf, and Ta are predicted to form stable hetero-dimers with late TMs including Fe, Co, Ni, Ru, Rh, Pd, Os, Ir, and Pt.
This preference follows from the ten-electron rule: combining an early and a late TM brings the total $d$-electron count close to ten and stabilizes the hetero-dimer, whereas the corresponding early--early and late--late homo-dimers lie further from this optimum and are therefore less stable.
In contrast, combinations of central 4$d$ and 5$d$ TMs often favor homo-dimer formation, limiting the number of stable hetero-dimers in these regions of the combinatorial space.
An important exception is provided by central 3$d$ elements, particularly Cr and Mn, whose aggregation tendencies are drastically reduced by spin polarization.
As a result, these elements do not prefer homo-dimer formation, and we instead find them forming hetero-dimers with very late TMs.
The experimentally realized systems CrPtAg(111)\cite{kress2023priori} and CrPdAg(111) are examples of this behavior.
Overall, our results open a sizable and largely unexplored compositional space of early--late TM combinations for targeted experimental exploration. 
Many of these pair an abundant central 3$d$ or early TM with a more expensive noble metal that might otherwise prefer to segregate into the bulk, anchoring it in the surface where it is accessible and thereby reducing the overall noble metal loading required.

Beyond catalysis, control over local dopant interactions may also provide new approaches to traditional metallurgy, where dopant pairing, clustering, and precipitation are exploited to tune material properties including radiation resistance, mechanical strength, thermal stability, grain structure, and corrosion resistance. 
In dilute Mg--Zn--Y alloys, for example, Zn--Y co-segregation stabilizes the solute-enriched stacking faults,\cite{egami2020thermodynamic} while in Al--Sc--Zr the two solutes co-precipitate into coarsening-resistant core--shell particles.\cite{booth2011coarsening} 
Additionally, a computational screening for radiation-resistant dilute Cu alloys pairs a strongly vacancy-binding solute with a partner that binds it, identifying early--late TM combinations such as Zr--Ni and Zr--Co,\cite{vasudevan2026density} for which our Atlas also predicts favorable dimer formation in Cu.
The ten-electron rule therefore rationalizes solute pairing beyond the context in which it was derived, and the principles developed here may help identify unexplored dilute alloy compositions.

The insights developed here can also be used to tune the electronic and magnetic properties of materials. 
Certain active site motifs may enable engineering of atomic sites with giant spin splitting\cite{ast2007giant} or the Kondo effect.\cite{knorr2002kondo} 
By exploiting the reactor--anchor concept, magnetic dopants that would otherwise segregate into the bulk can be thermodynamically stabilized in the surface through the formation of hetero-dimers with early TMs. 
Promising examples include ZrFeCu(111) and ZrFeCu(100).
In both systems, the Fe dopant is predicted to retain a pronounced local magnetic moment, with onsite magnetizations of 1.7 and 1.5~$\mu_\mathrm{B}$, respectively, while the surface hetero-dimer is stabilized by thermodynamics. 
Specifically, the corresponding aggregation energies are $-40$ and $-64$~kJ~mol$^{-1}$, and the segregation energies are $+13$ and $+11$~kJ~mol$^{-1}$, respectively.
The resulting surface accessibility of the magnetic dopants may increase their responsiveness to external stimuli and facilitate their manipulation, thereby enabling the rational tuning of surface-localized electronic and magnetic properties for future functional materials.

\section{Conclusion}

This work establishes a conceptual framework for understanding and predicting the structure and stability of active sites of transition metal dopants in dilute alloys.
The results reveal four distinct classes of dopant combinations: early--early and late--late pairs that favor formation of SAAs, and central--central as well as early--late pairs that favor dopant aggregation.
These trends are rationalized by a ten-electron rule, which generally explains when dopant--dopant bonding is maximized.
For a small subset of elements, spin polarization introduces an additional dimension, enabling certain 3$d$ elements to form surface-stable SAAs and selected trimetallic hetero-dimer sites in cases where their 4$d$ and 5$d$ analogues instead favor homo-aggregation and segregation into the bulk.

Beyond fundamental insights, the framework identifies practical design handles for engineering local atomic environments. 
The choice of dopant element, host metal, and surface facet, together with the reactor--anchor concept, allow active site preferences and stability to be tuned. 
Together, these findings establish a foundation for understanding and designing highly dilute alloys with desired active site structures.
A natural next step is to ask how the electron count that stabilizes a dopant dimer influences its reactivity, connecting the stability trends established here to the function of the resulting sites.

More broadly, by establishing a bottom-up understanding of the thermodynamic stability of a large number of functional sites, this work provides a foundation for engineering increasingly complex local atomic arrangements in multimetallic materials.
The same local interactions identified here are expected to govern the short-range order that emerges in high-entropy alloys and other dilute and concentrated alloys.
These insights therefore provide a general framework for designing materials with tailored catalytic, electronic, and magnetic properties.

\clearpage

\section{Methods}

\subsection{Experimental Methods}

\textbf{Low-Temperature Scanning Tunneling Microscopy (LT-STM).}  
Experiments were performed using an Omicron LT-STM system under ultra-high vacuum (UHV) conditions (base pressure $<10^{-11}$ mbar). 
The Ag(111) and Cu(100) single crystals were cleaned in an auxiliary preparation chamber (chamber pressure $< 2 \times 10^{-10}$ mbar) via repeated cycles of Ar$^{+}$ ion bombardment (1 keV, 10--20 $\mu$A) followed by thermal annealing at 750~K.

\textbf{Cu-Based Single-Atom Alloys (SAAs).}  
Sub-monolayer quantities of Rh, Ni, Pt, and Pd (Goodfellow, 99.9\% purity) were deposited using flux-monitored EFM 3 electron beam evaporators (Scienta Omicron).
During deposition, the Cu(100) substrate was held at 300~K with a calibrated metal flux of $\sim 0.01$~ML/min.

\textbf{PdCrAg(111) Dual-Atom Alloys (DAAs).}  
Cr was incorporated using Cr(CO)$_6$ (Sigma-Aldrich, 98\%) as a molecular precursor. 
The PdAg(111) surface was prepared by first heating the Ag(111) crystal to 323~K, followed by Pd deposition via electron beam evaporation. 
Following Pd incorporation, Cr was introduced using Cr(CO)$_6$. 
The Cr(CO)$_6$ was housed in a glass vial connected to the UHV system through a turbo-pumped gas manifold and a precision leak valve. 
After evacuating the manifold to $<10^{-4}$~mbar, the Cr(CO)$_6$ was dosed onto the pre-formed PdAg(111) surface at a substrate temperature of 323~K and a chamber pressure of $5\times10^{-7}$ mbar. 
Finally, the prepared surfaces were cryogenically cooled to 80~K for STM characterization. 
This process has been described in detail before.\cite{kress2023priori}

\subsection{Theoretical Methods}

The Cu and Ag metal surfaces were modeled using 4$\times$3$\times$6 supercells, cleaved along the (100) and (111) facets, with a vacuum height of 15~Å and the bottom two layers fixed.
SAAs contain one dopant atom, whereas DAAs contain two dopant atoms arranged such that at least two host atoms separate the dopants from their periodic images, reducing interactions mediated through directly neighboring host atoms.\cite{schumann2021periodic}
Dopants were placed either in the first layer (surface model) or in the third layer (bulk model).
The layers above and below the third layer were allowed to relax to accommodate the dopant.

Density functional theory (DFT) calculations with periodic boundary conditions (pbc) were performed with the optB86b-vdW functional,\cite{optB86b-vdW} as implemented in the Vienna Ab initio Simulation Package (VASP), version 6.3.0.\cite{kresse1993ab,kresse1996efficiency,kresse1996efficient,kresse1999ultrasoft}
A plane-wave basis set was employed for valence electrons, together with the projector augmented wave (PAW) method for core electrons.
PAW potentials recommended by VASP and the Materials Project\cite{jain2013commentary} were used; in cases where the recommendations were inconsistent, the potential treating more electrons explicitly was selected.
Spin polarization, first-order Methfessel--Paxton smearing (SIGMA = 0.1), and dipole corrections were applied.

For each SAA and DAA, we calculated the low-spin state and attempted to locate high-spin solutions using two approaches:
(1) initializing the on-site magnetic moment for each dopant to 1.5 times the maximum possible number of unpaired electrons, and
(2) restarting a spin-unrestricted calculation from the charge density of a spin-restricted calculation.
Both approaches were necessary to reliably identify the energetically most stable spin states.
For dopant dimers, both ferromagnetic and antiferromagnetic coupling of the local dopant spins were considered.
All quantities discussed in this work were calculated using the energetically most stable spin state for the respective SAA or DAA.
 
For structure optimizations, an energy cutoff of 600~eV was used, and the Brillouin zone was sampled with a Monkhorst--Pack mesh of 3$\times$3$\times$1 \textbf{k}-points.
Convergence thresholds were set to $10^{-7}$~eV for electronic energies and 0.01~eV~Å$^{-1}$ for ionic forces.

Unit cell vectors were determined using the equation-of-state method,\cite{alchagirov2003reply} as implemented in the Atomic Simulation Environment (ASE).
An energy cutoff of 800~eV was used to minimize the influence of volume changes on the basis set, and the Brillouin zone of the bulk, represented by a primitive one-atom unit cell, was sampled using a Monkhorst--Pack mesh of 31$\times$31$\times$31 \textbf{k}-points.

\begin{acknowledgement}
F.B. acknowledges support from the Alexander von Humboldt Foundation through a Feodor Lynen Research Fellowship, from the Isaac Newton Trust through an Early Career Fellowship, and from Churchill College, Cambridge, through a Postdoctoral By-Fellowship. 
This work has been funded by the European Union (ERC, n-AQUA, 101071937).
Views and opinions expressed are, however, those of the authors only and do not necessarily reflect those of the European Union or the European Research Council Executive Agency.
Neither the European Union nor the granting authority can be held responsible for them. 
E.C.H.S. thanks the US National Science Foundation for support under grant NSF CHE-2334970.
Y.W. thanks the US Department of Energy, BES, CPIMS program under contract DE-SC0004738.
This work was performed using resources provided by the Cambridge Service for Data Driven Discovery (CSD3) operated by the University of Cambridge Research Computing Service (www.csd3.cam.ac.uk), provided by Dell EMC and Intel using Tier-2 funding from the Engineering and Physical Sciences Research Council (capital grant EP/T022159/1), and DiRAC funding from the Science and Technology Facilities Council (https://www.dirac.ac.uk), with additional access through a University of Cambridge EPSRC Core Equipment Award (EP/X034712/1). 
We additionally acknowledge computational support and resources from the UK National High-Performance Computing Service, Advanced Research Computing High End Resource (ARCHER2).
Access for ARCHER2 was obtained via the Materials Chemistry Consortium (MCC), funded by EPSRC grant references EP/X035859 and EP/F067496. 
Further computational support and resources were provided by YOUNG, the Tier-2 High Performance Computing Hub in Materials and Molecular Modeling (MMM), which is partially funded by EPSRC grant reference EP/T022213. 
We also acknowledge the EuroHPC Joint Undertaking for awarding project ID EHPC-REG-2024R02-130 access to Leonardo at CINECA (Italy), and project ID EHPC-REG-2025R02-112 access to JUPITER at Jülich (Germany). 
\end{acknowledgement}


\clearpage

\bibliography{refs}

@article{hannagan2020single,
  title={Single-atom alloy catalysis},
  author={Hannagan, Ryan T and Giannakakis, Georgios and Flytzani-Stephanopoulos, Maria and Sykes, E Charles H},
  journal={Chem. Rev.},
  volume={120},
  number={21},
  pages={12044},
  year={2020},
  publisher={ACS Publications}
}

@article{cui2018bridging,
  title={Bridging homogeneous and heterogeneous catalysis by heterogeneous single-metal-site catalysts},
  author={Cui, Xinjiang and Li, Wu and Ryabchuk, Pavel and Junge, Kathrin and Beller, Matthias},
  journal={Nat. Catal.},
  volume={1},
  number={6},
  pages={385},
  year={2018},
  publisher={Nature Publishing Group UK London}
}

@article{yang2013single,
  title={Single-atom catalysts: a new frontier in heterogeneous catalysis},
  author={Yang, Xiao-Feng and Wang, Aiqin and Qiao, Botao and Li, JUN and Liu, Jingyue and Zhang, Tao},
  journal={Acc. Chem. Res.},
  volume={46},
  number={8},
  pages={1740},
  year={2013},
  publisher={ACS Publications}
}

@article{kyriakou2012isolated,
  title={Isolated metal atom geometries as a strategy for selective heterogeneous hydrogenations},
  author={Kyriakou, Georgios and Boucher, Matthew B and Jewell, April D and Lewis, Emily A and Lawton, Timothy J and Baber, Ashleigh E and Tierney, Heather L and Flytzani-Stephanopoulos, Maria and Sykes, E Charles H},
  journal={Science},
  volume={335},
  number={6073},
  pages={1209},
  year={2012},
  publisher={American Association for the Advancement of Science}
}

@article{norskov2002universality,
  title={Universality in heterogeneous catalysis},
  author={N{\o}rskov, Jens K and Bligaard, Thomas and Logadottir, Ashildur and Bahn, S and Hansen, Lars B and Bollinger, Mikkel and Bengaard, H and Hammer, Bj{\o}rk and Sljivancanin, Z and Mavrikakis, Manos and others},
  journal={J. Catal.},
  volume={209},
  number={2},
  pages={275},
  year={2002},
  publisher={Elsevier}
}

@article{perez2019strategies,
  title={Strategies to break linear scaling relationships},
  author={P{\'e}rez-Ram{\'\i}rez, Javier and L{\'o}pez, N{\'u}ria},
  journal={Nat. Catal.},
  volume={2},
  number={11},
  pages={971},
  year={2019},
  publisher={Nature Publishing Group UK London}
}

@article{wang2022observation,
  title={Observation and Characterization of Dicarbonyls on a RhCu Single-Atom Alloy},
  author={Wang, Yicheng and Schumann, Julia and Happel, Elizabeth E and {\c{C}}{\i}nar, Volkan and Sykes, E Charles H and Stamatakis, Michail and Michaelides, Angelos and Hannagan, Ryan T},
  journal={J. Phys. Chem. Lett.},
  volume={13},
  number={27},
  pages={6316},
  year={2022},
  publisher={ACS Publications}
}

@article{schumann2024ten,
  title={Ten-electron count rule for the binding of adsorbates on single-atom alloy catalysts}, 
  author={Schumann, Julia and Stamatakis, Michail and Michaelides, Angelos and R{\'e}ocreux, Romain},
  journal={Nat. Chem.},
  volume={16},
  pages={749},
  year={2024},
  publisher={Nature Publishing Group UK London}
}

@article{hammer1999improved,
  title={Improved adsorption energetics within density-functional theory using revised Perdew-Burke-Ernzerhof functionals},
  author={Hammer, Bj{\o}rk and Hansen, Lars Bruno and N{\o}rskov, Jens Kehlet},
  journal={Phys. Rev. B.},
  volume={59},
  number={11},
  pages={7413},
  year={1999},
  publisher={APS}
}

@article{optB86b-vdW,
  title = {Van der Waals density functionals applied to solids},
  author = {Klime{\v{s}}, Ji{\v{r}}{\'{\i}} and Bowler, David R. and Michaelides, Angelos},
  journal = {Phys. Rev. B},
  volume = {83},
  issue = {19},
  pages = {195131},
  numpages = {13},
  year = {2011},
  month = {May},
  publisher = {American Physical Society},
}

@article{PBE,
  title = {Generalized Gradient Approximation Made Simple},
  author = {Perdew, John P. and Burke, Kieron and Ernzerhof, Matthias},
  journal = {Phys. Rev. Lett.},
  volume = {77},
  issue = {18},
  pages = {3865},
  numpages = {0},
  year = {1996},
  month = {Oct},
  publisher = {American Physical Society}
}

@article{PBEsol,
  title = {Restoring the Density-Gradient Expansion for Exchange in Solids and Surfaces},
  author = {Perdew, John P. and Ruzsinszky, Adrienn and Csonka, G\'abor I. and Vydrov, Oleg A. and Scuseria, Gustavo E. and Constantin, Lucian A. and Zhou, Xiaolan and Burke, Kieron},
  journal = {Phys. Rev. Lett.},
  volume = {100},
  issue = {13},
  pages = {136406},
  numpages = {4},
  year = {2008},
  month = {Apr},
  publisher = {American Physical Society}
}

@article{PBEsol-er,
  title = {Erratum: Restoring the Density-Gradient Expansion for Exchange in Solids and Surfaces [Phys. Rev. Lett. 100, 136406 (2008)]},
  author = {Perdew, John P. and Ruzsinszky, Adrienn and Csonka, G\'abor I. and Vydrov, Oleg A. and Scuseria, Gustavo E. and Constantin, Lucian A. and Zhou, Xiaolan and Burke, Kieron},
  journal = {Phys. Rev. Lett.},
  volume = {102},
  issue = {3},
  pages = {039902},
  numpages = {1},
  year = {2009},
  month = {Jan},
  publisher = {American Physical Society}
}

@article{schimka2010accurate,
  title={Accurate surface and adsorption energies from many-body perturbation theory},
  author={Schimka, Laurids and Harl, J and Stroppa, A and Gr{\"u}neis, A and Marsman, M and Mittendorfer, F and Kresse, G},
  journal={Nat. Mater.},
  volume={9},
  number={9},
  pages={741},
  year={2010},
  publisher={Nature Publishing Group UK London}
}

@article{kresse1996efficiency,
  title={Efficiency of ab-initio total energy calculations for metals and semiconductors using a plane-wave basis set},
  author={Kresse, Georg and Furthm{\"u}ller, J{\"u}rgen},
  journal={Comput. Mater. Sci.},
  volume={6},
  number={1},
  pages={15},
  year={1996},
  publisher={Elsevier}
}

@article{kresse1996efficient,
  title={Efficient iterative schemes for ab initio total-energy calculations using a plane-wave basis set},
  author={Kresse, Georg and Furthm{\"u}ller, J{\"u}rgen},
  journal={Phys. Rev. B},
  volume={54},
  number={16},
  pages={11169},
  year={1996},
  publisher={APS}
}

@article{kresse1993ab,
  title={Ab initio molecular dynamics for liquid metals},
  author={Kresse, Georg and Hafner, J{\"u}rgen},
  journal={Phys. Rev. B},
  volume={47},
  number={1},
  pages={558},
  year={1993},
  publisher={APS}
}

@article{kresse1999ultrasoft,
  title={From ultrasoft pseudopotentials to the projector augmented-wave method},
  author={Kresse, Georg and Joubert, Daniel},
  journal={Phys. Rev. B},
  volume={59},
  number={3},
  pages={1758},
  year={1999},
  publisher={APS}
}

@article{alchagirov2003reply,
  title={Reply to “Comment on ‘Energy and pressure versus volume: Equations of state motivated by the stabilized jellium model’”},
  author={Alchagirov, Alim B and Perdew, John P and Boettger, Jonathan C and Albers, RC and Fiolhais, Carlos},
  journal={Phys. Rev. B},
  volume={67},
  number={2},
  pages={026103},
  year={2003},
  publisher={APS}
}

@article{michaelides2003identification,
  title={Identification of general linear relationships between activation energies and enthalpy changes for dissociation reactions at surfaces},
  author={Michaelides, Angelos and Liu, Z-P and Zhang, CJ and Alavi, Ali and King, David A and Hu, Peijun},
  journal={J. Am. Chem. Soc.},
  volume={125},
  number={13},
  pages={3704},
  year={2003},
  publisher={ACS Publications}
}

@article{sun2018breaking,
  title={Breaking the scaling relationship via thermally stable Pt/Cu single atom alloys for catalytic dehydrogenation},
  author={Sun, Guodong and Zhao, Zhi-Jian and Mu, Rentao and Zha, Shenjun and Li, Lulu and Chen, Sai and Zang, Ketao and Luo, Jun and Li, Zhenglong and Purdy, Stephen C and others},
  journal={Nat. Commun.},
  volume={9},
  number={1},
  pages={4454},
  year={2018},
  publisher={Nature Publishing Group UK London}
}

@article{greiner2018free,
  title={Free-atom-like d states in single-atom alloy catalysts},
  author={Greiner, Mark T and Jones, TE and Beeg, Sebastian and Zwiener, Leon and Scherzer, Michael and Girgsdies, Frank and Piccinin, S and Armbr{\"u}ster, Marc and Knop-Gericke, Axel and Schl{\"o}gl, Robert},
  journal={Nat. Chem.},
  volume={10},
  number={10},
  pages={1008},
  year={2018},
  publisher={Nature Publishing Group UK London}
}

@article{lucci2015selective,
  title={Selective hydrogenation of 1, 3-butadiene on platinum--copper alloys at the single-atom limit},
  author={Lucci, Felicia R and Liu, Jilei and Marcinkowski, Matthew D and Yang, Ming and Allard, Lawrence F and Flytzani-Stephanopoulos, Maria and Sykes, E Charles H},
  journal={Nat. Commun.},
  volume={6},
  number={1},
  pages={8550},
  year={2015},
  publisher={Nature Publishing Group UK London}
}

@article{jiang2020facet,
  title={Facet engineering accelerates spillover hydrogenation on highly diluted metal nanocatalysts},
  author={Jiang, Lizhi and Liu, Kunlong and Hung, Sung-Fu and Zhou, Lingyun and Qin, Ruixuan and Zhang, Qinghua and Liu, Pengxin and Gu, Lin and Chen, Hao Ming and Fu, Gang and others},
  journal={Nat. Nanotechnol.},
  volume={15},
  number={10},
  pages={848},
  year={2020},
  publisher={Nature Publishing Group UK London}
}

@article{muir2020adsorption,
  title={Adsorption of CO to characterize the structure of a Pd/Ag (111) single-atom alloy surface},
  author={Muir, Mark and Trenary, Michael},
  journal={J. Phys. Chem. C},
  volume={124},
  number={27},
  pages={14722},
  year={2020},
  publisher={ACS Publications}
}

@article{batatia2025foundation,
  title={A foundation model for atomistic materials chemistry},
  author={Batatia, Ilyes and Benner, Philipp and Chiang, Yuan and Elena, Alin M and Kov{\'a}cs, D{\'a}vid P and Riebesell, Janosh and Advincula, Xavier R and Asta, Mark and Avaylon, Matthew and Baldwin, William J and others},
  journal={J. Chem. Phys.},
  volume={163},
  number={18},
  pages={184110},
  year={2025},
  publisher={AIP Publishing}
}

@article{jain2013commentary,
  title={Commentary: The Materials Project: A materials genome approach to accelerating materials innovation},
  author={Jain, Anubhav and Ong, Shyue Ping and Hautier, Geoffroy and Chen, Wei and Richards, William Davidson and Dacek, Stephen and Cholia, Shreyas and Gunter, Dan and Skinner, David and Ceder, Gerbrand and others},
  journal={APL Mater.},
  volume={1},
  number={1},
  pages={011002},
  year={2013},
}

@article{reocreux2022stick,
  title={Stick or spill? Scaling relationships for the binding energies of adsorbates on single-atom alloy catalysts},
  author={R{\'e}ocreux, Romain and Sykes, E Charles H and Michaelides, Angelos and Stamatakis, Michail},
  journal={J. Phys. Chem. Lett.},
  volume={13},
  number={31},
  pages={7314},
  year={2022},
  publisher={ACS Publications}
}

@article{reocreux2021one,
  title={One decade of computational studies on single-atom alloys: is in silico design within reach?},
  author={R{\'e}ocreux, Romain and Stamatakis, Michail},
  journal={Acc. Chem. Res.},
  volume={55},
  number={1},
  pages={87},
  year={2021},
  publisher={ACS Publications}
}

@article{eren2018structure,
  title={Structure of copper--cobalt surface alloys in equilibrium with carbon monoxide gas},
  author={Eren, Baran and Torres, Daniel and Karsl{\i}o{\u{g}}lu, Osman and Liu, Zongyuan and Wu, Cheng Hao and Stacchiola, Dario and Bluhm, Hendrik and Somorjai, Gabor A and Salmeron, Miquel},
  journal={J. Am. Chem. Soc.},
  volume={140},
  number={21},
  pages={6575},
  year={2018},
  publisher={ACS Publications}
}

@article{christiansen2024single,
  title={Single-Atom Substituents in Copper Surfaces May Adsorb Multiple CO Molecules},
  author={Christiansen, Magnus AH and Pe{\~n}a-Torres, Alejandro and J{\'o}nsson, Elvar Ö and J{\'o}nsson, Hannes},
  journal={J. Phys. Chem. Lett.},
  volume={15},
  pages={5654},
  year={2024},
  publisher={ACS Publications}
}

@article{lee2022dilute,
  title={Dilute alloys based on Au, Ag, or Cu for efficient catalysis: from synthesis to active sites},
  author={Lee, Jennifer D and Miller, Jeffrey B and Shneidman, Anna V and Sun, Lixin and Weaver, Jason F and Aizenberg, Joanna and Biener, Juergen and Boscoboinik, J Anibal and Foucher, Alexandre C and Frenkel, Anatoly I and others},
  journal={Chem. Rev.},
  volume={122},
  number={9},
  pages={8758},
  year={2022},
  publisher={ACS Publications}
}

@article{spivey2021selective,
  title={Selective interactions between free-atom-like d-states in single-atom alloy catalysts and near-frontier molecular orbitals},
  author={Spivey, Taylor D and Holewinski, Adam},
  journal={J. Am. Chem. Soc.},
  volume={143},
  number={31},
  pages={11897},
  year={2021},
  publisher={ACS Publications}
}

@article{zhang2022tuning,
  title={Tuning reactivity in trimetallic dual-atom alloys: molecular-like electronic states and ensemble effects},
  author={Zhang, Shengjie and Sykes, E Charles H and Montemore, Matthew M},
  journal={Chem. Sci.},
  volume={13},
  number={47},
  pages={14070},
  year={2022},
  publisher={Royal Society of Chemistry}
}

@article{kress2023priori,
  title={A priori design of dual-atom alloy sites and experimental demonstration of ethanol dehydrogenation and dehydration on PtCrAg},
  author={Kress, Paul L and Zhang, Shengjie and Wang, Yicheng and {\c{C}}{\i}nar, Volkan and Friend, Cynthia M and Sykes, E Charles H and Montemore, Matthew M},
  journal={J. Am. Chem. Soc.},
  volume={145},
  number={15},
  pages={8401},
  year={2023},
  publisher={ACS Publications}
}

@inbook{demand_catalysis,
author = {de Jong, Krijn P.},
publisher = {John Wiley \& Sons, Ltd},
isbn = {9783527626854},
title = {Deposition Precipitation},
booktitle = {Synthesis of Solid Catalysts},
chapter = {6},
pages = {111},
url = {https://onlinelibrary.wiley.com/doi/abs/10.1002/9783527626854.ch6},
eprint = {https://onlinelibrary.wiley.com/doi/pdf/10.1002/9783527626854.ch6},
year = {2009}
}

@misc{ma2006heterogeneous,
  title={Heterogeneous catalysis by metals},
  author={Ma, Zhen and Zaera, Francisco},
  journal={Inorg. Chem.},
  volume={30},
  pages={27},
  year={2006},
  publisher={John Wiley \& Sons, Ltd: Chichester, UK}
}

@article{rosen2023free,
  title={Free-atom-like d states beyond the dilute limit of single-atom alloys},
  author={Rosen, Andrew S and Vijay, Sudarshan and Persson, Kristin A},
  journal={Chem. Sci.},
  volume={14},
  number={6},
  pages={1503},
  year={2023},
  publisher={Royal Society of Chemistry}
}

@article{weaver2025self,
  title={Self-Stabilized Heterometallic Pair Sites for Selective Ethanol Dehydrogenation on Pt--Cr--Ag Alloy Catalysts},
  author={Weaver, Jason F and Xiang, Shuting and Jamir, Jovenal and K{\"u}st, Ulrike and R{\"a}misch, Lisa and Grespi, Andrea and Wallander, Harald and Zetterberg, Johan and Arias, Steven and Fornero, Esteban L and others},
  journal={Angewandte Chemie},
  volume={137},
  number={49},
  pages={e202513844},
  year={2025},
  publisher={Wiley Online Library}
}

@article{chen2022electronic,
  title={Electronic structure of single-atom alloys and its impact on the catalytic activities},
  author={Chen, Ziyi and Zhang, Peng},
  journal={ACS omega},
  volume={7},
  number={2},
  pages={1585},
  year={2022},
  publisher={ACS Publications}
}

@article{berger2025dopant,
  title={When Are Dopant d-States Free-Atom-Like? Periodic Trends and Confinement Effects in Single-Atom Alloys},
  author={Berger, Fabian and Michaelides, Angelos},
  journal={J. Am. Chem. Soc.},
  volume={147},
  number={41},
  pages={37079},
  year={2025},
  publisher={ACS Publications}
}

@article{bronsted1928acid,
  title={Acid and Basic Catalysis.},
  author={Bronsted, JN},
  journal={Chem. Rev.},
  volume={5},
  number={3},
  pages={231},
  year={1928},
  publisher={ACS Publications}
}

@article{bell1936theory,
  title={The theory of reactions involving proton transfers},
  author={Bell, Ronald Percy},
  journal={Proc. R. Soc. Lond. A. Math. Phys. Sci.},
  volume={154},
  number={882},
  pages={414},
  year={1936},
  publisher={The Royal Society London}
}

@article{evans1937introduction,
  title={On the introduction of thermodynamic variables into reaction kinetics},
  author={Evans, MG and Polanyi, M},
  journal={Trans. Faraday Soc.},
  volume={33},
  pages={448},
  year={1937},
  publisher={Royal Society of Chemistry}
}

@article{evans1938inertia,
  title={Inertia and driving force of chemical reactions},
  author={Evans, MG and Polanyi, Michael},
  journal={Trans. Faraday Soc.},
  volume={34},
  pages={11},
  year={1938},
  publisher={Royal Society of Chemistry}
}

@book{sabatier1920catalyse,
  title={La catalyse en chimie organique},
  author={Sabatier, Paul},
  volume={3},
  year={1920},
  publisher={C. B{\'e}ranger}
}

@article{che2013nobel,
  title={Nobel Prize in chemistry 1912 to Sabatier: Organic chemistry or catalysis?},
  author={Che, Michel},
  journal={Catal. Today},
  volume={218},
  pages={162},
  year={2013},
  publisher={Elsevier}
}

@article{stratton2023addressing,
  title={Addressing complexity in catalyst design: From volcanos and scaling to more sophisticated design strategies},
  author={Stratton, Sarah M and Zhang, Shengjie and Montemore, Matthew M},
  journal={Surf. Sci. Rep.},
  volume={78},
  number={3},
  pages={100597},
  year={2023},
  publisher={Elsevier}
}

@article{nwaokorie2022alloy,
  title={Alloy catalyst design beyond the volcano plot by breaking scaling relations},
  author={Nwaokorie, Chukwudi F and Montemore, Matthew M},
  journal={J. Phys. Chem. C},
  volume={126},
  number={8},
  pages={3993},
  year={2022},
  publisher={ACS Publications}
}

@article{monasterial2020more,
  title={When more is less: Nonmonotonic trends in adsorption on clusters in alloy surfaces},
  author={Monasterial, Abigale P and Hinderks, Calla A and Viriyavaree, Songkun and Montemore, Matthew M},
  journal={J. Chem. Phys.},
  volume={153},
  number={11},
  pages={111102},
  year={2020},
  publisher={AIP Publishing}
}

@article{hannagan2021first,
  title={First-principles design of a single-atom--alloy propane dehydrogenation catalyst},
  author={Hannagan, Ryan T and Giannakakis, Georgios and R{\'e}ocreux, Romain and Schumann, Julia and Finzel, Jordan and Wang, Yicheng and Michaelides, Angelos and Deshlahra, Prashant and Christopher, Phillip and Flytzani-Stephanopoulos, Maria and others},
  journal={Science},
  volume={372},
  number={6549},
  pages={1444},
  year={2021},
  publisher={American Association for the Advancement of Science}
}

@article{jalil2025nickel,
  title={Nickel promotes selective ethylene epoxidation on silver},
  author={Jalil, Anika and Happel, Elizabeth E and Cramer, Laura and Hunt, Adrian and Hoffman, Adam S and Waluyo, Iradwikanari and Montemore, Matthew M and Christopher, Phillip and Sykes, E Charles H},
  journal={Science},
  volume={387},
  number={6736},
  pages={869},
  year={2025},
  publisher={American Association for the Advancement of Science}
}

@article{inderwildi2007adding,
  title={When adding an unreactive metal enhances catalytic activity: NOx decomposition over silver--rhodium bimetallic surfaces},
  author={Inderwildi, OR and Jenkins, SJ and King, DA},
  journal={Surf. Sci.},
  volume={601},
  number={17},
  pages={L103},
  year={2007},
  publisher={Elsevier}
}

@article{zhang2019catalytic,
  title={Catalytic activity of palladium-doped silver dilute nanoalloys for formate oxidation from a theoretical perspective},
  author={Zhang, Nan and Chen, Fuyi and Guo, Longfei},
  journal={Phys. Chem. Chem. Phys.},
  volume={21},
  number={40},
  pages={22598},
  year={2019},
  publisher={Royal Society of Chemistry}
}

@article{fung2020electronic,
  title={Electronic band contraction induced low temperature methane activation on metal alloys},
  author={Fung, Victor and Hu, Guoxiang and Sumpter, Bobby},
  journal={J. Mater. Chem. A},
  volume={8},
  number={12},
  pages={6057},
  year={2020},
  publisher={Royal Society of Chemistry}
}

@article{he2024selective,
  title={Selective orbital coupling: an adsorption mechanism in single-atom catalysis},
  author={He, Chen and Lee, Chih-Heng and Meng, Lei and Chen, Hsin-Yi Tiffany and Li, Zhe},
  journal={J. Am. Chem. Soc.},
  volume={146},
  number={18},
  pages={12395},
  year={2024},
  publisher={ACS Publications}
}

@article{michaelides2007unhappy,
  title={The unhappy marriage of transition and noble metal atoms: A new way to enhance catalytic activity?(A perspective on:“When adding an unreactive metal enhances catalytic activity: NO x decomposition over silver rhodium bimetallic surfaces” by OR Inderwildi, SJ Jenkins, DA King)},
  author={Michaelides, Angelos},
  journal={Surf. Sci.},
  volume={601},
  number={17},
  pages={3529},
  year={2007}
}

@article{kayode2021factors,
  title={Factors controlling oxophilicity and carbophilicity of transition metals and main group metals},
  author={Kayode, Gbolade O and Montemore, Matthew M},
  journal={J. Mater. Chem. A},
  volume={9},
  number={39},
  pages={22325},
  year={2021},
  publisher={Royal Society of Chemistry}
}

@article{he2022atomically,
  title={Atomically dispersed heteronuclear dual-atom catalysts: A new rising star in atomic catalysis},
  author={He, Tianwei and Santiago, Alain R Puente and Kong, Youchao and Ahsan, Md Ariful and Luque, Rafael and Du, Aijun and Pan, Hui},
  journal={Small},
  volume={18},
  number={12},
  pages={2106091},
  year={2022},
  publisher={Wiley Online Library}
}

@article{lin2024machine,
  title={Machine learning-assisted dual-atom sites design with interpretable descriptors unifying electrocatalytic reactions},
  author={Lin, Xiaoyun and Du, Xiaowei and Wu, Shican and Zhen, Shiyu and Liu, Wei and Pei, Chunlei and Zhang, Peng and Zhao, Zhi-Jian and Gong, Jinlong},
  journal={Nat. Commun.},
  volume={15},
  number={1},
  pages={8169},
  year={2024},
  publisher={Nature Publishing Group UK London}
}

@article{rao2020extendable,
  title={Extendable machine learning model for the stability of single atom alloys},
  author={Rao, Karun K and Do, Quan K and Pham, Khoa and Maiti, Debtanu and Grabow, Lars C},
  journal={Top. Catal.},
  volume={63},
  number={7},
  pages={728},
  year={2020},
  publisher={Springer}
}

@article{he2025stability,
  title={Stability of Single Atom Alloys Catalyst: A Theoretical and Experimental Perspective},
  author={He, Tianwei and Shi, Ran and Zhou, Tong and Puente Santiago, Alain Rafael and Liu, Qingju},
  journal={ACS Catal.},
  volume={15},
  pages={10005},
  year={2025},
  publisher={ACS Publications}
}

@article{berger2024bringing,
  title={Bringing molecules together: Synergistic coadsorption at dopant sites of single atom alloys},
  author={Berger, Fabian and Schumann, Julia and R{\'e}ocreux, Romain and Stamatakis, Michail and Michaelides, Angelos},
  journal={J. Am. Chem. Soc.},
  volume={146},
  number={41},
  pages={28119},
  year={2024},
  publisher={ACS Publications}
}

@article{papanikolaou2019co,
  title={CO-induced aggregation and segregation of highly dilute alloys: a density functional theory study},
  author={Papanikolaou, Konstantinos G and Darby, Matthew T and Stamatakis, Michail},
  journal={J. Phys. Chem. C},
  volume={123},
  number={14},
  pages={9128},
  year={2019},
  publisher={ACS Publications}
}

@article{darby2018carbon,
  title={Carbon monoxide poisoning resistance and structural stability of single atom alloys},
  author={Darby, Matthew T and Sykes, E Charles H and Michaelides, Angelos and Stamatakis, Michail},
  journal={Top. Catal.},
  volume={61},
  number={5},
  pages={428},
  year={2018},
  publisher={Springer}
}

@article{papanikolaou2019engineering,
  title={Engineering the surface architecture of highly dilute alloys: an ab initio Monte Carlo approach},
  author={Papanikolaou, Konstantinos G and Darby, Matthew T and Stamatakis, Michail},
  journal={ACS Catal.},
  volume={10},
  number={2},
  pages={1224},
  year={2019},
  publisher={ACS Publications}
}

@article{kothakonda2025discovering,
  title={Discovering Ni/Cu Single-Atom Alloy as a Highly Active and Selective Catalyst for Direct Methane Conversion to Ethylene: A First-Principles Kinetic Study},
  author={Kothakonda, Manish and LaCroix, Sarah and Zhou, Chengyu and Yang, Ji and Su, Ji and Zhao, Qing},
  journal={ACS Catal.},
  volume={15},
  pages={11608},
  year={2025},
  publisher={ACS Publications}
}

@article{ouyang2021directing,
  title={Directing reaction pathways via in situ control of active site geometries in PdAu single-atom alloy catalysts},
  author={Ouyang, Mengyao and Papanikolaou, Konstantinos G and Boubnov, Alexey and Hoffman, Adam S and Giannakakis, Georgios and Bare, Simon R and Stamatakis, Michail and Flytzani-Stephanopoulos, Maria and Sykes, E Charles H},
  journal={Nat. Commun.},
  volume={12},
  number={1},
  pages={1549},
  year={2021},
  publisher={Nature Publishing Group UK London}
}

@article{zhang2020alloying,
  title={Alloying effect in silver-based dilute nanoalloy catalysts for oxygen reduction reactions},
  author={Zhang, Nan and Chen, Fuyi and Jin, Yachao and Wang, Jiali and Jin, Tao and Kou, Bo},
  journal={J. Catal.},
  volume={384},
  pages={37},
  year={2020},
  publisher={Elsevier}
}

@article{nilekar2009surface,
  title={Surface segregation energies in low-index open surfaces of bimetallic transition metal alloys},
  author={Nilekar, Anand Udaykumar and Ruban, Andrei V and Mavrikakis, Manos},
  journal={Surf. Sci.},
  volume={603},
  number={1},
  pages={91},
  year={2009},
  publisher={Elsevier}
}

@article{pei2015ag,
  title={Ag alloyed Pd single-atom catalysts for efficient selective hydrogenation of acetylene to ethylene in excess ethylene},
  author={Pei, Guang Xian and Liu, Xiao Yan and Wang, Aiqin and Lee, Adam F and Isaacs, Mark A and Li, Lin and Pan, Xiaoli and Yang, Xiaofeng and Wang, Xiaodong and Tai, Zhijun and others},
  journal={ACS Catal.},
  volume={5},
  number={6},
  pages={3717},
  year={2015},
  publisher={ACS Publications}
}

@article{zhou2020light,
  title={Light-driven methane dry reforming with single atomic site antenna-reactor plasmonic photocatalysts},
  author={Zhou, Linan and Martirez, John Mark P and Finzel, Jordan and Zhang, Chao and Swearer, Dayne F and Tian, Shu and Robatjazi, Hossein and Lou, Minhan and Dong, Liangliang and Henderson, Luke and others},
  journal={Nat. Energy},
  volume={5},
  number={1},
  pages={61},
  year={2020},
  publisher={Nature Publishing Group UK London}
}

@article{ro2022bifunctional,
  title={Bifunctional hydroformylation on heterogeneous Rh-WO x pair site catalysts},
  author={Ro, Insoo and Qi, Ji and Lee, Seungyeon and Xu, Mingjie and Yan, Xingxu and Xie, Zhenhua and Zakem, Gregory and Morales, Austin and Chen, Jingguang G and Pan, Xiaoqing and others},
  journal={Nature},
  volume={609},
  number={7926},
  pages={287},
  year={2022},
  publisher={Nature Publishing Group UK London}
}

@article{bunting2023reactivity,
  title={Reactivity of single-atom alloy nanoparticles: modeling the dehydrogenation of propane},
  author={Bunting, Rhys J and Wodaczek, Felix and Torabi, Tina and Cheng, Bingqing},
  journal={J. Am. Chem. Soc.},
  volume={145},
  number={27},
  pages={14894},
  year={2023},
  publisher={ACS Publications}
}

@article{meng2024coverage,
  title={Coverage-dependent activation of CO over Ni/Cu (100) single atom alloys (SAAs)},
  author={Meng, Weiwen and Li, Ling and Zhao, Rui and Liu, Yu and Wang, Xuan and Qiu, Hengshan},
  journal={J. Chem. Phys.},
  volume={161},
  number={1},
  pages={014712},
  year={2024},
  publisher={AIP Publishing}
}

@article{patel2019elucidating,
  title={Elucidating the composition of PtAg surface alloys with atomic-scale imaging and spectroscopy},
  author={Patel, Dipna A and Kress, Paul L and Cramer, Laura A and Larson, Amanda M and Sykes, E Charles H},
  journal={J. Chem. Phys.},
  volume={151},
  number={16},
  pages={164705},
  year={2019},
  publisher={AIP Publishing}
}

@article{mohrhusen2024modifying,
  title={Modifying the Reactivity of Single Pd Sites in a Trimetallic Sn-Pd-Ag Surface Alloy: Tuning CO Binding Strength},
  author={Mohrhusen, Lars and Zhang, Shengjie and Montemore, Matthew M and Madix, Robert J},
  journal={Small},
  volume={20},
  number={48},
  pages={2405715},
  year={2024},
  publisher={Wiley Online Library}
}

@article{sun2025full,
  title={Full utilization of noble metals by atom abstraction for propane dehydrogenation},
  author={Sun, Guodong and Luo, Ran and Fu, Donglong and Wu, Kexin and Wang, Xianhui and Bian, Xiaoqing and Lu, Zhenpu and Chang, Xin and Wang, Zhi and Huang, Siwei and others},
  journal={Science},
  volume={390},
  number={6776},
  pages={eadw3053},
  year={2025},
  publisher={American Association for the Advancement of Science}
}

@article{karageorgiou2026mechanisms,
  title={Mechanisms for the formation of active sites in single-atom alloys},
  author={Karageorgiou, Ioannis and Michaelides, Angelos and Berger, Fabian},
  journal={Nanoscale},
  volume={18},
  pages={9709},
  year={2026},
  publisher={Royal Society of Chemistry}
}

@article{weinhold2007high,
  title={High bond orders in metal-metal bonding},
  author={Weinhold, Frank and Landis, Clark R},
  journal={Science},
  volume={316},
  number={5821},
  pages={61},
  year={2007},
  publisher={American Association for the Advancement of Science}
}

@article{nguyen2005synthesis,
  title={Synthesis of a stable compound with fivefold bonding between two chromium (I) centers},
  author={Nguyen, Tailuan and Sutton, Andrew D and Brynda, Marcin and Fettinger, James C and Long, Gary J and Power, Philip P},
  journal={Science},
  volume={310},
  number={5749},
  pages={844},
  year={2005},
  publisher={American Association for the Advancement of Science}
}

@article{brynda2006quantum,
  title={A Quantum Chemical Study of the Quintuple Bond between Two Chromium Centers in [PhCrCrPh]: trans-Bent versus Linear Geometry},
  author={Brynda, Marcin and Gagliardi, Laura and Widmark, Per-Olof and Power, Philip P and Roos, Bj{\"o}rn O},
  journal={Angew. Chem. Int. Ed.},
  volume={118},
  number={23},
  pages={3888},
  year={2006},
  publisher={Wiley Online Library}
}

@article{schumann2021periodic,
  title={Periodic trends in adsorption energies around single-atom alloy active sites},
  author={Schumann, Julia and Bao, Yutian and Hannagan, Ryan T and Sykes, E Charles H and Stamatakis, Michail and Michaelides, Angelos},
  journal={J. Phys. Chem. Lett.},
  volume={12},
  number={41},
  pages={10060},
  year={2021},
  publisher={ACS Publications}
}

@article{lucci2014atomic,
  title={Atomic scale surface structure of Pt/Cu (111) surface alloys},
  author={Lucci, Felicia R and Lawton, Timothy J and Pronschinske, Alex and Sykes, E Charles H},
  journal={J. Phys. Chem. C.},
  volume={118},
  number={6},
  pages={3015},
  year={2014},
  publisher={ACS Publications}
}

@article{chen2026dual,
  title={Dual-atom Rh-Co catalysts for synergistically boosting nitrile hydrogenation},
  author={Chen, Jiawei and Chen, Hongqiu and Cai, Xiangbin and Wang, Yue and Peng, Mi and Sun, Bo and Diao, Jiangyong and Sun, Geng and Ma, Ding and Liu, Hongyang},
  journal={Nat. Commun.},
  year={2026},
  publisher={Nature Publishing Group UK London}
}

@article{ngan2026early,
  title={Early transition metal Cu-based single-atom alloys for selective propane dehydrogenation to propylene},
  author={Ngan, Hio Tong and Sautet, Philippe},
  journal={Chem Catal.},
  volume={6},
  number={5},
  pages={101691},
  year={2026},
  publisher={Elsevier}
}

@article{christiansen2025multiple,
  title={Multiple Adsorption of CO Molecules on Transition Metal Substitutional Impurities in Copper Surfaces},
  author={Christiansen, Magnus AH and Wang, Wei and J{\'o}nsson, Elvar {\"O} and Cicero, Giancarlo and J{\'o}nsson, Hannes},
  journal={ChemCatChem},
  volume={17},
  number={18},
  pages={e00765},
  year={2025},
  publisher={Wiley Online Library}
}

@article{ast2007giant,
  title={Giant spin splitting through surface alloying},
  author={Ast, Christian R and Henk, J{\"u}rgen and Ernst, Arthur and Moreschini, Luca and Falub, Mihaela C and Pacil{\'e}, Daniela and Bruno, Patrick and Kern, Klaus and Grioni, Marco},
  journal={Phys. Rev. Lett.},
  volume={98},
  number={18},
  pages={186807},
  year={2007},
  publisher={APS}
}

@article{knorr2002kondo,
  title={Kondo effect of single Co adatoms on Cu surfaces},
  author={Knorr, Nikolaus and Schneider, M Alexander and Diekh{\"o}ner, Lars and Wahl, Peter and Kern, Klaus},
  journal={Phys. Rev. Lett.},
  volume={88},
  number={9},
  pages={096804},
  year={2002},
  publisher={APS}
}

@article{li2026synthesis,
  title={Synthesis of 4H-phase high-entropy alloys for electrocatalysis},
  author={Li, Zijian and Zhang, An and Chen, Changsheng and Yang, Hua and Sun, Mingzi and Zhang, Qinghua and Xi, Shibo and Zhai, Li and Long, Xinyue and Li, Lujiang and others},
  journal={Nat. Mater.},
  volume={25},
  pages={972},
  year={2026},
  publisher={Nature Publishing Group UK London}
}

@article{ding2026differentiable,
  title={Differentiable inverse design of short-range order in high-entropy alloys: from target sro to target property},
  author={Ding, Tiancheng and Feugmo, Conrard Giresse Tetsassi},
  journal={arXiv preprint arXiv:2607.02219},
  year={2026}
}

@article{mazitov2024surface,
  title={Surface segregation in high-entropy alloys from alchemical machine learning},
  author={Mazitov, Arslan and Springer, Maximilian A and Lopanitsyna, Nataliya and Fraux, Guillaume and De, Sandip and Ceriotti, Michele},
  journal={J. Phys. Mater.},
  volume={7},
  number={2},
  pages={025007},
  year={2024},
  publisher={IOP Publishing}
}

@article{greeley2004alloy,
  title={Alloy catalysts designed from first principles},
  author={Greeley, Jeff and Mavrikakis, Manos},
  journal={Nat. Mater.},
  volume={3},
  number={11},
  pages={810},
  year={2004},
  publisher={Nature Publishing Group UK London}
}

@article{greeley2006computational,
  title={Computational high-throughput screening of electrocatalytic materials for hydrogen evolution},
  author={Greeley, Jeff and Jaramillo, Thomas F and Bonde, Jacob and Chorkendorff, IB and N{\o}rskov, Jens K},
  journal={Nat. Mater.},
  volume={5},
  number={11},
  pages={909},
  year={2006},
  publisher={Nature Publishing Group UK London}
}

@article{han2021single,
  title={Single-atom alloy catalysts designed by first-principles calculations and artificial intelligence},
  author={Han, Zhong-Kang and Sarker, Debalaya and Ouyang, Runhai and Mazheika, Aliaksei and Gao, Yi and Levchenko, Sergey V},
  journal={Nat. Commun.},
  volume={12},
  number={1},
  pages={1833},
  year={2021},
  publisher={Nature Publishing Group UK London}
}

@article{marcinkowski2013controlling,
  title={Controlling a spillover pathway with the molecular cork effect},
  author={Marcinkowski, Matthew D and Jewell, April D and Stamatakis, Michail and Boucher, Matthew B and Lewis, Emily A and Murphy, Colin J and Kyriakou, Georgios and Sykes, E Charles H},
  journal={Nat. Mater.},
  volume={12},
  number={6},
  pages={523},
  year={2013},
  publisher={Nature Publishing Group UK London}
}

@article{finzel2026metal,
  title={Metal hybridization in dilute-alloy catalysts promotes sintering resistance by decreasing surface mobility},
  author={Finzel, Jordan and Dannar, Audrey and Sun, Shoutian and Hoffman, Adam S and Soni, Yogita and Wang, Bin and Bare, Simon R and Sykes, E Charles H and Christopher, Phillip},
  journal={Nat. Mater.},
  volume={25},
  number={6},
  pages={964},
  year={2026},
  publisher={Nature Publishing Group UK London}
}

@article{van2021unlocking,
  title={Unlocking synergy in bimetallic catalysts by core--shell design},
  author={van der Hoeven, Jessi ES and Jelic, Jelena and Olthof, Liselotte A and Totarella, Giorgio and van Dijk-Moes, Relinde JA and Krafft, Jean-Marc and Louis, Catherine and Studt, Felix and van Blaaderen, Alfons and de Jongh, Petra E},
  journal={Nat. Mater.},
  volume={20},
  number={9},
  pages={1216},
  year={2021},
  publisher={Nature Publishing Group UK London}
}

@article{zugic2017dynamic,
  title={Dynamic restructuring drives catalytic activity on nanoporous gold--silver alloy catalysts},
  author={Zugic, Branko and Wang, Lucun and Heine, Christian and Zakharov, Dmitri N and Lechner, Barbara AJ and Stach, Eric A and Biener, Juergen and Salmeron, Miquel and Madix, Robert J and Friend, Cynthia M},
  journal={Nat. Mater.},
  volume={16},
  number={5},
  pages={558},
  year={2017},
  publisher={Nature Publishing Group UK London}
}

@article{han2022single,
  title={A single-atom library for guided monometallic and concentration-complex multimetallic designs},
  author={Han, Lili and Cheng, Hao and Liu, Wei and Li, Haoqiang and Ou, Pengfei and Lin, Ruoqian and Wang, Hsiao-Tsu and Pao, Chih-Wen and Head, Ashley R and Wang, Chia-Hsin and others},
  journal={Nat. Mater.},
  volume={21},
  number={6},
  pages={681},
  year={2022},
  publisher={Nature Publishing Group UK London}
}

@article{xie2021percolation,
  title={A percolation theory for designing corrosion-resistant alloys},
  author={Xie, Yusi and Artymowicz, Dorota M and Lopes, Pietro P and Aiello, Ashlee and Wang, Duo and Hart, James L and Anber, Elaf and Taheri, Mitra L and Zhuang, Houlong and Newman, Roger C and others},
  journal={Nat. Mater.},
  volume={20},
  number={6},
  pages={789},
  year={2021},
  publisher={Nature Publishing Group UK London}
}

@article{ruban1999surface,
  title={Surface segregation energies in transition-metal alloys},
  author={Ruban, AV and Skriver, Hans Lomholt and N{\o}rskov, Jens Kehlet},
  journal={Phys. Rev. B},
  volume={59},
  number={24},
  pages={15990},
  year={1999},
  publisher={APS}
}

@article{huang2025decoding,
  title={Decoding the stability of transition-metal alloys with theory-infused deep learning},
  author={Huang, Yang and Wang, Shih-Han and Cao, Shuyi and Achenie, Luke EK and Xin, Hongliang},
  journal={arXiv preprint arXiv:2506.03031},
  year={2025}
}

@article{friedel1969physics,
  title={The physics of metals},
  author={Friedel, J and others},
  journal={Cambridge University Press, Cambridge, 1969) p},
  volume={340},
  year={1969}
}

@article{morse1986clusters,
  title={Clusters of transition-metal atoms},
  author={Morse, Michael D},
  journal={Chem. Rev.},
  volume={86},
  number={6},
  pages={1049},
  year={1986},
  publisher={ACS Publications}
}

@article{christensen1997phase,
  title={Phase diagrams for surface alloys},
  author={Christensen, Asbj{\o}rn and Ruban, AV and Stoltze, Per and Jacobsen, Karsten Wedel and Skriver, Hans Lomholt and N{\o}rskov, Jens Kehlet and Besenbacher, Flemming},
  journal={Phys. Rev. B},
  volume={56},
  number={10},
  pages={5822},
  year={1997},
  publisher={APS}
}

@article{zhang2020short,
  title={Short-range order and its impact on the CrCoNi medium-entropy alloy},
  author={Zhang, Ruopeng and Zhao, Shiteng and Ding, Jun and Chong, Yan and Jia, Tao and Ophus, Colin and Asta, Mark and Ritchie, Robert O and Minor, Andrew M},
  journal={Nature},
  volume={581},
  number={7808},
  pages={283},
  year={2020},
  publisher={Nature Publishing Group UK London}
}

@article{chen2021direct,
  title={Direct observation of chemical short-range order in a medium-entropy alloy},
  author={Chen, Xuefei and Wang, Qi and Cheng, Zhiying and Zhu, Mingliu and Zhou, Hao and Jiang, Ping and Zhou, Lingling and Xue, Qiqi and Yuan, Fuping and Zhu, Jing and others},
  journal={Nature},
  volume={592},
  number={7856},
  pages={712},
  year={2021},
  publisher={Nature Publishing Group UK London}
}

@article{darby2018lonely,
  title={Lonely atoms with special gifts: breaking linear scaling relationships in heterogeneous catalysis with single-atom alloys},
  author={Darby, Matthew T and Stamatakis, Michail and Michaelides, Angelos and Sykes, E Charles H},
  journal={J. Phys. Chem. Lett.},
  volume={9},
  number={18},
  pages={5636},
  year={2018},
  publisher={ACS Publications}
}

@article{laltrimetallic,
  title={Trimetallic Single-Atom Alloys: Active Site Synergy or Most Reactive Atom Dominance? The Case of RhPtCu (111)},
  author={Lal, Vinita and H. Sykes, E Charles},
  journal={J. Phys. Chem. C.},
  year={2026},
  publisher={ACS Publications}
}

@article{zhangmagnetic,
  title={Magnetic Moments in Single-Atom Alloys: Trends, Mechanisms, and Implications for Tunable Adsorption},
  author={Zhang, Shengjie and Riviere, Collette I and Montemore, Matthew M},
  journal={ACS Phys. Chem. Au},
  year={2026},
  publisher={ACS Publications}
}

@article{vasudevan2026density,
  title={Density functional theory-informed design of radiation-resistant dilute ternary Cu alloys},
  author={Vasudevan, Vaibhav and Schuler, Thomas and Bellon, Pascal and Averback, Robert},
  journal={Phys. Rev. Mater.},
  volume={10},
  number={3},
  pages={033602},
  year={2026},
  publisher={APS}
}

@article{egami2020thermodynamic,
  title={Thermodynamic origin of solute-enriched stacking-fault in dilute Mg-Zn-Y alloys},
  author={Egami, M and Ohnuma, I and Enoki, M and Ohtani, H and Abe, Eiji},
  journal={Mater. Des.},
  volume={188},
  pages={108452},
  year={2020},
  publisher={Elsevier}
}

@article{booth2011coarsening,
  title={Coarsening resistance at 400 C of precipitation-strengthened Al--Zr--Sc--Er alloys},
  author={Booth-Morrison, Christopher and Dunand, David C and Seidman, David N},
  journal={Acta Mater.},
  volume={59},
  number={18},
  pages={7029},
  year={2011},
  publisher={Elsevier}
}

@article{wang2020surface,
  title={Surface facet dependence of competing alloying mechanisms},
  author={Wang, Yicheng and Papanikolaou, Konstantinos G and Hannagan, Ryan T and Patel, Dipna A and Balema, Tedros A and Cramer, Laura A and Kress, Paul L and Stamatakis, Michail and Sykes, E Charles H},
  journal={J. Chem. Phys.},
  volume={153},
  number={24},
  year={2020},
  publisher={AIP Publishing}
}

\end{document}


\clearpage

\tableofcontents

\clearpage

\section{DFT Exchange-Correlation Functional Dependence of Segregation and Aggregation Energies}

To assess whether our findings are sensitive to the chosen exchange–correlation functional, we perform additional density functional theory (DFT) calculations for aggregation and segregation energies.
In addition to the optB86b-vdW functional employed throughout the main text, we test three alternative generalized gradient approximation (GGA) functionals: PBEsol, PBE and RPBE. 
These functionals are selected for their known differences in describing lattice constants, surface energetics, and electronic localization.\cite{schimka2010accurate}
We also assess the performance of the foundation machine learning interatomic potential (MLIP) MACE-MP-0.

\begin{figure}[htbp]
    \includegraphics[width=14cm,height=\textheight,keepaspectratio]{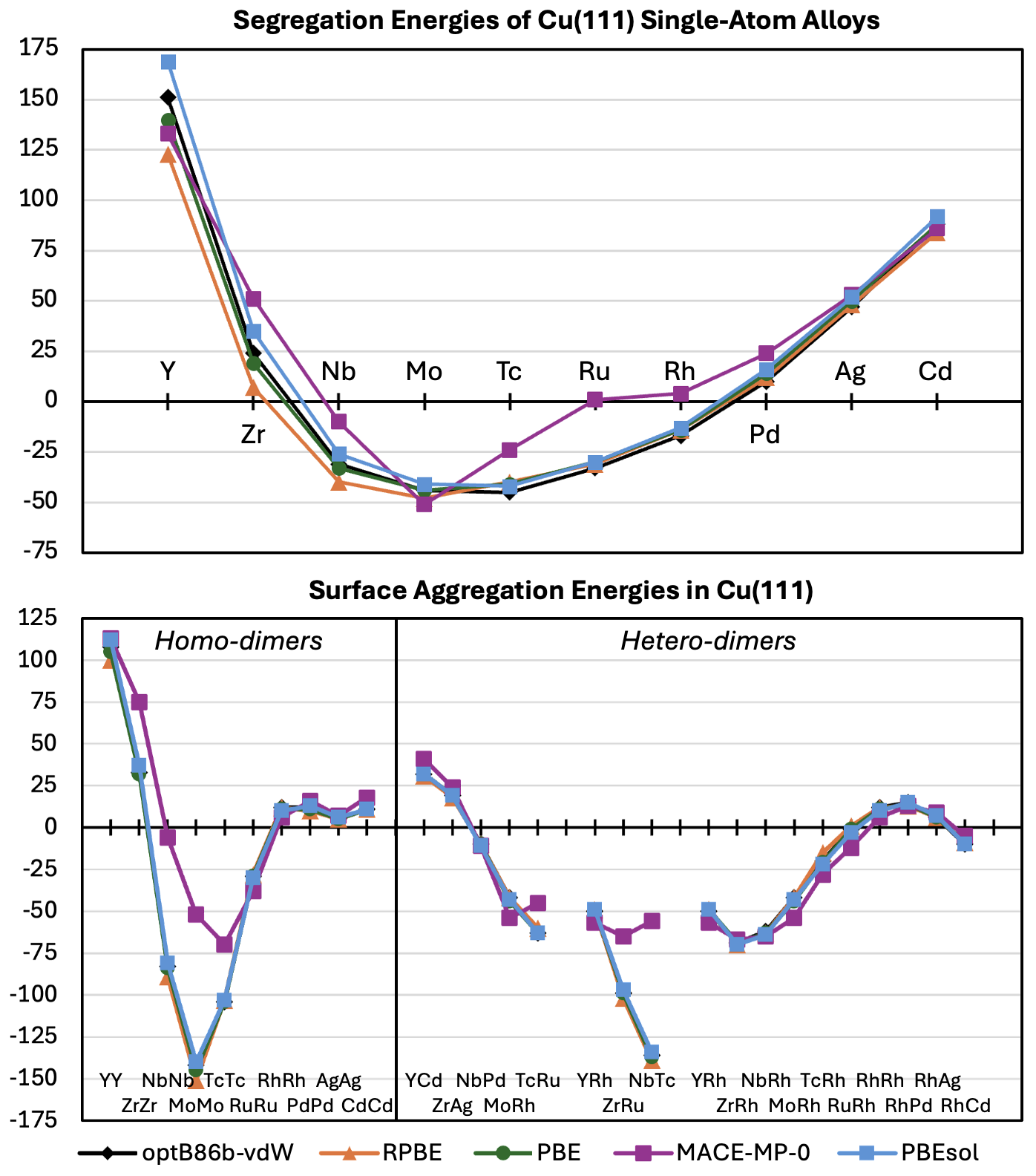}
    \caption{
        Comparison of segregation and aggregation energies predicted by different exchange--correlation functionals and the foundation model MACE-MP-0.
        Top: Segregation energies of 4$d$ transition-metal dopants in Cu(111) calculated with the optB86b-vdW (black diamonds), RPBE (orange triangles), PBE (green circles), and PBEsol (blue squares) functionals, together with the foundation MLIP MACE-MP-0 (purple triangles).
        Bottom: Aggregation energies of selected homo- and hetero-dimers in Cu(111) obtained with the same methods.
        The four DFT functionals are in good quantitative agreement, while MACE-MP-0 reproduces the overall trends but exhibits larger deviations.
        All energies are reported in kJ~mol$^{-1}$.
    }
    \label{fig:dft}
\end{figure}

Figure~\ref{fig:dft} summarizes the functional dependence of the segregation energies for the series of 4$d$ transition-metal dopants in Cu(111), while Table~\ref{tab:seg_energies_cu111_4d} lists the corresponding numerical values. 
All four DFT functionals predict the same segregation trend and are in close quantitative agreement. 
The largest deviations occur for the earliest transition metals, which exhibit a strong thermodynamic preference for surface incorporation.
MACE-MP-0 reproduces the same overall trends but exhibits somewhat larger deviations.

The aggregation energies show even better agreement across the investigated DFT functionals.
The corresponding data points are nearly indistinguishable in Figure~\ref{fig:dft}, reflecting the excellent quantitative agreement between the functionals. 
MACE-MP-0 again captures the overall trends but exhibits larger deviations, particularly for systems containing early transition metals.

Overall, the excellent agreement between the four DFT functionals, despite their distinct behavior, demonstrates that the predicted segregation and aggregation energies are robust with respect to the choice of functional. 
While MACE-MP-0 reproduces the qualitative trends, its larger deviations indicate that it is not yet sufficiently accurate to replace DFT for the quantitative prediction of these properties.

\begin{table}[ht]
  \centering
  \renewcommand{\arraystretch}{1.2}
  \caption{Segregation energies for 4$d$ transition metal dopants in Cu(111) calculated using the functionals optB86b-vdW, RPBE, PBE, and PBEsol, as well as the foundation MLIP MACE-MP-0. All energies are reported in kJ~mol$^{-1}$.}
  \vspace{8pt}
  \begin{tabular}{@{\hspace{8pt}}*{6}{c@{\hspace{8pt}}}}
    \hline\hline
    TM & optB86b-vdW & RPBE & PBE & MACE-MP-0 & PBEsol \\
    \hline
    Y  & $151$  & $123$  & $140$  & $133$  & $169$ \\
    Zr & $24$   & $7$    & $19$   & $51$   & $35$  \\
    Nb & $-31$  & $-40$  & $-33$  & $-10$  & $-26$ \\
    Mo & $-44$  & $-48$  & $-44$  & $-51$  & $-41$ \\
    Tc & $-45$  & $-40$  & $-41$  & $-24$  & $-42$ \\
    Ru & $-33$  & $-31$  & $-30$  & $1$    & $-30$ \\
    Rh & $-17$  & $-14$  & $-14$  & $4$    & $-13$ \\
    Pd & $10$   & $12$   & $14$   & $24$   & $16$  \\
    Ag & $47$   & $48$   & $50$   & $53$   & $52$  \\
    Cd & $88$   & $84$   & $88$   & $86$   & $92$  \\
    \hline\hline
  \end{tabular}
  \label{tab:seg_energies_cu111_4d}
\end{table}

\begin{table}[ht]
  \centering
  \renewcommand{\arraystretch}{1.2}
  \caption{Surface aggregation energies for selected 4$d$ transition metal dopant combinations in Cu(111) calculated using the functionals optB86b-vdW, RPBE, PBE, and PBEsol, as well as the foundation MLIP MACE-MP-0. All energies are reported in kJ~mol$^{-1}$. The aggregation energy calculation for Zr--Zr with RPBE did not converge.}
  \vspace{8pt}
  \begin{tabular}{@{\hspace{8pt}}*{6}{c@{\hspace{8pt}}}}
    \hline\hline
    Pair & optB86b-vdW & RPBE & PBE & MACE-MP-0 & PBEsol \\
    \hline
    Y--Y   & $108$  & $100$ & $105$ & $113$ & $112$ \\
    Zr--Zr & $33$   &    -   & $32$  & $75$  & $37$  \\
    Nb--Nb & $-83$  & $-89$ & $-84$ & $-6$  & $-81$ \\
    Mo--Mo & $-142$ & $-151$& $-145$& $-52$ & $-140$ \\
    Tc--Tc & $-104$ & $-103$& $-104$& $-70$ & $-103$ \\
    Ru--Ru & $-29$  & $-28$ & $-29$ & $-38$ & $-30$ \\
    Rh--Rh & $12$   & $12$  & $11$  & $6$   & $10$  \\
    Pd--Pd & $12$   & $10$  & $11$  & $16$  & $13$  \\
    Ag--Ag & $5$    & $5$   & $5$   & $7$   & $6$   \\
    Cd--Cd & $11$   & $11$  & $11$  & $18$  & $11$  \\
    \hline
    Y--Cd  & $32$   & $31$  & $32$  & $41$  & $32$  \\
    Zr--Ag & $19$   & $18$  & $19$  & $24$  & $19$  \\
    Nb--Pd & $-10$  & $-10$ & $-10$ & $-11$ & $-11$ \\
    Mo--Rh & $-42$  & $-42$ & $-44$ & $-54$ & $-43$ \\
    Tc--Ru & $-63$  & $-60$ & $-63$ & $-45$ & $-63$ \\
    \hline
    Y--Rh  & $-50$  & $-49$ & $-49$ & $-57$ & $-49$ \\
    Zr--Ru & $-99$  & $-102$& $-99$ & $-65$ & $-97$ \\
    Nb--Tc & $-136$ & $-139$& $-137$& $-56$ & $-134$ \\
    \hline
    Y--Rh  & $-50$  & $-49$ & $-49$ & $-57$ & $-49$ \\
    Zr--Rh & $-69$  & $-70$ & $-69$ & $-67$ & $-70$ \\
    Nb--Rh & $-62$  & $-63$ & $-63$ & $-65$ & $-64$ \\
    Mo--Rh & $-42$  & $-42$ & $-44$ & $-54$ & $-43$ \\
    Tc--Rh & $-20$  & $-15$ & $-21$ & $-28$ & $-22$ \\
    Ru--Rh & $-1$   & $1$   & $-1$  & $-12$ & $-3$  \\
    Rh--Rh & $12$   & $12$  & $11$  & $6$   & $10$  \\
    Rh--Pd & $15$   & $13$  & $14$  & $13$  & $15$  \\
    Rh--Ag & $6$    & $6$   & $6$   & $9$   & $7$   \\
    Rh--Cd & $-10$  & $-9$  & $-9$  & $-5$  & $-10$ \\
    \hline\hline
  \end{tabular}
  \label{tab:agg_energies_cu111}
\end{table}

\clearpage

\section{Aggregation Energies}

\subsection{Surface Aggregation}


\begin{figure}[htbp]
    \includegraphics[width=12cm,height=\textheight,keepaspectratio]{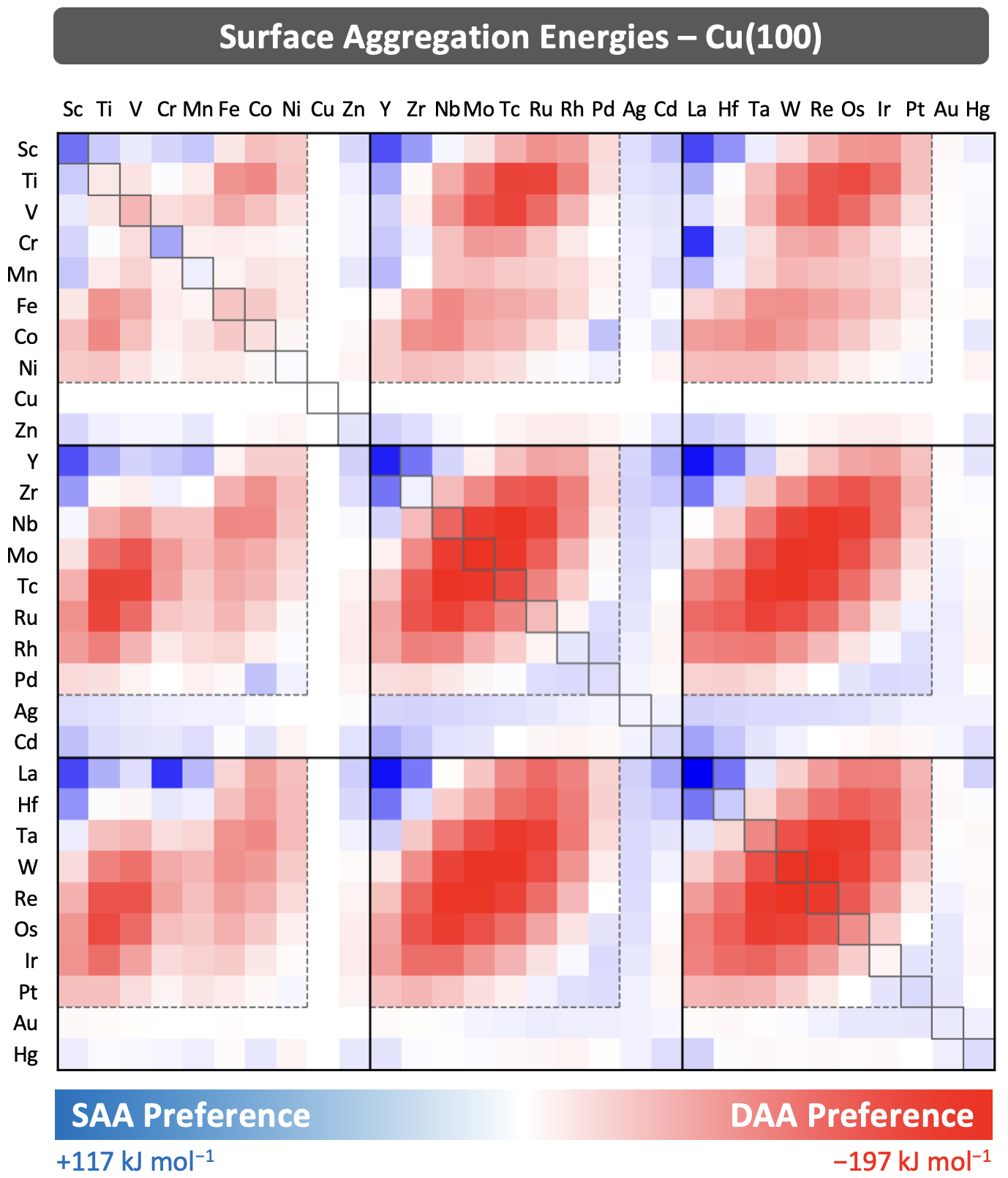}
    \caption{
        Aggregation energies for the formation of all possible transition metal dopant dimers in the Cu(100) surface, analogous to Fig.~2 in the main text. 
        Preferences for single-atom sites are represented in blue, while preferences for dopant dimers are shown in red. 
    }
    \label{fig:Cu100-surf-aggr}
\end{figure}

\begin{figure}[htbp]
    \includegraphics[width=9.5cm,height=\textheight,keepaspectratio]{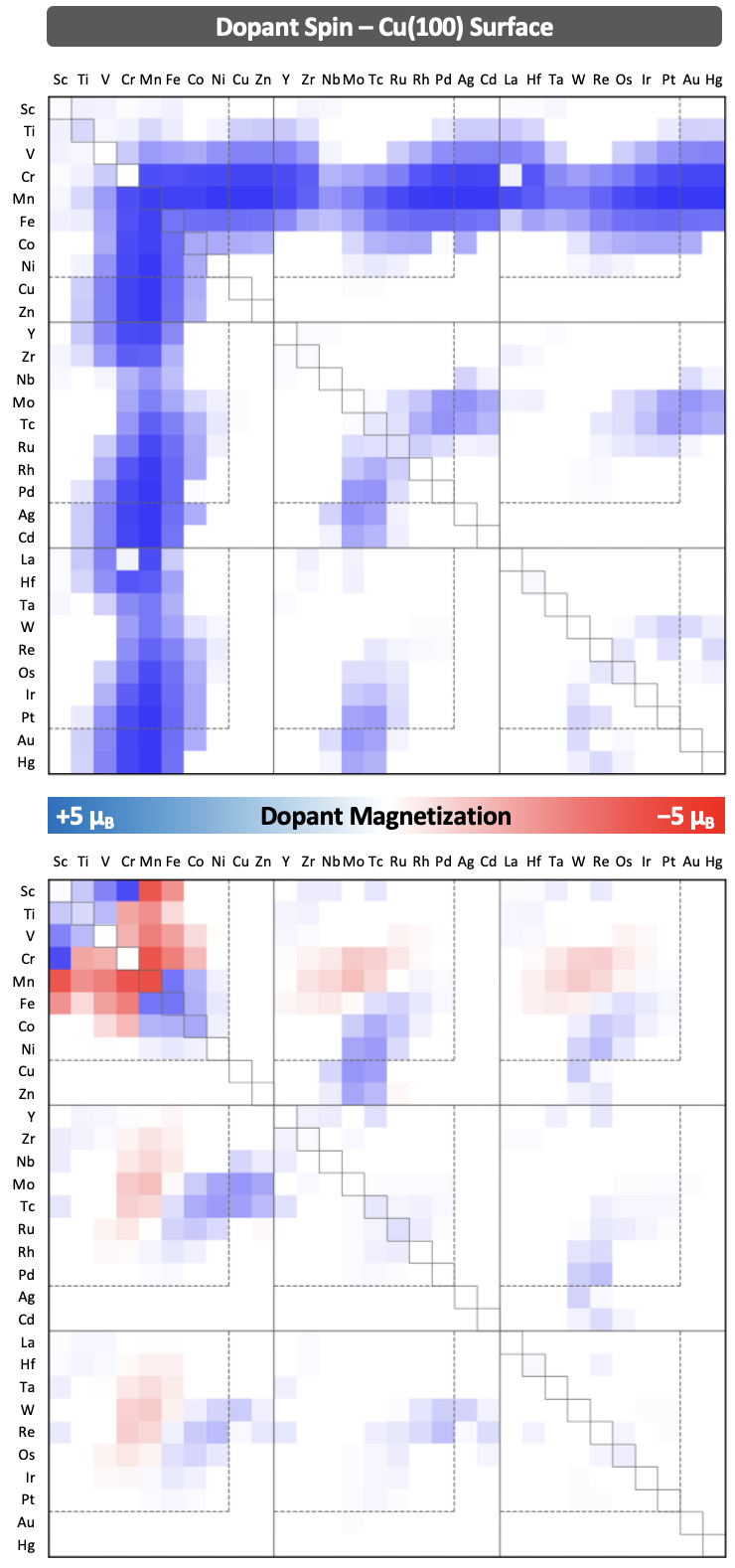}
    \caption{
        Local dopant magnetizations in dual-atom sites located in the surface layer of Cu(100).
        Top: The local magnetization of the dopant with the larger magnetic moment is shown in blue.
        Bottom: The local magnetization of the second dopant is shown in blue when its magnetic moment is aligned parallel to that of the first dopant (ferromagnetic coupling) and in red when it is aligned antiparallel (antiferromagnetic coupling).
    }
    \label{fig:Cu100-surf-spin}
\end{figure}

\begin{table}[ht]
  \centering
  \renewcommand{\arraystretch}{1.2}
  \caption{\textbf{Aggregation energies} for \textbf{dual-atom} sites in the \textbf{surface} layer of \textbf{Cu(100)} for all combinations of \textbf{3\textit{d}} transition metals (columns) and all other transition metals (rows), as obtained from initial \textbf{high-spin} configurations. Energies are reported in kJ~mol$^{-1}$.}
  
  \vspace{8pt}
  \begin{tabular}{@{\hspace{8pt}}*{11}{c@{\hspace{8pt}}}}
    \hline\hline
    TM & Sc & Ti & V & Cr & Mn & Fe & Co & Ni & Cu & Zn \\
    \hline
    Sc & $64$ & $24$ & $9$ & $18$ & $26$ & $-19$ & $-52$ & $-43$ & $0$ & $18$ \\
    Ti & $24$ & $-18$ & $-23$ & $2$ & $-16$ & $-88$ & $-97$ & $-46$ & $0$ & $8$ \\
    V  & $9$ & $-23$ & $-58$ & $-28$ & $-36$ & $-68$ & $-51$ & $-22$ & $0$ & $4$ \\
    Cr & $18$ & $2$ & $-28$ & $40$ & $-13$ & $-16$ & $-12$ & $-7$ & $0$ & $5$ \\
    Mn & $26$ & $-16$ & $-36$ & $-13$ & $8$ & $-8$ & $-21$ & $-18$ & $0$ & $10$ \\
    Fe & $-19$ & $-88$ & $-68$ & $-16$ & $-8$ & $-46$ & $-44$ & $-18$ & $0$ & $-2$ \\
    Co & $-52$ & $-97$ & $-51$ & $-12$ & $-21$ & $-44$ & $-26$ & $-6$ & $0$ & $-6$ \\
    Ni & $-43$ & $-46$ & $-22$ & $-7$ & $-18$ & $-18$ & $-6$ & $3$ & $0$ & $-8$ \\
    Cu & $0$ & $0$ & $0$ & $0$ & $0$ & $0$ & $0$ & $0$ & $0$ & $0$ \\
    Zn & $18$ & $8$ & $4$ & $5$ & $10$ & $-2$ & $-6$ & $-8$ & $0$ & $13$ \\
    \hline
    Y  & $80$ & $38$ & $20$ & $25$ & $33$ & $-10$ & $-39$ & $-40$ & $0$ & $21$ \\
    Zr & $46$ & $-6$ & $-13$ & $7$ & $0$ & $-64$ & $-90$ & $-54$ & $0$ & $15$ \\
    Nb & $5$ & $-66$ & $-88$ & $-49$ & $-49$ & $-97$ & $-94$ & $-47$ & $0$ & $3$ \\
    Mo & $-25$ & $-115$ & $-141$ & $-82$ & $-55$ & $-77$ & $-66$ & $-33$ & $0$ & $1$ \\
    Tc & $-64$ & $-165$ & $-161$ & $-79$ & $-43$ & $-69$ & $-58$ & $-25$ & $0$ & $-10$ \\
    Ru & $-90$ & $-162$ & $-122$ & $-45$ & $-38$ & $-58$ & $-37$ & $-7$ & $0$ & $-15$ \\
    Rh & $-80$ & $-105$ & $-60$ & $-17$ & $-29$ & $-34$ & $-12$ & $2$ & $0$ & $-16$ \\
    Pd & $-30$ & $-26$ & $-9$ & $-1$ & $-11$ & $-6$ & $28$ & $7$ & $0$ & $-9$ \\
    Ag & $15$ & $13$ & $10$ & $7$ & $7$ & $6$ & $3$ & $1$ & $0$ & $2$ \\
    Cd & $29$ & $16$ & $12$ & $11$ & $15$ & $1$ & $12$ & $-8$ & $0$ & $13$ \\
    \hline
    La & $85$ & $36$ & $15$ & $94$ & $33$ & $-33$ & $-79$ & $-53$ & $0$ & $22$ \\
    Hf & $50$ & $2$ & $-7$ & $10$ & $8$ & $-52$ & $-83$ & $-55$ & $0$ & $18$ \\
    Ta & $9$ & $-52$ & $-59$ & $-27$ & $-34$ & $-89$ & $-97$ & $-55$ & $0$ & $6$ \\
    W  & $-28$ & $-103$ & $-118$ & $-69$ & $-57$ & $-90$ & $-81$ & $-42$ & $0$ & $-3$ \\
    Re & $-62$ & $-145$ & $-145$ & $-76$ & $-52$ & $-78$ & $-64$ & $-30$ & $0$ & $-11$ \\
    Os & $-86$ & $-155$ & $-123$ & $-54$ & $-44$ & $-64$ & $-44$ & $-13$ & $0$ & $-14$ \\
    Ir & $-88$ & $-120$ & $-77$ & $-28$ & $-36$ & $-43$ & $-20$ & $-3$ & $0$ & $-16$ \\
    Pt & $-52$ & $-51$ & $-26$ & $-10$ & $-20$ & $-16$ & $-5$ & $4$ & $0$ & $-10$ \\
    Au & $-6$ & $-4$ & $-2$ & $0$ & $-1$ & $-2$ & $0$ & $0$ & $0$ & $1$ \\
    Hg & $9$ & $2$ & $3$ & $4$ & $8$ & $-3$ & $11$ & $-8$ & $0$ & $11$ \\
    \hline\hline
  \end{tabular}
  \label{tab:surf_agg_cu100_3d}
\end{table}

\begin{table}[ht]
  \centering
  \renewcommand{\arraystretch}{1.2}
  \caption{\textbf{Aggregation energies} for \textbf{dual-atom} sites in the \textbf{surface} layer of \textbf{Cu(100)} for all combinations of \textbf{4\textit{d}} transition metals (columns) and all other transition metals (rows), as obtained from initial \textbf{high-spin} configurations. Energies are reported in kJ~mol$^{-1}$.}
  \vspace{8pt}
  \begin{tabular}{@{\hspace{8pt}}*{11}{c@{\hspace{8pt}}}}
    \hline\hline
    TM & Y & Zr & Nb & Mo & Tc & Ru & Rh & Pd & Ag & Cd \\
    \hline
    Sc & $80$ & $46$ & $5$ & $-25$ & $-64$ & $-90$ & $-80$ & $-30$ & $15$ & $29$ \\
    Ti & $38$ & $-6$ & $-66$ & $-115$ & $-165$ & $-162$ & $-105$ & $-26$ & $13$ & $16$ \\
    V  & $20$ & $-13$ & $-88$ & $-141$ & $-161$ & $-122$ & $-60$ & $-9$ & $10$ & $12$ \\
    Cr & $25$ & $7$ & $-49$ & $-82$ & $-79$ & $-45$ & $-17$ & $-1$ & $7$ & $11$ \\
    Mn & $33$ & $0$ & $-49$ & $-55$ & $-43$ & $-38$ & $-29$ & $-11$ & $7$ & $15$ \\
    Fe & $-10$ & $-64$ & $-97$ & $-77$ & $-69$ & $-58$ & $-34$ & $-6$ & $6$ & $1$ \\
    Co & $-39$ & $-90$ & $-94$ & $-66$ & $-58$ & $-37$ & $-12$ & $28$ & $3$ & $12$ \\
    Ni & $-40$ & $-54$ & $-47$ & $-33$ & $-25$ & $-7$ & $2$ & $7$ & $1$ & $-8$ \\
    Cu & $0$ & $0$ & $0$ & $0$ & $0$ & $0$ & $0$ & $0$ & $0$ & $0$ \\
    Zn & $21$ & $15$ & $3$ & $1$ & $-10$ & $-15$ & $-16$ & $-9$ & $2$ & $13$ \\
    \hline
    Y  & $102$ & $64$ & $19$ & $-12$ & $-48$ & $-71$ & $-68$ & $-28$ & $19$ & $38$ \\
    Zr & $64$ & $7$ & $-54$ & $-95$ & $-138$ & $-144$ & $-104$ & $-31$ & $20$ & $26$ \\
    Nb & $19$ & $-54$ & $-128$ & $-173$ & $-197$ & $-165$ & $-99$ & $-19$ & $18$ & $13$ \\
    Mo & $-12$ & $-95$ & $-173$ & $-197$ & $-184$ & $-135$ & $-65$ & $-7$ & $16$ & $11$ \\
    Tc & $-48$ & $-138$ & $-197$ & $-184$ & $-156$ & $-101$ & $-40$ & $2$ & $14$ & $-1$ \\
    Ru & $-71$ & $-144$ & $-165$ & $-135$ & $-101$ & $-54$ & $-9$ & $15$ & $11$ & $-6$ \\
    Rh & $-68$ & $-104$ & $-99$ & $-65$ & $-40$ & $-9$ & $11$ & $17$ & $8$ & $-10$ \\
    Pd & $-28$ & $-31$ & $-19$ & $-7$ & $2$ & $15$ & $17$ & $15$ & $5$ & $-5$ \\
    Ag & $19$ & $20$ & $18$ & $16$ & $14$ & $11$ & $8$ & $5$ & $5$ & $6$ \\
    Cd & $38$ & $26$ & $13$ & $11$ & $-1$ & $-6$ & $-10$ & $-5$ & $6$ & $18$ \\
    \hline
    La & $110$ & $61$ & $-2$ & $-47$ & $-98$ & $-122$ & $-102$ & $-36$ & $22$ & $43$ \\
    Hf & $64$ & $15$ & $-40$ & $-77$ & $-118$ & $-134$ & $-107$ & $-37$ & $19$ & $26$ \\
    Ta & $21$ & $-44$ & $-109$ & $-150$ & $-182$ & $-166$ & $-109$ & $-30$ & $17$ & $12$ \\
    W  & $-17$ & $-92$ & $-162$ & $-193$ & $-192$ & $-153$ & $-87$ & $-16$ & $17$ & $6$ \\
    Re & $-50$ & $-129$ & $-190$ & $-187$ & $-166$ & $-120$ & $-57$ & $0$ & $16$ & $-1$ \\
    Os & $-72$ & $-144$ & $-173$ & $-149$ & $-120$ & $-73$ & $-22$ & $12$ & $14$ & $-4$ \\
    Ir & $-78$ & $-120$ & $-119$ & $-88$ & $-60$ & $-24$ & $3$ & $17$ & $10$ & $-8$ \\
    Pt & $-48$ & $-57$ & $-46$ & $-26$ & $-12$ & $7$ & $14$ & $16$ & $7$ & $-6$ \\
    Au & $-4$ & $-2$ & $1$ & $5$ & $7$ & $8$ & $8$ & $8$ & $6$ & $4$ \\
    Hg & $12$ & $2$ & $-2$ & $2$ & $-6$ & $-7$ & $-8$ & $-3$ & $6$ & $15$ \\
    \hline\hline
  \end{tabular}
  \label{tab:surf_agg_cu100_4d}
\end{table}

\begin{table}[ht]
  \centering
  \renewcommand{\arraystretch}{1.2}
  \caption{\textbf{Aggregation energies} for \textbf{dual-atom} sites in the \textbf{surface} layer of \textbf{Cu(100)} for all combinations of \textbf{5\textit{d}} transition metals (columns) and all other transition metals (rows), as obtained from initial \textbf{high-spin} configurations. Energies are reported in kJ~mol$^{-1}$.}
  \vspace{8pt}
  \begin{tabular}{@{\hspace{8pt}}*{11}{c@{\hspace{8pt}}}}
    \hline\hline
    TM & La & Hf & Ta & W & Re & Os & Ir & Pt & Au & Hg \\
    \hline
    Sc & $85$ & $50$ & $9$ & $-28$ & $-62$ & $-86$ & $-88$ & $-52$ & $-6$ & $9$ \\
    Ti & $36$ & $2$ & $-52$ & $-103$ & $-145$ & $-155$ & $-120$ & $-51$ & $-4$ & $2$ \\
    V  & $15$ & $-7$ & $-59$ & $-118$ & $-145$ & $-123$ & $-77$ & $-26$ & $-2$ & $3$ \\
    Cr & $94$ & $10$ & $-27$ & $-69$ & $-76$ & $-54$ & $-28$ & $-10$ & $0$ & $4$ \\
    Mn & $33$ & $8$ & $-34$ & $-57$ & $-52$ & $-44$ & $-36$ & $-20$ & $-1$ & $8$ \\
    Fe & $-33$ & $-52$ & $-89$ & $-90$ & $-78$ & $-64$ & $-43$ & $-16$ & $-2$ & $-3$ \\
    Co & $-79$ & $-83$ & $-97$ & $-81$ & $-64$ & $-44$ & $-20$ & $-5$ & $0$ & $11$ \\
    Ni & $-53$ & $-55$ & $-55$ & $-42$ & $-30$ & $-13$ & $-3$ & $4$ & $0$ & $-8$ \\
    Cu & $0$ & $0$ & $0$ & $0$ & $0$ & $0$ & $0$ & $0$ & $0$ & $0$ \\
    Zn & $22$ & $18$ & $6$ & $-3$ & $-11$ & $-14$ & $-16$ & $-10$ & $1$ & $11$ \\
    \hline
    Y  & $110$ & $64$ & $21$ & $-17$ & $-50$ & $-72$ & $-78$ & $-48$ & $-4$ & $12$ \\
    Zr & $61$ & $15$ & $-44$ & $-92$ & $-129$ & $-144$ & $-120$ & $-57$ & $-2$ & $2$ \\
    Nb & $-2$ & $-40$ & $-109$ & $-162$ & $-190$ & $-173$ & $-119$ & $-46$ & $1$ & $-2$ \\
    Mo & $-47$ & $-77$ & $-150$ & $-193$ & $-187$ & $-149$ & $-88$ & $-26$ & $5$ & $2$ \\
    Tc & $-98$ & $-118$ & $-182$ & $-192$ & $-166$ & $-120$ & $-60$ & $-12$ & $7$ & $-6$ \\
    Ru & $-122$ & $-134$ & $-166$ & $-153$ & $-120$ & $-73$ & $-24$ & $7$ & $8$ & $-7$ \\
    Rh & $-102$ & $-107$ & $-109$ & $-87$ & $-57$ & $-22$ & $3$ & $14$ & $8$ & $-8$ \\
    Pd & $-36$ & $-37$ & $-30$ & $-16$ & $0$ & $12$ & $17$ & $16$ & $8$ & $-3$ \\
    Ag & $22$ & $19$ & $17$ & $17$ & $16$ & $14$ & $10$ & $7$ & $6$ & $6$ \\
    Cd & $43$ & $26$ & $12$ & $6$ & $-1$ & $-4$ & $-8$ & $-6$ & $4$ & $15$ \\
    \hline
    La & $117$ & $63$ & $11$ & $-38$ & $-80$ & $-106$ & $-103$ & $-59$ & $-3$ & $21$ \\
    Hf & $63$ & $24$ & $-32$ & $-78$ & $-115$ & $-135$ & $-120$ & $-63$ & $-5$ & $1$ \\
    Ta & $11$ & $-32$ & $-96$ & $-147$ & $-179$ & $-173$ & $-128$ & $-57$ & $-2$ & $-5$ \\
    W  & $-38$ & $-78$ & $-147$ & $-192$ & $-196$ & $-166$ & $-109$ & $-39$ & $2$ & $-4$ \\
    Re & $-80$ & $-115$ & $-179$ & $-196$ & $-178$ & $-138$ & $-79$ & $-18$ & $6$ & $-5$ \\
    Os & $-106$ & $-135$ & $-173$ & $-166$ & $-138$ & $-92$ & $-41$ & $1$ & $10$ & $-4$ \\
    Ir & $-103$ & $-120$ & $-128$ & $-109$ & $-79$ & $-41$ & $-8$ & $12$ & $11$ & $-5$ \\
    Pt & $-59$ & $-63$ & $-57$ & $-39$ & $-18$ & $1$ & $12$ & $17$ & $11$ & $-1$ \\
    Au & $-3$ & $-5$ & $-2$ & $2$ & $6$ & $10$ & $11$ & $11$ & $10$ & $8$ \\
    Hg & $21$ & $1$ & $-5$ & $-4$ & $-5$ & $-4$ & $-5$ & $-1$ & $8$ & $16$ \\
    \hline\hline
  \end{tabular}
  \label{tab:surf_agg_cu100_5d}
\end{table}

\begin{table}[ht]
  \centering
  \renewcommand{\arraystretch}{1.2}
  \caption{\textbf{Aggregation energies} for \textbf{dual-atom} sites in the \textbf{surface} layer of \textbf{Cu(100)} for all combinations of \textbf{3\textit{d}} transition metals (columns) and all other transition metals (rows), as obtained from initial \textbf{low-spin} configurations. Energies are reported in kJ~mol$^{-1}$.}
  \vspace{8pt}
  \begin{tabular}{@{\hspace{8pt}}*{11}{c@{\hspace{8pt}}}}
    \hline\hline
    TM & Sc & Ti & V & Cr & Mn & Fe & Co & Ni & Cu & Zn \\
    \hline
    Sc & $65$ & $31$ & $-4$ & $-22$ & $-49$ & $-70$ & $-74$ & $-43$ & $0$ & $18$ \\
    Ti & $31$ & $-20$ & $-85$ & $-133$ & $-169$ & $-170$ & $-124$ & $-51$ & $0$ & $9$ \\
    V  & $-4$ & $-85$ & $-176$ & $-223$ & $-230$ & $-190$ & $-118$ & $-45$ & $-3$ & $-2$ \\
    Cr & $-22$ & $-133$ & $-223$ & $-242$ & $-207$ & $-153$ & $-90$ & $-28$ & $0$ & $-3$ \\
    Mn & $-49$ & $-169$ & $-230$ & $-207$ & $-164$ & $-112$ & $-60$ & $-21$ & $-3$ & $-8$ \\
    Fe & $-70$ & $-170$ & $-190$ & $-153$ & $-112$ & $-68$ & $-36$ & $-10$ & $0$ & $-8$ \\
    Co & $-74$ & $-124$ & $-118$ & $-90$ & $-60$ & $-36$ & $-18$ & $-2$ & $0$ & $-12$ \\
    Ni & $-43$ & $-51$ & $-45$ & $-28$ & $-21$ & $-10$ & $-2$ & $3$ & $0$ & $-8$ \\
    Cu & $0$ & $0$ & $-3$ & $0$ & $-3$ & $0$ & $0$ & $0$ & $0$ & $0$ \\
    Zn & $18$ & $9$ & $-2$ & $-3$ & $-8$ & $-8$ & $-12$ & $-8$ & $0$ & $13$ \\
    \hline
    Y  & $80$ & $42$ & $6$ & $-11$ & $-35$ & $-50$ & $-61$ & $-40$ & $0$ & $21$ \\
    Zr & $47$ & $-8$ & $-62$ & $-96$ & $-126$ & $-135$ & $-112$ & $-54$ & $0$ & $15$ \\
    Nb & $3$ & $-73$ & $-148$ & $-189$ & $-200$ & $-174$ & $-118$ & $-49$ & $0$ & $2$ \\
    Mo & $-37$ & $-132$ & $-212$ & $-228$ & $-204$ & $-155$ & $-96$ & $-34$ & $0$ & $-9$ \\
    Tc & $-69$ & $-175$ & $-226$ & $-205$ & $-168$ & $-113$ & $-63$ & $-19$ & $0$ & $-13$ \\
    Ru & $-90$ & $-167$ & $-176$ & $-186$ & $-107$ & $-63$ & $-30$ & $-5$ & $0$ & $-15$ \\
    Rh & $-80$ & $-110$ & $-101$ & $-74$ & $-47$ & $-23$ & $-7$ & $2$ & $0$ & $-16$ \\
    Pd & $-30$ & $-30$ & $-24$ & $-10$ & $-6$ & $2$ & $6$ & $7$ & $0$ & $-9$ \\
    Ag & $15$ & $13$ & $10$ & $10$ & $6$ & $5$ & $3$ & $1$ & $0$ & $2$ \\
    Cd & $29$ & $17$ & $3$ & $3$ & $-4$ & $-6$ & $-10$ & $-8$ & $0$ & $13$ \\
    \hline
    La & $85$ & $40$ & $-12$ & $-47$ & $-85$ & $-108$ & $-101$ & $-53$ & $0$ & $22$ \\
    Hf & $50$ & $1$ & $-45$ & $-72$ & $-100$ & $-116$ & $-105$ & $-55$ & $0$ & $18$ \\
    Ta & $10$ & $-57$ & $-116$ & $-152$ & $-172$ & $-161$ & $-119$ & $-55$ & $0$ & $6$ \\
    W  & $-31$ & $-110$ & $-179$ & $-202$ & $-192$ & $-155$ & $-105$ & $-43$ & $0$ & $-5$ \\
    Re & $-62$ & $-150$ & $-204$ & $-193$ & $-167$ & $-120$ & $-73$ & $-27$ & $0$ & $-10$ \\
    Os & $-86$ & $-160$ & $-178$ & $-157$ & $-121$ & $-77$ & $-44$ & $-13$ & $0$ & $-14$ \\
    Ir & $-88$ & $-125$ & $-120$ & $-95$ & $-64$ & $-38$ & $-18$ & $-3$ & $0$ & $-16$ \\
    Pt & $-52$ & $-56$ & $-47$ & $-29$ & $-21$ & $-8$ & $0$ & $4$ & $0$ & $-10$ \\
    Au & $-6$ & $-5$ & $-7$ & $-1$ & $-2$ & $-1$ & $0$ & $0$ & $0$ & $1$ \\
    Hg & $9$ & $0$ & $-9$ & $-6$ & $-9$ & $-9$ & $-11$ & $-8$ & $0$ & $11$ \\
    \hline\hline
  \end{tabular}
  \label{tab:surf_agg_cu100_ls_3d}
\end{table}

\begin{table}[ht]
  \centering
  \renewcommand{\arraystretch}{1.2}
  \caption{\textbf{Aggregation energies} for \textbf{dual-atom} sites in the \textbf{surface} layer of \textbf{Cu(100)} for all combinations of \textbf{4\textit{d}} transition metals (columns) and all other transition metals (rows), as obtained from initial \textbf{low-spin} configurations. Energies are reported in kJ~mol$^{-1}$.}
  \vspace{8pt}
  \begin{tabular}{@{\hspace{8pt}}*{11}{c@{\hspace{8pt}}}}
    \hline\hline
    TM & Y & Zr & Nb & Mo & Tc & Ru & Rh & Pd & Ag & Cd \\
    \hline
    Sc & $80$ & $47$ & $3$ & $-37$ & $-69$ & $-90$ & $-80$ & $-30$ & $15$ & $29$ \\
    Ti & $42$ & $-8$ & $-73$ & $-132$ & $-175$ & $-167$ & $-110$ & $-30$ & $13$ & $17$ \\
    V  & $6$ & $-62$ & $-148$ & $-212$ & $-226$ & $-176$ & $-101$ & $-24$ & $10$ & $3$ \\
    Cr & $-11$ & $-96$ & $-189$ & $-228$ & $-205$ & $-186$ & $-74$ & $-10$ & $10$ & $3$ \\
    Mn & $-35$ & $-126$ & $-200$ & $-204$ & $-168$ & $-107$ & $-47$ & $-6$ & $6$ & $-4$ \\
    Fe & $-50$ & $-135$ & $-174$ & $-155$ & $-113$ & $-63$ & $-23$ & $2$ & $5$ & $-6$ \\
    Co & $-61$ & $-112$ & $-118$ & $-96$ & $-63$ & $-30$ & $-7$ & $6$ & $3$ & $-10$ \\
    Ni & $-40$ & $-54$ & $-49$ & $-34$ & $-19$ & $-5$ & $2$ & $7$ & $1$ & $-8$ \\
    Cu & $0$ & $0$ & $0$ & $0$ & $0$ & $0$ & $0$ & $0$ & $0$ & $0$ \\
    Zn & $21$ & $15$ & $2$ & $-9$ & $-13$ & $-15$ & $-16$ & $-9$ & $2$ & $13$ \\
    \hline
    Y  & $102$ & $64$ & $17$ & $-24$ & $-53$ & $-71$ & $-68$ & $-28$ & $19$ & $38$ \\
    Zr & $64$ & $7$ & $-56$ & $-107$ & $-143$ & $-144$ & $-104$ & $-31$ & $20$ & $26$ \\
    Nb & $17$ & $-56$ & $-131$ & $-188$ & $-204$ & $-167$ & $-101$ & $-21$ & $19$ & $11$ \\
    Mo & $-24$ & $-107$ & $-188$ & $-222$ & $-202$ & $-147$ & $-74$ & $-6$ & $16$ & $2$ \\
    Tc & $-53$ & $-143$ & $-204$ & $-202$ & $-167$ & $-105$ & $-39$ & $7$ & $14$ & $-3$ \\
    Ru & $-71$ & $-144$ & $-167$ & $-147$ & $-105$ & $-50$ & $-8$ & $16$ & $11$ & $-6$ \\
    Rh & $-68$ & $-104$ & $-101$ & $-74$ & $-39$ & $-8$ & $11$ & $17$ & $8$ & $-10$ \\
    Pd & $-28$ & $-31$ & $-21$ & $-6$ & $7$ & $16$ & $17$ & $15$ & $5$ & $-5$ \\
    Ag & $19$ & $20$ & $19$ & $16$ & $14$ & $11$ & $8$ & $5$ & $5$ & $6$ \\
    Cd & $38$ & $26$ & $11$ & $2$ & $-3$ & $-6$ & $-10$ & $-5$ & $6$ & $18$ \\
    \hline
    La & $110$ & $62$ & $-3$ & $-59$ & $-104$ & $-122$ & $-102$ & $-36$ & $22$ & $43$ \\
    Hf & $64$ & $16$ & $-42$ & $-89$ & $-124$ & $-134$ & $-107$ & $-37$ & $19$ & $26$ \\
    Ta & $22$ & $-44$ & $-111$ & $-162$ & $-188$ & $-166$ & $-109$ & $-30$ & $17$ & $12$ \\
    W  & $-20$ & $-94$ & $-166$ & $-208$ & $-200$ & $-155$ & $-89$ & $-15$ & $16$ & $3$ \\
    Re & $-49$ & $-129$ & $-191$ & $-199$ & $-172$ & $-118$ & $-55$ & $0$ & $16$ & $0$ \\
    Os & $-72$ & $-144$ & $-175$ & $-160$ & $-123$ & $-70$ & $-22$ & $12$ & $14$ & $-4$ \\
    Ir & $-78$ & $-120$ & $-121$ & $-97$ & $-61$ & $-24$ & $3$ & $17$ & $10$ & $-8$ \\
    Pt & $-48$ & $-57$ & $-47$ & $-28$ & $-9$ & $8$ & $14$ & $16$ & $7$ & $-6$ \\
    Au & $-4$ & $-2$ & $1$ & $3$ & $7$ & $8$ & $8$ & $8$ & $6$ & $4$ \\
    Hg & $12$ & $2$ & $-4$ & $-7$ & $-7$ & $-7$ & $-8$ & $-3$ & $6$ & $15$ \\
    \hline\hline
  \end{tabular}
  \label{tab:surf_agg_cu100_ls_4d}
\end{table}

\begin{table}[ht]
  \centering
  \renewcommand{\arraystretch}{1.2}
  \caption{\textbf{Aggregation energies} for \textbf{dual-atom} sites in the \textbf{surface} layer of \textbf{Cu(100)} for all combinations of \textbf{5\textit{d}} transition metals (columns) and all other transition metals (rows), as obtained from initial \textbf{low-spin} configurations. Energies are reported in kJ~mol$^{-1}$.}
  \vspace{8pt}
  \begin{tabular}{@{\hspace{8pt}}*{11}{c@{\hspace{8pt}}}}
    \hline\hline
    TM & La & Hf & Ta & W & Re & Os & Ir & Pt & Au & Hg \\
    \hline
    Sc & $85$ & $50$ & $10$ & $-31$ & $-62$ & $-86$ & $-88$ & $-52$ & $-6$ & $9$ \\
    Ti & $40$ & $1$ & $-57$ & $-110$ & $-150$ & $-160$ & $-125$ & $-56$ & $-5$ & $0$ \\
    V  & $-12$ & $-45$ & $-116$ & $-179$ & $-204$ & $-178$ & $-120$ & $-47$ & $-7$ & $-9$ \\
    Cr & $-47$ & $-72$ & $-152$ & $-202$ & $-193$ & $-157$ & $-95$ & $-29$ & $-1$ & $-6$ \\
    Mn & $-85$ & $-100$ & $-172$ & $-192$ & $-167$ & $-121$ & $-64$ & $-21$ & $-2$ & $-9$ \\
    Fe & $-108$ & $-116$ & $-161$ & $-155$ & $-120$ & $-77$ & $-38$ & $-8$ & $-1$ & $-9$ \\
    Co & $-101$ & $-105$ & $-119$ & $-105$ & $-73$ & $-44$ & $-18$ & $0$ & $0$ & $-11$ \\
    Ni & $-53$ & $-55$ & $-55$ & $-43$ & $-27$ & $-13$ & $-3$ & $4$ & $0$ & $-8$ \\
    Cu & $0$ & $0$ & $0$ & $0$ & $0$ & $0$ & $0$ & $0$ & $0$ & $0$ \\
    Zn & $22$ & $18$ & $6$ & $-5$ & $-10$ & $-14$ & $-16$ & $-10$ & $1$ & $11$ \\
    \hline
    Y  & $110$ & $64$ & $22$ & $-20$ & $-49$ & $-72$ & $-78$ & $-48$ & $-4$ & $12$ \\
    Zr & $62$ & $16$ & $-44$ & $-94$ & $-129$ & $-144$ & $-120$ & $-57$ & $-2$ & $2$ \\
    Nb & $-3$ & $-42$ & $-111$ & $-166$ & $-191$ & $-175$ & $-121$ & $-47$ & $1$ & $-4$ \\
    Mo & $-59$ & $-89$ & $-162$ & $-208$ & $-199$ & $-160$ & $-97$ & $-28$ & $3$ & $-7$ \\
    Tc & $-104$ & $-124$ & $-188$ & $-200$ & $-172$ & $-123$ & $-61$ & $-9$ & $7$ & $-7$ \\
    Ru & $-122$ & $-134$ & $-166$ & $-155$ & $-118$ & $-70$ & $-24$ & $8$ & $8$ & $-7$ \\
    Rh & $-102$ & $-107$ & $-109$ & $-89$ & $-55$ & $-22$ & $3$ & $14$ & $8$ & $-8$ \\
    Pd & $-36$ & $-37$ & $-30$ & $-15$ & $0$ & $12$ & $17$ & $16$ & $8$ & $-3$ \\
    Ag & $22$ & $19$ & $17$ & $16$ & $16$ & $14$ & $10$ & $7$ & $6$ & $6$ \\
    Cd & $43$ & $26$ & $12$ & $3$ & $0$ & $-4$ & $-8$ & $-6$ & $4$ & $15$ \\
    \hline
    La & $117$ & $63$ & $11$ & $-41$ & $-80$ & $-106$ & $-103$ & $-59$ & $-3$ & $21$ \\
    Hf & $63$ & $24$ & $-32$ & $-81$ & $-115$ & $-135$ & $-120$ & $-63$ & $-5$ & $1$ \\
    Ta & $11$ & $-32$ & $-96$ & $-149$ & $-179$ & $-173$ & $-128$ & $-57$ & $-2$ & $-5$ \\
    W  & $-41$ & $-81$ & $-149$ & $-197$ & $-199$ & $-168$ & $-112$ & $-39$ & $1$ & $-6$ \\
    Re & $-80$ & $-115$ & $-179$ & $-199$ & $-177$ & $-136$ & $-77$ & $-18$ & $6$ & $-4$ \\
    Os & $-106$ & $-135$ & $-173$ & $-168$ & $-136$ & $-91$ & $-41$ & $1$ & $10$ & $-4$ \\
    Ir & $-103$ & $-120$ & $-128$ & $-112$ & $-77$ & $-41$ & $-8$ & $12$ & $11$ & $-5$ \\
    Pt & $-59$ & $-63$ & $-57$ & $-39$ & $-18$ & $1$ & $12$ & $17$ & $11$ & $-1$ \\
    Au & $-3$ & $-5$ & $-2$ & $1$ & $6$ & $10$ & $11$ & $11$ & $10$ & $8$ \\
    Hg & $21$ & $1$ & $-5$ & $-6$ & $-4$ & $-4$ & $-5$ & $-1$ & $8$ & $16$ \\
    \hline\hline
  \end{tabular}
  \label{tab:surf_agg_cu100_ls_5d}
\end{table}

\clearpage


\begin{figure}[htbp]
    \includegraphics[width=12cm,height=\textheight,keepaspectratio]{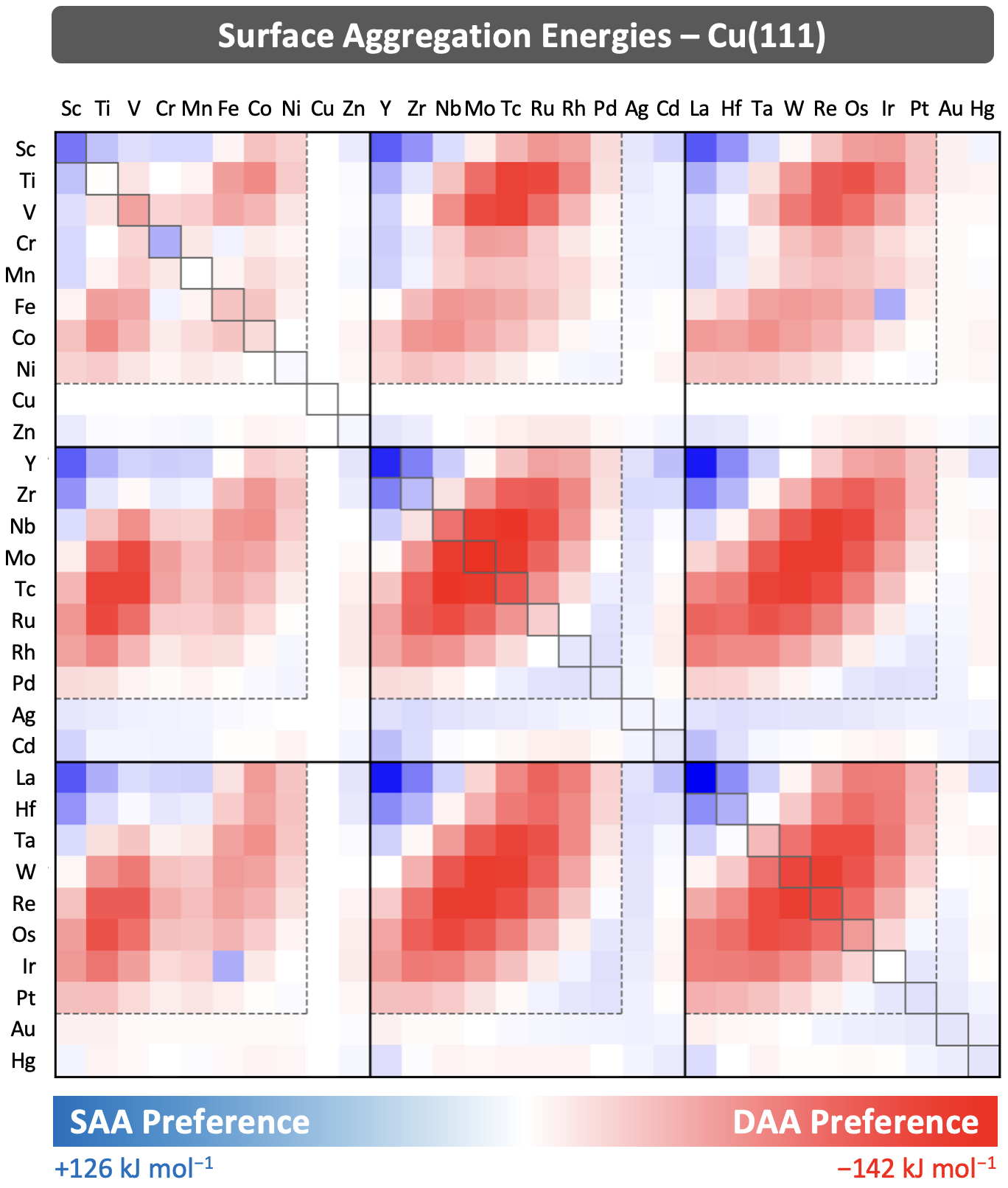}
    \caption{
        Aggregation energies for the formation of all possible transition metal dopant dimers in the Cu(111) surface, analogous to Fig.~2 in the main text.
        Preferences for single-atom sites are represented in blue, while preferences for dopant dimers are shown in red.  
        }
    \label{fig:Cu111-surf-aggr}
\end{figure}

\begin{figure}[htbp]
    \includegraphics[width=9.5cm,height=\textheight,keepaspectratio]{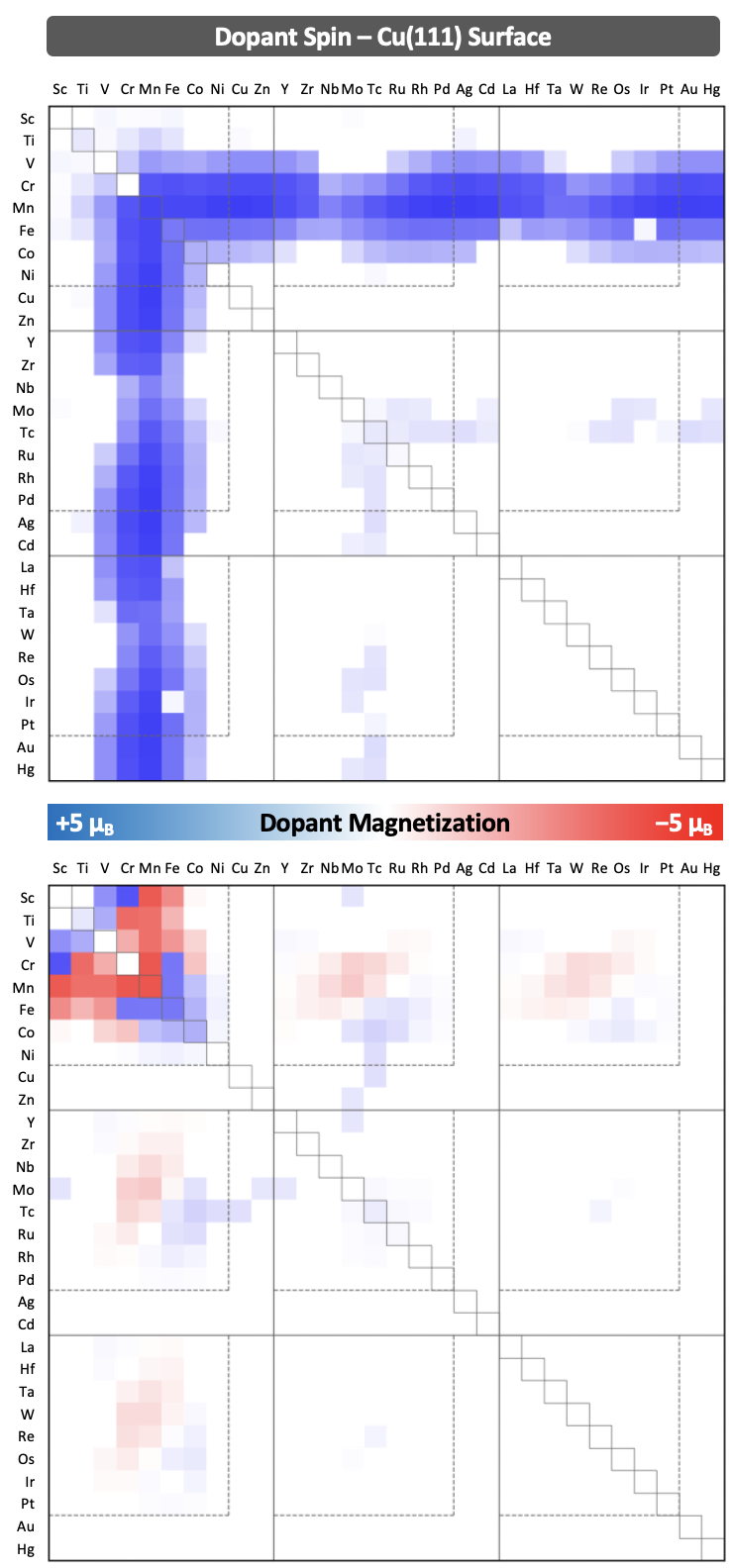}
    \caption{
        Local dopant magnetizations in dual-atom sites located in the surface layer of Cu(111).
        Top: The local magnetization of the dopant with the larger magnetic moment is shown in blue.
        Bottom: The local magnetization of the second dopant is shown in blue when its magnetic moment is aligned parallel to that of the first dopant (ferromagnetic coupling) and in red when it is aligned antiparallel (antiferromagnetic coupling).    }
    \label{fig:Cu111-surf-spin}
\end{figure}

\begin{table}[ht]
  \centering
  \renewcommand{\arraystretch}{1.2}
  \caption{\textbf{Aggregation energies} for \textbf{dual-atom} sites in the \textbf{surface} layer of \textbf{Cu(111)} for all combinations of \textbf{3\textit{d}} transition metals (columns) and all other transition metals (rows), as obtained from initial \textbf{high-spin} configurations. Energies are reported in kJ~mol$^{-1}$.}
  \vspace{8pt}
  \begin{tabular}{@{\hspace{8pt}}*{11}{c@{\hspace{8pt}}}}
    \hline\hline
    TM & Sc & Ti & V & Cr & Mn & Fe & Co & Ni & Cu & Zn \\
    \hline
    Sc & $67$ & $31$ & $16$ & $19$ & $19$ & $-7$ & $-36$ & $-26$ & $0$ & $10$ \\
    Ti & $31$ & $-1$ & $-16$ & $0$ & $-6$ & $-56$ & $-69$ & $-31$ & $0$ & $2$ \\
    V  & $16$ & $-16$ & $-55$ & $-24$ & $-29$ & $-51$ & $-41$ & $-15$ & $0$ & $2$ \\
    Cr & $19$ & $0$ & $-24$ & $40$ & $-14$ & $6$ & $-11$ & $-6$ & $0$ & $3$ \\
    Mn & $19$ & $-6$ & $-29$ & $-14$ & $0$ & $-6$ & $-21$ & $-13$ & $0$ & $4$ \\
    Fe & $-7$ & $-56$ & $-51$ & $6$ & $-6$ & $-34$ & $-34$ & $-8$ & $0$ & $-2$ \\
    Co & $-36$ & $-69$ & $-41$ & $-11$ & $-21$ & $-34$ & $-21$ & $-1$ & $0$ & $-7$ \\
    Ni & $-26$ & $-31$ & $-15$ & $-6$ & $-13$ & $-8$ & $-1$ & $4$ & $0$ & $-6$ \\
    Cu & $0$ & $0$ & $0$ & $0$ & $0$ & $0$ & $0$ & $0$ & $0$ & $0$ \\
    Zn & $10$ & $2$ & $2$ & $3$ & $4$ & $-2$ & $-7$ & $-6$ & $0$ & $4$ \\
    \hline
    Y  & $80$ & $39$ & $22$ & $24$ & $24$ & $-2$ & $-30$ & $-25$ & $0$ & $13$ \\
    Zr & $54$ & $12$ & $-4$ & $10$ & $6$ & $-40$ & $-62$ & $-35$ & $0$ & $9$ \\
    Nb & $17$ & $-36$ & $-66$ & $-28$ & $-27$ & $-61$ & $-65$ & $-29$ & $0$ & $0$ \\
    Mo & $-10$ & $-86$ & $-111$ & $-57$ & $-38$ & $-57$ & $-51$ & $-20$ & $0$ & $-4$ \\
    Tc & $-42$ & $-119$ & $-119$ & $-54$ & $-35$ & $-49$ & $-38$ & $-11$ & $0$ & $-10$ \\
    Ru & $-62$ & $-113$ & $-86$ & $-31$ & $-29$ & $-37$ & $-22$ & $-1$ & $0$ & $-13$ \\
    Rh & $-54$ & $-72$ & $-42$ & $-13$ & $-22$ & $-20$ & $-5$ & $5$ & $0$ & $-13$ \\
    Pd & $-21$ & $-19$ & $-7$ & $-3$ & $-8$ & $-1$ & $3$ & $6$ & $0$ & $-5$ \\
    Ag & $12$ & $10$ & $8$ & $6$ & $6$ & $3$ & $1$ & $0$ & $0$ & $2$ \\
    Cd & $21$ & $5$ & $5$ & $7$ & $7$ & $-1$ & $-1$ & $-6$ & $0$ & $6$ \\
    \hline
    La & $83$ & $40$ & $17$ & $22$ & $23$ & $-17$ & $-59$ & $-34$ & $0$ & $14$ \\
    Hf & $52$ & $16$ & $3$ & $13$ & $10$ & $-30$ & $-55$ & $-35$ & $0$ & $11$ \\
    Ta & $18$ & $-18$ & $-34$ & $-9$ & $-13$ & $-55$ & $-65$ & $-33$ & $0$ & $3$ \\
    W  & $-5$ & $-60$ & $-80$ & $-37$ & $-32$ & $-59$ & $-55$ & $-26$ & $0$ & $0$ \\
    Re & $-36$ & $-99$ & $-99$ & $-48$ & $-38$ & $-55$ & $-45$ & $-19$ & $0$ & $-7$ \\
    Os & $-58$ & $-105$ & $-85$ & $-37$ & $-34$ & $-43$ & $-29$ & $-7$ & $0$ & $-11$ \\
    Ir & $-60$ & $-82$ & $-55$ & $-22$ & $-28$ & $41$ & $-13$ & $0$ & $0$ & $-11$ \\
    Pt & $-37$ & $-38$ & $-21$ & $-10$ & $-15$ & $-9$ & $-2$ & $3$ & $0$ & $-5$ \\
    Au & $-8$ & $-8$ & $-4$ & $-3$ & $-2$ & $-2$ & $-2$ & $-2$ & $0$ & $2$ \\
    Hg & $5$ & $-6$ & $-4$ & $0$ & $1$ & $-3$ & $-6$ & $-4$ & $0$ & $4$ \\
    \hline\hline
  \end{tabular}
  \label{tab:surf_agg_cu111_3d}
\end{table}

\begin{table}[ht]
  \centering
  \renewcommand{\arraystretch}{1.2}
  \caption{\textbf{Aggregation energies} for \textbf{dual-atom} sites in the \textbf{surface} layer of \textbf{Cu(111)} for all combinations of \textbf{4\textit{d}} transition metals (columns) and all other transition metals (rows), as obtained from initial \textbf{high-spin} configurations. Energies are reported in kJ~mol$^{-1}$.}
  \vspace{8pt}
  \begin{tabular}{@{\hspace{8pt}}*{11}{c@{\hspace{8pt}}}}
    \hline\hline
    TM & Y & Zr & Nb & Mo & Tc & Ru & Rh & Pd & Ag & Cd \\
    \hline
    Sc & $80$ & $54$ & $17$ & $-10$ & $-42$ & $-62$ & $-54$ & $-21$ & $12$ & $21$ \\
    Ti & $39$ & $12$ & $-36$ & $-86$ & $-119$ & $-113$ & $-72$ & $-19$ & $10$ & $5$ \\
    V  & $22$ & $-4$ & $-66$ & $-111$ & $-119$ & $-86$ & $-42$ & $-7$ & $8$ & $5$ \\
    Cr & $24$ & $10$ & $-28$ & $-57$ & $-54$ & $-31$ & $-13$ & $-3$ & $6$ & $7$ \\
    Mn & $24$ & $6$ & $-27$ & $-38$ & $-35$ & $-29$ & $-22$ & $-8$ & $6$ & $7$ \\
    Fe & $-2$ & $-40$ & $-61$ & $-57$ & $-49$ & $-37$ & $-20$ & $-1$ & $3$ & $-1$ \\
    Co & $-30$ & $-62$ & $-65$ & $-51$ & $-38$ & $-22$ & $-5$ & $3$ & $1$ & $-1$ \\
    Ni & $-25$ & $-35$ & $-29$ & $-20$ & $-11$ & $-1$ & $5$ & $6$ & $0$ & $-6$ \\
    Cu & $0$ & $0$ & $0$ & $0$ & $0$ & $0$ & $0$ & $0$ & $0$ & $0$ \\
    Zn & $13$ & $9$ & $0$ & $-4$ & $-10$ & $-13$ & $-13$ & $-5$ & $2$ & $6$ \\
    \hline
    Y  & $108$ & $63$ & $24$ & $-3$ & $-34$ & $-53$ & $-50$ & $-20$ & $15$ & $32$ \\
    Zr & $63$ & $33$ & $-17$ & $-64$ & $-97$ & $-99$ & $-69$ & $-19$ & $19$ & $17$ \\
    Nb & $24$ & $-17$ & $-83$ & $-125$ & $-136$ & $-110$ & $-62$ & $-10$ & $14$ & $2$ \\
    Mo & $-3$ & $-64$ & $-125$ & $-142$ & $-128$ & $-91$ & $-42$ & $0$ & $12$ & $-1$ \\
    Tc & $-34$ & $-97$ & $-136$ & $-128$ & $-104$ & $-63$ & $-20$ & $8$ & $10$ & $-5$ \\
    Ru & $-53$ & $-99$ & $-110$ & $-91$ & $-63$ & $-29$ & $-1$ & $14$ & $8$ & $-9$ \\
    Rh & $-50$ & $-69$ & $-62$ & $-42$ & $-20$ & $-1$ & $12$ & $15$ & $6$ & $-10$ \\
    Pd & $-20$ & $-19$ & $-10$ & $0$ & $8$ & $14$ & $15$ & $12$ & $4$ & $-3$ \\
    Ag & $15$ & $19$ & $14$ & $12$ & $10$ & $8$ & $6$ & $4$ & $5$ & $6$ \\
    Cd & $32$ & $17$ & $2$ & $-1$ & $-5$ & $-9$ & $-10$ & $-3$ & $6$ & $11$ \\
    \hline
    La & $114$ & $64$ & $21$ & $-25$ & $-70$ & $-92$ & $-76$ & $-28$ & $15$ & $32$ \\
    Hf & $58$ & $36$ & $-6$ & $-46$ & $-79$ & $-88$ & $-69$ & $-24$ & $16$ & $16$ \\
    Ta & $23$ & $-6$ & $-58$ & $-100$ & $-118$ & $-105$ & $-68$ & $-15$ & $14$ & $5$ \\
    W  & $0$ & $-46$ & $-103$ & $-127$ & $-122$ & $-96$ & $-52$ & $-6$ & $13$ & $3$ \\
    Re & $-31$ & $-85$ & $-123$ & $-124$ & $-109$ & $-75$ & $-34$ & $3$ & $13$ & $-1$ \\
    Os & $-53$ & $-96$ & $-113$ & $-99$ & $-77$ & $-44$ & $-11$ & $12$ & $12$ & $-5$ \\
    Ir & $-57$ & $-79$ & $-76$ & $-59$ & $-37$ & $-13$ & $6$ & $15$ & $10$ & $-7$ \\
    Pt & $-37$ & $-38$ & $-30$ & $-17$ & $-4$ & $7$ & $12$ & $14$ & $8$ & $-1$ \\
    Au & $-9$ & $-2$ & $-2$ & $1$ & $3$ & $5$ & $5$ & $6$ & $8$ & $5$ \\
    Hg & $16$ & $2$ & $-7$ & $-5$ & $-6$ & $-8$ & $-6$ & $1$ & $6$ & $10$ \\
    \hline\hline
  \end{tabular}
  \label{tab:surf_agg_cu111_4d}
\end{table}

\begin{table}[ht]
  \centering
  \renewcommand{\arraystretch}{1.2}
  \caption{\textbf{Aggregation energies} for \textbf{dual-atom} sites in the \textbf{surface} layer of \textbf{Cu(111)} for all combinations of \textbf{5\textit{d}} transition metals (columns) and all other transition metals (rows), as obtained from initial \textbf{high-spin} configurations. Energies are reported in kJ~mol$^{-1}$.}
  \vspace{8pt}
  \begin{tabular}{@{\hspace{8pt}}*{11}{c@{\hspace{8pt}}}}
    \hline\hline
    TM & La & Hf & Ta & W & Re & Os & Ir & Pt & Au & Hg \\
    \hline
    Sc & $83$ & $52$ & $18$ & $-5$ & $-36$ & $-58$ & $-60$ & $-37$ & $-8$ & $5$ \\
    Ti & $40$ & $16$ & $-18$ & $-60$ & $-99$ & $-105$ & $-82$ & $-38$ & $-8$ & $-6$ \\
    V  & $17$ & $3$ & $-34$ & $-80$ & $-99$ & $-85$ & $-55$ & $-21$ & $-4$ & $-4$ \\
    Cr & $22$ & $13$ & $-9$ & $-37$ & $-48$ & $-37$ & $-22$ & $-10$ & $-3$ & $0$ \\
    Mn & $23$ & $10$ & $-13$ & $-32$ & $-38$ & $-34$ & $-28$ & $-15$ & $-2$ & $1$ \\
    Fe & $-17$ & $-30$ & $-55$ & $-59$ & $-55$ & $-43$ & $41$ & $-9$ & $-2$ & $-3$ \\
    Co & $-59$ & $-55$ & $-65$ & $-55$ & $-45$ & $-29$ & $-13$ & $-2$ & $-2$ & $-6$ \\
    Ni & $-34$ & $-35$ & $-33$ & $-26$ & $-19$ & $-7$ & $0$ & $3$ & $-2$ & $-4$ \\
    Cu & $0$ & $0$ & $0$ & $0$ & $0$ & $0$ & $0$ & $0$ & $0$ & $0$ \\
    Zn & $14$ & $11$ & $3$ & $0$ & $-7$ & $-11$ & $-11$ & $-5$ & $2$ & $4$ \\
    \hline
    Y  & $114$ & $58$ & $23$ & $0$ & $-31$ & $-53$ & $-57$ & $-37$ & $-9$ & $16$ \\
    Zr & $64$ & $36$ & $-6$ & $-46$ & $-85$ & $-96$ & $-79$ & $-38$ & $-2$ & $2$ \\
    Nb & $21$ & $-6$ & $-58$ & $-103$ & $-123$ & $-113$ & $-76$ & $-30$ & $-2$ & $-7$ \\
    Mo & $-25$ & $-46$ & $-100$ & $-127$ & $-124$ & $-99$ & $-59$ & $-17$ & $1$ & $-5$ \\
    Tc & $-70$ & $-79$ & $-118$ & $-122$ & $-109$ & $-77$ & $-37$ & $-4$ & $3$ & $-6$ \\
    Ru & $-92$ & $-88$ & $-105$ & $-96$ & $-75$ & $-44$ & $-13$ & $7$ & $5$ & $-8$ \\
    Rh & $-76$ & $-69$ & $-68$ & $-52$ & $-34$ & $-11$ & $6$ & $12$ & $5$ & $-6$ \\
    Pd & $-28$ & $-24$ & $-15$ & $-6$ & $3$ & $12$ & $15$ & $14$ & $6$ & $1$ \\
    Ag & $15$ & $16$ & $14$ & $13$ & $13$ & $12$ & $10$ & $8$ & $8$ & $6$ \\
    Cd & $32$ & $16$ & $5$ & $3$ & $-1$ & $-5$ & $-7$ & $-1$ & $5$ & $10$ \\
    \hline
    La & $126$ & $56$ & $23$ & $-7$ & $-50$ & $-77$ & $-76$ & $-46$ & $-10$ & $18$ \\
    Hf & $56$ & $39$ & $2$ & $-32$ & $-70$ & $-87$ & $-77$ & $-41$ & $-5$ & $0$ \\
    Ta & $23$ & $2$ & $-41$ & $-84$ & $-111$ & $-109$ & $-81$ & $-36$ & $-4$ & $-6$ \\
    W  & $-7$ & $-32$ & $-84$ & $-115$ & $-122$ & $-104$ & $-68$ & $-24$ & $1$ & $-2$ \\
    Re & $-50$ & $-70$ & $-111$ & $-122$ & $-113$ & $-87$ & $-49$ & $-11$ & $5$ & $-2$ \\
    Os & $-77$ & $-87$ & $-109$ & $-104$ & $-87$ & $-58$ & $-24$ & $4$ & $8$ & $-3$ \\
    Ir & $-76$ & $-77$ & $-81$ & $-68$ & $-49$ & $-24$ & $-1$ & $12$ & $9$ & $-2$ \\
    Pt & $-46$ & $-41$ & $-36$ & $-24$ & $-11$ & $4$ & $12$ & $16$ & $11$ & $5$ \\
    Au & $-10$ & $-5$ & $-4$ & $1$ & $5$ & $8$ & $9$ & $11$ & $14$ & $9$ \\
    Hg & $18$ & $0$ & $-6$ & $-2$ & $-2$ & $-3$ & $-2$ & $5$ & $9$ & $13$ \\
    \hline\hline
  \end{tabular}
  \label{tab:surf_agg_cu111_5d}
\end{table}

\begin{table}[ht]
  \centering
  \renewcommand{\arraystretch}{1.2}
    \caption{\textbf{Aggregation energies} for \textbf{dual-atom} sites in the \textbf{surface} layer of \textbf{Cu(111)} for all combinations of \textbf{3\textit{d}} transition metals (columns) and all other transition metals (rows), as obtained from initial \textbf{low-spin} configurations. Energies are reported in kJ~mol$^{-1}$.}

  \vspace{8pt}
  \begin{tabular}{@{\hspace{8pt}}*{11}{c@{\hspace{8pt}}}}
    \hline\hline
    TM & Sc & Ti & V & Cr & Mn & Fe & Co & Ni & Cu & Zn \\
    \hline
    Sc & $67$ & $31$ & $7$ & $-9$ & $-28$ & $-47$ & $-46$ & $-26$ & $0$ & $10$ \\
    Ti & $31$ & $-1$ & $-44$ & $-93$ & $-118$ & $-118$ & $-78$ & $-31$ & $0$ & $2$ \\
    V  & $7$ & $-44$ & $-120$ & $-163$ & $-164$ & $-131$ & $-71$ & $-24$ & $0$ & $-4$ \\
    Cr & $-9$ & $-93$ & $-163$ & $-182$ & $-150$ & $-107$ & $-58$ & $-16$ & $-1$ & $-5$ \\
    Mn & $-28$ & $-118$ & $-164$ & $-150$ & $-116$ & $-77$ & $-37$ & $-10$ & $0$ & $-7$ \\
    Fe & $-47$ & $-118$ & $-131$ & $-107$ & $-77$ & $-48$ & $-21$ & $-3$ & $0$ & $-10$ \\
    Co & $-46$ & $-78$ & $-71$ & $-58$ & $-37$ & $-21$ & $-5$ & $3$ & $0$ & $-10$ \\
    Ni & $-26$ & $-31$ & $-24$ & $-16$ & $-10$ & $-3$ & $3$ & $4$ & $0$ & $-6$ \\
    Cu & $0$ & $0$ & $0$ & $-1$ & $0$ & $0$ & $0$ & $0$ & $0$ & $0$ \\
    Zn & $10$ & $2$ & $-4$ & $-5$ & $-7$ & $-10$ & $-10$ & $-6$ & $0$ & $4$ \\
    \hline
    Y  & $80$ & $39$ & $14$ & $-3$ & $-21$ & $-36$ & $-39$ & $-25$ & $0$ & $13$ \\
    Zr & $54$ & $12$ & $-28$ & $-65$ & $-88$ & $-94$ & $-71$ & $-35$ & $0$ & $9$ \\
    Nb & $17$ & $-36$ & $-98$ & $-135$ & $-139$ & $-118$ & $-74$ & $-29$ & $0$ & $0$ \\
    Mo & $-10$ & $-86$ & $-144$ & $-157$ & $-135$ & $-102$ & $-58$ & $-20$ & $0$ & $-4$ \\
    Tc & $-42$ & $-120$ & $-152$ & $-140$ & $-111$ & $-75$ & $-39$ & $-11$ & $0$ & $-10$ \\
    Ru & $-62$ & $-113$ & $-116$ & $-97$ & $-70$ & $-40$ & $-17$ & $-1$ & $0$ & $-13$ \\
    Rh & $-54$ & $-72$ & $-60$ & $-45$ & $-26$ & $-12$ & $0$ & $5$ & $0$ & $-13$ \\
    Pd & $-21$ & $-19$ & $-10$ & $-5$ & $1$ & $4$ & $7$ & $6$ & $0$ & $-5$ \\
    Ag & $12$ & $10$ & $8$ & $6$ & $4$ & $3$ & $1$ & $0$ & $0$ & $2$ \\
    Cd & $21$ & $5$ & $-2$ & $-3$ & $-6$ & $-9$ & $-11$ & $-6$ & $0$ & $6$ \\
    \hline
    La & $83$ & $40$ & $13$ & $-20$ & $-56$ & $-80$ & $-68$ & $-34$ & $0$ & $14$ \\
    Hf & $52$ & $16$ & $-14$ & $-43$ & $-66$ & $-77$ & $-65$ & $-35$ & $0$ & $11$ \\
    Ta & $18$ & $-18$ & $-66$ & $-101$ & $-112$ & $-105$ & $-74$ & $-33$ & $0$ & $3$ \\
    W  & $-5$ & $-60$ & $-112$ & $-131$ & $-123$ & $-100$ & $-64$ & $-26$ & $0$ & $0$ \\
    Re & $-36$ & $-99$ & $-131$ & $-130$ & $-112$ & $-84$ & $-50$ & $-19$ & $0$ & $-7$ \\
    Os & $-58$ & $-105$ & $-115$ & $-104$ & $-82$ & $-54$ & $-28$ & $-7$ & $0$ & $-11$ \\
    Ir & $-60$ & $-82$ & $-75$ & $-62$ & $-43$ & $-25$ & $-9$ & $0$ & $0$ & $-11$ \\
    Pt & $-37$ & $-38$ & $-30$ & $-21$ & $-12$ & $-5$ & $1$ & $3$ & $0$ & $-5$ \\
    Au & $-8$ & $-8$ & $-7$ & $-5$ & $-4$ & $-3$ & $-3$ & $-2$ & $0$ & $2$ \\
    Hg & $5$ & $-6$ & $-9$ & $-8$ & $-9$ & $-11$ & $-11$ & $-4$ & $0$ & $4$ \\
    \hline\hline
  \end{tabular}
  \label{tab:surf_agg_cu111_ls_3d}
\end{table}

\begin{table}[ht]
  \centering
  \renewcommand{\arraystretch}{1.2}
  \caption{\textbf{Aggregation energies} for \textbf{dual-atom} sites in the \textbf{surface} layer of \textbf{Cu(111)} for all combinations of \textbf{4\textit{d}} transition metals (columns) and all other transition metals (rows), as obtained from initial \textbf{low-spin} configurations. Energies are reported in kJ~mol$^{-1}$.}
  \vspace{8pt}
  \begin{tabular}{@{\hspace{8pt}}*{11}{c@{\hspace{8pt}}}}
    \hline\hline
    TM & Y & Zr & Nb & Mo & Tc & Ru & Rh & Pd & Ag & Cd \\
    \hline
    Sc & $80$ & $54$ & $17$ & $-10$ & $-42$ & $-62$ & $-54$ & $-21$ & $12$ & $21$ \\
    Ti & $39$ & $12$ & $-36$ & $-86$ & $-120$ & $-113$ & $-72$ & $-19$ & $10$ & $5$ \\
    V  & $14$ & $-28$ & $-98$ & $-144$ & $-152$ & $-116$ & $-60$ & $-10$ & $8$ & $-2$ \\
    Cr & $-3$ & $-65$ & $-135$ & $-157$ & $-140$ & $-97$ & $-45$ & $-5$ & $6$ & $-3$ \\
    Mn & $-21$ & $-88$ & $-139$ & $-135$ & $-111$ & $-70$ & $-26$ & $1$ & $4$ & $-6$ \\
    Fe & $-36$ & $-94$ & $-118$ & $-102$ & $-75$ & $-40$ & $-12$ & $4$ & $3$ & $-9$ \\
    Co & $-39$ & $-71$ & $-74$ & $-58$ & $-39$ & $-17$ & $0$ & $7$ & $1$ & $-11$ \\
    Ni & $-25$ & $-35$ & $-29$ & $-20$ & $-11$ & $-1$ & $5$ & $6$ & $0$ & $-6$ \\
    Cu & $0$ & $0$ & $0$ & $0$ & $0$ & $0$ & $0$ & $0$ & $0$ & $0$ \\
    Zn & $13$ & $9$ & $0$ & $-4$ & $-10$ & $-13$ & $-13$ & $-5$ & $2$ & $6$ \\
    \hline
    Y  & $108$ & $63$ & $24$ & $-3$ & $-34$ & $-53$ & $-50$ & $-20$ & $15$ & $32$ \\
    Zr & $63$ & $33$ & $-17$ & $-64$ & $-98$ & $-99$ & $-69$ & $-19$ & $19$ & $17$ \\
    Nb & $24$ & $-17$ & $-83$ & $-125$ & $-136$ & $-110$ & $-62$ & $-10$ & $14$ & $2$ \\
    Mo & $-3$ & $-64$ & $-125$ & $-142$ & $-128$ & $-91$ & $-42$ & $0$ & $12$ & $-1$ \\
    Tc & $-34$ & $-98$ & $-136$ & $-128$ & $-103$ & $-63$ & $-21$ & $8$ & $11$ & $-6$ \\
    Ru & $-53$ & $-99$ & $-110$ & $-91$ & $-63$ & $-29$ & $-1$ & $14$ & $8$ & $-9$ \\
    Rh & $-50$ & $-69$ & $-62$ & $-42$ & $-21$ & $-1$ & $12$ & $15$ & $6$ & $-10$ \\
    Pd & $-20$ & $-19$ & $-10$ & $0$ & $8$ & $14$ & $15$ & $12$ & $4$ & $-3$ \\
    Ag & $15$ & $19$ & $14$ & $12$ & $11$ & $8$ & $6$ & $4$ & $5$ & $6$ \\
    Cd & $32$ & $17$ & $2$ & $-1$ & $-6$ & $-9$ & $-10$ & $-3$ & $6$ & $11$ \\
    \hline
    La & $114$ & $64$ & $21$ & $-25$ & $-71$ & $-92$ & $-76$ & $-28$ & $15$ & $32$ \\
    Hf & $58$ & $36$ & $-6$ & $-46$ & $-79$ & $-88$ & $-69$ & $-24$ & $16$ & $16$ \\
    Ta & $23$ & $-6$ & $-58$ & $-100$ & $-118$ & $-105$ & $-68$ & $-15$ & $14$ & $5$ \\
    W  & $0$ & $-46$ & $-103$ & $-127$ & $-123$ & $-96$ & $-52$ & $-6$ & $13$ & $3$ \\
    Re & $-31$ & $-85$ & $-123$ & $-124$ & $-108$ & $-75$ & $-34$ & $3$ & $13$ & $-1$ \\
    Os & $-53$ & $-96$ & $-113$ & $-98$ & $-76$ & $-44$ & $-11$ & $12$ & $12$ & $-5$ \\
    Ir & $-57$ & $-79$ & $-76$ & $-59$ & $-37$ & $-13$ & $6$ & $15$ & $10$ & $-7$ \\
    Pt & $-37$ & $-38$ & $-30$ & $-17$ & $-5$ & $7$ & $12$ & $14$ & $8$ & $-1$ \\
    Au & $-9$ & $-3$ & $-2$ & $1$ & $4$ & $5$ & $5$ & $6$ & $8$ & $5$ \\
    Hg & $16$ & $2$ & $-7$ & $-5$ & $-6$ & $-8$ & $-6$ & $1$ & $6$ & $10$ \\
    \hline\hline
  \end{tabular}
  \label{tab:surf_agg_cu111_ls_4d}
\end{table}

\begin{table}[ht]
  \centering
  \renewcommand{\arraystretch}{1.2}
  \caption{\textbf{Aggregation energies} for \textbf{dual-atom} sites in the \textbf{surface} layer of \textbf{Cu(111)} for all combinations of \textbf{5\textit{d}} transition metals (columns) and all other transition metals (rows), as obtained from initial \textbf{low-spin} configurations. Energies are reported in kJ~mol$^{-1}$.}

  \vspace{8pt}
  \begin{tabular}{@{\hspace{8pt}}*{11}{c@{\hspace{8pt}}}}
    \hline\hline
    TM & La & Hf & Ta & W & Re & Os & Ir & Pt & Au & Hg \\
    \hline
    Sc & $83$ & $52$ & $18$ & $-5$ & $-36$ & $-58$ & $-60$ & $-37$ & $-8$ & $5$ \\
    Ti & $40$ & $16$ & $-18$ & $-60$ & $-99$ & $-105$ & $-82$ & $-38$ & $-8$ & $-6$ \\
    V  & $13$ & $-14$ & $-66$ & $-112$ & $-131$ & $-115$ & $-75$ & $-30$ & $-7$ & $-9$ \\
    Cr & $-20$ & $-43$ & $-101$ & $-131$ & $-130$ & $-104$ & $-62$ & $-21$ & $-5$ & $-8$ \\
    Mn & $-56$ & $-66$ & $-112$ & $-123$ & $-112$ & $-82$ & $-43$ & $-12$ & $-4$ & $-9$ \\
    Fe & $-80$ & $-77$ & $-105$ & $-100$ & $-84$ & $-54$ & $-25$ & $-5$ & $-3$ & $-11$ \\
    Co & $-68$ & $-65$ & $-74$ & $-64$ & $-50$ & $-28$ & $-9$ & $1$ & $-3$ & $-11$ \\
    Ni & $-34$ & $-35$ & $-33$ & $-26$ & $-19$ & $-7$ & $0$ & $3$ & $-2$ & $-4$ \\
    Cu & $0$ & $0$ & $0$ & $0$ & $0$ & $0$ & $0$ & $0$ & $0$ & $0$ \\
    Zn & $14$ & $11$ & $3$ & $0$ & $-7$ & $-11$ & $-11$ & $-5$ & $2$ & $4$ \\
    \hline
    Y  & $114$ & $58$ & $23$ & $0$ & $-31$ & $-53$ & $-57$ & $-37$ & $-9$ & $16$ \\
    Zr & $64$ & $36$ & $-6$ & $-46$ & $-85$ & $-96$ & $-79$ & $-38$ & $-3$ & $2$ \\
    Nb & $21$ & $-6$ & $-58$ & $-103$ & $-123$ & $-113$ & $-76$ & $-30$ & $-2$ & $-7$ \\
    Mo & $-25$ & $-46$ & $-100$ & $-127$ & $-124$ & $-98$ & $-59$ & $-17$ & $1$ & $-5$ \\
    Tc & $-71$ & $-79$ & $-118$ & $-123$ & $-108$ & $-76$ & $-37$ & $-5$ & $4$ & $-6$ \\
    Ru & $-92$ & $-88$ & $-105$ & $-96$ & $-75$ & $-44$ & $-13$ & $7$ & $5$ & $-8$ \\
    Rh & $-76$ & $-69$ & $-68$ & $-52$ & $-34$ & $-11$ & $6$ & $12$ & $5$ & $-6$ \\
    Pd & $-28$ & $-24$ & $-15$ & $-6$ & $3$ & $12$ & $15$ & $14$ & $6$ & $1$ \\
    Ag & $15$ & $16$ & $14$ & $13$ & $13$ & $12$ & $10$ & $8$ & $8$ & $6$ \\
    Cd & $32$ & $16$ & $5$ & $3$ & $-1$ & $-5$ & $-7$ & $-1$ & $5$ & $10$ \\
    \hline
    La & $127$ & $56$ & $23$ & $-7$ & $-50$ & $-77$ & $-76$ & $-46$ & $-10$ & $18$ \\
    Hf & $56$ & $39$ & $2$ & $-32$ & $-70$ & $-87$ & $-77$ & $-41$ & $-5$ & $0$ \\
    Ta & $23$ & $2$ & $-41$ & $-84$ & $-111$ & $-109$ & $-81$ & $-36$ & $-4$ & $-6$ \\
    W  & $-7$ & $-32$ & $-84$ & $-115$ & $-122$ & $-104$ & $-68$ & $-24$ & $1$ & $-2$ \\
    Re & $-50$ & $-70$ & $-111$ & $-122$ & $-113$ & $-87$ & $-49$ & $-11$ & $5$ & $-2$ \\
    Os & $-77$ & $-87$ & $-109$ & $-104$ & $-87$ & $-58$ & $-24$ & $4$ & $8$ & $-3$ \\
    Ir & $-76$ & $-77$ & $-81$ & $-68$ & $-49$ & $-24$ & $-1$ & $12$ & $9$ & $-2$ \\
    Pt & $-46$ & $-41$ & $-36$ & $-24$ & $-11$ & $4$ & $12$ & $16$ & $11$ & $5$ \\
    Au & $-10$ & $-5$ & $-4$ & $1$ & $5$ & $8$ & $9$ & $11$ & $14$ & $9$ \\
    Hg & $18$ & $0$ & $-6$ & $-2$ & $-2$ & $-3$ & $-2$ & $5$ & $9$ & $13$ \\
    \hline\hline
  \end{tabular}
  \label{tab:surf_agg_cu111_ls_5d}
\end{table}

\clearpage


\begin{figure}[htbp]
    \includegraphics[width=12cm,height=\textheight,keepaspectratio]{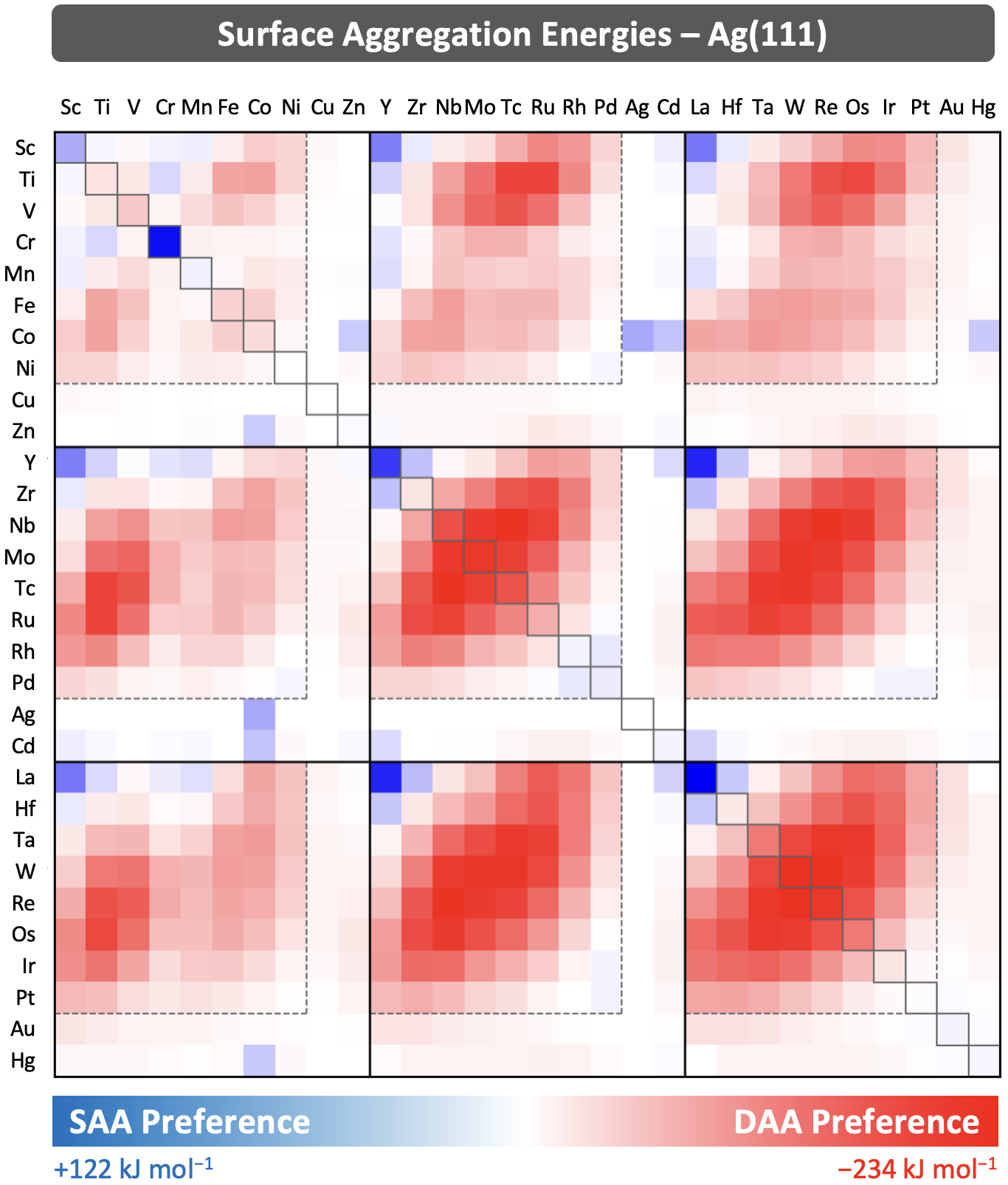}
    \caption{
        Aggregation energies for the formation of all possible transition metal dopant dimers in the Ag(111) surface, as also shown in Fig.~2 of the main text.
        Preferences for single-atom sites are represented in blue, while preferences for dopant dimers are shown in red.     }
    \label{fig:Ag111-surf-aggr}
\end{figure}

\begin{figure}[htbp]
    \includegraphics[width=9.5cm,height=\textheight,keepaspectratio]{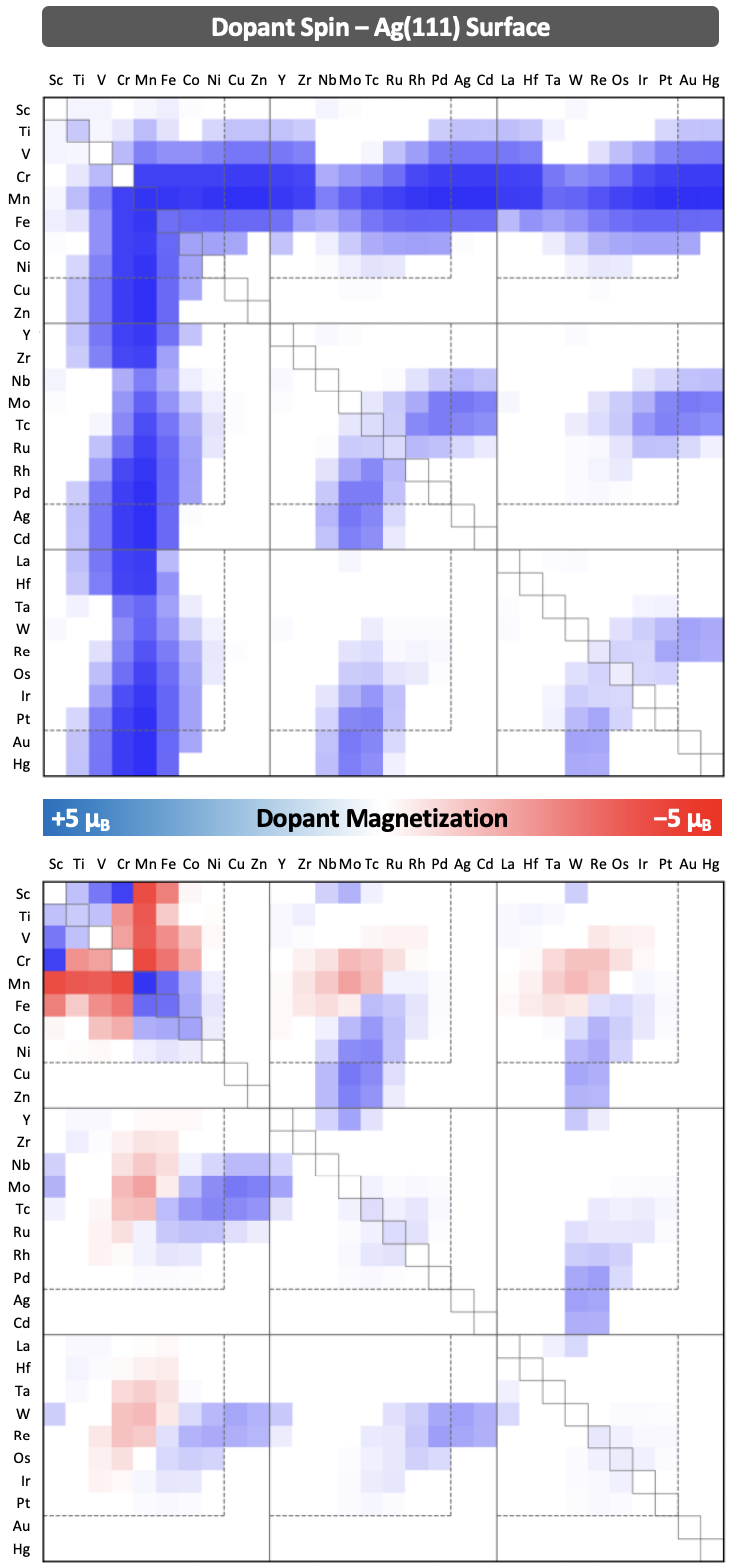}
    \caption{
        Local dopant magnetizations in dual-atom sites located in the surface layer of Ag(111).
        Top: The local magnetization of the dopant with the larger magnetic moment is shown in blue.
        Bottom: The local magnetization of the second dopant is shown in blue when its magnetic moment is aligned parallel to that of the first dopant (ferromagnetic coupling) and in red when it is aligned antiparallel (antiferromagnetic coupling).    }
    \label{fig:Ag111-surf-spin}
\end{figure}

\begin{table}[ht]
  \centering
  \renewcommand{\arraystretch}{1.2}
  \caption{\textbf{Aggregation energies} for \textbf{dual-atom} sites in the \textbf{surface} layer of \textbf{Ag(111)} for all combinations of \textbf{3\textit{d}} transition metals (columns) and all other transition metals (rows), as obtained from initial \textbf{high-spin} configurations. Energies are reported in kJ~mol$^{-1}$.}

  \vspace{8pt}
  \begin{tabular}{@{\hspace{8pt}}*{11}{c@{\hspace{8pt}}}}
    \hline\hline
    TM & Sc & Ti & V & Cr & Mn & Fe & Co & Ni & Cu & Zn \\
    \hline
    Sc & $26$ & $3$ & $-7$ & $4$ & $6$ & $-18$ & $-49$ & $-41$ & $-6$ & $0$ \\
    Ti & $3$ & $-26$ & $-23$ & $12$ & $-19$ & $-85$ & $-91$ & $-41$ & $-5$ & $-1$ \\
    V  & $-7$ & $-23$ & $-53$ & $-12$ & $-34$ & $-57$ & $-44$ & $-17$ & $-3$ & $-3$ \\
    Cr & $4$ & $12$ & $-12$ & $75$ & $122$ & $-12$ & $-9$ & $-8$ & $-1$ & $0$ \\
    Mn & $6$ & $-19$ & $-34$ & $122$ & $5$ & $-9$ & $-23$ & $-20$ & $-2$ & $1$ \\
    Fe & $-18$ & $-85$ & $-57$ & $-12$ & $-9$ & $-42$ & $-46$ & $-18$ & $-2$ & $-2$ \\
    Co & $-49$ & $-91$ & $-44$ & $-9$ & $-23$ & $-46$ & $-35$ & $-9$ & $-1$ & $16$ \\
    Ni & $-41$ & $-41$ & $-17$ & $-8$ & $-20$ & $-18$ & $-9$ & $0$ & $0$ & $-7$ \\
    Cu & $-6$ & $-5$ & $-3$ & $-1$ & $-2$ & $-2$ & $-1$ & $0$ & $0$ & $0$ \\
    Zn & $0$ & $-1$ & $-3$ & $0$ & $1$ & $-2$ & $16$ & $-7$ & $0$ & $2$ \\
    \hline
    Y  & $40$ & $14$ & $1$ & $8$ & $11$ & $-11$ & $-35$ & $-39$ & $-8$ & $2$ \\
    Zr & $7$ & $-25$ & $-26$ & $-6$ & $-12$ & $-62$ & $-87$ & $-53$ & $-8$ & $-5$ \\
    Nb & $-20$ & $-89$ & $-106$ & $-53$ & $-57$ & $-94$ & $-91$ & $-49$ & $-8$ & $-8$ \\
    Mo & $-32$ & $-137$ & $-149$ & $-76$ & $-51$ & $-65$ & $-60$ & $-33$ & $-6$ & $-7$ \\
    Tc & $-78$ & $-191$ & $-170$ & $-74$ & $-46$ & $-69$ & $-65$ & $-32$ & $-6$ & $-13$ \\
    Ru & $-116$ & $-191$ & $-136$ & $-48$ & $-50$ & $-69$ & $-53$ & $-17$ & $-6$ & $-21$ \\
    Rh & $-97$ & $-113$ & $-63$ & $-20$ & $-38$ & $-39$ & $-20$ & $-1$ & $-3$ & $-17$ \\
    Pd & $-39$ & $-30$ & $-12$ & $-9$ & $-16$ & $-8$ & $-2$ & $3$ & $0$ & $-8$ \\
    Ag & $0$ & $0$ & $0$ & $0$ & $0$ & $0$ & $27$ & $0$ & $0$ & $0$ \\
    Cd & $5$ & $2$ & $-1$ & $2$ & $2$ & $-1$ & $18$ & $-6$ & $-1$ & $2$ \\
    \hline
    La & $42$ & $11$ & $-6$ & $6$ & $9$ & $-32$ & $-84$ & $-58$ & $-12$ & $1$ \\
    Hf & $6$ & $-19$ & $-22$ & $-5$ & $-8$ & $-51$ & $-81$ & $-55$ & $-8$ & $-3$ \\
    Ta & $-21$ & $-67$ & $-68$ & $-26$ & $-42$ & $-89$ & $-96$ & $-58$ & $-10$ & $-8$ \\
    W  & $-47$ & $-129$ & $-135$ & $-75$ & $-68$ & $-94$ & $-88$ & $-50$ & $-10$ & $-12$ \\
    Re & $-80$ & $-174$ & $-159$ & $-81$ & $-64$ & $-84$ & $-78$ & $-42$ & $-10$ & $-16$ \\
    Os & $-113$ & $-185$ & $-139$ & $-61$ & $-60$ & $-79$ & $-63$ & $-27$ & $-10$ & $-23$ \\
    Ir & $-108$ & $-134$ & $-85$ & $-33$ & $-50$ & $-52$ & $-33$ & $-11$ & $-6$ & $-19$ \\
    Pt & $-67$ & $-61$ & $-32$ & $-20$ & $-29$ & $-22$ & $-12$ & $-4$ & $-3$ & $-12$ \\
    Au & $-24$ & $-19$ & $-13$ & $-9$ & $-10$ & $-7$ & $-4$ & $-3$ & $-1$ & $-3$ \\
    Hg & $-8$ & $-8$ & $-8$ & $-4$ & $-3$ & $-4$ & $17$ & $-6$ & $-2$ & $0$ \\
    \hline\hline
  \end{tabular}
  \label{tab:surf_agg_ag111_3d}
\end{table}

\begin{table}[ht]
  \centering
  \renewcommand{\arraystretch}{1.2}
  \caption{\textbf{Aggregation energies} for \textbf{dual-atom} sites in the \textbf{surface} layer of \textbf{Ag(111)} for all combinations of \textbf{4\textit{d}} transition metals (columns) and all other transition metals (rows), as obtained from initial \textbf{high-spin} configurations. Energies are reported in kJ~mol$^{-1}$.}

  \vspace{8pt}
  \begin{tabular}{@{\hspace{8pt}}*{11}{c@{\hspace{8pt}}}}
    \hline\hline
    TM & Y & Zr & Nb & Mo & Tc & Ru & Rh & Pd & Ag & Cd \\
    \hline
    Sc & $40$ & $7$ & $-20$ & $-32$ & $-78$ & $-116$ & $-97$ & $-39$ & $0$ & $5$ \\
    Ti & $14$ & $-25$ & $-89$ & $-137$ & $-191$ & $-191$ & $-113$ & $-30$ & $0$ & $2$ \\
    V  & $1$ & $-26$ & $-106$ & $-149$ & $-170$ & $-136$ & $-63$ & $-12$ & $0$ & $-1$ \\
    Cr & $8$ & $-6$ & $-53$ & $-76$ & $-74$ & $-48$ & $-20$ & $-9$ & $0$ & $2$ \\
    Mn & $11$ & $-12$ & $-57$ & $-51$ & $-46$ & $-50$ & $-38$ & $-16$ & $0$ & $2$ \\
    Fe & $-11$ & $-62$ & $-94$ & $-65$ & $-69$ & $-69$ & $-39$ & $-8$ & $0$ & $-1$ \\
    Co & $-35$ & $-87$ & $-91$ & $-60$ & $-65$ & $-53$ & $-20$ & $-2$ & $27$ & $18$ \\
    Ni & $-39$ & $-53$ & $-49$ & $-33$ & $-32$ & $-17$ & $-1$ & $3$ & $0$ & $-6$ \\
    Cu & $-8$ & $-8$ & $-8$ & $-6$ & $-6$ & $-6$ & $-3$ & $0$ & $0$ & $-1$ \\
    Zn & $2$ & $-5$ & $-8$ & $-7$ & $-13$ & $-21$ & $-17$ & $-8$ & $0$ & $2$ \\
    \hline
    Y  & $61$ & $19$ & $-8$ & $-20$ & $-56$ & $-94$ & $-85$ & $-41$ & $0$ & $12$ \\
    Zr & $19$ & $-25$ & $-84$ & $-122$ & $-169$ & $-181$ & $-124$ & $-42$ & $0$ & $0$ \\
    Nb & $-8$ & $-84$ & $-175$ & $-209$ & $-231$ & $-197$ & $-115$ & $-31$ & $0$ & $-3$ \\
    Mo & $-20$ & $-122$ & $-209$ & $-214$ & $-199$ & $-155$ & $-72$ & $-18$ & $0$ & $-3$ \\
    Tc & $-56$ & $-169$ & $-231$ & $-199$ & $-173$ & $-118$ & $-53$ & $-11$ & $0$ & $-7$ \\
    Ru & $-94$ & $-181$ & $-197$ & $-155$ & $-118$ & $-77$ & $-27$ & $2$ & $0$ & $-14$ \\
    Rh & $-85$ & $-124$ & $-115$ & $-72$ & $-53$ & $-27$ & $4$ & $7$ & $0$ & $-13$ \\
    Pd & $-41$ & $-42$ & $-31$ & $-18$ & $-11$ & $2$ & $7$ & $6$ & $0$ & $-7$ \\
    Ag & $0$ & $0$ & $0$ & $0$ & $0$ & $0$ & $0$ & $0$ & $0$ & $0$ \\
    Cd & $12$ & $0$ & $-3$ & $-3$ & $-7$ & $-14$ & $-13$ & $-7$ & $0$ & $4$ \\
    \hline
    La & $68$ & $21$ & $-24$ & $-60$ & $-122$ & $-162$ & $-133$ & $-54$ & $0$ & $14$ \\
    Hf & $18$ & $-23$ & $-66$ & $-95$ & $-143$ & $-167$ & $-127$ & $-49$ & $0$ & $2$ \\
    Ta & $-11$ & $-69$ & $-146$ & $-179$ & $-211$ & $-198$ & $-127$ & $-40$ & $0$ & $-3$ \\
    W  & $-35$ & $-124$ & $-205$ & $-220$ & $-219$ & $-186$ & $-105$ & $-28$ & $0$ & $-6$ \\
    Re & $-63$ & $-164$ & $-227$ & $-207$ & $-193$ & $-147$ & $-73$ & $-16$ & $0$ & $-9$ \\
    Os & $-95$ & $-180$ & $-206$ & $-174$ & $-143$ & $-102$ & $-41$ & $-3$ & $0$ & $-14$ \\
    Ir & $-96$ & $-145$ & $-143$ & $-102$ & $-77$ & $-48$ & $-9$ & $3$ & $0$ & $-13$ \\
    Pt & $-66$ & $-77$ & $-66$ & $-41$ & $-31$ & $-12$ & $0$ & $4$ & $0$ & $-8$ \\
    Au & $-26$ & $-26$ & $-20$ & $-14$ & $-11$ & $-6$ & $-3$ & $0$ & $0$ & $-2$ \\
    Hg & $-4$ & $-13$ & $-12$ & $-9$ & $-10$ & $-14$ & $-11$ & $-4$ & $0$ & $2$ \\
    \hline\hline
  \end{tabular}
  \label{tab:surf_agg_ag111_4d}
\end{table}

\begin{table}[ht]
  \centering
  \renewcommand{\arraystretch}{1.2}
  \caption{\textbf{Aggregation energies} for \textbf{dual-atom} sites in the \textbf{surface} layer of \textbf{Ag(111)} for all combinations of \textbf{5\textit{d}} transition metals (columns) and all other transition metals (rows), as obtained from initial \textbf{high-spin} configurations. Energies are reported in kJ~mol$^{-1}$.}
  \vspace{8pt}
  \begin{tabular}{@{\hspace{8pt}}*{11}{c@{\hspace{8pt}}}}
    \hline\hline
    TM & La & Hf & Ta & W & Re & Os & Ir & Pt & Au & Hg \\
    \hline
    Sc & $42$ & $6$ & $-21$ & $-47$ & $-80$ & $-113$ & $-108$ & $-67$ & $-24$ & $-8$ \\
    Ti & $11$ & $-19$ & $-67$ & $-129$ & $-174$ & $-185$ & $-134$ & $-61$ & $-19$ & $-8$ \\
    V  & $-6$ & $-22$ & $-68$ & $-135$ & $-159$ & $-139$ & $-85$ & $-32$ & $-13$ & $-8$ \\
    Cr & $6$ & $-5$ & $-26$ & $-75$ & $-81$ & $-61$ & $-33$ & $-20$ & $-9$ & $-4$ \\
    Mn & $9$ & $-8$ & $-42$ & $-68$ & $-64$ & $-60$ & $-50$ & $-29$ & $-10$ & $-3$ \\
    Fe & $-32$ & $-51$ & $-89$ & $-94$ & $-84$ & $-79$ & $-52$ & $-22$ & $-7$ & $-4$ \\
    Co & $-84$ & $-81$ & $-96$ & $-88$ & $-78$ & $-63$ & $-33$ & $-12$ & $-4$ & $17$ \\
    Ni & $-58$ & $-55$ & $-58$ & $-50$ & $-42$ & $-27$ & $-11$ & $-4$ & $-3$ & $-6$ \\
    Cu & $-12$ & $-8$ & $-10$ & $-10$ & $-10$ & $-10$ & $-6$ & $-3$ & $-1$ & $-2$ \\
    Zn & $1$ & $-3$ & $-8$ & $-12$ & $-16$ & $-23$ & $-19$ & $-12$ & $-3$ & $0$ \\
    \hline
    Y  & $68$ & $18$ & $-11$ & $-35$ & $-63$ & $-95$ & $-96$ & $-66$ & $-26$ & $-4$ \\
    Zr & $21$ & $-23$ & $-69$ & $-124$ & $-164$ & $-180$ & $-145$ & $-77$ & $-26$ & $-13$ \\
    Nb & $-24$ & $-66$ & $-146$ & $-205$ & $-227$ & $-206$ & $-143$ & $-66$ & $-20$ & $-12$ \\
    Mo & $-60$ & $-95$ & $-179$ & $-220$ & $-207$ & $-174$ & $-102$ & $-41$ & $-14$ & $-9$ \\
    Tc & $-122$ & $-143$ & $-211$ & $-219$ & $-193$ & $-143$ & $-77$ & $-31$ & $-11$ & $-10$ \\
    Ru & $-162$ & $-167$ & $-198$ & $-186$ & $-147$ & $-102$ & $-48$ & $-12$ & $-6$ & $-14$ \\
    Rh & $-133$ & $-127$ & $-127$ & $-105$ & $-73$ & $-41$ & $-9$ & $0$ & $-3$ & $-11$ \\
    Pd & $-54$ & $-49$ & $-40$ & $-28$ & $-16$ & $-3$ & $3$ & $4$ & $0$ & $-4$ \\
    Ag & $0$ & $0$ & $0$ & $0$ & $0$ & $0$ & $0$ & $0$ & $0$ & $0$ \\
    Cd & $14$ & $2$ & $-3$ & $-6$ & $-9$ & $-14$ & $-13$ & $-8$ & $-2$ & $2$ \\
    \hline
    La & $79$ & $17$ & $-16$ & $-57$ & $-107$ & $-144$ & $-134$ & $-82$ & $-28$ & $0$ \\
    Hf & $17$ & $-21$ & $-59$ & $-104$ & $-146$ & $-170$ & $-147$ & $-84$ & $-28$ & $-11$ \\
    Ta & $-16$ & $-59$ & $-128$ & $-186$ & $-217$ & $-211$ & $-154$ & $-77$ & $-24$ & $-11$ \\
    W  & $-57$ & $-104$ & $-186$ & $-233$ & $-234$ & $-206$ & $-137$ & $-59$ & $-18$ & $-11$ \\
    Re & $-107$ & $-146$ & $-217$ & $-234$ & $-216$ & $-173$ & $-104$ & $-41$ & $-12$ & $-10$ \\
    Os & $-144$ & $-170$ & $-211$ & $-206$ & $-173$ & $-129$ & $-65$ & $-20$ & $-7$ & $-12$ \\
    Ir & $-134$ & $-147$ & $-154$ & $-137$ & $-104$ & $-65$ & $-25$ & $-6$ & $-3$ & $-9$ \\
    Pt & $-82$ & $-84$ & $-77$ & $-59$ & $-41$ & $-20$ & $-6$ & $1$ & $1$ & $-2$ \\
    Au & $-28$ & $-28$ & $-24$ & $-18$ & $-12$ & $-7$ & $-3$ & $1$ & $3$ & $1$ \\
    Hg & $0$ & $-11$ & $-11$ & $-11$ & $-10$ & $-12$ & $-9$ & $-2$ & $1$ & $3$ \\
    \hline\hline
  \end{tabular}
  \label{tab:surf_agg_ag111_5d}
\end{table}

\begin{table}[ht]
  \centering
  \renewcommand{\arraystretch}{1.2}
  \caption{\textbf{Aggregation energies} for \textbf{dual-atom} sites in the \textbf{surface} layer of \textbf{Ag(111)} for all combinations of \textbf{3\textit{d}} transition metals (columns) and all other transition metals (rows), as obtained from initial \textbf{low-spin} configurations. Energies are reported in kJ~mol$^{-1}$.}
  \vspace{8pt}
  \begin{tabular}{@{\hspace{8pt}}*{11}{c@{\hspace{8pt}}}}
    \hline\hline
    TM & Sc & Ti & V & Cr & Mn & Fe & Co & Ni & Cu & Zn \\
    \hline
    Sc & $26$ & $-1$ & $-15$ & $-25$ & $-50$ & $-74$ & $-76$ & $-41$ & $-6$ & $0$ \\
    Ti & $-1$ & $-39$ & $-105$ & $-156$ & $-187$ & $-190$ & $-130$ & $-49$ & $-6$ & $-7$ \\
    V  & $-15$ & $-105$ & $-200$ & $-245$ & $-246$ & $-225$ & $-144$ & $-43$ & $-5$ & $-6$ \\
    Cr & $-25$ & $-156$ & $-245$ & $-263$ & $-361$ & $-195$ & $-93$ & $-30$ & $-4$ & $-4$ \\
    Mn & $-50$ & $-187$ & $-246$ & $-361$ & $-353$ & $-117$ & $-64$ & $-23$ & $-3$ & $-6$ \\
    Fe & $-74$ & $-190$ & $-225$ & $-195$ & $-117$ & $-144$ & $-45$ & $-14$ & $-2$ & $-9$ \\
    Co & $-76$ & $-130$ & $-144$ & $-93$ & $-64$ & $-45$ & $-22$ & $-4$ & $-2$ & $-11$ \\
    Ni & $-41$ & $-49$ & $-43$ & $-30$ & $-23$ & $-14$ & $-4$ & $0$ & $0$ & $-7$ \\
    Cu & $-6$ & $-6$ & $-5$ & $-4$ & $-3$ & $-2$ & $-2$ & $0$ & $0$ & $0$ \\
    Zn & $0$ & $-7$ & $-6$ & $-4$ & $-6$ & $-9$ & $-11$ & $-7$ & $0$ & $2$ \\
    \hline
    Y  & $40$ & $9$ & $-6$ & $-15$ & $-33$ & $-50$ & $-58$ & $-39$ & $-8$ & $2$ \\
    Zr & $7$ & $-32$ & $-78$ & $-113$ & $-139$ & $-147$ & $-114$ & $-53$ & $-8$ & $-5$ \\
    Nb & $-21$ & $-104$ & $-183$ & $-219$ & $-220$ & $-185$ & $-119$ & $-51$ & $-9$ & $-11$ \\
    Mo & $-57$ & $-179$ & $-253$ & $-260$ & $-222$ & $-190$ & $-105$ & $-43$ & $-9$ & $-16$ \\
    Tc & $-98$ & $-223$ & $-264$ & $-263$ & $-245$ & $-184$ & $-75$ & $-29$ & $-8$ & $-22$ \\
    Ru & $-117$ & $-203$ & $-210$ & $-217$ & $-121$ & $-76$ & $-41$ & $-12$ & $-6$ & $-21$ \\
    Rh & $-97$ & $-125$ & $-110$ & $-79$ & $-50$ & $-31$ & $-11$ & $-1$ & $-3$ & $-17$ \\
    Pd & $-39$ & $-33$ & $-25$ & $-16$ & $-11$ & $-4$ & $3$ & $3$ & $0$ & $-8$ \\
    Ag & $0$ & $0$ & $0$ & $0$ & $0$ & $0$ & $0$ & $0$ & $0$ & $0$ \\
    Cd & $5$ & $-3$ & $-3$ & $-2$ & $-4$ & $-6$ & $-9$ & $-6$ & $-1$ & $2$ \\
    \hline
    La & $42$ & $8$ & $-16$ & $-52$ & $-89$ & $-120$ & $-111$ & $-58$ & $-12$ & $1$ \\
    Hf & $6$ & $-28$ & $-56$ & $-82$ & $-108$ & $-125$ & $-108$ & $-55$ & $-8$ & $-3$ \\
    Ta & $-21$ & $-79$ & $-141$ & $-175$ & $-187$ & $-173$ & $-122$ & $-58$ & $-10$ & $-8$ \\
    W  & $-49$ & $-147$ & $-215$ & $-230$ & $-210$ & $-173$ & $-116$ & $-54$ & $-11$ & $-13$ \\
    Re & $-89$ & $-195$ & $-241$ & $-224$ & $-191$ & $-146$ & $-92$ & $-42$ & $-12$ & $-22$ \\
    Os & $-113$ & $-197$ & $-203$ & $-229$ & $-141$ & $-97$ & $-58$ & $-24$ & $-10$ & $-23$ \\
    Ir & $-108$ & $-146$ & $-134$ & $-109$ & $-75$ & $-51$ & $-27$ & $-11$ & $-6$ & $-19$ \\
    Pt & $-67$ & $-67$ & $-57$ & $-39$ & $-30$ & $-18$ & $-7$ & $-4$ & $-3$ & $-12$ \\
    Au & $-24$ & $-22$ & $-16$ & $-13$ & $-10$ & $-7$ & $-6$ & $-3$ & $-1$ & $-3$ \\
    Hg & $-8$ & $-12$ & $-10$ & $-8$ & $-8$ & $-8$ & $-10$ & $-6$ & $-2$ & $0$ \\
    \hline\hline
  \end{tabular}
  \label{tab:surf_agg_ag111_ls_3d}
\end{table}

\begin{table}[ht]
  \centering
  \renewcommand{\arraystretch}{1.2}
  \caption{\textbf{Aggregation energies} for \textbf{dual-atom} sites in the \textbf{surface} layer of \textbf{Ag(111)} for all combinations of \textbf{4\textit{d}} transition metals (columns) and all other transition metals (rows), as obtained from initial \textbf{low-spin} configurations. Energies are reported in kJ~mol$^{-1}$.}
  \vspace{8pt}
  \begin{tabular}{@{\hspace{8pt}}*{11}{c@{\hspace{8pt}}}}
    \hline\hline
    TM & Y & Zr & Nb & Mo & Tc & Ru & Rh & Pd & Ag & Cd \\
    \hline
    Sc & $40$ & $7$ & $-21$ & $-57$ & $-98$ & $-117$ & $-97$ & $-39$ & $0$ & $5$ \\
    Ti & $9$ & $-32$ & $-104$ & $-179$ & $-223$ & $-203$ & $-125$ & $-33$ & $0$ & $-3$ \\
    V  & $-6$ & $-78$ & $-183$ & $-253$ & $-264$ & $-210$ & $-110$ & $-25$ & $0$ & $-3$ \\
    Cr & $-15$ & $-113$ & $-219$ & $-260$ & $-263$ & $-217$ & $-79$ & $-16$ & $0$ & $-2$ \\
    Mn & $-33$ & $-139$ & $-220$ & $-222$ & $-245$ & $-121$ & $-50$ & $-11$ & $0$ & $-4$ \\
    Fe & $-50$ & $-147$ & $-185$ & $-190$ & $-184$ & $-76$ & $-31$ & $-4$ & $0$ & $-6$ \\
    Co & $-58$ & $-114$ & $-119$ & $-105$ & $-75$ & $-41$ & $-11$ & $3$ & $0$ & $-9$ \\
    Ni & $-39$ & $-53$ & $-51$ & $-43$ & $-29$ & $-12$ & $-1$ & $3$ & $0$ & $-6$ \\
    Cu & $-8$ & $-8$ & $-9$ & $-9$ & $-8$ & $-6$ & $-3$ & $0$ & $0$ & $-1$ \\
    Zn & $2$ & $-5$ & $-11$ & $-16$ & $-22$ & $-21$ & $-17$ & $-8$ & $0$ & $2$ \\
    \hline
    Y  & $61$ & $19$ & $-10$ & $-40$ & $-76$ & $-95$ & $-85$ & $-41$ & $0$ & $12$ \\
    Zr & $19$ & $-25$ & $-87$ & $-152$ & $-190$ & $-181$ & $-124$ & $-42$ & $0$ & $0$ \\
    Nb & $-10$ & $-87$ & $-181$ & $-243$ & $-255$ & $-200$ & $-115$ & $-31$ & $0$ & $-4$ \\
    Mo & $-40$ & $-152$ & $-243$ & $-276$ & $-247$ & $-180$ & $-92$ & $-20$ & $0$ & $-9$ \\
    Tc & $-76$ & $-190$ & $-255$ & $-247$ & $-208$ & $-136$ & $-57$ & $-5$ & $0$ & $-14$ \\
    Ru & $-95$ & $-181$ & $-200$ & $-180$ & $-136$ & $-73$ & $-19$ & $5$ & $0$ & $-15$ \\
    Rh & $-85$ & $-124$ & $-115$ & $-92$ & $-57$ & $-19$ & $4$ & $7$ & $0$ & $-13$ \\
    Pd & $-41$ & $-42$ & $-31$ & $-20$ & $-5$ & $5$ & $7$ & $6$ & $0$ & $-7$ \\
    Ag & $0$ & $0$ & $0$ & $0$ & $0$ & $0$ & $0$ & $0$ & $0$ & $0$ \\
    Cd & $12$ & $0$ & $-4$ & $-9$ & $-14$ & $-15$ & $-13$ & $-7$ & $0$ & $4$ \\
    \hline
    La & $68$ & $21$ & $-27$ & $-91$ & $-143$ & $-162$ & $-133$ & $-54$ & $0$ & $14$ \\
    Hf & $18$ & $-23$ & $-69$ & $-126$ & $-164$ & $-167$ & $-127$ & $-49$ & $0$ & $2$ \\
    Ta & $-11$ & $-69$ & $-149$ & $-210$ & $-232$ & $-198$ & $-127$ & $-40$ & $0$ & $-3$ \\
    W  & $-37$ & $-130$ & $-215$ & $-257$ & $-245$ & $-191$ & $-109$ & $-29$ & $0$ & $-6$ \\
    Re & $-71$ & $-173$ & $-239$ & $-246$ & $-217$ & $-156$ & $-78$ & $-16$ & $0$ & $-12$ \\
    Os & $-95$ & $-180$ & $-210$ & $-197$ & $-161$ & $-99$ & $-39$ & $-2$ & $0$ & $-14$ \\
    Ir & $-96$ & $-145$ & $-143$ & $-124$ & $-87$ & $-42$ & $-9$ & $3$ & $0$ & $-13$ \\
    Pt & $-66$ & $-77$ & $-67$ & $-51$ & $-28$ & $-8$ & $0$ & $4$ & $0$ & $-8$ \\
    Au & $-26$ & $-26$ & $-22$ & $-16$ & $-10$ & $-6$ & $-3$ & $0$ & $0$ & $-2$ \\
    Hg & $-4$ & $-13$ & $-13$ & $-13$ & $-16$ & $-14$ & $-11$ & $-4$ & $0$ & $2$ \\
    \hline\hline
  \end{tabular}
  \label{tab:surf_agg_ag111_ls_4d}
\end{table}

\begin{table}[ht]
  \centering
  \renewcommand{\arraystretch}{1.2}
  \caption{\textbf{Aggregation energies} for \textbf{dual-atom} sites in the \textbf{surface} layer of \textbf{Ag(111)} for all combinations of \textbf{5\textit{d}} transition metals (columns) and all other transition metals (rows), as obtained from initial \textbf{low-spin} configurations. Energies are reported in kJ~mol$^{-1}$.}
  \vspace{8pt}
  \begin{tabular}{@{\hspace{8pt}}*{11}{c@{\hspace{8pt}}}}
    \hline\hline
    TM & La & Hf & Ta & W & Re & Os & Ir & Pt & Au & Hg \\
    \hline
    Sc & $42$ & $6$ & $-21$ & $-49$ & $-89$ & $-113$ & $-108$ & $-67$ & $-24$ & $-8$ \\
    Ti & $8$ & $-28$ & $-79$ & $-147$ & $-195$ & $-197$ & $-146$ & $-67$ & $-22$ & $-12$ \\
    V  & $-16$ & $-56$ & $-141$ & $-215$ & $-241$ & $-203$ & $-134$ & $-57$ & $-16$ & $-10$ \\
    Cr & $-52$ & $-82$ & $-175$ & $-230$ & $-224$ & $-229$ & $-109$ & $-39$ & $-13$ & $-8$ \\
    Mn & $-89$ & $-108$ & $-187$ & $-210$ & $-191$ & $-141$ & $-75$ & $-30$ & $-10$ & $-8$ \\
    Fe & $-120$ & $-125$ & $-173$ & $-173$ & $-146$ & $-97$ & $-51$ & $-18$ & $-7$ & $-8$ \\
    Co & $-111$ & $-108$ & $-122$ & $-116$ & $-92$ & $-58$ & $-27$ & $-7$ & $-6$ & $-10$ \\
    Ni & $-58$ & $-55$ & $-58$ & $-54$ & $-42$ & $-24$ & $-11$ & $-4$ & $-3$ & $-6$ \\
    Cu & $-12$ & $-8$ & $-10$ & $-11$ & $-12$ & $-10$ & $-6$ & $-3$ & $-1$ & $-2$ \\
    Zn & $1$ & $-3$ & $-8$ & $-13$ & $-22$ & $-23$ & $-19$ & $-12$ & $-3$ & $0$ \\
    \hline
    Y  & $68$ & $18$ & $-11$ & $-37$ & $-71$ & $-95$ & $-96$ & $-66$ & $-26$ & $-4$ \\
    Zr & $21$ & $-23$ & $-69$ & $-130$ & $-173$ & $-180$ & $-145$ & $-77$ & $-26$ & $-13$ \\
    Nb & $-27$ & $-69$ & $-149$ & $-215$ & $-239$ & $-210$ & $-143$ & $-67$ & $-22$ & $-13$ \\
    Mo & $-91$ & $-126$ & $-210$ & $-257$ & $-246$ & $-197$ & $-124$ & $-51$ & $-16$ & $-13$ \\
    Tc & $-143$ & $-164$ & $-232$ & $-245$ & $-217$ & $-161$ & $-87$ & $-28$ & $-10$ & $-16$ \\
    Ru & $-162$ & $-167$ & $-198$ & $-191$ & $-156$ & $-99$ & $-42$ & $-8$ & $-6$ & $-14$ \\
    Rh & $-133$ & $-127$ & $-127$ & $-109$ & $-78$ & $-39$ & $-9$ & $0$ & $-3$ & $-11$ \\
    Pd & $-54$ & $-49$ & $-40$ & $-29$ & $-16$ & $-2$ & $3$ & $4$ & $0$ & $-4$ \\
    Ag & $0$ & $0$ & $0$ & $0$ & $0$ & $0$ & $0$ & $0$ & $0$ & $0$ \\
    Cd & $14$ & $2$ & $-3$ & $-6$ & $-12$ & $-14$ & $-13$ & $-8$ & $-2$ & $2$ \\
    \hline
    La & $79$ & $17$ & $-16$ & $-62$ & $-116$ & $-144$ & $-134$ & $-82$ & $-28$ & $0$ \\
    Hf & $17$ & $-21$ & $-59$ & $-111$ & $-155$ & $-170$ & $-147$ & $-84$ & $-28$ & $-11$ \\
    Ta & $-16$ & $-59$ & $-128$ & $-192$ & $-225$ & $-211$ & $-154$ & $-77$ & $-24$ & $-11$ \\
    W  & $-62$ & $-111$ & $-192$ & $-246$ & $-249$ & $-209$ & $-142$ & $-65$ & $-18$ & $-9$ \\
    Re & $-116$ & $-155$ & $-225$ & $-249$ & $-229$ & $-180$ & $-110$ & $-42$ & $-12$ & $-13$ \\
    Os & $-144$ & $-170$ & $-211$ & $-209$ & $-180$ & $-127$ & $-64$ & $-18$ & $-7$ & $-12$ \\
    Ir & $-134$ & $-147$ & $-154$ & $-142$ & $-110$ & $-64$ & $-25$ & $-6$ & $-3$ & $-9$ \\
    Pt & $-82$ & $-84$ & $-77$ & $-65$ & $-42$ & $-18$ & $-6$ & $1$ & $1$ & $-2$ \\
    Au & $-28$ & $-28$ & $-24$ & $-18$ & $-12$ & $-7$ & $-3$ & $1$ & $3$ & $1$ \\
    Hg & $0$ & $-11$ & $-11$ & $-9$ & $-13$ & $-12$ & $-9$ & $-2$ & $1$ & $3$ \\
    \hline\hline
  \end{tabular}
  \label{tab:surf_agg_ag111_ls_5d}
\end{table}

\clearpage

\subsection{Bulk Aggregation}


\begin{figure}[htbp]
    \includegraphics[width=12cm,height=\textheight,keepaspectratio]{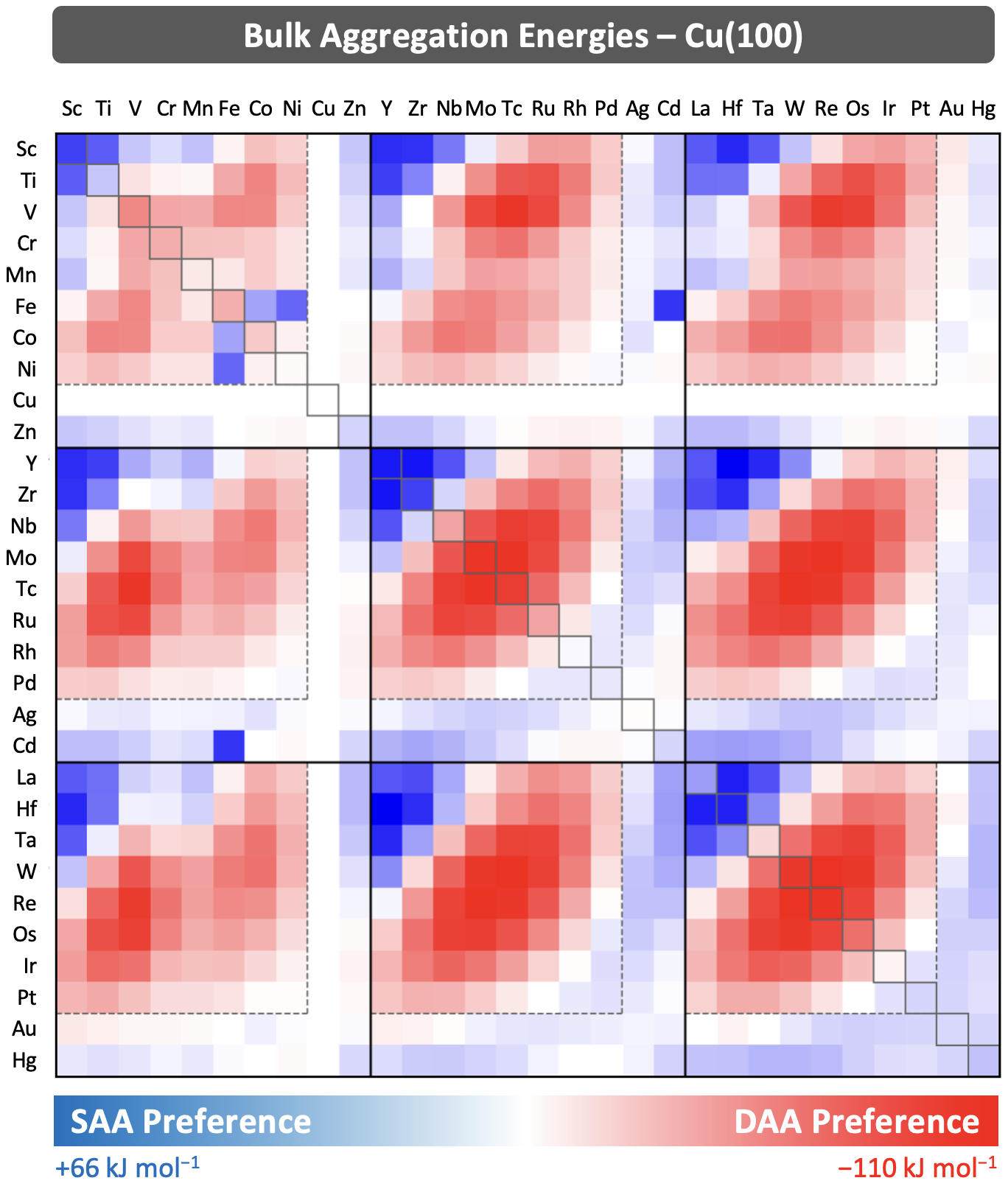}
    \caption{
        Aggregation energies for the formation of all possible transition metal dopant dimers in a bulk layer of the Cu(100) system. 
        Preferences for single-atom sites are represented in blue, while preferences for dopant dimers are shown in red. 
    }
    \label{fig:Cu100-bulk-aggr}
\end{figure}

\begin{table}[ht]
  \centering
  \renewcommand{\arraystretch}{1.2}
  \caption{\textbf{Aggregation energies} for \textbf{dual-atom} sites in the \textbf{bulk} layer of \textbf{Cu(100)} for all combinations of \textbf{3\textit{d}} transition metals (columns) and all other transition metals (rows), as obtained from initial \textbf{high-spin} configurations. Energies are reported in kJ~mol$^{-1}$.}
  \vspace{8pt}
  \begin{tabular}{@{\hspace{8pt}}*{11}{c@{\hspace{8pt}}}}
    \hline\hline
    TM & Sc & Ti & V & Cr & Mn & Fe & Co & Ni & Cu & Zn \\
    \hline
    Sc & $49$ & $42$ & $15$ & $9$ & $16$ & $-5$ & $-28$ & $-21$ & $0$ & $15$ \\
    Ti & $42$ & $15$ & $-13$ & $-4$ & $-4$ & $-38$ & $-55$ & $-30$ & $0$ & $12$ \\
    V  & $15$ & $-13$ & $-54$ & $-39$ & $-38$ & $-55$ & $-54$ & $-24$ & $0$ & $8$ \\
    Cr & $9$ & $-4$ & $-39$ & $-37$ & $-28$ & $-27$ & $-23$ & $-12$ & $0$ & $5$ \\
    Mn & $16$ & $-4$ & $-38$ & $-28$ & $-10$ & $-11$ & $-23$ & $-13$ & $0$ & $6$ \\
    Fe & $-5$ & $-38$ & $-55$ & $-27$ & $-11$ & $-35$ & $24$ & $39$ & $0$ & $0$ \\
    Co & $-28$ & $-55$ & $-54$ & $-23$ & $-23$ & $24$ & $-23$ & $-7$ & $0$ & $-2$ \\
    Ni & $-21$ & $-30$ & $-24$ & $-12$ & $-13$ & $39$ & $-7$ & $-2$ & $0$ & $-4$ \\
    Cu & $0$ & $0$ & $0$ & $0$ & $0$ & $0$ & $0$ & $0$ & $0$ & $0$ \\
    Zn & $15$ & $12$ & $8$ & $5$ & $6$ & $0$ & $-2$ & $-4$ & $0$ & $11$ \\
    \hline
    Y  & $54$ & $50$ & $22$ & $14$ & $20$ & $2$ & $-21$ & $-18$ & $0$ & $16$ \\
    Zr & $53$ & $32$ & $0$ & $3$ & $10$ & $-24$ & $-46$ & $-29$ & $0$ & $16$ \\
    Nb & $34$ & $-7$ & $-46$ & $-26$ & $-25$ & $-51$ & $-62$ & $-33$ & $0$ & $10$ \\
    Mo & $5$ & $-50$ & $-88$ & $-57$ & $-42$ & $-55$ & $-57$ & $-27$ & $0$ & $4$ \\
    Tc & $-22$ & $-78$ & $-104$ & $-64$ & $-40$ & $-47$ & $-42$ & $-18$ & $0$ & $-1$ \\
    Ru & $-43$ & $-83$ & $-88$ & $-46$ & $-31$ & $-37$ & $-27$ & $-9$ & $0$ & $-5$ \\
    Rh & $-43$ & $-60$ & $-52$ & $-24$ & $-22$ & $-22$ & $-11$ & $-3$ & $0$ & $-7$ \\
    Pd & $-23$ & $-24$ & $-14$ & $-9$ & $-9$ & $-5$ & $0$ & $2$ & $0$ & $-4$ \\
    Ag & $2$ & $6$ & $6$ & $3$ & $3$ & $4$ & $7$ & $1$ & $0$ & $2$ \\
    Cd & $17$ & $17$ & $12$ & $7$ & $9$ & $52$ & $0$ & $-3$ & $0$ & $11$ \\
    \hline
    La & $43$ & $37$ & $12$ & $7$ & $16$ & $-6$ & $-35$ & $-24$ & $0$ & $18$ \\
    Hf & $55$ & $37$ & $4$ & $5$ & $11$ & $-23$ & $-45$ & $-30$ & $0$ & $18$ \\
    Ta & $43$ & $5$ & $-34$ & $-17$ & $-19$ & $-49$ & $-64$ & $-36$ & $0$ & $14$ \\
    W  & $15$ & $-40$ & $-79$ & $-52$ & $-42$ & $-60$ & $-64$ & $-32$ & $0$ & $8$ \\
    Re & $-14$ & $-70$ & $-98$ & $-63$ & $-43$ & $-53$ & $-51$ & $-23$ & $0$ & $3$ \\
    Os & $-39$ & $-84$ & $-94$ & $-55$ & $-38$ & $-44$ & $-35$ & $-15$ & $0$ & $-2$ \\
    Ir & $-45$ & $-69$ & $-65$ & $-32$ & $-28$ & $-29$ & $-18$ & $-6$ & $0$ & $-5$ \\
    Pt & $-32$ & $-38$ & $-28$ & $-16$ & $-15$ & $-12$ & $-2$ & $-1$ & $0$ & $-3$ \\
    Au & $-10$ & $-7$ & $-4$ & $-4$ & $-2$ & $-1$ & $4$ & $1$ & $0$ & $1$ \\
    Hg & $6$ & $8$ & $7$ & $3$ & $5$ & $2$ & $0$ & $-2$ & $0$ & $10$ \\
    \hline\hline
  \end{tabular}
  \label{tab:bulk_agg_cu100_3d}
\end{table}

\begin{table}[ht]
  \centering
  \renewcommand{\arraystretch}{1.2}
  \caption{\textbf{Aggregation energies} for \textbf{dual-atom} sites in the \textbf{bulk} layer of \textbf{Cu(100)} for all combinations of \textbf{4\textit{d}} transition metals (columns) and all other transition metals (rows), as obtained from initial \textbf{high-spin} configurations. Energies are reported in kJ~mol$^{-1}$.}
  \vspace{8pt}
  \begin{tabular}{@{\hspace{8pt}}*{11}{c@{\hspace{8pt}}}}
    \hline\hline
    TM & Y & Zr & Nb & Mo & Tc & Ru & Rh & Pd & Ag & Cd \\
    \hline
    Sc & $54$ & $53$ & $34$ & $5$ & $-22$ & $-43$ & $-43$ & $-23$ & $2$ & $17$ \\
    Ti & $50$ & $32$ & $-7$ & $-50$ & $-78$ & $-83$ & $-60$ & $-24$ & $6$ & $17$ \\
    V  & $22$ & $0$ & $-46$ & $-88$ & $-104$ & $-88$ & $-52$ & $-14$ & $6$ & $12$ \\
    Cr & $14$ & $3$ & $-26$ & $-57$ & $-64$ & $-46$ & $-24$ & $-9$ & $3$ & $7$ \\
    Mn & $20$ & $10$ & $-25$ & $-42$ & $-40$ & $-31$ & $-22$ & $-9$ & $3$ & $9$ \\
    Fe & $2$ & $-24$ & $-51$ & $-55$ & $-47$ & $-37$ & $-22$ & $-5$ & $4$ & $52$ \\
    Co & $-21$ & $-46$ & $-62$ & $-57$ & $-42$ & $-27$ & $-11$ & $0$ & $7$ & $0$ \\
    Ni & $-18$ & $-29$ & $-33$ & $-27$ & $-18$ & $-9$ & $-3$ & $2$ & $1$ & $-3$ \\
    Cu & $0$ & $0$ & $0$ & $0$ & $0$ & $0$ & $0$ & $0$ & $0$ & $0$ \\
    Zn & $16$ & $16$ & $10$ & $4$ & $-1$ & $-5$ & $-7$ & $-4$ & $2$ & $11$ \\
    \hline
    Y  & $59$ & $61$ & $44$ & $15$ & $-10$ & $-32$ & $-35$ & $-21$ & $3$ & $20$ \\
    Zr & $61$ & $49$ & $10$ & $-29$ & $-56$ & $-67$ & $-54$ & $-24$ & $7$ & $23$ \\
    Nb & $44$ & $10$ & $-41$ & $-79$ & $-96$ & $-90$ & $-60$ & $-20$ & $11$ & $20$ \\
    Mo & $15$ & $-29$ & $-79$ & $-106$ & $-107$ & $-87$ & $-50$ & $-9$ & $12$ & $14$ \\
    Tc & $-10$ & $-56$ & $-96$ & $-107$ & $-96$ & $-69$ & $-30$ & $0$ & $12$ & $9$ \\
    Ru & $-32$ & $-67$ & $-90$ & $-87$ & $-69$ & $-41$ & $-10$ & $6$ & $9$ & $1$ \\
    Rh & $-35$ & $-54$ & $-60$ & $-50$ & $-30$ & $-10$ & $2$ & $7$ & $5$ & $-3$ \\
    Pd & $-21$ & $-24$ & $-20$ & $-9$ & $0$ & $6$ & $7$ & $6$ & $1$ & $-4$ \\
    Ag & $3$ & $7$ & $11$ & $12$ & $12$ & $9$ & $5$ & $1$ & $-1$ & $1$ \\
    Cd & $20$ & $23$ & $20$ & $14$ & $9$ & $1$ & $-3$ & $-4$ & $1$ & $11$ \\
    \hline
    La & $44$ & $46$ & $22$ & $-8$ & $-36$ & $-50$ & $-44$ & $-22$ & $6$ & $24$ \\
    Hf & $66$ & $54$ & $19$ & $-23$ & $-52$ & $-66$ & $-55$ & $-25$ & $8$ & $26$ \\
    Ta & $56$ & $24$ & $-29$ & $-71$ & $-92$ & $-90$ & $-64$ & $-23$ & $12$ & $25$ \\
    W  & $29$ & $-17$ & $-71$ & $-103$ & $-109$ & $-93$ & $-57$ & $-13$ & $15$ & $21$ \\
    Re & $2$ & $-47$ & $-92$ & $-107$ & $-102$ & $-78$ & $-39$ & $-1$ & $16$ & $16$ \\
    Os & $-23$ & $-65$ & $-94$ & $-95$ & $-81$ & $-53$ & $-18$ & $6$ & $13$ & $8$ \\
    Ir & $-33$ & $-59$ & $-71$ & $-63$ & $-44$ & $-20$ & $-1$ & $8$ & $9$ & $2$ \\
    Pt & $-25$ & $-35$ & $-34$ & $-23$ & $-10$ & $0$ & $5$ & $8$ & $5$ & $1$ \\
    Au & $-7$ & $-5$ & $-1$ & $4$ & $7$ & $7$ & $6$ & $4$ & $3$ & $4$ \\
    Hg & $9$ & $13$ & $14$ & $11$ & $9$ & $3$ & $0$ & $0$ & $3$ & $11$ \\
    \hline\hline
  \end{tabular}
  \label{tab:bulk_agg_cu100_4d}
\end{table}

\begin{table}[ht]
  \centering
  \renewcommand{\arraystretch}{1.2}
  \caption{\textbf{Aggregation energies} for \textbf{dual-atom} sites in the \textbf{bulk} layer of \textbf{Cu(100)} for all combinations of \textbf{5\textit{d}} transition metals (columns) and all other transition metals (rows), as obtained from initial \textbf{high-spin} configurations. Energies are reported in kJ~mol$^{-1}$.}
  \vspace{8pt}
  \begin{tabular}{@{\hspace{8pt}}*{11}{c@{\hspace{8pt}}}}
    \hline\hline
    TM & La & Hf & Ta & W & Re & Os & Ir & Pt & Au & Hg \\
    \hline
    Sc & $43$ & $55$ & $43$ & $15$ & $-14$ & $-39$ & $-45$ & $-32$ & $-10$ & $6$ \\
    Ti & $37$ & $37$ & $5$ & $-40$ & $-70$ & $-84$ & $-69$ & $-38$ & $-7$ & $8$ \\
    V  & $12$ & $4$ & $-34$ & $-79$ & $-98$ & $-94$ & $-65$ & $-28$ & $-4$ & $7$ \\
    Cr & $7$ & $5$ & $-17$ & $-52$ & $-63$ & $-55$ & $-32$ & $-16$ & $-4$ & $3$ \\
    Mn & $16$ & $11$ & $-19$ & $-42$ & $-43$ & $-38$ & $-28$ & $-15$ & $-2$ & $5$ \\
    Fe & $-6$ & $-23$ & $-49$ & $-60$ & $-53$ & $-44$ & $-29$ & $-12$ & $-1$ & $2$ \\
    Co & $-35$ & $-45$ & $-64$ & $-64$ & $-51$ & $-35$ & $-18$ & $-2$ & $4$ & $0$ \\
    Ni & $-24$ & $-30$ & $-36$ & $-32$ & $-23$ & $-15$ & $-6$ & $-1$ & $1$ & $-2$ \\
    Cu & $0$ & $0$ & $0$ & $0$ & $0$ & $0$ & $0$ & $0$ & $0$ & $0$ \\
    Zn & $18$ & $18$ & $14$ & $8$ & $3$ & $-2$ & $-5$ & $-3$ & $1$ & $10$ \\
    \hline
    Y  & $44$ & $66$ & $56$ & $29$ & $2$ & $-23$ & $-33$ & $-25$ & $-7$ & $9$ \\
    Zr & $46$ & $54$ & $24$ & $-17$ & $-47$ & $-65$ & $-59$ & $-35$ & $-5$ & $13$ \\
    Nb & $22$ & $19$ & $-29$ & $-71$ & $-92$ & $-94$ & $-71$ & $-34$ & $-1$ & $14$ \\
    Mo & $-8$ & $-23$ & $-71$ & $-103$ & $-107$ & $-95$ & $-63$ & $-23$ & $4$ & $11$ \\
    Tc & $-36$ & $-52$ & $-92$ & $-109$ & $-102$ & $-81$ & $-44$ & $-10$ & $7$ & $9$ \\
    Ru & $-50$ & $-66$ & $-90$ & $-93$ & $-78$ & $-53$ & $-20$ & $0$ & $7$ & $3$ \\
    Rh & $-44$ & $-55$ & $-64$ & $-57$ & $-39$ & $-18$ & $-1$ & $5$ & $6$ & $0$ \\
    Pd & $-22$ & $-25$ & $-23$ & $-13$ & $-1$ & $6$ & $8$ & $8$ & $4$ & $0$ \\
    Ag & $6$ & $8$ & $12$ & $15$ & $16$ & $13$ & $9$ & $5$ & $3$ & $3$ \\
    Cd & $24$ & $26$ & $25$ & $21$ & $16$ & $8$ & $2$ & $1$ & $4$ & $11$ \\
    \hline
    La & $26$ & $58$ & $45$ & $18$ & $-8$ & $-29$ & $-32$ & $-21$ & $0$ & $15$ \\
    Hf & $58$ & $57$ & $30$ & $-12$ & $-44$ & $-64$ & $-61$ & $-38$ & $-6$ & $16$ \\
    Ta & $45$ & $30$ & $-18$ & $-64$ & $-88$ & $-95$ & $-75$ & $-38$ & $0$ & $19$ \\
    W  & $18$ & $-12$ & $-64$ & $-100$ & $-110$ & $-102$ & $-71$ & $-28$ & $6$ & $19$ \\
    Re & $-8$ & $-44$ & $-88$ & $-110$ & $-107$ & $-90$ & $-52$ & $-12$ & $10$ & $17$ \\
    Os & $-29$ & $-64$ & $-95$ & $-102$ & $-90$ & $-66$ & $-29$ & $-1$ & $12$ & $12$ \\
    Ir & $-32$ & $-61$ & $-75$ & $-71$ & $-52$ & $-29$ & $-6$ & $8$ & $11$ & $8$ \\
    Pt & $-21$ & $-38$ & $-38$ & $-28$ & $-12$ & $-1$ & $8$ & $10$ & $11$ & $8$ \\
    Au & $0$ & $-6$ & $0$ & $6$ & $10$ & $12$ & $11$ & $11$ & $10$ & $10$ \\
    Hg & $15$ & $16$ & $19$ & $19$ & $17$ & $12$ & $8$ & $8$ & $10$ & $16$ \\
    \hline\hline
  \end{tabular}
  \label{tab:bulk_agg_cu100_5d}
\end{table}

\begin{table}[ht]
  \centering
  \renewcommand{\arraystretch}{1.2}
  \caption{\textbf{Aggregation energies} for \textbf{dual-atom} sites in the \textbf{bulk} layer of \textbf{Cu(100)} for all combinations of \textbf{3\textit{d}} transition metals (columns) and all other transition metals (rows), as obtained from initial \textbf{low-spin} configurations. Energies are reported in kJ~mol$^{-1}$.}
  \vspace{8pt}
  \begin{tabular}{@{\hspace{8pt}}*{11}{c@{\hspace{8pt}}}}
    \hline\hline
    TM & Sc & Ti & V & Cr & Mn & Fe & Co & Ni & Cu & Zn \\
    \hline
    Sc & $49$ & $42$ & $26$ & $2$ & $-15$ & $-29$ & $-31$ & $-21$ & $0$ & $15$ \\
    Ti & $42$ & $22$ & $-17$ & $-52$ & $-74$ & $-77$ & $-59$ & $-30$ & $0$ & $12$ \\
    V  & $26$ & $-17$ & $-72$ & $-112$ & $-121$ & $-102$ & $-66$ & $-29$ & $0$ & $6$ \\
    Cr & $2$ & $-52$ & $-112$ & $-138$ & $-124$ & $-92$ & $-56$ & $-23$ & $139$ & $2$ \\
    Mn & $-15$ & $-74$ & $-121$ & $-124$ & $-101$ & $-71$ & $-41$ & $-17$ & $0$ & $0$ \\
    Fe & $-29$ & $-77$ & $-102$ & $-92$ & $-71$ & $-49$ & $-28$ & $-11$ & $0$ & $-2$ \\
    Co & $-31$ & $-59$ & $-66$ & $-56$ & $-41$ & $-28$ & $-16$ & $-7$ & $0$ & $-4$ \\
    Ni & $-21$ & $-30$ & $-29$ & $-23$ & $-17$ & $-11$ & $-7$ & $-2$ & $0$ & $-4$ \\
    Cu & $0$ & $0$ & $0$ & $139$ & $0$ & $0$ & $0$ & $0$ & $0$ & $0$ \\
    Zn & $15$ & $12$ & $6$ & $2$ & $0$ & $-2$ & $-4$ & $-4$ & $0$ & $11$ \\
    \hline
    Y  & $54$ & $50$ & $33$ & $11$ & $-6$ & $-20$ & $-25$ & $-18$ & $0$ & $16$ \\
    Zr & $53$ & $35$ & $0$ & $-31$ & $-50$ & $-58$ & $-49$ & $-29$ & $0$ & $16$ \\
    Nb & $35$ & $-7$ & $-55$ & $-86$ & $-96$ & $-89$ & $-66$ & $-33$ & $0$ & $10$ \\
    Mo & $5$ & $-50$ & $-97$ & $-117$ & $-111$ & $-91$ & $-61$ & $-27$ & $0$ & $4$ \\
    Tc & $-22$ & $-78$ & $-113$ & $-114$ & $-98$ & $-74$ & $-45$ & $-18$ & $0$ & $-1$ \\
    Ru & $-43$ & $-83$ & $-97$ & $-87$ & $-68$ & $-48$ & $-26$ & $-9$ & $0$ & $-5$ \\
    Rh & $-43$ & $-60$ & $-59$ & $-46$ & $-31$ & $-19$ & $-9$ & $-3$ & $0$ & $-7$ \\
    Pd & $-23$ & $-24$ & $-18$ & $-10$ & $-4$ & $-1$ & $1$ & $2$ & $0$ & $-4$ \\
    Ag & $2$ & $6$ & $8$ & $8$ & $7$ & $6$ & $4$ & $1$ & $0$ & $2$ \\
    Cd & $17$ & $17$ & $12$ & $8$ & $5$ & $2$ & $-2$ & $-3$ & $0$ & $11$ \\
    \hline
    La & $43$ & $37$ & $11$ & $-15$ & $-37$ & $-47$ & $-39$ & $-24$ & $0$ & $18$ \\
    Hf & $55$ & $39$ & $7$ & $-24$ & $-44$ & $-54$ & $-49$ & $-30$ & $0$ & $18$ \\
    Ta & $43$ & $5$ & $-43$ & $-75$ & $-88$ & $-86$ & $-67$ & $-36$ & $0$ & $14$ \\
    W  & $15$ & $-40$ & $-88$ & $-110$ & $-110$ & $-96$ & $-68$ & $-32$ & $0$ & $8$ \\
    Re & $-14$ & $-70$ & $-107$ & $-114$ & $-103$ & $-83$ & $-54$ & $-23$ & $0$ & $3$ \\
    Os & $-39$ & $-84$ & $-103$ & $-98$ & $-81$ & $-62$ & $-37$ & $-15$ & $0$ & $-2$ \\
    Ir & $-45$ & $-69$ & $-73$ & $-62$ & $-46$ & $-31$ & $-17$ & $-6$ & $0$ & $-5$ \\
    Pt & $-32$ & $-38$ & $-33$ & $-24$ & $-16$ & $-9$ & $-5$ & $-1$ & $0$ & $-3$ \\
    Au & $-10$ & $-7$ & $-4$ & $-1$ & $0$ & $1$ & $0$ & $1$ & $0$ & $1$ \\
    Hg & $6$ & $8$ & $6$ & $4$ & $3$ & $0$ & $-2$ & $-2$ & $0$ & $10$ \\
    \hline\hline
  \end{tabular}
  \label{tab:bulk_agg_cu100_ls_3d}
\end{table}

\begin{table}[ht]
  \centering
  \renewcommand{\arraystretch}{1.2}
  \caption{\textbf{Aggregation energies} for \textbf{dual-atom} sites in the \textbf{bulk} layer of \textbf{Cu(100)} for all combinations of \textbf{4\textit{d}} transition metals (columns) and all other transition metals (rows), as obtained from initial \textbf{low-spin} configurations. Energies are reported in kJ~mol$^{-1}$.}

  \vspace{8pt}
  \begin{tabular}{@{\hspace{8pt}}*{11}{c@{\hspace{8pt}}}}
    \hline\hline
    TM & Y & Zr & Nb & Mo & Tc & Ru & Rh & Pd & Ag & Cd \\
    \hline
    Sc & $54$ & $53$ & $35$ & $5$ & $-22$ & $-43$ & $-43$ & $-23$ & $2$ & $17$ \\
    Ti & $50$ & $35$ & $-7$ & $-50$ & $-78$ & $-83$ & $-60$ & $-24$ & $6$ & $17$ \\
    V  & $33$ & $0$ & $-55$ & $-97$ & $-113$ & $-97$ & $-59$ & $-18$ & $8$ & $12$ \\
    Cr & $11$ & $-31$ & $-86$ & $-117$ & $-114$ & $-87$ & $-46$ & $-10$ & $8$ & $8$ \\
    Mn & $-6$ & $-50$ & $-96$ & $-111$ & $-98$ & $-68$ & $-31$ & $-4$ & $7$ & $5$ \\
    Fe & $-20$ & $-58$ & $-89$ & $-91$ & $-74$ & $-48$ & $-19$ & $-1$ & $6$ & $2$ \\
    Co & $-25$ & $-49$ & $-66$ & $-61$ & $-45$ & $-26$ & $-9$ & $1$ & $4$ & $-2$ \\
    Ni & $-18$ & $-29$ & $-33$ & $-27$ & $-18$ & $-9$ & $-3$ & $2$ & $1$ & $-3$ \\
    Cu & $0$ & $0$ & $0$ & $0$ & $0$ & $0$ & $0$ & $0$ & $0$ & $0$ \\
    Zn & $16$ & $16$ & $10$ & $4$ & $-1$ & $-5$ & $-7$ & $-4$ & $2$ & $11$ \\
    \hline
    Y  & $59$ & $61$ & $44$ & $15$ & $-10$ & $-32$ & $-35$ & $-21$ & $3$ & $20$ \\
    Zr & $61$ & $49$ & $10$ & $-29$ & $-56$ & $-67$ & $-54$ & $-24$ & $7$ & $23$ \\
    Nb & $44$ & $10$ & $-41$ & $-79$ & $-96$ & $-90$ & $-60$ & $-20$ & $11$ & $20$ \\
    Mo & $15$ & $-29$ & $-79$ & $-106$ & $-107$ & $-87$ & $-50$ & $-9$ & $12$ & $14$ \\
    Tc & $-10$ & $-56$ & $-96$ & $-107$ & $-96$ & $-69$ & $-30$ & $0$ & $12$ & $9$ \\
    Ru & $-32$ & $-67$ & $-90$ & $-87$ & $-69$ & $-41$ & $-10$ & $6$ & $9$ & $1$ \\
    Rh & $-35$ & $-54$ & $-60$ & $-50$ & $-30$ & $-10$ & $2$ & $7$ & $5$ & $-3$ \\
    Pd & $-21$ & $-24$ & $-20$ & $-9$ & $0$ & $6$ & $7$ & $6$ & $1$ & $-4$ \\
    Ag & $3$ & $7$ & $11$ & $12$ & $12$ & $9$ & $5$ & $1$ & $-1$ & $1$ \\
    Cd & $20$ & $23$ & $20$ & $14$ & $9$ & $1$ & $-3$ & $-4$ & $1$ & $11$ \\
    \hline
    La & $44$ & $46$ & $22$ & $-8$ & $-36$ & $-50$ & $-44$ & $-22$ & $6$ & $24$ \\
    Hf & $66$ & $54$ & $19$ & $-23$ & $-52$ & $-66$ & $-55$ & $-25$ & $8$ & $26$ \\
    Ta & $56$ & $24$ & $-29$ & $-71$ & $-92$ & $-90$ & $-64$ & $-23$ & $12$ & $25$ \\
    W  & $29$ & $-17$ & $-71$ & $-103$ & $-109$ & $-93$ & $-57$ & $-13$ & $15$ & $21$ \\
    Re & $2$ & $-47$ & $-92$ & $-107$ & $-102$ & $-78$ & $-39$ & $-1$ & $16$ & $16$ \\
    Os & $-23$ & $-65$ & $-94$ & $-95$ & $-81$ & $-53$ & $-18$ & $6$ & $13$ & $8$ \\
    Ir & $-33$ & $-59$ & $-71$ & $-63$ & $-44$ & $-20$ & $-1$ & $8$ & $9$ & $2$ \\
    Pt & $-25$ & $-35$ & $-34$ & $-23$ & $-10$ & $0$ & $5$ & $8$ & $5$ & $1$ \\
    Au & $-7$ & $-5$ & $-1$ & $4$ & $7$ & $7$ & $6$ & $4$ & $3$ & $4$ \\
    Hg & $9$ & $13$ & $14$ & $11$ & $9$ & $3$ & $0$ & $0$ & $3$ & $11$ \\
    \hline\hline
  \end{tabular}
  \label{tab:bulk_agg_cu100_ls_4d}
\end{table}

\begin{table}[ht]
  \centering
  \renewcommand{\arraystretch}{1.2}
  \caption{\textbf{Aggregation energies} for \textbf{dual-atom} sites in the \textbf{bulk} layer of \textbf{Cu(100)} for all combinations of \textbf{5\textit{d}} transition metals (columns) and all other transition metals (rows), as obtained from initial \textbf{low-spin} configurations. Energies are reported in kJ~mol$^{-1}$.}

  \vspace{8pt}
  \begin{tabular}{@{\hspace{8pt}}*{11}{c@{\hspace{8pt}}}}
    \hline\hline
    TM & La & Hf & Ta & W & Re & Os & Ir & Pt & Au & Hg \\
    \hline
    Sc & $43$ & $55$ & $43$ & $15$ & $-14$ & $-39$ & $-45$ & $-32$ & $-10$ & $6$ \\
    Ti & $37$ & $39$ & $5$ & $-40$ & $-70$ & $-84$ & $-69$ & $-38$ & $-7$ & $8$ \\
    V  & $11$ & $7$ & $-43$ & $-88$ & $-107$ & $-103$ & $-73$ & $-33$ & $-4$ & $6$ \\
    Cr & $-15$ & $-24$ & $-75$ & $-110$ & $-114$ & $-98$ & $-62$ & $-24$ & $-1$ & $4$ \\
    Mn & $-37$ & $-44$ & $-88$ & $-110$ & $-103$ & $-81$ & $-46$ & $-16$ & $0$ & $3$ \\
    Fe & $-47$ & $-54$ & $-86$ & $-96$ & $-83$ & $-62$ & $-31$ & $-9$ & $1$ & $0$ \\
    Co & $-39$ & $-49$ & $-67$ & $-68$ & $-54$ & $-37$ & $-17$ & $-5$ & $0$ & $-2$ \\
    Ni & $-24$ & $-30$ & $-36$ & $-32$ & $-23$ & $-15$ & $-6$ & $-1$ & $1$ & $-2$ \\
    Cu & $0$ & $0$ & $0$ & $0$ & $0$ & $0$ & $0$ & $0$ & $0$ & $0$ \\
    Zn & $18$ & $18$ & $14$ & $8$ & $3$ & $-2$ & $-5$ & $-3$ & $1$ & $10$ \\
    \hline
    Y  & $44$ & $66$ & $56$ & $29$ & $2$ & $-23$ & $-33$ & $-25$ & $-7$ & $9$ \\
    Zr & $46$ & $54$ & $24$ & $-17$ & $-47$ & $-65$ & $-59$ & $-35$ & $-5$ & $13$ \\
    Nb & $22$ & $19$ & $-29$ & $-71$ & $-92$ & $-94$ & $-71$ & $-34$ & $-1$ & $14$ \\
    Mo & $-8$ & $-23$ & $-71$ & $-103$ & $-107$ & $-95$ & $-63$ & $-23$ & $4$ & $11$ \\
    Tc & $-36$ & $-52$ & $-92$ & $-109$ & $-102$ & $-81$ & $-44$ & $-10$ & $7$ & $9$ \\
    Ru & $-50$ & $-66$ & $-90$ & $-93$ & $-78$ & $-53$ & $-20$ & $0$ & $7$ & $3$ \\
    Rh & $-44$ & $-55$ & $-64$ & $-57$ & $-39$ & $-18$ & $-1$ & $5$ & $6$ & $0$ \\
    Pd & $-22$ & $-25$ & $-23$ & $-13$ & $-1$ & $6$ & $8$ & $8$ & $4$ & $0$ \\
    Ag & $6$ & $8$ & $12$ & $15$ & $16$ & $13$ & $9$ & $5$ & $3$ & $3$ \\
    Cd & $24$ & $26$ & $25$ & $21$ & $16$ & $8$ & $2$ & $1$ & $4$ & $11$ \\
    \hline
    La & $26$ & $58$ & $45$ & $18$ & $-8$ & $-29$ & $-32$ & $-21$ & $0$ & $15$ \\
    Hf & $58$ & $57$ & $30$ & $-12$ & $-44$ & $-64$ & $-61$ & $-38$ & $-6$ & $16$ \\
    Ta & $45$ & $30$ & $-18$ & $-64$ & $-88$ & $-95$ & $-75$ & $-38$ & $0$ & $19$ \\
    W  & $18$ & $-12$ & $-64$ & $-100$ & $-110$ & $-102$ & $-71$ & $-28$ & $6$ & $19$ \\
    Re & $-8$ & $-44$ & $-88$ & $-110$ & $-107$ & $-90$ & $-52$ & $-12$ & $10$ & $17$ \\
    Os & $-29$ & $-64$ & $-95$ & $-102$ & $-90$ & $-66$ & $-29$ & $-1$ & $12$ & $12$ \\
    Ir & $-32$ & $-61$ & $-75$ & $-71$ & $-52$ & $-29$ & $-6$ & $8$ & $11$ & $8$ \\
    Pt & $-21$ & $-38$ & $-38$ & $-28$ & $-12$ & $-1$ & $8$ & $10$ & $11$ & $8$ \\
    Au & $0$ & $-6$ & $0$ & $6$ & $10$ & $12$ & $11$ & $11$ & $10$ & $10$ \\
    Hg & $15$ & $16$ & $19$ & $19$ & $17$ & $12$ & $8$ & $8$ & $10$ & $16$ \\
    \hline\hline
  \end{tabular}
  \label{tab:bulk_agg_cu100_ls_5d}
\end{table}

\clearpage


\begin{figure}[htbp]
    \includegraphics[width=12cm,height=\textheight,keepaspectratio]{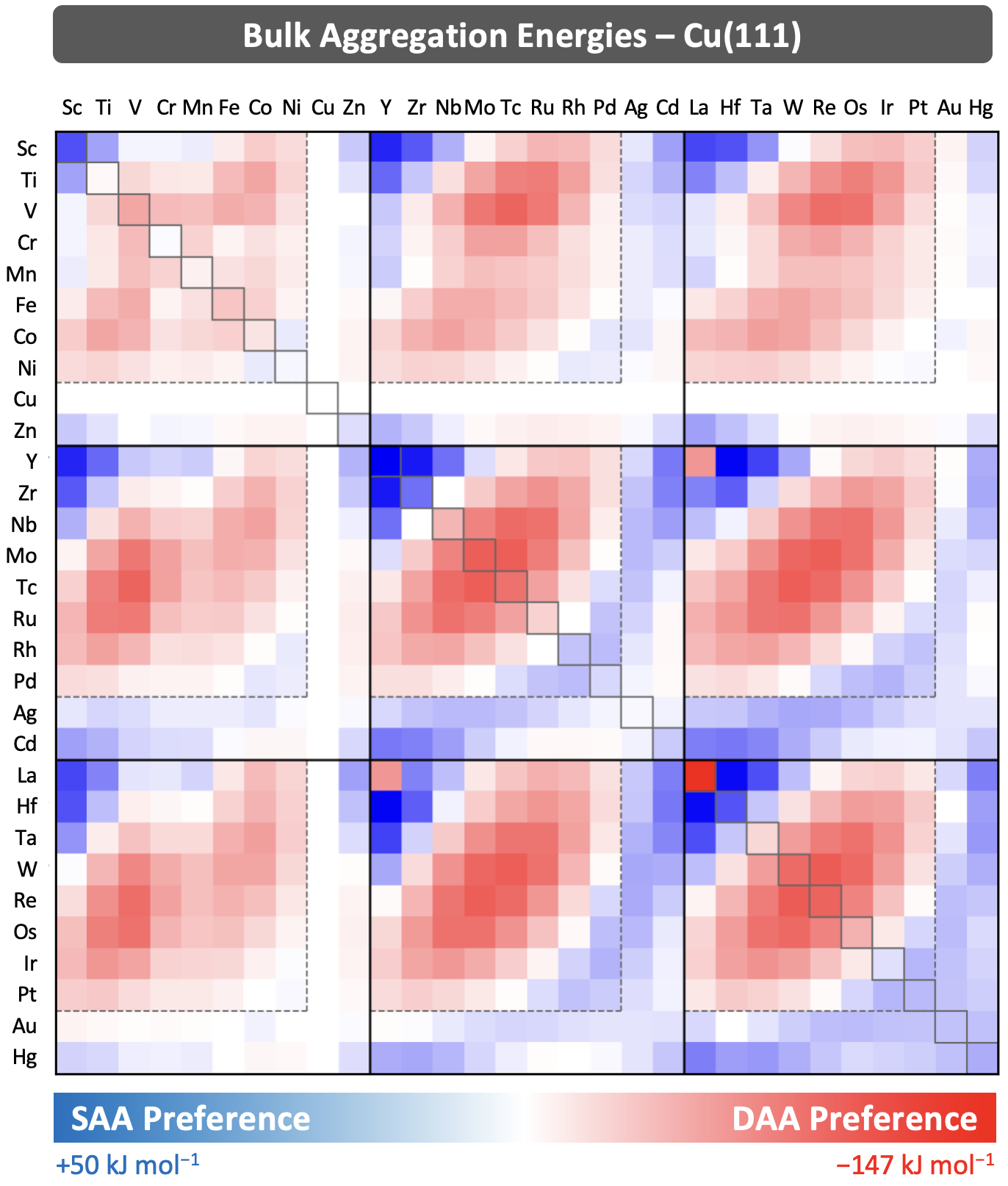}
    \caption{
        Aggregation energies for the formation of all possible transition metal dopant dimers in a bulk layer of the Cu(111) system. 
        Preferences for single-atom sites are represented in blue, while preferences for dopant dimers are shown in red. 
    }
    \label{fig:Cu111-bulk-aggr}
\end{figure}

\begin{table}[ht]
  \centering
  \renewcommand{\arraystretch}{1.2}
  \caption{\textbf{Aggregation energies} for \textbf{dual-atom} sites in the \textbf{bulk} layer of \textbf{Cu(111)} for all combinations of \textbf{3\textit{d}} transition metals (columns) and all other transition metals (rows), as obtained from initial \textbf{high-spin} configurations. Energies are reported in kJ~mol$^{-1}$.}

  \vspace{8pt}
  \begin{tabular}{@{\hspace{8pt}}*{11}{c@{\hspace{8pt}}}}
    \hline\hline
    TM & Sc & Ti & V & Cr & Mn & Fe & Co & Ni & Cu & Zn \\
    \hline
    Sc & $35$ & $18$ & $2$ & $2$ & $4$ & $-12$ & $-30$ & $-21$ & $0$ & $11$ \\
    Ti & $18$ & $-4$ & $-24$ & $-15$ & $-13$ & $-40$ & $-53$ & $-26$ & $0$ & $6$ \\
    V  & $2$ & $-24$ & $-53$ & $-40$ & $-39$ & $-50$ & $-46$ & $-18$ & $0$ & $0$ \\
    Cr & $2$ & $-15$ & $-40$ & $1$ & $-27$ & $-6$ & $-18$ & $-10$ & $0$ & $2$ \\
    Mn & $4$ & $-13$ & $-39$ & $-27$ & $-9$ & $-17$ & $-24$ & $-12$ & $0$ & $2$ \\
    Fe & $-12$ & $-40$ & $-50$ & $-6$ & $-17$ & $-34$ & $-29$ & $-8$ & $0$ & $-4$ \\
    Co & $-30$ & $-53$ & $-46$ & $-18$ & $-24$ & $-29$ & $-15$ & $4$ & $0$ & $-8$ \\
    Ni & $-21$ & $-26$ & $-18$ & $-10$ & $-12$ & $-8$ & $4$ & $2$ & $0$ & $-6$ \\
    Cu & $0$ & $0$ & $0$ & $0$ & $0$ & $0$ & $0$ & $0$ & $0$ & $0$ \\
    Zn & $11$ & $6$ & $0$ & $2$ & $2$ & $-4$ & $-8$ & $-6$ & $0$ & $6$ \\
    \hline
    Y  & $43$ & $30$ & $11$ & $8$ & $10$ & $-5$ & $-25$ & $-20$ & $0$ & $15$ \\
    Zr & $33$ & $11$ & $-12$ & $-6$ & $-2$ & $-29$ & $-46$ & $-27$ & $0$ & $11$ \\
    Nb & $16$ & $-19$ & $-47$ & $-29$ & $-27$ & $-48$ & $-55$ & $-25$ & $0$ & $3$ \\
    Mo & $-6$ & $-53$ & $-83$ & $-55$ & $-38$ & $-49$ & $-46$ & $-18$ & $0$ & $-5$ \\
    Tc & $-28$ & $-76$ & $-96$ & $-57$ & $-35$ & $-41$ & $-31$ & $-9$ & $0$ & $-9$ \\
    Ru & $-44$ & $-80$ & $-82$ & $-39$ & $-30$ & $-32$ & $-17$ & $-2$ & $0$ & $-11$ \\
    Rh & $-41$ & $-56$ & $-43$ & $-19$ & $-20$ & $-17$ & $-2$ & $4$ & $0$ & $-9$ \\
    Pd & $-21$ & $-18$ & $-9$ & $-6$ & $-7$ & $-2$ & $5$ & $4$ & $0$ & $-6$ \\
    Ag & $5$ & $8$ & $7$ & $4$ & $4$ & $4$ & $5$ & $1$ & $0$ & $1$ \\
    Cd & $19$ & $15$ & $8$ & $7$ & $6$ & $1$ & $-5$ & $-5$ & $0$ & $7$ \\
    \hline
    La & $37$ & $24$ & $5$ & $5$ & $9$ & $-14$ & $-41$ & $-26$ & $0$ & $19$ \\
    Hf & $34$ & $13$ & $-10$ & $-5$ & $-2$ & $-27$ & $-46$ & $-28$ & $0$ & $13$ \\
    Ta & $21$ & $-11$ & $-37$ & $-22$ & $-22$ & $-47$ & $-58$ & $-29$ & $0$ & $7$ \\
    W  & $1$ & $-43$ & $-73$ & $-50$ & $-39$ & $-53$ & $-54$ & $-23$ & $0$ & $-1$ \\
    Re & $-20$ & $-67$ & $-89$ & $-56$ & $-38$ & $-46$ & $-38$ & $-14$ & $0$ & $-6$ \\
    Os & $-39$ & $-78$ & $-85$ & $-45$ & $-34$ & $-37$ & $-24$ & $-6$ & $0$ & $-9$ \\
    Ir & $-42$ & $-64$ & $-55$ & $-27$ & $-25$ & $-23$ & $-8$ & $0$ & $0$ & $-7$ \\
    Pt & $-29$ & $-33$ & $-21$ & $-13$ & $-13$ & $-9$ & $-1$ & $1$ & $0$ & $-5$ \\
    Au & $-7$ & $-4$ & $-2$ & $-3$ & $-2$ & $0$ & $2$ & $0$ & $0$ & $1$ \\
    Hg & $9$ & $8$ & $3$ & $3$ & $4$ & $0$ & $-5$ & $-4$ & $0$ & $7$ \\
    \hline\hline
  \end{tabular}
  \label{tab:bulk_agg_cu111_3d}
\end{table}

\begin{table}[ht]
  \centering
  \renewcommand{\arraystretch}{1.2}
  \caption{\textbf{Aggregation energies} for \textbf{dual-atom} sites in the \textbf{bulk} layer of \textbf{Cu(111)} for all combinations of \textbf{4\textit{d}} transition metals (columns) and all other transition metals (rows), as obtained from initial \textbf{high-spin} configurations. Energies are reported in kJ~mol$^{-1}$.}
  \vspace{8pt}
  \begin{tabular}{@{\hspace{8pt}}*{11}{c@{\hspace{8pt}}}}
    \hline\hline
    TM & Y & Zr & Nb & Mo & Tc & Ru & Rh & Pd & Ag & Cd \\
    \hline
    Sc & $43$ & $33$ & $16$ & $-6$ & $-28$ & $-44$ & $-41$ & $-21$ & $5$ & $19$ \\
    Ti & $30$ & $11$ & $-19$ & $-53$ & $-76$ & $-80$ & $-56$ & $-18$ & $8$ & $15$ \\
    V  & $11$ & $-12$ & $-47$ & $-83$ & $-96$ & $-82$ & $-43$ & $-9$ & $7$ & $8$ \\
    Cr & $8$ & $-6$ & $-29$ & $-55$ & $-57$ & $-39$ & $-19$ & $-6$ & $4$ & $7$ \\
    Mn & $10$ & $-2$ & $-27$ & $-38$ & $-35$ & $-30$ & $-20$ & $-7$ & $4$ & $6$ \\
    Fe & $-5$ & $-29$ & $-48$ & $-49$ & $-41$ & $-32$ & $-17$ & $-2$ & $4$ & $1$ \\
    Co & $-25$ & $-46$ & $-55$ & $-46$ & $-31$ & $-17$ & $-2$ & $5$ & $5$ & $-5$ \\
    Ni & $-20$ & $-27$ & $-25$ & $-18$ & $-9$ & $-2$ & $4$ & $4$ & $1$ & $-5$ \\
    Cu & $0$ & $0$ & $0$ & $0$ & $0$ & $0$ & $0$ & $0$ & $0$ & $0$ \\
    Zn & $15$ & $11$ & $3$ & $-5$ & $-9$ & $-11$ & $-9$ & $-6$ & $1$ & $7$ \\
    \hline
    Y  & $50$ & $45$ & $28$ & $6$ & $-14$ & $-33$ & $-34$ & $-19$ & $8$ & $26$ \\
    Zr & $45$ & $28$ & $-1$ & $-32$ & $-55$ & $-65$ & $-51$ & $-19$ & $12$ & $25$ \\
    Nb & $28$ & $-1$ & $-43$ & $-75$ & $-91$ & $-85$ & $-53$ & $-11$ & $13$ & $19$ \\
    Mo & $6$ & $-32$ & $-75$ & $-99$ & $-101$ & $-79$ & $-38$ & $-2$ & $13$ & $10$ \\
    Tc & $-14$ & $-55$ & $-91$ & $-101$ & $-87$ & $-56$ & $-18$ & $7$ & $12$ & $3$ \\
    Ru & $-33$ & $-65$ & $-85$ & $-79$ & $-56$ & $-27$ & $0$ & $11$ & $9$ & $-4$ \\
    Rh & $-34$ & $-51$ & $-53$ & $-38$ & $-18$ & $0$ & $12$ & $13$ & $5$ & $-4$ \\
    Pd & $-19$ & $-19$ & $-11$ & $-2$ & $7$ & $11$ & $13$ & $8$ & $2$ & $-3$ \\
    Ag & $8$ & $12$ & $13$ & $13$ & $12$ & $9$ & $5$ & $2$ & $1$ & $3$ \\
    Cd & $26$ & $25$ & $19$ & $10$ & $3$ & $-4$ & $-4$ & $-3$ & $3$ & $11$ \\
    \hline
    La & $-63$ & $25$ & $13$ & $-14$ & $-35$ & $-47$ & $-42$ & $-18$ & $11$ & $25$ \\
    Hf & $49$ & $32$ & $2$ & $-30$ & $-53$ & $-64$ & $-53$ & $-21$ & $11$ & $26$ \\
    Ta & $37$ & $9$ & $-32$ & $-67$ & $-84$ & $-84$ & $-56$ & $-14$ & $15$ & $24$ \\
    W  & $17$ & $-21$ & $-66$ & $-94$ & $-100$ & $-85$ & $-44$ & $-3$ & $17$ & $16$ \\
    Re & $-3$ & $-45$ & $-84$ & $-99$ & $-92$ & $-64$ & $-23$ & $8$ & $17$ & $10$ \\
    Os & $-23$ & $-60$ & $-86$ & $-85$ & $-66$ & $-36$ & $-4$ & $13$ & $14$ & $4$ \\
    Ir & $-29$ & $-54$ & $-62$ & $-50$ & $-29$ & $-8$ & $9$ & $15$ & $10$ & $3$ \\
    Pt & $-22$ & $-29$ & $-24$ & $-13$ & $-1$ & $7$ & $12$ & $10$ & $7$ & $2$ \\
    Au & $-1$ & $1$ & $4$ & $7$ & $8$ & $8$ & $6$ & $5$ & $5$ & $6$ \\
    Hg & $16$ & $17$ & $14$ & $8$ & $4$ & $-1$ & $0$ & $1$ & $5$ & $11$ \\
    \hline\hline
  \end{tabular}
  \label{tab:bulk_agg_cu111_4d}
\end{table}

\begin{table}[ht]
  \centering
  \renewcommand{\arraystretch}{1.2}
  \caption{\textbf{Aggregation energies} for \textbf{dual-atom} sites in the \textbf{bulk} layer of \textbf{Cu(111)} for all combinations of \textbf{5\textit{d}} transition metals (columns) and all other transition metals (rows), as obtained from initial \textbf{high-spin} configurations. Energies are reported in kJ~mol$^{-1}$.}
  \vspace{8pt}
  \begin{tabular}{@{\hspace{8pt}}*{11}{c@{\hspace{8pt}}}}
    \hline\hline
    TM & La & Hf & Ta & W & Re & Os & Ir & Pt & Au & Hg \\
    \hline
    Sc & $37$ & $34$ & $21$ & $1$ & $-20$ & $-39$ & $-42$ & $-29$ & $-7$ & $9$ \\
    Ti & $24$ & $13$ & $-11$ & $-43$ & $-67$ & $-78$ & $-64$ & $-33$ & $-4$ & $8$ \\
    V  & $5$ & $-10$ & $-37$ & $-73$ & $-89$ & $-85$ & $-55$ & $-21$ & $-2$ & $3$ \\
    Cr & $5$ & $-5$ & $-22$ & $-50$ & $-56$ & $-45$ & $-27$ & $-13$ & $-3$ & $3$ \\
    Mn & $9$ & $-2$ & $-22$ & $-39$ & $-38$ & $-34$ & $-25$ & $-13$ & $-2$ & $4$ \\
    Fe & $-14$ & $-27$ & $-47$ & $-53$ & $-46$ & $-37$ & $-23$ & $-9$ & $0$ & $0$ \\
    Co & $-41$ & $-46$ & $-58$ & $-54$ & $-38$ & $-24$ & $-8$ & $-1$ & $2$ & $-5$ \\
    Ni & $-26$ & $-28$ & $-29$ & $-23$ & $-14$ & $-6$ & $0$ & $1$ & $0$ & $-4$ \\
    Cu & $0$ & $0$ & $0$ & $0$ & $0$ & $0$ & $0$ & $0$ & $0$ & $0$ \\
    Zn & $19$ & $13$ & $7$ & $-1$ & $-6$ & $-9$ & $-7$ & $-5$ & $1$ & $7$ \\
    \hline
    Y  & $-63$ & $49$ & $37$ & $17$ & $-3$ & $-23$ & $-29$ & $-22$ & $-1$ & $16$ \\
    Zr & $25$ & $32$ & $9$ & $-21$ & $-45$ & $-60$ & $-54$ & $-29$ & $1$ & $17$ \\
    Nb & $13$ & $2$ & $-32$ & $-66$ & $-84$ & $-86$ & $-62$ & $-24$ & $4$ & $14$ \\
    Mo & $-14$ & $-30$ & $-67$ & $-94$ & $-99$ & $-85$ & $-50$ & $-13$ & $7$ & $8$ \\
    Tc & $-35$ & $-53$ & $-84$ & $-100$ & $-92$ & $-66$ & $-29$ & $-1$ & $8$ & $4$ \\
    Ru & $-47$ & $-64$ & $-84$ & $-85$ & $-64$ & $-36$ & $-8$ & $7$ & $8$ & $-1$ \\
    Rh & $-42$ & $-53$ & $-56$ & $-44$ & $-23$ & $-4$ & $9$ & $12$ & $6$ & $0$ \\
    Pd & $-18$ & $-21$ & $-14$ & $-3$ & $8$ & $13$ & $15$ & $10$ & $5$ & $1$ \\
    Ag & $11$ & $11$ & $15$ & $17$ & $17$ & $14$ & $10$ & $7$ & $5$ & $5$ \\
    Cd & $25$ & $26$ & $24$ & $16$ & $10$ & $4$ & $3$ & $2$ & $6$ & $11$ \\
    \hline
    La & $-147$ & $48$ & $35$ & $13$ & $-7$ & $-25$ & $-28$ & $-15$ & $8$ & $25$ \\
    Hf & $48$ & $34$ & $11$ & $-18$ & $-43$ & $-60$ & $-56$ & $-32$ & $0$ & $19$ \\
    Ta & $35$ & $11$ & $-24$ & $-59$ & $-80$ & $-86$ & $-66$ & $-28$ & $5$ & $20$ \\
    W  & $13$ & $-18$ & $-59$ & $-91$ & $-101$ & $-92$ & $-58$ & $-17$ & $9$ & $16$ \\
    Re & $-7$ & $-43$ & $-80$ & $-101$ & $-96$ & $-75$ & $-37$ & $-3$ & $13$ & $11$ \\
    Os & $-25$ & $-60$ & $-86$ & $-92$ & $-75$ & $-47$ & $-13$ & $8$ & $13$ & $7$ \\
    Ir & $-28$ & $-56$ & $-66$ & $-58$ & $-37$ & $-13$ & $6$ & $14$ & $12$ & $9$ \\
    Pt & $-15$ & $-32$ & $-28$ & $-17$ & $-3$ & $8$ & $14$ & $14$ & $12$ & $10$ \\
    Au & $8$ & $0$ & $5$ & $9$ & $13$ & $13$ & $12$ & $12$ & $12$ & $12$ \\
    Hg & $25$ & $19$ & $20$ & $16$ & $11$ & $7$ & $9$ & $10$ & $12$ & $16$ \\
    \hline\hline
  \end{tabular}
  \label{tab:bulk_agg_cu111_5d}
\end{table}

\begin{table}[ht]
  \centering
  \renewcommand{\arraystretch}{1.2}
  \caption{\textbf{Aggregation energies} for \textbf{dual-atom} sites in the \textbf{bulk} layer of \textbf{Cu(111)} for all combinations of \textbf{3\textit{d}} transition metals (columns) and all other transition metals (rows), as obtained from initial \textbf{low-spin} configurations. Energies are reported in kJ~mol$^{-1}$.}
  \vspace{8pt}
  \begin{tabular}{@{\hspace{8pt}}*{11}{c@{\hspace{8pt}}}}
    \hline\hline
    TM & Sc & Ti & V & Cr & Mn & Fe & Co & Ni & Cu & Zn \\
    \hline
    Sc & $35$ & $18$ & $4$ & $-11$ & $-24$ & $-34$ & $-33$ & $-21$ & $0$ & $11$ \\
    Ti & $18$ & $-3$ & $-29$ & $-56$ & $-74$ & $-76$ & $-56$ & $-26$ & $0$ & $6$ \\
    V  & $4$ & $-29$ & $-73$ & $-108$ & $-115$ & $-96$ & $-58$ & $-22$ & $0$ & $0$ \\
    Cr & $-11$ & $-56$ & $-108$ & $-131$ & $-118$ & $-86$ & $-46$ & $-15$ & $0$ & $-6$ \\
    Mn & $-24$ & $-74$ & $-115$ & $-118$ & $-94$ & $-61$ & $-29$ & $-9$ & $0$ & $-10$ \\
    Fe & $-34$ & $-76$ & $-96$ & $-86$ & $-61$ & $-36$ & $-15$ & $-2$ & $0$ & $-10$ \\
    Co & $-33$ & $-56$ & $-58$ & $-46$ & $-29$ & $-15$ & $-4$ & $1$ & $0$ & $-8$ \\
    Ni & $-21$ & $-26$ & $-22$ & $-15$ & $-9$ & $-2$ & $1$ & $2$ & $0$ & $-6$ \\
    Cu & $0$ & $0$ & $0$ & $0$ & $0$ & $0$ & $0$ & $0$ & $0$ & $0$ \\
    Zn & $11$ & $6$ & $0$ & $-6$ & $-10$ & $-10$ & $-8$ & $-6$ & $0$ & $6$ \\
    \hline
    Y  & $43$ & $30$ & $14$ & $-3$ & $-16$ & $-27$ & $-28$ & $-20$ & $0$ & $15$ \\
    Zr & $33$ & $11$ & $-15$ & $-39$ & $-54$ & $-60$ & $-49$ & $-27$ & $0$ & $11$ \\
    Nb & $16$ & $-19$ & $-57$ & $-84$ & $-92$ & $-85$ & $-58$ & $-25$ & $0$ & $3$ \\
    Mo & $-6$ & $-53$ & $-93$ & $-111$ & $-105$ & $-84$ & $-49$ & $-18$ & $0$ & $-5$ \\
    Tc & $-28$ & $-76$ & $-107$ & $-109$ & $-91$ & $-62$ & $-31$ & $-9$ & $0$ & $-9$ \\
    Ru & $-44$ & $-80$ & $-92$ & $-81$ & $-58$ & $-34$ & $-14$ & $-2$ & $0$ & $-11$ \\
    Rh & $-41$ & $-56$ & $-52$ & $-36$ & $-20$ & $-7$ & $1$ & $4$ & $0$ & $-9$ \\
    Pd & $-21$ & $-18$ & $-10$ & $-3$ & $1$ & $6$ & $7$ & $4$ & $0$ & $-6$ \\
    Ag & $5$ & $8$ & $9$ & $8$ & $6$ & $5$ & $2$ & $1$ & $0$ & $1$ \\
    Cd & $19$ & $15$ & $9$ & $2$ & $-2$ & $-5$ & $-5$ & $-5$ & $0$ & $7$ \\
    \hline
    La & $37$ & $24$ & $-2$ & $-25$ & $-41$ & $-49$ & $-44$ & $-26$ & $0$ & $19$ \\
    Hf & $34$ & $13$ & $-11$ & $-34$ & $-49$ & $-56$ & $-49$ & $-28$ & $0$ & $13$ \\
    Ta & $21$ & $-11$ & $-47$ & $-73$ & $-84$ & $-82$ & $-61$ & $-29$ & $0$ & $7$ \\
    W  & $1$ & $-43$ & $-83$ & $-103$ & $-103$ & $-89$ & $-56$ & $-23$ & $0$ & $-1$ \\
    Re & $-20$ & $-67$ & $-99$ & $-107$ & $-95$ & $-71$ & $-40$ & $-14$ & $0$ & $-6$ \\
    Os & $-39$ & $-78$ & $-95$ & $-89$ & $-70$ & $-46$ & $-23$ & $-6$ & $0$ & $-9$ \\
    Ir & $-42$ & $-64$ & $-65$ & $-52$ & $-34$ & $-18$ & $-7$ & $0$ & $0$ & $-7$ \\
    Pt & $-29$ & $-33$ & $-25$ & $-16$ & $-9$ & $-2$ & $1$ & $1$ & $0$ & $-5$ \\
    Au & $-7$ & $-4$ & $-1$ & $0$ & $0$ & $0$ & $0$ & $0$ & $0$ & $1$ \\
    Hg & $9$ & $8$ & $4$ & $-1$ & $-4$ & $-7$ & $-6$ & $-4$ & $0$ & $7$ \\
    \hline\hline
  \end{tabular}
  \label{tab:bulk_agg_cu111_ls_3d}
\end{table}

\begin{table}[ht]
  \centering
  \renewcommand{\arraystretch}{1.2}
  \caption{\textbf{Aggregation energies} for \textbf{dual-atom} sites in the \textbf{bulk} layer of \textbf{Cu(111)} for all combinations of \textbf{4\textit{d}} transition metals (columns) and all other transition metals (rows), as obtained from initial \textbf{low-spin} configurations. Energies are reported in kJ~mol$^{-1}$.}
  \vspace{8pt}
  \begin{tabular}{@{\hspace{8pt}}*{11}{c@{\hspace{8pt}}}}
    \hline\hline
    TM & Y & Zr & Nb & Mo & Tc & Ru & Rh & Pd & Ag & Cd \\
    \hline
    Sc & $43$ & $33$ & $16$ & $-6$ & $-28$ & $-44$ & $-41$ & $-21$ & $5$ & $19$ \\
    Ti & $30$ & $11$ & $-19$ & $-53$ & $-76$ & $-80$ & $-56$ & $-18$ & $8$ & $15$ \\
    V  & $14$ & $-15$ & $-57$ & $-93$ & $-107$ & $-92$ & $-52$ & $-10$ & $9$ & $9$ \\
    Cr & $-3$ & $-39$ & $-84$ & $-111$ & $-109$ & $-81$ & $-36$ & $-3$ & $8$ & $2$ \\
    Mn & $-16$ & $-54$ & $-92$ & $-105$ & $-91$ & $-58$ & $-20$ & $1$ & $6$ & $-2$ \\
    Fe & $-27$ & $-60$ & $-85$ & $-84$ & $-62$ & $-34$ & $-7$ & $6$ & $5$ & $-5$ \\
    Co & $-28$ & $-49$ & $-58$ & $-49$ & $-31$ & $-14$ & $1$ & $7$ & $2$ & $-5$ \\
    Ni & $-20$ & $-27$ & $-25$ & $-18$ & $-9$ & $-2$ & $4$ & $4$ & $1$ & $-5$ \\
    Cu & $0$ & $0$ & $0$ & $0$ & $0$ & $0$ & $0$ & $0$ & $0$ & $0$ \\
    Zn & $15$ & $11$ & $3$ & $-5$ & $-9$ & $-11$ & $-9$ & $-6$ & $1$ & $7$ \\
    \hline
    Y  & $50$ & $45$ & $28$ & $6$ & $-14$ & $-33$ & $-34$ & $-19$ & $8$ & $26$ \\
    Zr & $45$ & $28$ & $-1$ & $-32$ & $-55$ & $-65$ & $-51$ & $-19$ & $12$ & $25$ \\
    Nb & $28$ & $-1$ & $-43$ & $-75$ & $-91$ & $-85$ & $-53$ & $-11$ & $13$ & $19$ \\
    Mo & $6$ & $-32$ & $-75$ & $-99$ & $-101$ & $-79$ & $-38$ & $-2$ & $13$ & $10$ \\
    Tc & $-14$ & $-55$ & $-91$ & $-101$ & $-87$ & $-56$ & $-18$ & $7$ & $12$ & $3$ \\
    Ru & $-33$ & $-65$ & $-85$ & $-79$ & $-56$ & $-27$ & $0$ & $11$ & $9$ & $-4$ \\
    Rh & $-34$ & $-51$ & $-53$ & $-38$ & $-18$ & $0$ & $12$ & $13$ & $5$ & $-4$ \\
    Pd & $-19$ & $-19$ & $-11$ & $-2$ & $7$ & $11$ & $13$ & $8$ & $2$ & $-3$ \\
    Ag & $8$ & $12$ & $13$ & $13$ & $12$ & $9$ & $5$ & $2$ & $1$ & $3$ \\
    Cd & $26$ & $25$ & $19$ & $10$ & $3$ & $-4$ & $-4$ & $-3$ & $3$ & $11$ \\
    \hline
    La & $-63$ & $25$ & $13$ & $-14$ & $-35$ & $-47$ & $-42$ & $-18$ & $11$ & $25$ \\
    Hf & $49$ & $32$ & $2$ & $-30$ & $-53$ & $-64$ & $-53$ & $-21$ & $11$ & $26$ \\
    Ta & $37$ & $9$ & $-32$ & $-67$ & $-84$ & $-84$ & $-56$ & $-14$ & $15$ & $24$ \\
    W  & $17$ & $-21$ & $-66$ & $-94$ & $-100$ & $-85$ & $-44$ & $-3$ & $17$ & $16$ \\
    Re & $-3$ & $-45$ & $-84$ & $-99$ & $-92$ & $-64$ & $-23$ & $8$ & $17$ & $10$ \\
    Os & $-23$ & $-60$ & $-86$ & $-85$ & $-66$ & $-36$ & $-4$ & $13$ & $14$ & $4$ \\
    Ir & $-29$ & $-54$ & $-62$ & $-50$ & $-29$ & $-8$ & $9$ & $15$ & $10$ & $3$ \\
    Pt & $-22$ & $-29$ & $-24$ & $-13$ & $-1$ & $7$ & $12$ & $10$ & $7$ & $2$ \\
    Au & $-1$ & $1$ & $4$ & $7$ & $8$ & $8$ & $6$ & $5$ & $5$ & $6$ \\
    Hg & $16$ & $17$ & $14$ & $8$ & $4$ & $-1$ & $0$ & $1$ & $5$ & $11$ \\
    \hline\hline
  \end{tabular}
\end{table}

\begin{table}[ht]
  \centering
  \renewcommand{\arraystretch}{1.2}
  \caption{\textbf{Aggregation energies} for \textbf{dual-atom} sites in the \textbf{bulk} layer of \textbf{Cu(111)} for all combinations of \textbf{5\textit{d}} transition metals (columns) and all other transition metals (rows), as obtained from initial \textbf{low-spin} configurations. Energies are reported in kJ~mol$^{-1}$.}
  \vspace{8pt}
  \begin{tabular}{@{\hspace{8pt}}*{11}{c@{\hspace{8pt}}}}
    \hline\hline
    TM & La & Hf & Ta & W & Re & Os & Ir & Pt & Au & Hg \\
    \hline
    Sc & $37$ & $34$ & $21$ & $1$ & $-20$ & $-39$ & $-42$ & $-29$ & $-7$ & $9$ \\
    Ti & $24$ & $13$ & $-11$ & $-43$ & $-67$ & $-78$ & $-64$ & $-33$ & $-4$ & $8$ \\
    V  & $-2$ & $-11$ & $-47$ & $-83$ & $-99$ & $-95$ & $-65$ & $-25$ & $-1$ & $4$ \\
    Cr & $-25$ & $-34$ & $-73$ & $-103$ & $-107$ & $-89$ & $-52$ & $-16$ & $0$ & $-1$ \\
    Mn & $-41$ & $-49$ & $-84$ & $-103$ & $-95$ & $-70$ & $-34$ & $-9$ & $0$ & $-4$ \\
    Fe & $-49$ & $-56$ & $-82$ & $-89$ & $-71$ & $-46$ & $-18$ & $-2$ & $0$ & $-7$ \\
    Co & $-44$ & $-49$ & $-61$ & $-56$ & $-40$ & $-23$ & $-7$ & $1$ & $0$ & $-6$ \\
    Ni & $-26$ & $-28$ & $-29$ & $-23$ & $-14$ & $-6$ & $0$ & $1$ & $0$ & $-4$ \\
    Cu & $0$ & $0$ & $0$ & $0$ & $0$ & $0$ & $0$ & $0$ & $0$ & $0$ \\
    Zn & $19$ & $13$ & $7$ & $-1$ & $-6$ & $-9$ & $-7$ & $-5$ & $1$ & $7$ \\
    \hline
    Y  & $-63$ & $49$ & $37$ & $17$ & $-3$ & $-23$ & $-29$ & $-22$ & $-1$ & $16$ \\
    Zr & $25$ & $32$ & $9$ & $-21$ & $-45$ & $-60$ & $-54$ & $-29$ & $1$ & $17$ \\
    Nb & $13$ & $2$ & $-32$ & $-66$ & $-84$ & $-86$ & $-62$ & $-24$ & $4$ & $14$ \\
    Mo & $-14$ & $-30$ & $-67$ & $-94$ & $-99$ & $-85$ & $-50$ & $-13$ & $7$ & $8$ \\
    Tc & $-35$ & $-53$ & $-84$ & $-100$ & $-92$ & $-66$ & $-29$ & $-1$ & $8$ & $4$ \\
    Ru & $-47$ & $-64$ & $-84$ & $-85$ & $-64$ & $-36$ & $-8$ & $7$ & $8$ & $-1$ \\
    Rh & $-42$ & $-53$ & $-56$ & $-44$ & $-23$ & $-4$ & $9$ & $12$ & $6$ & $0$ \\
    Pd & $-18$ & $-21$ & $-14$ & $-3$ & $8$ & $13$ & $15$ & $10$ & $5$ & $1$ \\
    Ag & $11$ & $11$ & $15$ & $17$ & $17$ & $14$ & $10$ & $7$ & $5$ & $5$ \\
    Cd & $25$ & $26$ & $24$ & $16$ & $10$ & $4$ & $3$ & $2$ & $6$ & $11$ \\
    \hline
    La & $-147$ & $48$ & $35$ & $13$ & $-7$ & $-25$ & $-28$ & $-15$ & $8$ & $25$ \\
    Hf & $48$ & $34$ & $11$ & $-18$ & $-43$ & $-60$ & $-56$ & $-32$ & $0$ & $19$ \\
    Ta & $35$ & $11$ & $-24$ & $-59$ & $-80$ & $-86$ & $-66$ & $-28$ & $5$ & $20$ \\
    W  & $13$ & $-18$ & $-59$ & $-91$ & $-101$ & $-92$ & $-58$ & $-17$ & $9$ & $16$ \\
    Re & $-7$ & $-43$ & $-80$ & $-101$ & $-96$ & $-76$ & $-37$ & $-3$ & $13$ & $11$ \\
    Os & $-25$ & $-60$ & $-86$ & $-92$ & $-76$ & $-47$ & $-13$ & $8$ & $13$ & $7$ \\
    Ir & $-28$ & $-56$ & $-66$ & $-58$ & $-37$ & $-13$ & $6$ & $14$ & $12$ & $9$ \\
    Pt & $-15$ & $-32$ & $-28$ & $-17$ & $-3$ & $8$ & $14$ & $14$ & $12$ & $10$ \\
    Au & $8$ & $0$ & $5$ & $9$ & $13$ & $13$ & $12$ & $12$ & $12$ & $12$ \\
    Hg & $25$ & $19$ & $20$ & $16$ & $11$ & $7$ & $9$ & $10$ & $12$ & $16$ \\
    \hline\hline
  \end{tabular}
\end{table}

\clearpage


\begin{figure}[htbp]
    \includegraphics[width=12cm,height=\textheight,keepaspectratio]{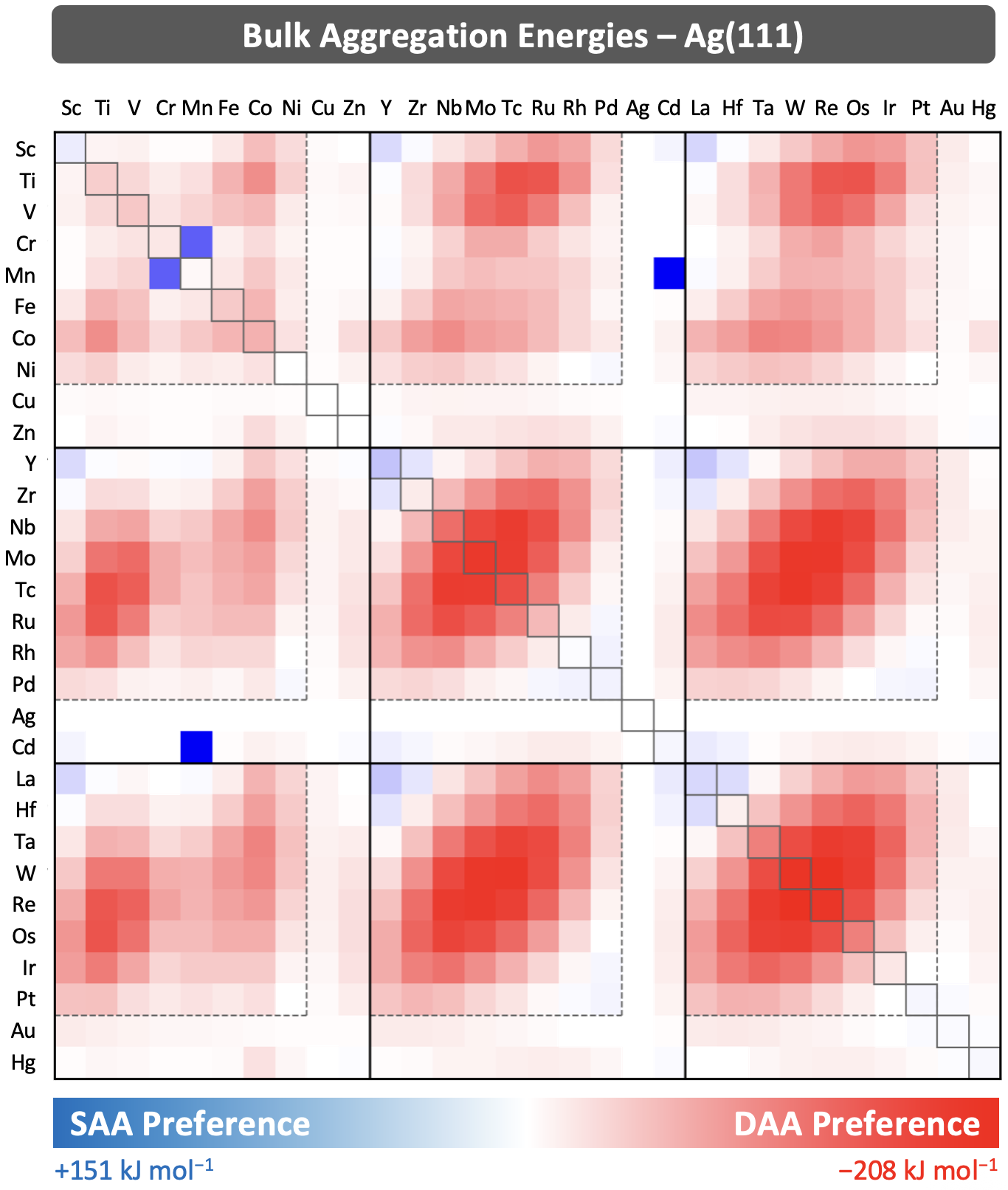}
    \caption{
        Aggregation energies for the formation of all possible transition metal dopant dimers in a bulk layer of the Ag(111) system. 
        Preferences for single-atom sites are represented in blue, while preferences for dopant dimers are shown in red.     }
    \label{fig:Ag111-bulk-aggr}
\end{figure}

\begin{table}[ht]
  \centering
  \renewcommand{\arraystretch}{1.2}
  \caption{\textbf{Aggregation energies} for \textbf{dual-atom} sites in the \textbf{bulk} layer of \textbf{Ag(111)} for all combinations of \textbf{3\textit{d}} transition metals (columns) and all other transition metals (rows), as obtained from initial \textbf{high-spin} configurations. Energies are reported in kJ~mol$^{-1}$.}
  \vspace{8pt}
  \begin{tabular}{@{\hspace{8pt}}*{11}{c@{\hspace{8pt}}}}
    \hline\hline
    TM & Sc & Ti & V & Cr & Mn & Fe & Co & Ni & Cu & Zn \\
    \hline
    Sc & $11$ & $-9$ & $-12$ & $-2$ & $-3$ & $-20$ & $-56$ & $-31$ & $-5$ & $-1$ \\
    Ti & $-9$ & $-42$ & $-34$ & $-16$ & $-27$ & $-64$ & $-96$ & $-39$ & $-6$ & $-8$ \\
    V  & $-12$ & $-34$ & $-47$ & $-24$ & $-36$ & $-51$ & $-60$ & $-18$ & $-4$ & $-6$ \\
    Cr & $-2$ & $-16$ & $-24$ & $-20$ & $95$ & $-14$ & $-30$ & $-10$ & $-2$ & $-2$ \\
    Mn & $-3$ & $-27$ & $-36$ & $95$ & $-6$ & $-20$ & $-47$ & $-18$ & $-3$ & $-2$ \\
    Fe & $-20$ & $-64$ & $-51$ & $-14$ & $-20$ & $-44$ & $-62$ & $-16$ & $-4$ & $-7$ \\
    Co & $-56$ & $-96$ & $-60$ & $-30$ & $-47$ & $-62$ & $-66$ & $-26$ & $-4$ & $-30$ \\
    Ni & $-31$ & $-39$ & $-18$ & $-10$ & $-18$ & $-16$ & $-26$ & $0$ & $-3$ & $-12$ \\
    Cu & $-5$ & $-6$ & $-4$ & $-2$ & $-3$ & $-4$ & $-4$ & $-3$ & $-1$ & $-1$ \\
    Zn & $-1$ & $-8$ & $-6$ & $-2$ & $-2$ & $-7$ & $-30$ & $-12$ & $-1$ & $1$ \\
    \hline
    Y  & $22$ & $2$ & $-5$ & $2$ & $3$ & $-12$ & $-46$ & $-28$ & $-5$ & $1$ \\
    Zr & $3$ & $-30$ & $-29$ & $-11$ & $-13$ & $-43$ & $-81$ & $-41$ & $-8$ & $-6$ \\
    Nb & $-22$ & $-73$ & $-77$ & $-37$ & $-46$ & $-76$ & $-99$ & $-45$ & $-11$ & $-18$ \\
    Mo & $-40$ & $-119$ & $-129$ & $-70$ & $-56$ & $-71$ & $-83$ & $-33$ & $-9$ & $-19$ \\
    Tc & $-67$ & $-155$ & $-143$ & $-71$ & $-50$ & $-66$ & $-76$ & $-25$ & $-8$ & $-23$ \\
    Ru & $-86$ & $-151$ & $-113$ & $-43$ & $-48$ & $-58$ & $-58$ & $-11$ & $-8$ & $-27$ \\
    Rh & $-73$ & $-95$ & $-54$ & $-22$ & $-36$ & $-33$ & $-34$ & $-1$ & $-4$ & $-23$ \\
    Pd & $-31$ & $-27$ & $-12$ & $-9$ & $-14$ & $-8$ & $-19$ & $4$ & $-2$ & $-12$ \\
    Ag & $0$ & $0$ & $0$ & $0$ & $0$ & $0$ & $0$ & $0$ & $0$ & $0$ \\
    Cd & $6$ & $-1$ & $-1$ & $1$ & $151$ & $-3$ & $-12$ & $-8$ & $0$ & $3$ \\
    \hline
    La & $24$ & $1$ & $-7$ & $1$ & $3$ & $-15$ & $-65$ & $-36$ & $-8$ & $-1$ \\
    Hf & $2$ & $-28$ & $-28$ & $-12$ & $-13$ & $-42$ & $-82$ & $-45$ & $-9$ & $-3$ \\
    Ta & $-21$ & $-66$ & $-60$ & $-32$ & $-43$ & $-78$ & $-106$ & $-53$ & $-13$ & $-16$ \\
    W  & $-46$ & $-115$ & $-116$ & $-70$ & $-65$ & $-87$ & $-103$ & $-48$ & $-14$ & $-24$ \\
    Re & $-72$ & $-149$ & $-137$ & $-78$ & $-64$ & $-80$ & $-88$ & $-37$ & $-14$ & $-28$ \\
    Os & $-90$ & $-152$ & $-122$ & $-57$ & $-58$ & $-69$ & $-69$ & $-22$ & $-11$ & $-29$ \\
    Ir & $-84$ & $-114$ & $-75$ & $-35$ & $-45$ & $-45$ & $-45$ & $-9$ & $-8$ & $-25$ \\
    Pt & $-51$ & $-53$ & $-28$ & $-19$ & $-25$ & $-19$ & $-27$ & $-1$ & $-4$ & $-15$ \\
    Au & $-16$ & $-15$ & $-10$ & $-7$ & $-7$ & $-5$ & $-4$ & $-2$ & $-2$ & $-3$ \\
    Hg & $-3$ & $-7$ & $-6$ & $-3$ & $-2$ & $-5$ & $-25$ & $-7$ & $-1$ & $1$ \\
    \hline\hline
  \end{tabular}
  \label{tab:bulk_agg_ag111_3d}
\end{table}

\begin{table}[ht]
  \centering
  \renewcommand{\arraystretch}{1.2}
  \caption{\textbf{Aggregation energies} for \textbf{dual-atom} sites in the \textbf{bulk} layer of \textbf{Ag(111)} for all combinations of \textbf{4\textit{d}} transition metals (columns) and all other transition metals (rows), as obtained from initial \textbf{high-spin} configurations. Energies are reported in kJ~mol$^{-1}$.}
  \vspace{8pt}
  \begin{tabular}{@{\hspace{8pt}}*{11}{c@{\hspace{8pt}}}}
    \hline\hline
    TM & Y & Zr & Nb & Mo & Tc & Ru & Rh & Pd & Ag & Cd \\
    \hline
    Sc & $22$ & $3$ & $-22$ & $-40$ & $-67$ & $-86$ & $-73$ & $-31$ & $0$ & $6$ \\
    Ti & $2$ & $-30$ & $-73$ & $-119$ & $-155$ & $-151$ & $-95$ & $-27$ & $0$ & $-1$ \\
    V  & $-5$ & $-29$ & $-77$ & $-129$ & $-143$ & $-113$ & $-54$ & $-12$ & $0$ & $-1$ \\
    Cr & $2$ & $-11$ & $-37$ & $-70$ & $-71$ & $-43$ & $-22$ & $-9$ & $0$ & $1$ \\
    Mn & $3$ & $-13$ & $-46$ & $-56$ & $-50$ & $-48$ & $-36$ & $-14$ & $0$ & $151$ \\
    Fe & $-12$ & $-43$ & $-76$ & $-71$ & $-66$ & $-58$ & $-33$ & $-8$ & $0$ & $-3$ \\
    Co & $-46$ & $-81$ & $-99$ & $-83$ & $-76$ & $-58$ & $-34$ & $-19$ & $0$ & $-12$ \\
    Ni & $-28$ & $-41$ & $-45$ & $-33$ & $-25$ & $-11$ & $-1$ & $4$ & $0$ & $-8$ \\
    Cu & $-5$ & $-8$ & $-11$ & $-9$ & $-8$ & $-8$ & $-4$ & $-2$ & $0$ & $0$ \\
    Zn & $1$ & $-6$ & $-18$ & $-19$ & $-23$ & $-27$ & $-23$ & $-12$ & $0$ & $3$ \\
    \hline
    Y  & $35$ & $16$ & $-9$ & $-27$ & $-49$ & $-68$ & $-63$ & $-32$ & $0$ & $10$ \\
    Zr & $16$ & $-16$ & $-59$ & $-93$ & $-124$ & $-129$ & $-94$ & $-35$ & $0$ & $5$ \\
    Nb & $-9$ & $-59$ & $-126$ & $-167$ & $-182$ & $-159$ & $-98$ & $-28$ & $0$ & $-5$ \\
    Mo & $-27$ & $-93$ & $-167$ & $-190$ & $-181$ & $-141$ & $-69$ & $-14$ & $0$ & $-7$ \\
    Tc & $-49$ & $-124$ & $-182$ & $-181$ & $-158$ & $-107$ & $-43$ & $-7$ & $0$ & $-13$ \\
    Ru & $-68$ & $-129$ & $-159$ & $-141$ & $-107$ & $-60$ & $-17$ & $5$ & $0$ & $-18$ \\
    Rh & $-63$ & $-94$ & $-98$ & $-69$ & $-43$ & $-17$ & $2$ & $8$ & $0$ & $-16$ \\
    Pd & $-32$ & $-35$ & $-28$ & $-14$ & $-7$ & $5$ & $8$ & $8$ & $0$ & $-10$ \\
    Ag & $0$ & $0$ & $0$ & $0$ & $0$ & $0$ & $0$ & $0$ & $0$ & $0$ \\
    Cd & $10$ & $5$ & $-5$ & $-7$ & $-13$ & $-18$ & $-16$ & $-10$ & $0$ & $5$ \\
    \hline
    La & $34$ & $14$ & $-22$ & $-51$ & $-81$ & $-99$ & $-83$ & $-38$ & $0$ & $12$ \\
    Hf & $16$ & $-15$ & $-54$ & $-87$ & $-116$ & $-128$ & $-99$ & $-40$ & $0$ & $8$ \\
    Ta & $-6$ & $-53$ & $-115$ & $-154$ & $-175$ & $-163$ & $-109$ & $-35$ & $0$ & $-3$ \\
    W  & $-29$ & $-95$ & $-165$ & $-192$ & $-194$ & $-162$ & $-93$ & $-23$ & $0$ & $-10$ \\
    Re & $-53$ & $-125$ & $-185$ & $-191$ & $-176$ & $-132$ & $-62$ & $-11$ & $0$ & $-15$ \\
    Os & $-70$ & $-134$ & $-171$ & $-160$ & $-131$ & $-83$ & $-31$ & $0$ & $0$ & $-17$ \\
    Ir & $-71$ & $-110$ & $-121$ & $-96$ & $-65$ & $-35$ & $-8$ & $5$ & $0$ & $-15$ \\
    Pt & $-48$ & $-60$ & $-54$ & $-35$ & $-22$ & $-6$ & $3$ & $6$ & $0$ & $-10$ \\
    Au & $-17$ & $-18$ & $-15$ & $-11$ & $-7$ & $-4$ & $-2$ & $1$ & $0$ & $-2$ \\
    Hg & $-2$ & $-4$ & $-11$ & $-10$ & $-13$ & $-16$ & $-13$ & $-7$ & $0$ & $3$ \\
    \hline\hline
  \end{tabular}
  \label{tab:bulk_agg_ag111_4d}
\end{table}

\begin{table}[ht]
  \centering
  \renewcommand{\arraystretch}{1.2}
  \caption{\textbf{Aggregation energies} for \textbf{dual-atom} sites in the \textbf{bulk} layer of \textbf{Ag(111)} for all combinations of \textbf{5\textit{d}} transition metals (columns) and all other transition metals (rows), as obtained from initial \textbf{high-spin} configurations. Energies are reported in kJ~mol$^{-1}$.}
  \vspace{8pt}
  \begin{tabular}{@{\hspace{8pt}}*{11}{c@{\hspace{8pt}}}}
    \hline\hline
    TM & La & Hf & Ta & W & Re & Os & Ir & Pt & Au & Hg \\
    \hline
    Sc & $24$ & $2$ & $-21$ & $-46$ & $-72$ & $-90$ & $-84$ & $-51$ & $-16$ & $-3$ \\
    Ti & $1$ & $-28$ & $-66$ & $-115$ & $-149$ & $-152$ & $-114$ & $-53$ & $-15$ & $-7$ \\
    V  & $-7$ & $-28$ & $-60$ & $-116$ & $-137$ & $-122$ & $-75$ & $-28$ & $-10$ & $-6$ \\
    Cr & $1$ & $-12$ & $-32$ & $-70$ & $-78$ & $-57$ & $-35$ & $-19$ & $-7$ & $-3$ \\
    Mn & $3$ & $-13$ & $-43$ & $-65$ & $-64$ & $-58$ & $-45$ & $-25$ & $-7$ & $-2$ \\
    Fe & $-15$ & $-42$ & $-78$ & $-87$ & $-80$ & $-69$ & $-45$ & $-19$ & $-5$ & $-5$ \\
    Co & $-65$ & $-82$ & $-106$ & $-103$ & $-88$ & $-69$ & $-45$ & $-27$ & $-4$ & $-25$ \\
    Ni & $-36$ & $-45$ & $-53$ & $-48$ & $-37$ & $-22$ & $-9$ & $-1$ & $-2$ & $-7$ \\
    Cu & $-8$ & $-9$ & $-13$ & $-14$ & $-14$ & $-11$ & $-8$ & $-4$ & $-2$ & $-1$ \\
    Zn & $-1$ & $-3$ & $-16$ & $-24$ & $-28$ & $-29$ & $-25$ & $-15$ & $-3$ & $1$ \\
    \hline
    Y  & $34$ & $16$ & $-6$ & $-29$ & $-53$ & $-70$ & $-71$ & $-48$ & $-17$ & $-2$ \\
    Zr & $14$ & $-15$ & $-53$ & $-95$ & $-125$ & $-134$ & $-110$ & $-60$ & $-18$ & $-4$ \\
    Nb & $-22$ & $-54$ & $-115$ & $-165$ & $-185$ & $-171$ & $-121$ & $-54$ & $-15$ & $-11$ \\
    Mo & $-51$ & $-87$ & $-154$ & $-192$ & $-191$ & $-160$ & $-96$ & $-35$ & $-11$ & $-10$ \\
    Tc & $-81$ & $-116$ & $-175$ & $-194$ & $-176$ & $-131$ & $-65$ & $-22$ & $-7$ & $-13$ \\
    Ru & $-99$ & $-128$ & $-163$ & $-162$ & $-132$ & $-83$ & $-35$ & $-6$ & $-4$ & $-16$ \\
    Rh & $-83$ & $-99$ & $-109$ & $-93$ & $-62$ & $-31$ & $-8$ & $3$ & $-2$ & $-13$ \\
    Pd & $-38$ & $-40$ & $-35$ & $-23$ & $-11$ & $0$ & $5$ & $6$ & $1$ & $-7$ \\
    Ag & $0$ & $0$ & $0$ & $0$ & $0$ & $0$ & $0$ & $0$ & $0$ & $0$ \\
    Cd & $12$ & $8$ & $-3$ & $-10$ & $-15$ & $-17$ & $-15$ & $-10$ & $-2$ & $3$ \\
    \hline
    La & $23$ & $21$ & $-7$ & $-39$ & $-67$ & $-85$ & $-81$ & $-51$ & $-17$ & $-1$ \\
    Hf & $21$ & $-14$ & $-51$ & $-92$ & $-122$ & $-136$ & $-116$ & $-66$ & $-19$ & $0$ \\
    Ta & $-7$ & $-51$ & $-109$ & $-159$ & $-184$ & $-177$ & $-133$ & $-65$ & $-16$ & $-7$ \\
    W  & $-39$ & $-92$ & $-159$ & $-201$ & $-208$ & $-181$ & $-122$ & $-50$ & $-13$ & $-12$ \\
    Re & $-67$ & $-122$ & $-184$ & $-208$ & $-196$ & $-157$ & $-91$ & $-31$ & $-8$ & $-14$ \\
    Os & $-85$ & $-136$ & $-177$ & $-181$ & $-157$ & $-109$ & $-53$ & $-13$ & $-4$ & $-13$ \\
    Ir & $-81$ & $-116$ & $-133$ & $-122$ & $-91$ & $-53$ & $-21$ & $-1$ & $-1$ & $-10$ \\
    Pt & $-51$ & $-66$ & $-65$ & $-50$ & $-31$ & $-13$ & $-1$ & $5$ & $3$ & $-5$ \\
    Au & $-17$ & $-19$ & $-16$ & $-13$ & $-8$ & $-4$ & $-1$ & $3$ & $3$ & $1$ \\
    Hg & $-1$ & $0$ & $-7$ & $-12$ & $-14$ & $-13$ & $-10$ & $-5$ & $1$ & $4$ \\
    \hline\hline
  \end{tabular}
  \label{tab:bulk_agg_ag111_5d}
\end{table}

\begin{table}[ht]
  \centering
  \renewcommand{\arraystretch}{1.2}
  \caption{\textbf{Aggregation energies} for \textbf{dual-atom} sites in the \textbf{bulk} layer of \textbf{Ag(111)} for all combinations of \textbf{3\textit{d}} transition metals (columns) and all other transition metals (rows), as obtained from initial \textbf{low-spin} configurations. Energies are reported in kJ~mol$^{-1}$.}
  \vspace{8pt}
  \begin{tabular}{@{\hspace{8pt}}*{11}{c@{\hspace{8pt}}}}
    \hline\hline
    TM & Sc & Ti & V & Cr & Mn & Fe & Co & Ni & Cu & Zn \\
    \hline
    Sc & $11$ & $-7$ & $-24$ & $-31$ & $-43$ & $-54$ & $-55$ & $-31$ & $-5$ & $-1$ \\
    Ti & $-7$ & $-36$ & $-76$ & $-117$ & $-63$ & $-143$ & $-97$ & $-41$ & $-7$ & $-8$ \\
    V  & $-24$ & $-76$ & $-160$ & $-204$ & $-205$ & $-163$ & $-99$ & $-36$ & $-7$ & $-14$ \\
    Cr & $-31$ & $-117$ & $-204$ & $-223$ & $-189$ & $-141$ & $-78$ & $-26$ & $-6$ & $-14$ \\
    Mn & $-43$ & $-63$ & $-205$ & $-189$ & $-325$ & $-98$ & $-54$ & $-20$ & $-6$ & $-16$ \\
    Fe & $-54$ & $-143$ & $-163$ & $-141$ & $-98$ & $-63$ & $-34$ & $-11$ & $-5$ & $-17$ \\
    Co & $-55$ & $-97$ & $-99$ & $-78$ & $-54$ & $-34$ & $-17$ & $-4$ & $-4$ & $-18$ \\
    Ni & $-31$ & $-41$ & $-36$ & $-26$ & $-20$ & $-11$ & $-4$ & $0$ & $-3$ & $-12$ \\
    Cu & $-5$ & $-7$ & $-7$ & $-6$ & $-6$ & $-5$ & $-4$ & $-3$ & $-1$ & $-1$ \\
    Zn & $-1$ & $-8$ & $-14$ & $-14$ & $-16$ & $-17$ & $-18$ & $-12$ & $-1$ & $1$ \\
    \hline
    Y  & $22$ & $4$ & $-13$ & $-20$ & $-29$ & $-38$ & $-43$ & $-28$ & $-5$ & $1$ \\
    Zr & $3$ & $-27$ & $-56$ & $-77$ & $-96$ & $-101$ & $-81$ & $-41$ & $-8$ & $-6$ \\
    Nb & $-21$ & $-74$ & $-134$ & $-165$ & $-169$ & $-145$ & $-99$ & $-45$ & $-11$ & $-18$ \\
    Mo & $-48$ & $-128$ & $-193$ & $-204$ & $-181$ & $-140$ & $-89$ & $-37$ & $-12$ & $-24$ \\
    Tc & $-73$ & $-163$ & $-206$ & $-223$ & $-156$ & $-109$ & $-63$ & $-24$ & $-10$ & $-27$ \\
    Ru & $-86$ & $-152$ & $-162$ & $-140$ & $-97$ & $-61$ & $-32$ & $-11$ & $-8$ & $-27$ \\
    Rh & $-73$ & $-97$ & $-89$ & $-62$ & $-42$ & $-23$ & $-8$ & $-1$ & $-4$ & $-23$ \\
    Pd & $-31$ & $-29$ & $-19$ & $-12$ & $-8$ & $-1$ & $4$ & $4$ & $-2$ & $-12$ \\
    Ag & $0$ & $0$ & $0$ & $0$ & $0$ & $0$ & $0$ & $0$ & $0$ & $0$ \\
    Cd & $6$ & $0$ & $-6$ & $-7$ & $-8$ & $-10$ & $-12$ & $-8$ & $0$ & $3$ \\
    \hline
    La & $24$ & $3$ & $-20$ & $-37$ & $-57$ & $-71$ & $-65$ & $-36$ & $-8$ & $-1$ \\
    Hf & $2$ & $-26$ & $-51$ & $-68$ & $-88$ & $-95$ & $-82$ & $-45$ & $-9$ & $-3$ \\
    Ta & $-21$ & $-67$ & $-116$ & $-142$ & $-153$ & $-142$ & $-106$ & $-53$ & $-13$ & $-16$ \\
    W  & $-46$ & $-117$ & $-173$ & $-189$ & $-179$ & $-148$ & $-103$ & $-49$ & $-15$ & $-24$ \\
    Re & $-73$ & $-151$ & $-194$ & $-224$ & $-224$ & $-124$ & $-80$ & $-36$ & $-14$ & $-29$ \\
    Os & $-90$ & $-154$ & $-170$ & $-155$ & $-118$ & $-82$ & $-50$ & $-22$ & $-11$ & $-29$ \\
    Ir & $-84$ & $-115$ & $-112$ & $-88$ & $-62$ & $-41$ & $-23$ & $-9$ & $-8$ & $-25$ \\
    Pt & $-51$ & $-54$ & $-43$ & $-30$ & $-23$ & $-13$ & $-5$ & $-1$ & $-4$ & $-15$ \\
    Au & $-16$ & $-15$ & $-12$ & $-9$ & $-8$ & $-6$ & $-4$ & $-2$ & $-2$ & $-3$ \\
    Hg & $-3$ & $-6$ & $-10$ & $-9$ & $-11$ & $-11$ & $-11$ & $-7$ & $-1$ & $1$ \\
    \hline\hline
  \end{tabular}
\end{table}

\begin{table}[ht]
  \centering
  \renewcommand{\arraystretch}{1.2}
  \caption{\textbf{Aggregation energies} for \textbf{dual-atom} sites in the \textbf{bulk} layer of \textbf{Ag(111)} for all combinations of \textbf{4\textit{d}} transition metals (columns) and all other transition metals (rows), as obtained from initial \textbf{low-spin} configurations. Energies are reported in kJ~mol$^{-1}$.}
  \vspace{8pt}
  \begin{tabular}{@{\hspace{8pt}}*{11}{c@{\hspace{8pt}}}}
    \hline\hline
    TM & Y & Zr & Nb & Mo & Tc & Ru & Rh & Pd & Ag & Cd \\
    \hline
    Sc & $22$ & $3$ & $-21$ & $-48$ & $-73$ & $-86$ & $-73$ & $-31$ & $0$ & $6$ \\
    Ti & $4$ & $-27$ & $-74$ & $-128$ & $-163$ & $-152$ & $-97$ & $-29$ & $0$ & $0$ \\
    V  & $-13$ & $-56$ & $-134$ & $-193$ & $-206$ & $-162$ & $-89$ & $-19$ & $0$ & $-6$ \\
    Cr & $-20$ & $-77$ & $-165$ & $-204$ & $-223$ & $-140$ & $-62$ & $-12$ & $0$ & $-7$ \\
    Mn & $-29$ & $-96$ & $-169$ & $-181$ & $-156$ & $-97$ & $-42$ & $-8$ & $0$ & $-8$ \\
    Fe & $-38$ & $-101$ & $-145$ & $-140$ & $-109$ & $-61$ & $-23$ & $-1$ & $0$ & $-10$ \\
    Co & $-43$ & $-81$ & $-99$ & $-89$ & $-63$ & $-32$ & $-8$ & $4$ & $0$ & $-12$ \\
    Ni & $-28$ & $-41$ & $-45$ & $-37$ & $-24$ & $-11$ & $-1$ & $4$ & $0$ & $-8$ \\
    Cu & $-5$ & $-8$ & $-11$ & $-12$ & $-10$ & $-8$ & $-4$ & $-2$ & $0$ & $0$ \\
    Zn & $1$ & $-6$ & $-18$ & $-24$ & $-27$ & $-27$ & $-23$ & $-12$ & $0$ & $3$ \\
    \hline
    Y  & $35$ & $16$ & $-8$ & $-32$ & $-55$ & $-68$ & $-63$ & $-32$ & $0$ & $10$ \\
    Zr & $16$ & $-16$ & $-59$ & $-101$ & $-130$ & $-129$ & $-94$ & $-35$ & $0$ & $5$ \\
    Nb & $-8$ & $-59$ & $-126$ & $-175$ & $-189$ & $-159$ & $-98$ & $-28$ & $0$ & $-5$ \\
    Mo & $-32$ & $-101$ & $-175$ & $-205$ & $-195$ & $-148$ & $-76$ & $-15$ & $0$ & $-12$ \\
    Tc & $-55$ & $-130$ & $-189$ & $-195$ & $-170$ & $-114$ & $-45$ & $-3$ & $0$ & $-17$ \\
    Ru & $-68$ & $-129$ & $-159$ & $-148$ & $-114$ & $-59$ & $-16$ & $5$ & $0$ & $-18$ \\
    Rh & $-63$ & $-94$ & $-98$ & $-76$ & $-45$ & $-16$ & $2$ & $8$ & $0$ & $-16$ \\
    Pd & $-32$ & $-35$ & $-28$ & $-15$ & $-3$ & $5$ & $8$ & $8$ & $0$ & $-10$ \\
    Ag & $0$ & $0$ & $0$ & $0$ & $0$ & $0$ & $0$ & $0$ & $0$ & $0$ \\
    Cd & $10$ & $5$ & $-5$ & $-12$ & $-17$ & $-18$ & $-16$ & $-10$ & $0$ & $5$ \\
    \hline
    La & $34$ & $14$ & $-22$ & $-59$ & $-87$ & $-99$ & $-83$ & $-38$ & $0$ & $12$ \\
    Hf & $16$ & $-15$ & $-54$ & $-94$ & $-122$ & $-128$ & $-99$ & $-40$ & $0$ & $8$ \\
    Ta & $-6$ & $-53$ & $-115$ & $-162$ & $-182$ & $-163$ & $-109$ & $-35$ & $0$ & $-3$ \\
    W  & $-30$ & $-96$ & $-165$ & $-201$ & $-201$ & $-162$ & $-93$ & $-23$ & $0$ & $-10$ \\
    Re & $-53$ & $-126$ & $-185$ & $-199$ & $-182$ & $-132$ & $-62$ & $-10$ & $0$ & $-15$ \\
    Os & $-70$ & $-134$ & $-171$ & $-166$ & $-138$ & $-83$ & $-31$ & $0$ & $0$ & $-17$ \\
    Ir & $-71$ & $-110$ & $-121$ & $-104$ & $-71$ & $-34$ & $-8$ & $5$ & $0$ & $-15$ \\
    Pt & $-48$ & $-60$ & $-54$ & $-38$ & $-20$ & $-6$ & $3$ & $6$ & $0$ & $-10$ \\
    Au & $-17$ & $-18$ & $-15$ & $-11$ & $-8$ & $-4$ & $-2$ & $1$ & $0$ & $-2$ \\
    Hg & $-2$ & $-4$ & $-11$ & $-15$ & $-17$ & $-16$ & $-13$ & $-7$ & $0$ & $3$ \\
    \hline\hline
  \end{tabular}
\end{table}

\begin{table}[ht]
  \centering
  \renewcommand{\arraystretch}{1.2}
  \caption{\textbf{Aggregation energies} for \textbf{dual-atom} sites in the \textbf{bulk} layer of \textbf{Ag(111)} for all combinations of \textbf{5\textit{d}} transition metals (columns) and all other transition metals (rows), as obtained from initial \textbf{low-spin} configurations. Energies are reported in kJ~mol$^{-1}$.}
  \vspace{8pt}
  \begin{tabular}{@{\hspace{8pt}}*{11}{c@{\hspace{8pt}}}}
    \hline\hline
    TM & La & Hf & Ta & W & Re & Os & Ir & Pt & Au & Hg \\
    \hline
    Sc & $24$ & $2$ & $-21$ & $-46$ & $-73$ & $-90$ & $-84$ & $-51$ & $-16$ & $-3$ \\
    Ti & $3$ & $-26$ & $-67$ & $-117$ & $-151$ & $-154$ & $-115$ & $-54$ & $-15$ & $-6$ \\
    V  & $-20$ & $-51$ & $-116$ & $-173$ & $-194$ & $-170$ & $-112$ & $-43$ & $-12$ & $-10$ \\
    Cr & $-37$ & $-68$ & $-142$ & $-189$ & $-224$ & $-155$ & $-88$ & $-30$ & $-9$ & $-9$ \\
    Mn & $-57$ & $-88$ & $-153$ & $-179$ & $-224$ & $-118$ & $-62$ & $-23$ & $-8$ & $-11$ \\
    Fe & $-71$ & $-95$ & $-142$ & $-148$ & $-124$ & $-82$ & $-41$ & $-13$ & $-6$ & $-11$ \\
    Co & $-65$ & $-82$ & $-106$ & $-103$ & $-80$ & $-50$ & $-23$ & $-5$ & $-4$ & $-11$ \\
    Ni & $-36$ & $-45$ & $-53$ & $-49$ & $-36$ & $-22$ & $-9$ & $-1$ & $-2$ & $-7$ \\
    Cu & $-8$ & $-9$ & $-13$ & $-15$ & $-14$ & $-11$ & $-8$ & $-4$ & $-2$ & $-1$ \\
    Zn & $-1$ & $-3$ & $-16$ & $-24$ & $-29$ & $-29$ & $-25$ & $-15$ & $-3$ & $1$ \\
    \hline
    Y  & $34$ & $16$ & $-6$ & $-30$ & $-53$ & $-70$ & $-71$ & $-48$ & $-17$ & $-2$ \\
    Zr & $14$ & $-15$ & $-53$ & $-96$ & $-126$ & $-134$ & $-110$ & $-60$ & $-18$ & $-4$ \\
    Nb & $-22$ & $-54$ & $-115$ & $-165$ & $-185$ & $-171$ & $-121$ & $-54$ & $-15$ & $-11$ \\
    Mo & $-59$ & $-94$ & $-162$ & $-201$ & $-199$ & $-166$ & $-104$ & $-38$ & $-11$ & $-15$ \\
    Tc & $-87$ & $-122$ & $-182$ & $-201$ & $-182$ & $-138$ & $-71$ & $-20$ & $-8$ & $-17$ \\
    Ru & $-99$ & $-128$ & $-163$ & $-162$ & $-132$ & $-83$ & $-34$ & $-6$ & $-4$ & $-16$ \\
    Rh & $-83$ & $-99$ & $-109$ & $-93$ & $-62$ & $-31$ & $-8$ & $3$ & $-2$ & $-13$ \\
    Pd & $-38$ & $-40$ & $-35$ & $-23$ & $-10$ & $0$ & $5$ & $6$ & $1$ & $-7$ \\
    Ag & $0$ & $0$ & $0$ & $0$ & $0$ & $0$ & $0$ & $0$ & $0$ & $0$ \\
    Cd & $12$ & $8$ & $-3$ & $-10$ & $-15$ & $-17$ & $-15$ & $-10$ & $-2$ & $3$ \\
    \hline
    La & $23$ & $21$ & $-7$ & $-39$ & $-67$ & $-85$ & $-81$ & $-51$ & $-17$ & $-1$ \\
    Hf & $21$ & $-14$ & $-51$ & $-92$ & $-123$ & $-136$ & $-116$ & $-66$ & $-19$ & $0$ \\
    Ta & $-7$ & $-51$ & $-109$ & $-160$ & $-184$ & $-177$ & $-133$ & $-65$ & $-16$ & $-7$ \\
    W  & $-39$ & $-92$ & $-160$ & $-202$ & $-209$ & $-182$ & $-122$ & $-51$ & $-13$ & $-12$ \\
    Re & $-67$ & $-123$ & $-184$ & $-209$ & $-197$ & $-158$ & $-92$ & $-31$ & $-9$ & $-14$ \\
    Os & $-85$ & $-136$ & $-177$ & $-182$ & $-158$ & $-109$ & $-53$ & $-13$ & $-4$ & $-13$ \\
    Ir & $-81$ & $-116$ & $-133$ & $-122$ & $-92$ & $-53$ & $-21$ & $-1$ & $-1$ & $-10$ \\
    Pt & $-51$ & $-66$ & $-65$ & $-51$ & $-31$ & $-13$ & $-1$ & $5$ & $3$ & $-5$ \\
    Au & $-17$ & $-19$ & $-16$ & $-13$ & $-9$ & $-4$ & $-1$ & $3$ & $3$ & $1$ \\
    Hg & $-1$ & $0$ & $-7$ & $-12$ & $-14$ & $-13$ & $-10$ & $-5$ & $1$ & $4$ \\
    \hline\hline
  \end{tabular}
  \label{tab:bulk_agg_ag111_ls_5d}
\end{table}

\clearpage

\section{Segregation Energies}

\subsection{Single-atom alloys}

\begin{figure}[htbp]
    \includegraphics[width=7.5cm,height=\textheight,keepaspectratio]{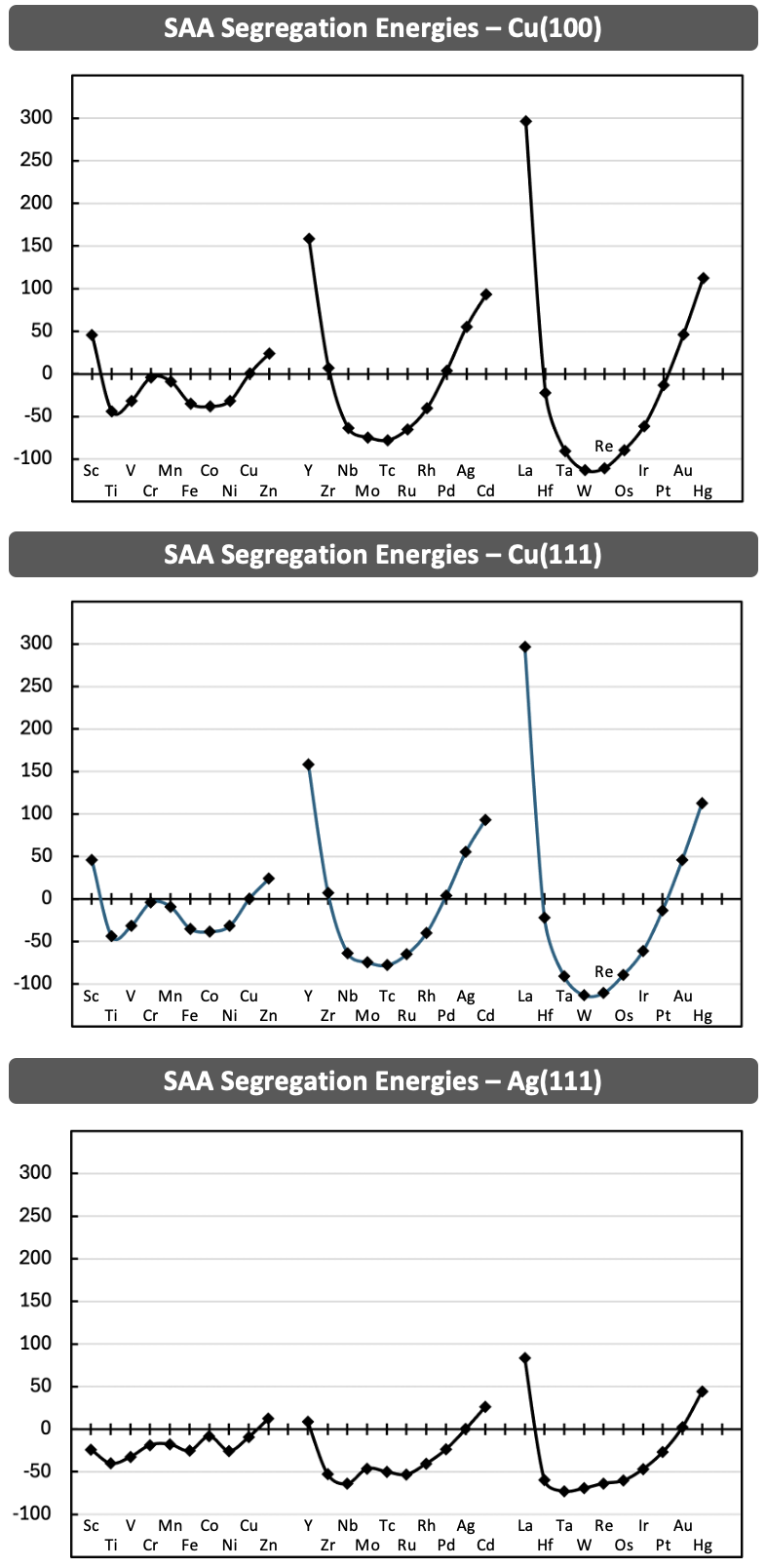}
    \caption{
        Segregation energies of all transition metal dopants in Cu(100), Cu(111), and Ag(111) single-atom alloys.
        Positive values indicate an energetic preference for the surface, whereas negative values indicate a preference for the bulk.
        All energies are reported in kJ~mol$^{-1}$.
    }
    \label{fig:SAA-seg}
\end{figure}

\begin{table}[ht]
  \centering
  \renewcommand{\arraystretch}{1.2}
  \caption{\textbf{Segregation energies} for \textbf{single-atom alloy} dopants in Cu(100), Cu(111), and Ag(111) for 3$d$, 4$d$, and 5$d$ transition metals. Results obtained from initial \textbf{low-spin} (LS) and \textbf{high-spin} (HS) configurations are reported. Energies are given in kJ~mol$^{-1}$.}
  \vspace{8pt}
  \begin{tabular}{@{\hspace{8pt}}*{7}{c@{\hspace{8pt}}}}
    \hline\hline
    TM & Cu(100) LS & Cu(100) HS & Cu(111) LS & Cu(111) HS & Ag(111) LS & Ag(111) HS \\
    \hline
    Sc & $46$  & $46$  & $49$  & $49$  & $-24$ & $-24$ \\
    Ti & $-49$ & $-44$ & $-22$ & $-22$ & $-51$ & $-40$ \\
    V  & $-81$ & $-32$ & $-43$ & $-21$ & $-49$ & $-32$ \\
    Cr & $-76$ & $-4$  & $-47$ & $-1$  & $-42$ & $-19$ \\
    Mn & $-71$ & $-9$  & $-44$ & $-3$  & $-37$ & $-18$ \\
    Fe & $-63$ & $-35$ & $-43$ & $-22$ & $-38$ & $-25$ \\
    Co & $-57$ & $-38$ & $-36$ & $-30$ & $-36$ & $-8$  \\
    Ni & $-32$ & $-32$ & $-21$ & $-21$ & $-26$ & $-26$ \\
    Cu & $0$   & $0$   & $0$   & $0$   & $-9$  & $-9$  \\
    Zn & $24$  & $24$  & $26$  & $26$  & $12$  & $12$  \\
    \hline
    Y  & $158$ & $158$ & $151$ & $151$ & $9$   & $9$   \\
    Zr & $7$   & $7$   & $24$  & $24$  & $-53$ & $-53$ \\
    Nb & $-65$ & $-64$ & $-31$ & $-31$ & $-67$ & $-64$ \\
    Mo & $-87$ & $-75$ & $-44$ & $-44$ & $-69$ & $-46$ \\
    Tc & $-84$ & $-78$ & $-45$ & $-45$ & $-64$ & $-50$ \\
    Ru & $-65$ & $-65$ & $-33$ & $-33$ & $-54$ & $-53$ \\
    Rh & $-40$ & $-40$ & $-17$ & $-17$ & $-41$ & $-41$ \\
    Pd & $4$   & $4$   & $10$  & $10$  & $-24$ & $-24$ \\
    Ag & $55$  & $55$  & $47$  & $47$  & $0$   & $0$   \\
    Cd & $93$  & $93$  & $88$  & $88$  & $26$  & $26$  \\
    \hline
    La & $296$ & $296$ & $285$ & $285$ & $84$  & $84$  \\
    Hf & $-22$ & $-22$ & $3$   & $3$   & $-60$ & $-60$ \\
    Ta & $-91$ & $-91$ & $-44$ & $-44$ & $-73$ & $-73$ \\
    W  & $-115$& $-113$& $-57$ & $-57$ & $-75$ & $-69$ \\
    Re & $-111$& $-111$& $-57$ & $-57$ & $-72$ & $-64$ \\
    Os & $-90$ & $-90$ & $-43$ & $-43$ & $-60$ & $-60$ \\
    Ir & $-61$ & $-61$ & $-24$ & $-24$ & $-47$ & $-47$ \\
    Pt & $-14$ & $-14$ & $8$   & $8$   & $-27$ & $-27$ \\
    Au & $46$  & $46$  & $48$  & $48$  & $2$   & $2$   \\
    Hg & $113$ & $113$ & $112$ & $112$ & $44$  & $44$  \\
    \hline\hline
  \end{tabular}
  \label{tab:segregation_energies_multisurface_spin}
\end{table}

\clearpage

\subsection{Dual-atom alloys}


\begin{figure}[htbp]
    \includegraphics[width=12cm,height=\textheight,keepaspectratio]{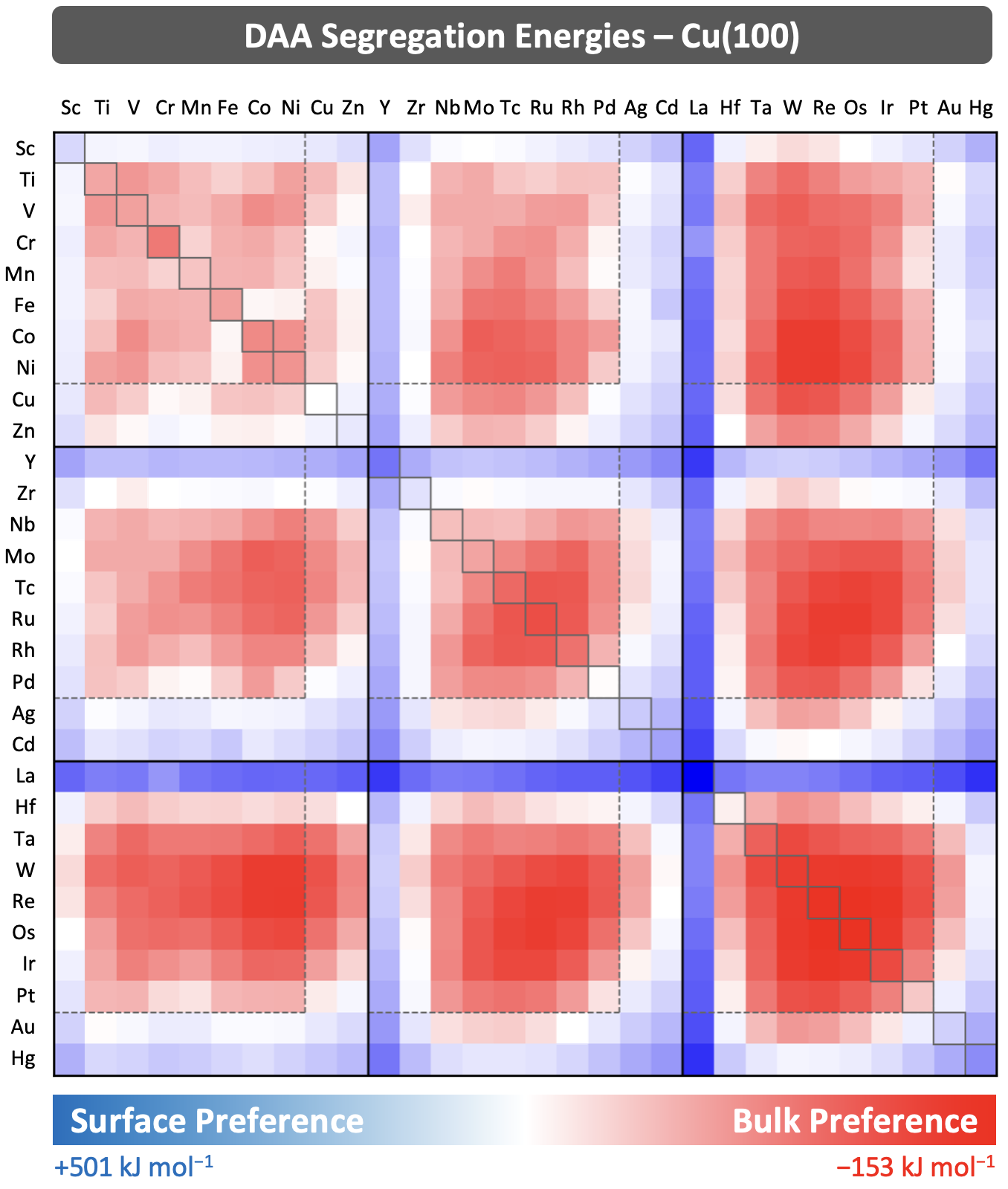}
    \caption{
        Segregation energies of all possible transition metal dopant dimers in Cu(100).
        Positive values (blue) indicate an energetic preference for the surface, whereas negative values (red) indicate a preference for bulk incorporation.
        All energies are reported in kJ~mol$^{-1}$.
           }
    \label{fig:Cu100-DAA-seg}
\end{figure}

\begin{table}[ht]
  \centering
  \renewcommand{\arraystretch}{1.2}
  \caption{\textbf{Segregation energies} for \textbf{dual-atom} sites in \textbf{Cu(100)} for all combinations of \textbf{3\textit{d}} transition metals (columns) and all other transition metals (rows), as obtained from initial \textbf{high-spin} configurations. Energies are reported in kJ~mol$^{-1}$.}
  \vspace{8pt}
  \begin{tabular}{@{\hspace{8pt}}*{11}{c@{\hspace{8pt}}}}
    \hline\hline
    TM & Sc & Ti & V & Cr & Mn & Fe & Co & Ni & Cu & Zn \\
    \hline
    Sc & $76$ & $20$ & $19$ & $32$ & $26$ & $24$ & $32$ & $36$ & $46$ & $67$ \\
    Ti & $20$ & $-55$ & $-66$ & $-54$ & $-41$ & $-30$ & $-40$ & $-60$ & $-44$ & $-16$ \\
    V & $19$ & $-66$ & $-59$ & $-47$ & $-43$ & $-53$ & $-73$ & $-65$ & $-31$ & $-4$ \\
    Cr & $32$ & $-54$ & $-47$ & $-86$ & $-28$ & $-50$ & $-53$ & $-41$ & $-4$ & $20$ \\
    Mn & $26$ & $-41$ & $-43$ & $-28$ & $-36$ & $-47$ & $-49$ & $-35$ & $-9$ & $11$ \\
    Fe & $24$ & $-30$ & $-53$ & $-50$ & $-47$ & $-59$ & $-6$ & $-9$ & $-35$ & $-9$ \\
    Co & $32$ & $-40$ & $-73$ & $-53$ & $-49$ & $-6$ & $-74$ & $-70$ & $-38$ & $-11$ \\
    Ni & $36$ & $-60$ & $-65$ & $-41$ & $-35$ & $-9$ & $-70$ & $-68$ & $-31$ & $-4$ \\
    Cu & $46$ & $-44$ & $-31$ & $-4$ & $-9$ & $-35$ & $-38$ & $-31$ & $1$ & $24$ \\
    Zn & $67$ & $-16$ & $-4$ & $20$ & $11$ & $-9$ & $-11$ & $-4$ & $24$ & $46$ \\
    \hline
    Y & $178$ & $126$ & $129$ & $143$ & $137$ & $135$ & $138$ & $148$ & $159$ & $177$ \\
    Zr & $59$ & $1$ & $-11$ & $-1$ & $7$ & $11$ & $13$ & $0$ & $7$ & $32$ \\
    Nb & $11$ & $-48$ & $-53$ & $-44$ & $-49$ & $-53$ & $-70$ & $-81$ & $-63$ & $-32$ \\
    Mo & $0$ & $-54$ & $-54$ & $-54$ & $-71$ & $-88$ & $-103$ & $-100$ & $-74$ & $-48$ \\
    Tc & $9$ & $-36$ & $-52$ & $-67$ & $-83$ & $-91$ & $-100$ & $-103$ & $-78$ & $-44$ \\
    Ru & $27$ & $-30$ & $-62$ & $-70$ & $-68$ & $-80$ & $-94$ & $-99$ & $-65$ & $-32$ \\
    Rh & $43$ & $-39$ & $-63$ & $-51$ & $-42$ & $-63$ & $-77$ & $-77$ & $-40$ & $-7$ \\
    Pd & $57$ & $-38$ & $-32$ & $-8$ & $-3$ & $-31$ & $-62$ & $-33$ & $4$ & $32$ \\
    Ag & $88$ & $4$ & $20$ & $47$ & $42$ & $18$ & $22$ & $24$ & $56$ & $79$ \\
    Cd & $126$ & $50$ & $62$ & $86$ & $78$ & $109$ & $43$ & $67$ & $94$ & $115$ \\
    \hline
    La & $300$ & $253$ & $262$ & $206$ & $271$ & $288$ & $301$ & $294$ & $297$ & $316$ \\
    Hf & $29$ & $-31$ & $-43$ & $-32$ & $-27$ & $-28$ & $-22$ & $-29$ & $-22$ & $2$ \\
    Ta & $-12$ & $-79$ & $-97$ & $-85$ & $-85$ & $-87$ & $-96$ & $-103$ & $-90$ & $-59$ \\
    W & $-24$ & $-94$ & $-105$ & $-100$ & $-107$ & $-118$ & $-134$ & $-135$ & $-113$ & $-78$ \\
    Re & $-17$ & $-80$ & $-95$ & $-103$ & $-111$ & $-120$ & $-135$ & $-136$ & $-110$ & $-73$ \\
    Os & $2$ & $-62$ & $-92$ & $-96$ & $-92$ & $-105$ & $-120$ & $-123$ & $-89$ & $-54$ \\
    Ir & $28$ & $-55$ & $-81$ & $-70$ & $-62$ & $-83$ & $-98$ & $-96$ & $-61$ & $-27$ \\
    Pt & $52$ & $-45$ & $-47$ & $-24$ & $-17$ & $-45$ & $-48$ & $-50$ & $-13$ & $17$ \\
    Au & $87$ & $-1$ & $12$ & $38$ & $35$ & $12$ & $12$ & $14$ & $46$ & $71$ \\
    Hg & $155$ & $74$ & $84$ & $108$ & $101$ & $82$ & $63$ & $87$ & $113$ & $136$ \\
    \hline\hline
  \end{tabular}
  \label{tab:daa_seg_cu100_hs_3d}
\end{table}

\begin{table}[ht]
  \centering
  \renewcommand{\arraystretch}{1.2}
  \caption{\textbf{Segregation energies} for \textbf{dual-atom} sites in \textbf{Cu(100)} for all combinations of \textbf{4\textit{d}} transition metals (columns) and all other transition metals (rows), as obtained from initial \textbf{high-spin} configurations. Energies are reported in kJ~mol$^{-1}$.}
  \vspace{8pt}
  \begin{tabular}{@{\hspace{8pt}}*{11}{c@{\hspace{8pt}}}}
    \hline\hline
    TM & Y & Zr & Nb & Mo & Tc & Ru & Rh & Pd & Ag & Cd \\
    \hline
    Sc & $178$ & $59$ & $11$ & $0$ & $9$ & $27$ & $43$ & $57$ & $88$ & $126$ \\
    Ti & $126$ & $1$ & $-48$ & $-54$ & $-36$ & $-30$ & $-39$ & $-38$ & $4$ & $50$ \\
    V  & $129$ & $-11$ & $-53$ & $-54$ & $-52$ & $-62$ & $-63$ & $-32$ & $20$ & $62$ \\
    Cr & $143$ & $-1$ & $-44$ & $-54$ & $-67$ & $-70$ & $-51$ & $-8$ & $47$ & $86$ \\
    Mn & $137$ & $7$ & $-49$ & $-71$ & $-83$ & $-68$ & $-42$ & $-3$ & $42$ & $78$ \\
    Fe & $135$ & $11$ & $-53$ & $-88$ & $-91$ & $-80$ & $-63$ & $-31$ & $18$ & $109$ \\
    Co & $138$ & $13$ & $-70$ & $-103$ & $-100$ & $-94$ & $-77$ & $-62$ & $22$ & $43$ \\
    Ni & $148$ & $0$ & $-81$ & $-100$ & $-103$ & $-99$ & $-77$ & $-33$ & $24$ & $67$ \\
    Cu & $159$ & $7$ & $-63$ & $-74$ & $-78$ & $-65$ & $-40$ & $4$ & $56$ & $94$ \\
    Zn & $177$ & $32$ & $-32$ & $-48$ & $-44$ & $-32$ & $-7$ & $32$ & $79$ & $115$ \\
    \hline
    Y  & $274$ & $162$ & $120$ & $111$ & $118$ & $133$ & $151$ & $169$ & $197$ & $234$ \\
    Zr & $162$ & $56$ & $8$ & $-2$ & $11$ & $19$ & $17$ & $17$ & $49$ & $97$ \\
    Nb & $120$ & $8$ & $-41$ & $-44$ & $-41$ & $-54$ & $-65$ & $-61$ & $-16$ & $36$ \\
    Mo & $111$ & $-2$ & $-44$ & $-57$ & $-75$ & $-91$ & $-100$ & $-73$ & $-22$ & $21$ \\
    Tc & $118$ & $11$ & $-41$ & $-75$ & $-96$ & $-111$ & $-108$ & $-75$ & $-25$ & $25$ \\
    Ru & $133$ & $19$ & $-54$ & $-91$ & $-111$ & $-117$ & $-107$ & $-71$ & $-12$ & $36$ \\
    Rh & $151$ & $17$ & $-65$ & $-100$ & $-108$ & $-107$ & $-90$ & $-47$ & $12$ & $59$ \\
    Pd & $169$ & $17$ & $-61$ & $-73$ & $-75$ & $-71$ & $-47$ & $-2$ & $55$ & $98$ \\
    Ag & $197$ & $49$ & $-16$ & $-22$ & $-25$ & $-12$ & $12$ & $55$ & $105$ & $144$ \\
    Cd & $234$ & $97$ & $36$ & $21$ & $25$ & $36$ & $59$ & $98$ & $144$ & $180$ \\
    \hline
    La & $389$ & $288$ & $257$ & $261$ & $281$ & $304$ & $314$ & $314$ & $336$ & $372$ \\
    Hf & $139$ & $24$ & $-26$ & $-42$ & $-33$ & $-18$ & $-11$ & $-6$ & $23$ & $71$ \\
    Ta & $102$ & $-15$ & $-74$ & $-86$ & $-78$ & $-81$ & $-86$ & $-80$ & $-41$ & $15$ \\
    W  & $92$ & $-31$ & $-86$ & $-97$ & $-108$ & $-119$ & $-124$ & $-107$ & $-60$ & $-5$ \\
    Re & $100$ & $-21$ & $-77$ & $-106$ & $-124$ & $-134$ & $-133$ & $-108$ & $-56$ & $-1$ \\
    Os & $118$ & $-3$ & $-74$ & $-111$ & $-129$ & $-135$ & $-125$ & $-92$ & $-35$ & $16$ \\
    Ir & $142$ & $7$ & $-77$ & $-111$ & $-123$ & $-122$ & $-106$ & $-66$ & $-7$ & $42$ \\
    Pt & $168$ & $15$ & $-66$ & $-85$ & $-89$ & $-85$ & $-63$ & $-18$ & $40$ & $86$ \\
    Au & $201$ & $49$ & $-20$ & $-30$ & $-32$ & $-21$ & $3$ & $47$ & $98$ & $139$ \\
    Hg & $268$ & $130$ & $65$ & $47$ & $49$ & $57$ & $81$ & $119$ & $165$ & $202$ \\
    \hline\hline
  \end{tabular}
  \label{tab:daa_seg_cu100_hs_4d}
\end{table}

\begin{table}[ht]
  \centering
  \renewcommand{\arraystretch}{1.2}
  \caption{\textbf{Segregation energies} for \textbf{dual-atom} sites in \textbf{Cu(100)} for all combinations of \textbf{5\textit{d}} transition metals (columns) and all other transition metals (rows), as obtained from initial \textbf{high-spin} configurations. Energies are reported in kJ~mol$^{-1}$.}
  \vspace{8pt}
  \begin{tabular}{@{\hspace{8pt}}*{11}{c@{\hspace{8pt}}}}
    \hline\hline
    TM & La & Hf & Ta & W & Re & Os & Ir & Pt & Au & Hg \\
    \hline
    Sc & $300$ & $29$ & $-12$ & $-24$ & $-17$ & $2$ & $28$ & $52$ & $87$ & $155$ \\
    Ti & $253$ & $-31$ & $-79$ & $-94$ & $-80$ & $-62$ & $-55$ & $-45$ & $-1$ & $74$ \\
    V  & $262$ & $-43$ & $-97$ & $-105$ & $-95$ & $-92$ & $-81$ & $-47$ & $12$ & $84$ \\
    Cr & $206$ & $-32$ & $-85$ & $-100$ & $-103$ & $-96$ & $-70$ & $-24$ & $38$ & $108$ \\
    Mn & $271$ & $-27$ & $-85$ & $-107$ & $-111$ & $-92$ & $-62$ & $-17$ & $35$ & $101$ \\
    Fe & $288$ & $-28$ & $-87$ & $-118$ & $-120$ & $-105$ & $-83$ & $-45$ & $12$ & $82$ \\
    Co & $301$ & $-22$ & $-96$ & $-134$ & $-135$ & $-120$ & $-98$ & $-48$ & $12$ & $63$ \\
    Ni & $294$ & $-29$ & $-103$ & $-135$ & $-136$ & $-123$ & $-96$ & $-50$ & $14$ & $87$ \\
    Cu & $297$ & $-22$ & $-90$ & $-113$ & $-110$ & $-89$ & $-61$ & $-13$ & $46$ & $113$ \\
    Zn & $316$ & $2$ & $-59$ & $-78$ & $-73$ & $-54$ & $-27$ & $17$ & $71$ & $136$ \\
    \hline
    Y  & $389$ & $139$ & $102$ & $92$ & $100$ & $118$ & $142$ & $168$ & $201$ & $268$ \\
    Zr & $288$ & $24$ & $-15$ & $-31$ & $-21$ & $-3$ & $7$ & $15$ & $49$ & $130$ \\
    Nb & $257$ & $-26$ & $-74$ & $-86$ & $-77$ & $-74$ & $-77$ & $-66$ & $-20$ & $65$ \\
    Mo & $261$ & $-42$ & $-86$ & $-97$ & $-106$ & $-111$ & $-111$ & $-85$ & $-30$ & $47$ \\
    Tc & $281$ & $-33$ & $-78$ & $-108$ & $-124$ & $-129$ & $-123$ & $-89$ & $-32$ & $49$ \\
    Ru & $304$ & $-18$ & $-81$ & $-119$ & $-134$ & $-135$ & $-122$ & $-85$ & $-21$ & $57$ \\
    Rh & $314$ & $-11$ & $-86$ & $-124$ & $-133$ & $-125$ & $-106$ & $-63$ & $3$ & $81$ \\
    Pd & $314$ & $-6$ & $-80$ & $-107$ & $-108$ & $-92$ & $-66$ & $-18$ & $47$ & $119$ \\
    Ag & $336$ & $23$ & $-41$ & $-60$ & $-56$ & $-35$ & $-7$ & $40$ & $98$ & $165$ \\
    Cd & $372$ & $71$ & $15$ & $-5$ & $-1$ & $16$ & $42$ & $86$ & $139$ & $202$ \\
    \hline
    La & $501$ & $270$ & $239$ & $239$ & $257$ & $283$ & $306$ & $321$ & $346$ & $404$ \\
    Hf & $270$ & $-11$ & $-51$ & $-69$ & $-62$ & $-41$ & $-24$ & $-10$ & $23$ & $106$ \\
    Ta & $239$ & $-51$ & $-103$ & $-121$ & $-111$ & $-102$ & $-99$ & $-85$ & $-43$ & $46$ \\
    W  & $239$ & $-69$ & $-121$ & $-134$ & $-138$ & $-139$ & $-136$ & $-115$ & $-64$ & $22$ \\
    Re & $257$ & $-62$ & $-111$ & $-138$ & $-151$ & $-152$ & $-145$ & $-118$ & $-61$ & $24$ \\
    Os & $283$ & $-41$ & $-102$ & $-139$ & $-152$ & $-153$ & $-139$ & $-105$ & $-42$ & $39$ \\
    Ir & $306$ & $-24$ & $-99$ & $-136$ & $-145$ & $-139$ & $-121$ & $-80$ & $-15$ & $65$ \\
    Pt & $321$ & $-10$ & $-85$ & $-115$ & $-118$ & $-105$ & $-80$ & $-34$ & $32$ & $108$ \\
    Au & $346$ & $23$ & $-43$ & $-64$ & $-61$ & $-42$ & $-15$ & $32$ & $92$ & $161$ \\
    Hg & $404$ & $106$ & $46$ & $22$ & $24$ & $39$ & $65$ & $108$ & $161$ & $225$ \\
    \hline\hline
  \end{tabular}
  \label{tab:daa_seg_cu100_hs_5d}
\end{table}


\begin{table}[ht]
  \centering
  \renewcommand{\arraystretch}{1.2}
  \caption{\textbf{Segregation energies} for \textbf{dual-atom} sites in \textbf{Cu(100)} for all combinations of \textbf{3\textit{d}} transition metals (columns) and all other transition metals (rows), as obtained from initial \textbf{low-spin} configurations. Energies are reported in kJ~mol$^{-1}$.}
  \vspace{8pt}
  \begin{tabular}{@{\hspace{8pt}}*{11}{c@{\hspace{8pt}}}}
    \hline\hline
    TM & Sc & Ti & V & Cr & Mn & Fe & Co & Ni & Cu & Zn \\
    \hline
    Sc & $76$ & $8$ & $-6$ & $-6$ & $10$ & $23$ & $32$ & $36$ & $46$ & $67$ \\
    Ti & $8$ & $-55$ & $-62$ & $-44$ & $-25$ & $-19$ & $-40$ & $-60$ & $-49$ & $-22$ \\
    V  & $-6$ & $-62$ & $-59$ & $-46$ & $-43$ & $-57$ & $-86$ & $-98$ & $-78$ & $-48$ \\
    Cr & $-6$ & $-44$ & $-46$ & $-48$ & $-63$ & $-78$ & $-98$ & $-102$ & $64$ & $-47$ \\
    Mn & $10$ & $-25$ & $-43$ & $-63$ & $-78$ & $-92$ & $-108$ & $-98$ & $-68$ & $-38$ \\
    Fe & $23$ & $-19$ & $-57$ & $-78$ & $-92$ & $-107$ & $-111$ & $-96$ & $-63$ & $-33$ \\
    Co & $32$ & $-40$ & $-86$ & $-98$ & $-108$ & $-111$ & $-111$ & $-93$ & $-56$ & $-25$ \\
    Ni & $36$ & $-60$ & $-98$ & $-102$ & $-98$ & $-96$ & $-93$ & $-68$ & $-31$ & $-4$ \\
    Cu & $46$ & $-49$ & $-78$ & $64$ & $-68$ & $-63$ & $-56$ & $-31$ & $1$ & $24$ \\
    Zn & $67$ & $-22$ & $-48$ & $-47$ & $-38$ & $-33$ & $-25$ & $-4$ & $24$ & $46$ \\
    \hline
    Y  & $178$ & $117$ & $104$ & $104$ & $118$ & $125$ & $138$ & $148$ & $159$ & $177$ \\
    Zr & $58$ & $1$ & $-13$ & $-4$ & $13$ & $21$ & $13$ & $0$ & $7$ & $32$ \\
    Nb & $12$ & $-48$ & $-53$ & $-38$ & $-32$ & $-44$ & $-70$ & $-81$ & $-65$ & $-33$ \\
    Mo & $0$ & $-54$ & $-54$ & $-52$ & $-64$ & $-86$ & $-108$ & $-112$ & $-87$ & $-51$ \\
    Tc & $9$ & $-36$ & $-52$ & $-69$ & $-84$ & $-108$ & $-123$ & $-114$ & $-83$ & $-47$ \\
    Ru & $27$ & $-30$ & $-67$ & $-42$ & $-97$ & $-113$ & $-117$ & $-101$ & $-65$ & $-32$ \\
    Rh & $43$ & $-39$ & $-79$ & $-89$ & $-95$ & $-99$ & $-99$ & $-77$ & $-40$ & $-7$ \\
    Pd & $57$ & $-39$ & $-71$ & $-71$ & $-65$ & $-62$ & $-58$ & $-33$ & $4$ & $32$ \\
    Ag & $88$ & $-1$ & $-29$ & $-22$ & $-14$ & $-7$ & $-1$ & $24$ & $56$ & $79$ \\
    Cd & $126$ & $44$ & $21$ & $23$ & $32$ & $38$ & $45$ & $67$ & $94$ & $115$ \\
    \hline
    La & $299$ & $244$ & $237$ & $253$ & $275$ & $295$ & $301$ & $294$ & $297$ & $316$ \\
    Hf & $29$ & $-33$ & $-51$ & $-50$ & $-37$ & $-23$ & $-22$ & $-29$ & $-22$ & $2$ \\
    Ta & $-12$ & $-78$ & $-99$ & $-90$ & $-78$ & $-79$ & $-96$ & $-103$ & $-90$ & $-59$ \\
    W  & $-24$ & $-94$ & $-105$ & $-100$ & $-104$ & $-119$ & $-135$ & $-137$ & $-115$ & $-78$ \\
    Re & $-17$ & $-80$ & $-95$ & $-107$ & $-117$ & $-137$ & $-148$ & $-139$ & $-110$ & $-73$ \\
    Os & $2$ & $-62$ & $-95$ & $-107$ & $-120$ & $-137$ & $-140$ & $-124$ & $-89$ & $-54$ \\
    Ir & $28$ & $-55$ & $-96$ & $-104$ & $-114$ & $-117$ & $-117$ & $-96$ & $-61$ & $-27$ \\
    Pt & $52$ & $-45$ & $-81$ & $-84$ & $-79$ & $-78$ & $-75$ & $-50$ & $-13$ & $17$ \\
    Au & $87$ & $-6$ & $-33$ & $-31$ & $-22$ & $-16$ & $-11$ & $14$ & $46$ & $71$ \\
    Hg & $155$ & $72$ & $46$ & $47$ & $53$ & $59$ & $65$ & $87$ & $113$ & $136$ \\
    \hline\hline
  \end{tabular}
  \label{tab:daa_seg_cu100_ls_3d}
\end{table}

\begin{table}[ht]
  \centering
  \renewcommand{\arraystretch}{1.2}
  \caption{\textbf{Segregation energies} for \textbf{dual-atom} sites in \textbf{Cu(100)} for all combinations of \textbf{4\textit{d}} transition metals (columns) and all other transition metals (rows), as obtained from initial \textbf{low-spin} configurations. Energies are reported in kJ~mol$^{-1}$.}
  \vspace{8pt}
  \begin{tabular}{@{\hspace{8pt}}*{11}{c@{\hspace{8pt}}}}
    \hline\hline
    TM & Y & Zr & Nb & Mo & Tc & Ru & Rh & Pd & Ag & Cd \\
    \hline
    Sc & $178$ & $58$ & $12$ & $0$ & $9$ & $27$ & $43$ & $57$ & $88$ & $126$ \\
    Ti & $117$ & $1$ & $-48$ & $-54$ & $-36$ & $-30$ & $-39$ & $-39$ & $-1$ & $44$ \\
    V  & $104$ & $-13$ & $-53$ & $-54$ & $-52$ & $-67$ & $-79$ & $-71$ & $-29$ & $21$ \\
    Cr & $104$ & $-4$ & $-38$ & $-52$ & $-69$ & $-42$ & $-89$ & $-71$ & $-22$ & $23$ \\
    Mn & $118$ & $13$ & $-32$ & $-64$ & $-84$ & $-97$ & $-95$ & $-65$ & $-14$ & $32$ \\
    Fe & $125$ & $21$ & $-44$ & $-86$ & $-108$ & $-113$ & $-99$ & $-62$ & $-7$ & $38$ \\
    Co & $138$ & $13$ & $-70$ & $-108$ & $-123$ & $-117$ & $-99$ & $-58$ & $-1$ & $45$ \\
    Ni & $148$ & $0$ & $-81$ & $-112$ & $-114$ & $-101$ & $-77$ & $-33$ & $24$ & $67$ \\
    Cu & $159$ & $7$ & $-65$ & $-87$ & $-83$ & $-65$ & $-40$ & $4$ & $56$ & $94$ \\
    Zn & $177$ & $32$ & $-33$ & $-51$ & $-47$ & $-32$ & $-7$ & $32$ & $79$ & $115$ \\
    \hline
    Y  & $274$ & $162$ & $120$ & $111$ & $117$ & $133$ & $151$ & $169$ & $197$ & $234$ \\
    Zr & $162$ & $56$ & $8$ & $-2$ & $11$ & $19$ & $17$ & $17$ & $49$ & $97$ \\
    Nb & $120$ & $8$ & $-41$ & $-44$ & $-41$ & $-54$ & $-65$ & $-61$ & $-18$ & $36$ \\
    Mo & $111$ & $-2$ & $-44$ & $-57$ & $-75$ & $-92$ & $-103$ & $-87$ & $-35$ & $19$ \\
    Tc & $117$ & $11$ & $-41$ & $-75$ & $-96$ & $-113$ & $-115$ & $-86$ & $-31$ & $21$ \\
    Ru & $133$ & $19$ & $-54$ & $-92$ & $-113$ & $-121$ & $-108$ & $-71$ & $-12$ & $35$ \\
    Rh & $151$ & $17$ & $-65$ & $-103$ & $-115$ & $-108$ & $-90$ & $-47$ & $12$ & $59$ \\
    Pd & $169$ & $17$ & $-61$ & $-87$ & $-86$ & $-71$ & $-47$ & $-2$ & $55$ & $98$ \\
    Ag & $197$ & $49$ & $-18$ & $-35$ & $-31$ & $-12$ & $12$ & $55$ & $105$ & $144$ \\
    Cd & $234$ & $97$ & $36$ & $19$ & $21$ & $35$ & $59$ & $98$ & $144$ & $180$ \\
    \hline
    La & $389$ & $288$ & $257$ & $261$ & $281$ & $304$ & $314$ & $314$ & $336$ & $372$ \\
    Hf & $139$ & $24$ & $-26$ & $-43$ & $-33$ & $-18$ & $-11$ & $-6$ & $23$ & $71$ \\
    Ta & $101$ & $-15$ & $-74$ & $-86$ & $-78$ & $-81$ & $-86$ & $-80$ & $-41$ & $15$ \\
    W  & $92$ & $-31$ & $-86$ & $-97$ & $-108$ & $-119$ & $-124$ & $-109$ & $-61$ & $-5$ \\
    Re & $99$ & $-21$ & $-77$ & $-106$ & $-124$ & $-136$ & $-135$ & $-108$ & $-56$ & $-2$ \\
    Os & $118$ & $-3$ & $-74$ & $-112$ & $-131$ & $-138$ & $-125$ & $-92$ & $-35$ & $16$ \\
    Ir & $142$ & $7$ & $-77$ & $-114$ & $-128$ & $-123$ & $-106$ & $-66$ & $-7$ & $42$ \\
    Pt & $168$ & $15$ & $-66$ & $-96$ & $-98$ & $-86$ & $-63$ & $-18$ & $40$ & $86$ \\
    Au & $201$ & $49$ & $-21$ & $-41$ & $-38$ & $-21$ & $3$ & $47$ & $98$ & $139$ \\
    Hg & $268$ & $130$ & $64$ & $44$ & $44$ & $57$ & $81$ & $119$ & $165$ & $202$ \\
    \hline\hline
  \end{tabular}
  \label{tab:daa_seg_cu100_ls_4d}
\end{table}

\begin{table}[ht]
  \centering
  \renewcommand{\arraystretch}{1.2}
  \caption{\textbf{Segregation energies} for \textbf{dual-atom} sites in \textbf{Cu(100)} for all combinations of \textbf{5\textit{d}} transition metals (columns) and all other transition metals (rows), as obtained from initial \textbf{low-spin} configurations. Energies are reported in kJ~mol$^{-1}$.}
  \vspace{8pt}
  \begin{tabular}{@{\hspace{8pt}}*{11}{c@{\hspace{8pt}}}}
    \hline\hline
    TM & La & Hf & Ta & W & Re & Os & Ir & Pt & Au & Hg \\
    \hline
    Sc & $299$ & $29$ & $-12$ & $-24$ & $-17$ & $2$ & $28$ & $52$ & $87$ & $155$ \\
    Ti & $244$ & $-33$ & $-78$ & $-94$ & $-80$ & $-62$ & $-55$ & $-45$ & $-6$ & $72$ \\
    V  & $237$ & $-51$ & $-99$ & $-105$ & $-95$ & $-95$ & $-96$ & $-81$ & $-33$ & $46$ \\
    Cr & $253$ & $-50$ & $-90$ & $-100$ & $-107$ & $-107$ & $-104$ & $-84$ & $-31$ & $47$ \\
    Mn & $275$ & $-37$ & $-78$ & $-104$ & $-117$ & $-120$ & $-114$ & $-79$ & $-22$ & $53$ \\
    Fe & $295$ & $-23$ & $-79$ & $-119$ & $-137$ & $-137$ & $-117$ & $-78$ & $-16$ & $59$ \\
    Co & $301$ & $-22$ & $-96$ & $-135$ & $-148$ & $-140$ & $-117$ & $-75$ & $-11$ & $65$ \\
    Ni & $294$ & $-29$ & $-103$ & $-137$ & $-139$ & $-124$ & $-96$ & $-50$ & $14$ & $87$ \\
    Cu & $297$ & $-22$ & $-90$ & $-115$ & $-110$ & $-89$ & $-61$ & $-13$ & $46$ & $113$ \\
    Zn & $316$ & $2$ & $-59$ & $-78$ & $-73$ & $-54$ & $-27$ & $17$ & $71$ & $136$ \\
    \hline
    Y  & $389$ & $139$ & $101$ & $92$ & $99$ & $118$ & $142$ & $168$ & $201$ & $268$ \\
    Zr & $288$ & $24$ & $-15$ & $-31$ & $-21$ & $-3$ & $7$ & $15$ & $49$ & $130$ \\
    Nb & $257$ & $-26$ & $-74$ & $-86$ & $-77$ & $-74$ & $-77$ & $-66$ & $-21$ & $64$ \\
    Mo & $261$ & $-43$ & $-86$ & $-97$ & $-106$ & $-112$ & $-114$ & $-96$ & $-41$ & $44$ \\
    Tc & $281$ & $-33$ & $-78$ & $-108$ & $-124$ & $-131$ & $-128$ & $-98$ & $-38$ & $44$ \\
    Ru & $304$ & $-18$ & $-81$ & $-119$ & $-136$ & $-138$ & $-123$ & $-86$ & $-21$ & $57$ \\
    Rh & $314$ & $-11$ & $-86$ & $-124$ & $-135$ & $-125$ & $-106$ & $-63$ & $3$ & $81$ \\
    Pd & $314$ & $-6$ & $-80$ & $-109$ & $-108$ & $-92$ & $-66$ & $-18$ & $47$ & $119$ \\
    Ag & $336$ & $23$ & $-41$ & $-61$ & $-56$ & $-35$ & $-7$ & $40$ & $98$ & $165$ \\
    Cd & $372$ & $71$ & $15$ & $-5$ & $-2$ & $16$ & $42$ & $86$ & $139$ & $202$ \\
    \hline
    La & $501$ & $270$ & $239$ & $239$ & $257$ & $283$ & $306$ & $321$ & $346$ & $404$ \\
    Hf & $270$ & $-11$ & $-51$ & $-69$ & $-62$ & $-41$ & $-24$ & $-10$ & $23$ & $106$ \\
    Ta & $239$ & $-51$ & $-103$ & $-121$ & $-111$ & $-102$ & $-99$ & $-85$ & $-43$ & $46$ \\
    W  & $239$ & $-69$ & $-121$ & $-134$ & $-138$ & $-139$ & $-136$ & $-117$ & $-65$ & $22$ \\
    Re & $257$ & $-62$ & $-111$ & $-138$ & $-151$ & $-155$ & $-147$ & $-118$ & $-61$ & $23$ \\
    Os & $283$ & $-41$ & $-102$ & $-139$ & $-155$ & $-155$ & $-139$ & $-105$ & $-42$ & $39$ \\
    Ir & $306$ & $-24$ & $-99$ & $-136$ & $-147$ & $-139$ & $-121$ & $-80$ & $-15$ & $65$ \\
    Pt & $321$ & $-10$ & $-85$ & $-117$ & $-118$ & $-105$ & $-80$ & $-34$ & $32$ & $108$ \\
    Au & $346$ & $23$ & $-43$ & $-65$ & $-61$ & $-42$ & $-15$ & $32$ & $92$ & $161$ \\
    Hg & $404$ & $106$ & $46$ & $22$ & $23$ & $39$ & $65$ & $108$ & $161$ & $225$ \\
    \hline\hline
  \end{tabular}
  \label{tab:daa_seg_cu100_ls_5d}
\end{table}

\clearpage


\begin{figure}[htbp]
    \includegraphics[width=12cm,height=\textheight,keepaspectratio]{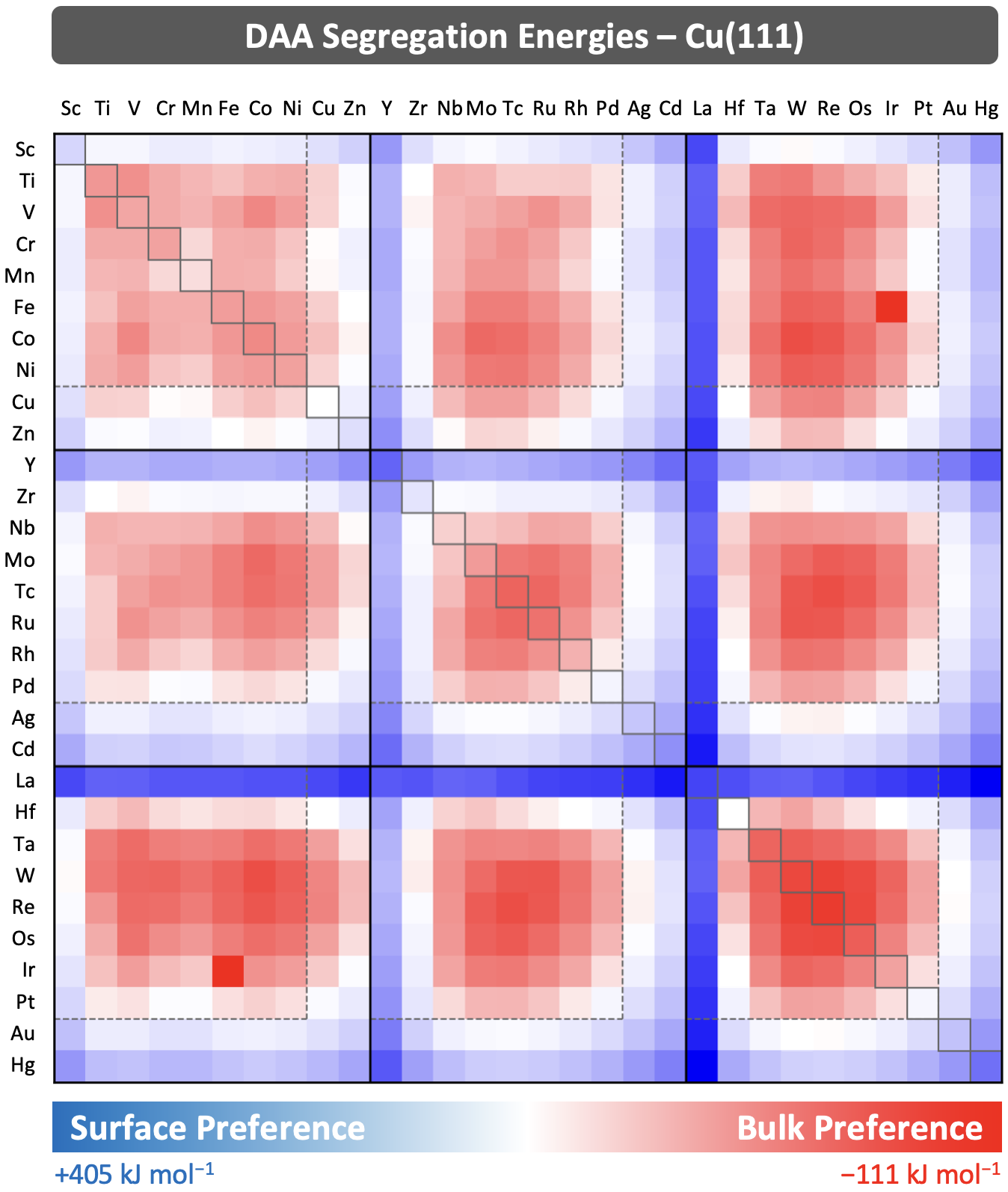}
    \caption{
        Segregation energies of all possible transition metal dopant dimers in Cu(111).
        Positive values (blue) indicate an energetic preference for the surface, whereas negative values (red) indicate a preference for bulk incorporation.
        All energies are reported in kJ~mol$^{-1}$.
    }
    \label{fig:Cu111-DAA-seg}
\end{figure}

\begin{table}[ht]
  \centering
  \renewcommand{\arraystretch}{1.2}
  \caption{\textbf{Segregation energies} for \textbf{dual-atom} sites in \textbf{Cu(111)} for all combinations of \textbf{3\textit{d}} transition metals (columns) and all other transition metals (rows), as obtained from initial \textbf{high-spin} configurations. Energies are reported in kJ~mol$^{-1}$.}
  \vspace{8pt}
  \begin{tabular}{@{\hspace{8pt}}*{11}{c@{\hspace{8pt}}}}
    \hline\hline
    TM & Sc & Ti & V & Cr & Mn & Fe & Co & Ni & Cu & Zn \\
    \hline
    Sc & $66$ & $14$ & $15$ & $31$ & $31$ & $21$ & $26$ & $33$ & $50$ & $75$ \\
    Ti & $14$ & $-47$ & $-51$ & $-38$ & $-33$ & $-28$ & $-36$ & $-38$ & $-21$ & $7$ \\
    V  & $15$ & $-51$ & $-40$ & $-38$ & $-34$ & $-43$ & $-55$ & $-44$ & $-21$ & $3$ \\
    Cr & $31$ & $-38$ & $-38$ & $-42$ & $-18$ & $-37$ & $-38$ & $-26$ & $-1$ & $23$ \\
    Mn & $31$ & $-33$ & $-34$ & $-18$ & $-15$ & $-37$ & $-36$ & $-23$ & $-3$ & $20$ \\
    Fe & $21$ & $-28$ & $-43$ & $-37$ & $-37$ & $-45$ & $-47$ & $-42$ & $-22$ & $2$ \\
    Co & $26$ & $-36$ & $-55$ & $-38$ & $-36$ & $-47$ & $-53$ & $-45$ & $-29$ & $-5$ \\
    Ni & $33$ & $-38$ & $-44$ & $-26$ & $-23$ & $-42$ & $-45$ & $-43$ & $-20$ & $4$ \\
    Cu & $50$ & $-21$ & $-21$ & $-1$ & $-3$ & $-22$ & $-29$ & $-20$ & $1$ & $26$ \\
    Zn & $75$ & $7$ & $3$ & $23$ & $20$ & $2$ & $-5$ & $4$ & $26$ & $53$ \\
    \hline
    Y  & $162$ & $120$ & $118$ & $134$ & $134$ & $125$ & $125$ & $135$ & $151$ & $178$ \\
    Zr & $53$ & $1$ & $-5$ & $7$ & $12$ & $13$ & $10$ & $12$ & $25$ & $51$ \\
    Nb & $17$ & $-36$ & $-33$ & $-33$ & $-34$ & $-40$ & $-51$ & $-47$ & $-30$ & $-2$ \\
    Mo & $9$ & $-33$ & $-37$ & $-44$ & $-48$ & $-59$ & $-69$ & $-62$ & $-44$ & $-19$ \\
    Tc & $19$ & $-23$ & $-43$ & $-50$ & $-48$ & $-59$ & $-67$ & $-63$ & $-44$ & $-18$ \\
    Ru & $34$ & $-23$ & $-50$ & $-42$ & $-37$ & $-50$ & $-58$ & $-54$ & $-33$ & $-6$ \\
    Rh & $45$ & $-23$ & $-39$ & $-25$ & $-19$ & $-37$ & $-44$ & $-39$ & $-17$ & $12$ \\
    Pd & $59$ & $-11$ & $-13$ & $5$ & $8$ & $-12$ & $-18$ & $-12$ & $11$ & $35$ \\
    Ag & $89$ & $23$ & $25$ & $43$ & $42$ & $26$ & $21$ & $27$ & $48$ & $72$ \\
    Cd & $135$ & $76$ & $70$ & $87$ & $85$ & $68$ & $55$ & $69$ & $89$ & $116$ \\
    \hline
    La & $289$ & $248$ & $252$ & $267$ & $267$ & $266$ & $274$ & $273$ & $286$ & $316$ \\
    Hf & $34$ & $-23$ & $-31$ & $-17$ & $-12$ & $-17$ & $-18$ & $-11$ & $3$ & $30$ \\
    Ta & $8$ & $-59$ & $-68$ & $-60$ & $-57$ & $-59$ & $-67$ & $-61$ & $-44$ & $-14$ \\
    W  & $-2$ & $-62$ & $-71$ & $-71$ & $-66$ & $-74$ & $-85$ & $-75$ & $-56$ & $-32$ \\
    Re & $8$ & $-48$ & $-69$ & $-67$ & $-60$ & $-71$ & $-80$ & $-73$ & $-57$ & $-31$ \\
    Os & $25$ & $-38$ & $-64$ & $-52$ & $-47$ & $-59$ & $-67$ & $-63$ & $-42$ & $-16$ \\
    Ir & $43$ & $-28$ & $-45$ & $-31$ & $-25$ & $-111$ & $-50$ & $-44$ & $-24$ & $6$ \\
    Pt & $64$ & $-9$ & $-14$ & $3$ & $6$ & $-14$ & $-21$ & $-14$ & $8$ & $33$ \\
    Au & $99$ & $31$ & $29$ & $47$ & $46$ & $28$ & $23$ & $30$ & $49$ & $73$ \\
    Hg & $165$ & $104$ & $98$ & $114$ & $111$ & $93$ & $83$ & $92$ & $112$ & $140$ \\
    \hline\hline
  \end{tabular}
  \label{tab:daa_seg_cu111_hs_3d}
\end{table}

\begin{table}[ht]
  \centering
  \renewcommand{\arraystretch}{1.2}
  \caption{\textbf{Segregation energies} for \textbf{dual-atom} sites in \textbf{Cu(111)} for all combinations of \textbf{4\textit{d}} transition metals (columns) and all other transition metals (rows), as obtained from initial \textbf{high-spin} configurations. Energies are reported in kJ~mol$^{-1}$.}
  \vspace{8pt}
  \begin{tabular}{@{\hspace{8pt}}*{11}{c@{\hspace{8pt}}}}
    \hline\hline
    TM & Y & Zr & Nb & Mo & Tc & Ru & Rh & Pd & Ag & Cd \\
    \hline
    Sc & $162$ & $53$ & $17$ & $9$ & $19$ & $34$ & $45$ & $59$ & $89$ & $135$ \\
    Ti & $120$ & $1$ & $-36$ & $-33$ & $-23$ & $-23$ & $-23$ & $-11$ & $23$ & $76$ \\
    V  & $118$ & $-5$ & $-33$ & $-37$ & $-43$ & $-50$ & $-39$ & $-13$ & $25$ & $70$ \\
    Cr & $134$ & $7$ & $-33$ & $-44$ & $-50$ & $-42$ & $-25$ & $5$ & $43$ & $87$ \\
    Mn & $134$ & $12$ & $-34$ & $-48$ & $-48$ & $-37$ & $-19$ & $8$ & $42$ & $85$ \\
    Fe & $125$ & $13$ & $-40$ & $-59$ & $-59$ & $-50$ & $-37$ & $-12$ & $26$ & $68$ \\
    Co & $125$ & $10$ & $-51$ & $-69$ & $-67$ & $-58$ & $-44$ & $-18$ & $21$ & $55$ \\
    Ni & $135$ & $12$ & $-47$ & $-62$ & $-63$ & $-54$ & $-39$ & $-12$ & $27$ & $69$ \\
    Cu & $151$ & $25$ & $-30$ & $-44$ & $-44$ & $-33$ & $-17$ & $11$ & $48$ & $89$ \\
    Zn & $178$ & $51$ & $-2$ & $-19$ & $-18$ & $-6$ & $12$ & $35$ & $72$ & $116$ \\
    \hline
    Y  & $244$ & $158$ & $123$ & $116$ & $126$ & $138$ & $149$ & $163$ & $191$ & $233$ \\
    Zr & $158$ & $44$ & $9$ & $12$ & $22$ & $24$ & $25$ & $34$ & $64$ & $120$ \\
    Nb & $123$ & $9$ & $-21$ & $-25$ & $-31$ & $-39$ & $-39$ & $-22$ & $15$ & $74$ \\
    Mo & $116$ & $12$ & $-25$ & $-46$ & $-62$ & $-65$ & $-57$ & $-36$ & $4$ & $54$ \\
    Tc & $126$ & $22$ & $-31$ & $-62$ & $-72$ & $-70$ & $-59$ & $-35$ & $4$ & $51$ \\
    Ru & $138$ & $24$ & $-39$ & $-65$ & $-70$ & $-64$ & $-50$ & $-25$ & $14$ & $60$ \\
    Rh & $149$ & $25$ & $-39$ & $-57$ & $-59$ & $-50$ & $-35$ & $-9$ & $29$ & $77$ \\
    Pd & $163$ & $34$ & $-22$ & $-36$ & $-35$ & $-25$ & $-9$ & $16$ & $55$ & $98$ \\
    Ag & $191$ & $64$ & $15$ & $4$ & $4$ & $14$ & $29$ & $55$ & $90$ & $132$ \\
    Cd & $233$ & $120$ & $74$ & $54$ & $51$ & $60$ & $77$ & $98$ & $132$ & $176$ \\
    \hline
    La & $260$ & $270$ & $247$ & $252$ & $276$ & $297$ & $302$ & $305$ & $329$ & $367$ \\
    Hf & $145$ & $23$ & $-20$ & $-25$ & $-16$ & $-7$ & $1$ & $15$ & $45$ & $101$ \\
    Ta & $120$ & $-6$ & $-49$ & $-56$ & $-55$ & $-57$ & $-50$ & $-33$ & $4$ & $63$ \\
    W  & $111$ & $-8$ & $-51$ & $-69$ & $-79$ & $-80$ & $-65$ & $-44$ & $-6$ & $45$ \\
    Re & $121$ & $7$ & $-49$ & $-76$ & $-84$ & $-78$ & $-63$ & $-42$ & $-6$ & $42$ \\
    Os & $137$ & $17$ & $-47$ & $-73$ & $-77$ & $-69$ & $-54$ & $-32$ & $6$ & $54$ \\
    Ir & $153$ & $26$ & $-41$ & $-60$ & $-61$ & $-52$ & $-39$ & $-14$ & $23$ & $74$ \\
    Pt & $174$ & $41$ & $-17$ & $-33$ & $-34$ & $-26$ & $-10$ & $14$ & $54$ & $99$ \\
    Au & $206$ & $76$ & $24$ & $10$ & $9$ & $18$ & $32$ & $58$ & $93$ & $137$ \\
    Hg & $263$ & $152$ & $102$ & $81$ & $77$ & $85$ & $101$ & $123$ & $159$ & $201$ \\
    \hline\hline
  \end{tabular}
  \label{tab:daa_seg_cu111_hs_4d}
\end{table}

\begin{table}[ht]
  \centering
  \renewcommand{\arraystretch}{1.2}
  \caption{\textbf{Segregation energies} for \textbf{dual-atom} sites in \textbf{Cu(111)} for all combinations of \textbf{5\textit{d}} transition metals (columns) and all other transition metals (rows), as obtained from initial \textbf{high-spin} configurations. Energies are reported in kJ~mol$^{-1}$.}
  \vspace{8pt}
  \begin{tabular}{@{\hspace{8pt}}*{11}{c@{\hspace{8pt}}}}
    \hline\hline
    TM & La & Hf & Ta & W & Re & Os & Ir & Pt & Au & Hg \\
    \hline
    Sc & $289$ & $34$ & $8$ & $-2$ & $8$ & $25$ & $43$ & $64$ & $99$ & $165$ \\
    Ti & $248$ & $-23$ & $-59$ & $-62$ & $-48$ & $-38$ & $-28$ & $-9$ & $31$ & $104$ \\
    V  & $252$ & $-31$ & $-68$ & $-71$ & $-69$ & $-64$ & $-45$ & $-14$ & $29$ & $98$ \\
    Cr & $267$ & $-17$ & $-60$ & $-71$ & $-67$ & $-52$ & $-31$ & $3$ & $47$ & $114$ \\
    Mn & $267$ & $-12$ & $-57$ & $-66$ & $-60$ & $-47$ & $-25$ & $6$ & $46$ & $111$ \\
    Fe & $266$ & $-17$ & $-59$ & $-74$ & $-71$ & $-59$ & $-111$ & $-14$ & $28$ & $93$ \\
    Co & $274$ & $-18$ & $-67$ & $-85$ & $-80$ & $-67$ & $-50$ & $-21$ & $23$ & $83$ \\
    Ni & $273$ & $-11$ & $-61$ & $-75$ & $-73$ & $-63$ & $-44$ & $-14$ & $30$ & $92$ \\
    Cu & $286$ & $3$ & $-44$ & $-56$ & $-57$ & $-42$ & $-24$ & $8$ & $49$ & $112$ \\
    Zn & $316$ & $30$ & $-14$ & $-32$ & $-31$ & $-16$ & $6$ & $33$ & $73$ & $140$ \\
    \hline
    Y  & $260$ & $145$ & $120$ & $111$ & $121$ & $137$ & $153$ & $174$ & $206$ & $263$ \\
    Zr & $270$ & $23$ & $-6$ & $-8$ & $7$ & $17$ & $26$ & $41$ & $76$ & $152$ \\
    Nb & $247$ & $-20$ & $-49$ & $-51$ & $-49$ & $-47$ & $-41$ & $-17$ & $24$ & $102$ \\
    Mo & $252$ & $-25$ & $-56$ & $-69$ & $-76$ & $-73$ & $-60$ & $-33$ & $10$ & $81$ \\
    Tc & $276$ & $-16$ & $-55$ & $-79$ & $-84$ & $-77$ & $-61$ & $-34$ & $9$ & $77$ \\
    Ru & $297$ & $-7$ & $-57$ & $-80$ & $-78$ & $-69$ & $-52$ & $-26$ & $18$ & $85$ \\
    Rh & $302$ & $1$ & $-50$ & $-65$ & $-63$ & $-54$ & $-39$ & $-10$ & $32$ & $101$ \\
    Pd & $305$ & $15$ & $-33$ & $-44$ & $-42$ & $-32$ & $-14$ & $14$ & $58$ & $123$ \\
    Ag & $329$ & $45$ & $4$ & $-6$ & $-6$ & $6$ & $23$ & $54$ & $93$ & $159$ \\
    Cd & $367$ & $101$ & $63$ & $45$ & $42$ & $54$ & $74$ & $99$ & $137$ & $201$ \\
    \hline
    La & $297$ & $280$ & $253$ & $249$ & $271$ & $294$ & $310$ & $324$ & $351$ & $405$ \\
    Hf & $280$ & $-1$ & $-32$ & $-41$ & $-27$ & $-14$ & $-1$ & $19$ & $56$ & $133$ \\
    Ta & $253$ & $-32$ & $-71$ & $-76$ & $-70$ & $-64$ & $-53$ & $-28$ & $14$ & $94$ \\
    W  & $249$ & $-41$ & $-76$ & $-89$ & $-93$ & $-87$ & $-71$ & $-42$ & $0$ & $73$ \\
    Re & $271$ & $-27$ & $-70$ & $-93$ & $-98$ & $-88$ & $-69$ & $-41$ & $-1$ & $68$ \\
    Os & $294$ & $-14$ & $-64$ & $-87$ & $-88$ & $-75$ & $-57$ & $-32$ & $10$ & $79$ \\
    Ir & $310$ & $-1$ & $-53$ & $-71$ & $-69$ & $-57$ & $-42$ & $-14$ & $27$ & $98$ \\
    Pt & $324$ & $19$ & $-28$ & $-42$ & $-41$ & $-32$ & $-14$ & $13$ & $56$ & $124$ \\
    Au & $351$ & $56$ & $14$ & $0$ & $-1$ & $10$ & $27$ & $56$ & $95$ & $164$ \\
    Hg & $405$ & $133$ & $94$ & $73$ & $68$ & $79$ & $98$ & $124$ & $164$ & $227$ \\
    \hline\hline
  \end{tabular}
  \label{tab:daa_seg_cu111_hs_5d}
\end{table}


\begin{table}[ht]
  \centering
  \renewcommand{\arraystretch}{1.2}
  \caption{\textbf{Segregation energies} for \textbf{dual-atom} sites in \textbf{Cu(111)} for all combinations of \textbf{3\textit{d}} transition metals (columns) and all other transition metals (rows), as obtained from initial \textbf{low-spin} configurations. Energies are reported in kJ~mol$^{-1}$.}
  \vspace{8pt}
  \begin{tabular}{@{\hspace{8pt}}*{11}{c@{\hspace{8pt}}}}
    \hline\hline
    TM & Sc & Ti & V & Cr & Mn & Fe & Co & Ni & Cu & Zn \\
    \hline
    Sc & $66$ & $14$ & $3$ & $1$ & $9$ & $19$ & $26$ & $33$ & $50$ & $75$ \\
    Ti & $14$ & $-46$ & $-50$ & $-32$ & $-22$ & $-24$ & $-36$ & $-38$ & $-21$ & $7$ \\
    V  & $3$ & $-50$ & $-40$ & $-35$ & $-39$ & $-52$ & $-67$ & $-62$ & $-43$ & $-15$ \\
    Cr & $1$ & $-32$ & $-35$ & $-42$ & $-58$ & $-68$ & $-71$ & $-66$ & $-46$ & $-22$ \\
    Mn & $9$ & $-22$ & $-39$ & $-58$ & $-66$ & $-71$ & $-73$ & $-64$ & $-44$ & $-21$ \\
    Fe & $19$ & $-24$ & $-52$ & $-68$ & $-71$ & $-73$ & $-74$ & $-63$ & $-42$ & $-17$ \\
    Co & $26$ & $-36$ & $-67$ & $-71$ & $-73$ & $-74$ & $-72$ & $-59$ & $-36$ & $-9$ \\
    Ni & $33$ & $-38$ & $-62$ & $-66$ & $-64$ & $-63$ & $-59$ & $-43$ & $-20$ & $4$ \\
    Cu & $50$ & $-21$ & $-43$ & $-46$ & $-44$ & $-42$ & $-36$ & $-20$ & $1$ & $26$ \\
    Zn & $75$ & $7$ & $-15$ & $-22$ & $-21$ & $-17$ & $-9$ & $4$ & $26$ & $53$ \\
    \hline
    Y  & $162$ & $120$ & $107$ & $104$ & $111$ & $117$ & $125$ & $135$ & $151$ & $178$ \\
    Zr & $53$ & $1$ & $-7$ & $3$ & $15$ & $16$ & $10$ & $12$ & $25$ & $51$ \\
    Nb & $17$ & $-36$ & $-33$ & $-26$ & $-28$ & $-41$ & $-51$ & $-47$ & $-30$ & $-2$ \\
    Mo & $8$ & $-33$ & $-37$ & $-44$ & $-59$ & $-70$ & $-72$ & $-62$ & $-44$ & $-19$ \\
    Tc & $19$ & $-23$ & $-43$ & $-60$ & $-69$ & $-75$ & $-74$ & $-63$ & $-44$ & $-18$ \\
    Ru & $34$ & $-23$ & $-53$ & $-63$ & $-65$ & $-70$ & $-66$ & $-54$ & $-33$ & $-6$ \\
    Rh & $45$ & $-23$ & $-52$ & $-55$ & $-56$ & $-56$ & $-53$ & $-39$ & $-17$ & $12$ \\
    Pd & $59$ & $-11$ & $-33$ & $-35$ & $-33$ & $-31$ & $-27$ & $-12$ & $11$ & $35$ \\
    Ag & $89$ & $23$ & $5$ & $2$ & $5$ & $6$ & $12$ & $27$ & $48$ & $72$ \\
    Cd & $135$ & $76$ & $56$ & $47$ & $48$ & $49$ & $57$ & $69$ & $89$ & $116$ \\
    \hline
    La & $289$ & $248$ & $227$ & $233$ & $257$ & $274$ & $274$ & $273$ & $286$ & $316$ \\
    Hf & $34$ & $-23$ & $-38$ & $-35$ & $-25$ & $-19$ & $-18$ & $-11$ & $3$ & $30$ \\
    Ta & $8$ & $-59$ & $-68$ & $-63$ & $-60$ & $-64$ & $-67$ & $-61$ & $-44$ & $-14$ \\
    W  & $-2$ & $-62$ & $-71$ & $-75$ & $-81$ & $-88$ & $-86$ & $-75$ & $-56$ & $-32$ \\
    Re & $8$ & $-48$ & $-69$ & $-80$ & $-85$ & $-88$ & $-84$ & $-73$ & $-57$ & $-31$ \\
    Os & $25$ & $-38$ & $-67$ & $-75$ & $-75$ & $-78$ & $-74$ & $-63$ & $-42$ & $-16$ \\
    Ir & $43$ & $-28$ & $-57$ & $-60$ & $-59$ & $-61$ & $-58$ & $-44$ & $-24$ & $6$ \\
    Pt & $64$ & $-9$ & $-32$ & $-35$ & $-33$ & $-32$ & $-29$ & $-14$ & $8$ & $33$ \\
    Au & $99$ & $31$ & $11$ & $7$ & $9$ & $9$ & $15$ & $30$ & $49$ & $73$ \\
    Hg & $165$ & $104$ & $82$ & $72$ & $73$ & $73$ & $81$ & $92$ & $112$ & $140$ \\
    \hline\hline
  \end{tabular}
  \label{tab:daa_seg_cu111_ls_3d}
\end{table}

\begin{table}[ht]
  \centering
  \renewcommand{\arraystretch}{1.2}
  \caption{\textbf{Segregation energies} for \textbf{dual-atom} sites in \textbf{Cu(111)} for all combinations of \textbf{4\textit{d}} transition metals (columns) and all other transition metals (rows), as obtained from initial \textbf{low-spin} configurations. Energies are reported in kJ~mol$^{-1}$.}
  \vspace{8pt}
  \begin{tabular}{@{\hspace{8pt}}*{11}{c@{\hspace{8pt}}}}
    \hline\hline
    TM & Y & Zr & Nb & Mo & Tc & Ru & Rh & Pd & Ag & Cd \\
    \hline
    Sc & $162$ & $53$ & $17$ & $8$ & $19$ & $34$ & $45$ & $59$ & $89$ & $135$ \\
    Ti & $120$ & $1$ & $-36$ & $-33$ & $-23$ & $-23$ & $-23$ & $-11$ & $23$ & $76$ \\
    V  & $107$ & $-7$ & $-33$ & $-37$ & $-43$ & $-53$ & $-52$ & $-33$ & $5$ & $56$ \\
    Cr & $104$ & $3$ & $-26$ & $-44$ & $-60$ & $-63$ & $-55$ & $-35$ & $2$ & $47$ \\
    Mn & $111$ & $15$ & $-28$ & $-59$ & $-69$ & $-65$ & $-56$ & $-33$ & $5$ & $48$ \\
    Fe & $117$ & $16$ & $-41$ & $-70$ & $-75$ & $-70$ & $-56$ & $-31$ & $6$ & $49$ \\
    Co & $125$ & $10$ & $-51$ & $-72$ & $-74$ & $-66$ & $-53$ & $-27$ & $12$ & $57$ \\
    Ni & $135$ & $12$ & $-47$ & $-62$ & $-63$ & $-54$ & $-39$ & $-12$ & $27$ & $69$ \\
    Cu & $151$ & $25$ & $-30$ & $-44$ & $-44$ & $-33$ & $-17$ & $11$ & $48$ & $89$ \\
    Zn & $178$ & $51$ & $-2$ & $-19$ & $-18$ & $-6$ & $12$ & $35$ & $72$ & $116$ \\
    \hline
    Y  & $244$ & $158$ & $123$ & $115$ & $126$ & $138$ & $149$ & $163$ & $191$ & $233$ \\
    Zr & $158$ & $44$ & $9$ & $12$ & $22$ & $24$ & $25$ & $34$ & $64$ & $120$ \\
    Nb & $123$ & $9$ & $-21$ & $-25$ & $-31$ & $-39$ & $-39$ & $-22$ & $15$ & $74$ \\
    Mo & $115$ & $12$ & $-25$ & $-46$ & $-62$ & $-65$ & $-58$ & $-36$ & $4$ & $54$ \\
    Tc & $126$ & $22$ & $-31$ & $-62$ & $-74$ & $-71$ & $-60$ & $-35$ & $3$ & $51$ \\
    Ru & $138$ & $24$ & $-39$ & $-65$ & $-71$ & $-64$ & $-50$ & $-25$ & $14$ & $60$ \\
    Rh & $149$ & $25$ & $-39$ & $-58$ & $-60$ & $-50$ & $-35$ & $-9$ & $29$ & $77$ \\
    Pd & $163$ & $34$ & $-22$ & $-36$ & $-35$ & $-25$ & $-9$ & $16$ & $55$ & $98$ \\
    Ag & $191$ & $64$ & $15$ & $4$ & $3$ & $14$ & $29$ & $55$ & $90$ & $132$ \\
    Cd & $233$ & $120$ & $74$ & $54$ & $51$ & $60$ & $77$ & $98$ & $132$ & $176$ \\
    \hline
    La & $260$ & $270$ & $247$ & $252$ & $276$ & $297$ & $302$ & $305$ & $329$ & $367$ \\
    Hf & $145$ & $23$ & $-20$ & $-25$ & $-16$ & $-7$ & $1$ & $15$ & $45$ & $101$ \\
    Ta & $120$ & $-6$ & $-49$ & $-56$ & $-55$ & $-57$ & $-50$ & $-33$ & $4$ & $63$ \\
    W  & $111$ & $-8$ & $-51$ & $-69$ & $-79$ & $-80$ & $-65$ & $-44$ & $-6$ & $45$ \\
    Re & $121$ & $7$ & $-49$ & $-76$ & $-86$ & $-78$ & $-63$ & $-42$ & $-6$ & $42$ \\
    Os & $137$ & $17$ & $-47$ & $-74$ & $-78$ & $-69$ & $-54$ & $-32$ & $6$ & $54$ \\
    Ir & $153$ & $26$ & $-41$ & $-60$ & $-61$ & $-52$ & $-39$ & $-14$ & $23$ & $74$ \\
    Pt & $174$ & $41$ & $-17$ & $-33$ & $-34$ & $-26$ & $-10$ & $14$ & $54$ & $99$ \\
    Au & $206$ & $76$ & $24$ & $10$ & $8$ & $18$ & $32$ & $58$ & $93$ & $137$ \\
    Hg & $263$ & $152$ & $102$ & $81$ & $77$ & $85$ & $101$ & $123$ & $159$ & $201$ \\
    \hline\hline
  \end{tabular}
  \label{tab:daa_seg_cu111_ls_4d}
\end{table}

\begin{table}[ht]
  \centering
  \renewcommand{\arraystretch}{1.2}
  \caption{\textbf{Segregation energies} for \textbf{dual-atom} sites in \textbf{Cu(111)} for all combinations of \textbf{5\textit{d}} transition metals (columns) and all other transition metals (rows), as obtained from initial \textbf{low-spin} configurations. Energies are reported in kJ~mol$^{-1}$.}
  \vspace{8pt}
  \begin{tabular}{@{\hspace{8pt}}*{11}{c@{\hspace{8pt}}}}
    \hline\hline
    TM & La & Hf & Ta & W & Re & Os & Ir & Pt & Au & Hg \\
    \hline
    Sc & $289$ & $34$ & $8$ & $-2$ & $8$ & $25$ & $43$ & $64$ & $99$ & $165$ \\
    Ti & $248$ & $-23$ & $-59$ & $-62$ & $-48$ & $-38$ & $-28$ & $-9$ & $31$ & $104$ \\
    V  & $227$ & $-38$ & $-68$ & $-71$ & $-69$ & $-67$ & $-57$ & $-32$ & $11$ & $82$ \\
    Cr & $233$ & $-35$ & $-63$ & $-75$ & $-80$ & $-75$ & $-60$ & $-35$ & $7$ & $72$ \\
    Mn & $257$ & $-25$ & $-60$ & $-81$ & $-85$ & $-75$ & $-59$ & $-33$ & $9$ & $73$ \\
    Fe & $274$ & $-19$ & $-64$ & $-88$ & $-88$ & $-78$ & $-61$ & $-32$ & $9$ & $73$ \\
    Co & $274$ & $-18$ & $-67$ & $-86$ & $-84$ & $-74$ & $-58$ & $-29$ & $15$ & $81$ \\
    Ni & $273$ & $-11$ & $-61$ & $-75$ & $-73$ & $-63$ & $-44$ & $-14$ & $30$ & $92$ \\
    Cu & $286$ & $3$ & $-44$ & $-56$ & $-57$ & $-42$ & $-24$ & $8$ & $49$ & $112$ \\
    Zn & $316$ & $30$ & $-14$ & $-32$ & $-31$ & $-16$ & $6$ & $33$ & $73$ & $140$ \\
    \hline
    Y  & $260$ & $145$ & $120$ & $111$ & $121$ & $137$ & $153$ & $174$ & $206$ & $263$ \\
    Zr & $270$ & $23$ & $-6$ & $-8$ & $7$ & $17$ & $26$ & $41$ & $76$ & $152$ \\
    Nb & $247$ & $-20$ & $-49$ & $-51$ & $-49$ & $-47$ & $-41$ & $-17$ & $24$ & $102$ \\
    Mo & $252$ & $-25$ & $-56$ & $-69$ & $-76$ & $-74$ & $-60$ & $-33$ & $10$ & $81$ \\
    Tc & $276$ & $-16$ & $-55$ & $-79$ & $-86$ & $-78$ & $-61$ & $-34$ & $8$ & $77$ \\
    Ru & $297$ & $-7$ & $-57$ & $-80$ & $-78$ & $-69$ & $-52$ & $-26$ & $18$ & $85$ \\
    Rh & $302$ & $1$ & $-50$ & $-65$ & $-63$ & $-54$ & $-39$ & $-10$ & $32$ & $101$ \\
    Pd & $305$ & $15$ & $-33$ & $-44$ & $-42$ & $-32$ & $-14$ & $14$ & $58$ & $123$ \\
    Ag & $329$ & $45$ & $4$ & $-6$ & $-6$ & $6$ & $23$ & $54$ & $93$ & $159$ \\
    Cd & $367$ & $101$ & $63$ & $45$ & $42$ & $54$ & $74$ & $99$ & $137$ & $201$ \\
    \hline
    La & $297$ & $280$ & $253$ & $249$ & $271$ & $294$ & $310$ & $324$ & $351$ & $405$ \\
    Hf & $280$ & $-1$ & $-32$ & $-41$ & $-27$ & $-14$ & $-1$ & $19$ & $56$ & $133$ \\
    Ta & $253$ & $-32$ & $-71$ & $-76$ & $-70$ & $-64$ & $-53$ & $-28$ & $14$ & $94$ \\
    W  & $249$ & $-41$ & $-76$ & $-89$ & $-93$ & $-87$ & $-71$ & $-42$ & $0$ & $73$ \\
    Re & $271$ & $-27$ & $-70$ & $-93$ & $-98$ & $-88$ & $-69$ & $-41$ & $-1$ & $68$ \\
    Os & $294$ & $-14$ & $-64$ & $-87$ & $-88$ & $-75$ & $-57$ & $-32$ & $10$ & $79$ \\
    Ir & $310$ & $-1$ & $-53$ & $-71$ & $-69$ & $-57$ & $-42$ & $-14$ & $27$ & $98$ \\
    Pt & $324$ & $19$ & $-28$ & $-42$ & $-41$ & $-32$ & $-14$ & $13$ & $56$ & $124$ \\
    Au & $351$ & $56$ & $14$ & $0$ & $-1$ & $10$ & $27$ & $56$ & $95$ & $164$ \\
    Hg & $405$ & $133$ & $94$ & $73$ & $68$ & $79$ & $98$ & $124$ & $164$ & $227$ \\
    \hline\hline
  \end{tabular}
  \label{tab:daa_seg_cu111_ls_5d}
\end{table}

\clearpage


\begin{figure}[htbp]
    \includegraphics[width=12cm,height=\textheight,keepaspectratio]{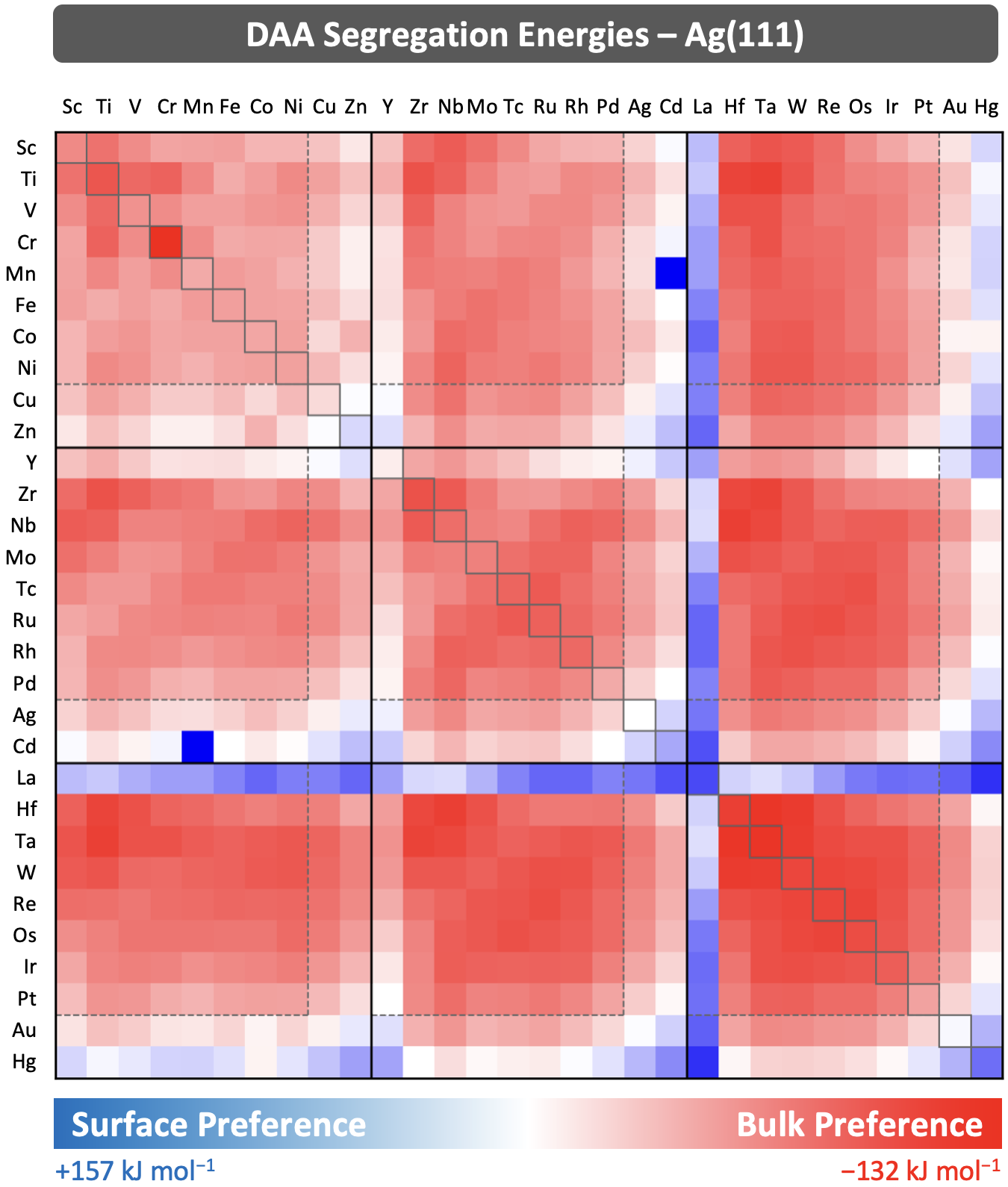}
    \caption{
        Segregation energies of all possible transition metal dopant dimers in Ag(111).
        Positive values (blue) indicate an energetic preference for the surface, whereas negative values (red) indicate a preference for bulk incorporation.
        All energies are reported in kJ~mol$^{-1}$.
    }
    \label{fig:Ag111-DAA-seg}
\end{figure}

\begin{table}[ht]
  \centering
  \renewcommand{\arraystretch}{1.2}
  \caption{\textbf{Segregation energies} for \textbf{dual-atom} sites in \textbf{Ag(111)} for all combinations of \textbf{3\textit{d}} transition metals (columns) and all other transition metals (rows), as obtained from initial \textbf{high-spin} configurations. Energies are reported in kJ~mol$^{-1}$.}
  \vspace{8pt}
  \begin{tabular}{@{\hspace{8pt}}*{11}{c@{\hspace{8pt}}}}
    \hline\hline
    TM & Sc & Ti & V & Cr & Mn & Fe & Co & Ni & Cu & Zn \\
    \hline
    Sc & $-64$ & $-77$ & $-62$ & $-49$ & $-50$ & $-52$ & $-39$ & $-39$ & $-32$ & $-13$ \\
    Ti & $-77$ & $-96$ & $-83$ & $-87$ & $-66$ & $-45$ & $-53$ & $-64$ & $-51$ & $-35$ \\
    V  & $-62$ & $-83$ & $-59$ & $-63$ & $-53$ & $-52$ & $-57$ & $-59$ & $-43$ & $-24$ \\
    Cr & $-49$ & $-87$ & $-63$ & $-132$ & $-63$ & $-46$ & $-48$ & $-46$ & $-29$ & $-9$ \\
    Mn & $-50$ & $-66$ & $-53$ & $-63$ & $-46$ & $-54$ & $-50$ & $-42$ & $-28$ & $-8$ \\
    Fe & $-52$ & $-45$ & $-52$ & $-46$ & $-54$ & $-52$ & $-50$ & $-49$ & $-36$ & $-18$ \\
    Co & $-39$ & $-53$ & $-57$ & $-48$ & $-50$ & $-50$ & $-48$ & $-51$ & $-21$ & $-42$ \\
    Ni & $-39$ & $-64$ & $-59$ & $-46$ & $-42$ & $-49$ & $-51$ & $-51$ & $-37$ & $-18$ \\
    Cu & $-32$ & $-51$ & $-43$ & $-29$ & $-28$ & $-36$ & $-21$ & $-37$ & $-20$ & $2$ \\
    Zn & $-13$ & $-35$ & $-24$ & $-9$ & $-8$ & $-18$ & $-42$ & $-18$ & $2$ & $24$ \\
    \hline
    Y  & $-33$ & $-44$ & $-29$ & $-16$ & $-17$ & $-18$ & $-11$ & $-6$ & $3$ & $20$ \\
    Zr & $-81$ & $-97$ & $-88$ & $-77$ & $-73$ & $-59$ & $-56$ & $-67$ & $-62$ & $-41$ \\
    Nb & $-91$ & $-88$ & $-68$ & $-67$ & $-71$ & $-72$ & $-81$ & $-86$ & $-76$ & $-61$ \\
    Mo & $-79$ & $-69$ & $-58$ & $-59$ & $-69$ & $-77$ & $-77$ & $-71$ & $-58$ & $-46$ \\
    Tc & $-64$ & $-55$ & $-55$ & $-66$ & $-72$ & $-73$ & $-69$ & $-69$ & $-62$ & $-48$ \\
    Ru & $-47$ & $-53$ & $-63$ & $-67$ & $-69$ & $-67$ & $-67$ & $-73$ & $-64$ & $-47$ \\
    Rh & $-41$ & $-63$ & $-64$ & $-62$ & $-56$ & $-60$ & $-63$ & $-66$ & $-51$ & $-34$ \\
    Pd & $-40$ & $-61$ & $-56$ & $-43$ & $-40$ & $-49$ & $-49$ & $-48$ & $-34$ & $-16$ \\
    Ag & $-24$ & $-40$ & $-32$ & $-19$ & $-18$ & $-25$ & $-36$ & $-26$ & $-9$ & $12$ \\
    Cd & $3$ & $-17$ & $-7$ & $7$ & $157$ & $-1$ & $-12$ & $-2$ & $18$ & $40$ \\
    \hline
    La & $41$ & $33$ & $50$ & $60$ & $60$ & $75$ & $95$ & $79$ & $78$ & $94$ \\
    Hf & $-88$ & $-109$ & $-98$ & $-86$ & $-82$ & $-76$ & $-69$ & $-75$ & $-70$ & $-47$ \\
    Ta & $-97$ & $-112$ & $-98$ & $-97$ & $-91$ & $-87$ & $-91$ & $-94$ & $-85$ & $-68$ \\
    W  & $-93$ & $-96$ & $-83$ & $-82$ & $-85$ & $-87$ & $-93$ & $-93$ & $-83$ & $-68$ \\
    Re & $-80$ & $-79$ & $-74$ & $-80$ & $-81$ & $-84$ & $-83$ & $-84$ & $-77$ & $-64$ \\
    Os & $-62$ & $-68$ & $-76$ & $-75$ & $-76$ & $-75$ & $-75$ & $-82$ & $-71$ & $-55$ \\
    Ir & $-47$ & $-67$ & $-70$ & $-67$ & $-60$ & $-65$ & $-67$ & $-70$ & $-57$ & $-40$ \\
    Pt & $-35$ & $-59$ & $-56$ & $-45$ & $-40$ & $-49$ & $-51$ & $-50$ & $-37$ & $-19$ \\
    Au & $-15$ & $-34$ & $-28$ & $-14$ & $-13$ & $-22$ & $-6$ & $-22$ & $-7$ & $14$ \\
    Hg & $25$ & $5$ & $14$ & $27$ & $28$ & $19$ & $-7$ & $17$ & $36$ & $58$ \\
    \hline\hline
  \end{tabular}
  \label{tab:daa_seg_ag111_hs_3d}
\end{table}

\begin{table}[ht]
  \centering
  \renewcommand{\arraystretch}{1.2}
  \caption{\textbf{Segregation energies} for \textbf{dual-atom} sites in \textbf{Ag(111)} for all combinations of \textbf{4\textit{d}} transition metals (columns) and all other transition metals (rows), as obtained from initial \textbf{high-spin} configurations. Energies are reported in kJ~mol$^{-1}$.}
  \vspace{8pt}
  \begin{tabular}{@{\hspace{8pt}}*{11}{c@{\hspace{8pt}}}}
    \hline\hline
    TM & Y & Zr & Nb & Mo & Tc & Ru & Rh & Pd & Ag & Cd \\
    \hline
    Sc & $-33$ & $-81$ & $-91$ & $-79$ & $-64$ & $-47$ & $-41$ & $-40$ & $-24$ & $3$ \\
    Ti & $-44$ & $-97$ & $-88$ & $-69$ & $-55$ & $-53$ & $-63$ & $-61$ & $-40$ & $-17$ \\
    V  & $-29$ & $-88$ & $-68$ & $-58$ & $-55$ & $-63$ & $-64$ & $-56$ & $-32$ & $-7$ \\
    Cr & $-16$ & $-77$ & $-67$ & $-59$ & $-66$ & $-67$ & $-62$ & $-43$ & $-19$ & $7$ \\
    Mn & $-17$ & $-73$ & $-71$ & $-69$ & $-72$ & $-69$ & $-56$ & $-40$ & $-18$ & $157$ \\
    Fe & $-18$ & $-59$ & $-72$ & $-77$ & $-73$ & $-67$ & $-60$ & $-49$ & $-25$ & $-1$ \\
    Co & $-11$ & $-56$ & $-81$ & $-77$ & $-69$ & $-67$ & $-63$ & $-49$ & $-36$ & $-12$ \\
    Ni & $-6$ & $-67$ & $-86$ & $-71$ & $-69$ & $-73$ & $-66$ & $-48$ & $-26$ & $-2$ \\
    Cu & $3$ & $-62$ & $-76$ & $-58$ & $-62$ & $-64$ & $-51$ & $-34$ & $-9$ & $18$ \\
    Zn & $20$ & $-41$ & $-61$ & $-46$ & $-48$ & $-47$ & $-34$ & $-16$ & $12$ & $40$ \\
    \hline
    Y  & $-9$ & $-48$ & $-56$ & $-44$ & $-34$ & $-18$ & $-10$ & $-6$ & $9$ & $34$ \\
    Zr & $-48$ & $-97$ & $-92$ & $-71$ & $-57$ & $-55$ & $-63$ & $-70$ & $-53$ & $-22$ \\
    Nb & $-56$ & $-92$ & $-80$ & $-68$ & $-66$ & $-80$ & $-88$ & $-84$ & $-64$ & $-40$ \\
    Mo & $-44$ & $-71$ & $-68$ & $-68$ & $-79$ & $-85$ & $-85$ & $-67$ & $-46$ & $-25$ \\
    Tc & $-34$ & $-57$ & $-66$ & $-79$ & $-85$ & $-92$ & $-80$ & $-70$ & $-50$ & $-29$ \\
    Ru & $-18$ & $-55$ & $-80$ & $-85$ & $-92$ & $-89$ & $-83$ & $-74$ & $-53$ & $-30$ \\
    Rh & $-10$ & $-63$ & $-88$ & $-85$ & $-80$ & $-83$ & $-83$ & $-64$ & $-41$ & $-18$ \\
    Pd & $-6$ & $-70$ & $-84$ & $-67$ & $-70$ & $-74$ & $-64$ & $-46$ & $-24$ & $-1$ \\
    Ag & $9$ & $-53$ & $-64$ & $-46$ & $-50$ & $-53$ & $-41$ & $-24$ & $0$ & $26$ \\
    Cd & $34$ & $-22$ & $-40$ & $-25$ & $-29$ & $-30$ & $-18$ & $-1$ & $26$ & $54$ \\
    \hline
    La & $58$ & $24$ & $21$ & $46$ & $75$ & $94$ & $93$ & $76$ & $84$ & $108$ \\
    Hf & $-53$ & $-105$ & $-112$ & $-98$ & $-82$ & $-74$ & $-73$ & $-75$ & $-60$ & $-28$ \\
    Ta & $-59$ & $-109$ & $-106$ & $-94$ & $-87$ & $-91$ & $-95$ & $-92$ & $-73$ & $-46$ \\
    W  & $-55$ & $-94$ & $-93$ & $-88$ & $-95$ & $-99$ & $-98$ & $-88$ & $-69$ & $-47$ \\
    Re & $-45$ & $-78$ & $-86$ & $-94$ & $-96$ & $-101$ & $-93$ & $-82$ & $-64$ & $-43$ \\
    Os & $-27$ & $-68$ & $-89$ & $-93$ & $-99$ & $-95$ & $-91$ & $-82$ & $-60$ & $-37$ \\
    Ir & $-13$ & $-65$ & $-90$ & $-87$ & $-86$ & $-87$ & $-86$ & $-69$ & $-47$ & $-23$ \\
    Pt & $0$ & $-63$ & $-80$ & $-67$ & $-68$ & $-74$ & $-65$ & $-49$ & $-27$ & $-3$ \\
    Au & $20$ & $-42$ & $-57$ & $-40$ & $-44$ & $-49$ & $-37$ & $-21$ & $2$ & $29$ \\
    Hg & $55$ & $0$ & $-18$ & $-3$ & $-9$ & $-11$ & $1$ & $18$ & $44$ & $72$ \\
    \hline\hline
  \end{tabular}
  \label{tab:daa_seg_ag111_hs_4d}
\end{table}

\begin{table}[ht]
  \centering
  \renewcommand{\arraystretch}{1.2}
  \caption{\textbf{Segregation energies} for \textbf{dual-atom} sites in \textbf{Ag(111)} for all combinations of \textbf{5\textit{d}} transition metals (columns) and all other transition metals (rows), as obtained from initial \textbf{high-spin} configurations. Energies are reported in kJ~mol$^{-1}$.}
  \vspace{8pt}
  \begin{tabular}{@{\hspace{8pt}}*{11}{c@{\hspace{8pt}}}}
    \hline\hline
    TM & La & Hf & Ta & W & Re & Os & Ir & Pt & Au & Hg \\
    \hline
    Sc & $41$ & $-88$ & $-97$ & $-93$ & $-80$ & $-62$ & $-47$ & $-35$ & $-15$ & $25$ \\
    Ti & $33$ & $-109$ & $-112$ & $-96$ & $-79$ & $-68$ & $-67$ & $-59$ & $-34$ & $5$ \\
    V  & $50$ & $-98$ & $-98$ & $-83$ & $-74$ & $-76$ & $-70$ & $-56$ & $-28$ & $14$ \\
    Cr & $60$ & $-86$ & $-97$ & $-82$ & $-80$ & $-75$ & $-67$ & $-45$ & $-14$ & $27$ \\
    Mn & $60$ & $-82$ & $-91$ & $-85$ & $-81$ & $-76$ & $-60$ & $-40$ & $-13$ & $28$ \\
    Fe & $75$ & $-76$ & $-87$ & $-87$ & $-84$ & $-75$ & $-65$ & $-49$ & $-22$ & $19$ \\
    Co & $95$ & $-69$ & $-91$ & $-93$ & $-83$ & $-75$ & $-67$ & $-51$ & $-6$ & $-7$ \\
    Ni & $79$ & $-75$ & $-94$ & $-93$ & $-84$ & $-82$ & $-70$ & $-50$ & $-22$ & $17$ \\
    Cu & $78$ & $-70$ & $-85$ & $-83$ & $-77$ & $-71$ & $-57$ & $-37$ & $-7$ & $36$ \\
    Zn & $94$ & $-47$ & $-68$ & $-68$ & $-64$ & $-55$ & $-40$ & $-19$ & $14$ & $58$ \\
    \hline
    Y  & $58$ & $-53$ & $-59$ & $-55$ & $-45$ & $-27$ & $-13$ & $0$ & $20$ & $55$ \\
    Zr & $24$ & $-105$ & $-109$ & $-94$ & $-78$ & $-68$ & $-65$ & $-63$ & $-42$ & $0$ \\
    Nb & $21$ & $-112$ & $-106$ & $-93$ & $-86$ & $-89$ & $-90$ & $-80$ & $-57$ & $-18$ \\
    Mo & $46$ & $-98$ & $-94$ & $-88$ & $-94$ & $-93$ & $-87$ & $-67$ & $-40$ & $-3$ \\
    Tc & $75$ & $-82$ & $-87$ & $-95$ & $-96$ & $-99$ & $-86$ & $-68$ & $-44$ & $-9$ \\
    Ru & $94$ & $-74$ & $-91$ & $-99$ & $-101$ & $-95$ & $-87$ & $-74$ & $-49$ & $-11$ \\
    Rh & $93$ & $-73$ & $-95$ & $-98$ & $-93$ & $-91$ & $-86$ & $-65$ & $-37$ & $1$ \\
    Pd & $76$ & $-75$ & $-92$ & $-88$ & $-82$ & $-82$ & $-69$ & $-49$ & $-21$ & $18$ \\
    Ag & $84$ & $-60$ & $-73$ & $-69$ & $-64$ & $-60$ & $-47$ & $-27$ & $2$ & $44$ \\
    Cd & $108$ & $-28$ & $-46$ & $-47$ & $-43$ & $-37$ & $-23$ & $-3$ & $29$ & $72$ \\
    \hline
    La & $111$ & $27$ & $20$ & $33$ & $60$ & $82$ & $90$ & $88$ & $97$ & $127$ \\
    Hf & $27$ & $-113$ & $-124$ & $-117$ & $-100$ & $-86$ & $-76$ & $-69$ & $-48$ & $-5$ \\
    Ta & $20$ & $-124$ & $-127$ & $-115$ & $-104$ & $-100$ & $-99$ & $-87$ & $-63$ & $-24$ \\
    W  & $33$ & $-117$ & $-115$ & $-107$ & $-107$ & $-105$ & $-101$ & $-88$ & $-62$ & $-25$ \\
    Re & $60$ & $-100$ & $-104$ & $-107$ & $-108$ & $-109$ & $-98$ & $-81$ & $-58$ & $-24$ \\
    Os & $82$ & $-86$ & $-100$ & $-105$ & $-109$ & $-101$ & $-95$ & $-81$ & $-56$ & $-17$ \\
    Ir & $90$ & $-76$ & $-99$ & $-101$ & $-98$ & $-95$ & $-89$ & $-69$ & $-43$ & $-3$ \\
    Pt & $88$ & $-69$ & $-87$ & $-88$ & $-81$ & $-81$ & $-69$ & $-50$ & $-24$ & $15$ \\
    Au & $97$ & $-48$ & $-63$ & $-62$ & $-58$ & $-56$ & $-43$ & $-24$ & $4$ & $47$ \\
    Hg & $127$ & $-5$ & $-24$ & $-25$ & $-24$ & $-17$ & $-3$ & $15$ & $47$ & $90$ \\
    \hline\hline
  \end{tabular}
  \label{tab:daa_seg_ag111_hs_5d}
\end{table}


\begin{table}[ht]
  \centering
  \renewcommand{\arraystretch}{1.2}
  \caption{\textbf{Segregation energies} for \textbf{dual-atom} sites in \textbf{Ag(111)} for all combinations of \textbf{3\textit{d}} transition metals (columns) and all other transition metals (rows), as obtained from initial \textbf{low} configurations. Energies are reported in kJ~mol$^{-1}$.}
  \vspace{8pt}
  \begin{tabular}{@{\hspace{8pt}}*{11}{c@{\hspace{8pt}}}}
    \hline\hline
    TM & Sc & Ti & V & Cr & Mn & Fe & Co & Ni & Cu & Zn \\
    \hline
    Sc & $-64$ & $-81$ & $-82$ & $-72$ & $-54$ & $-42$ & $-39$ & $-39$ & $-32$ & $-13$ \\
    Ti & $-81$ & $-98$ & $-71$ & $-54$ & $36$ & $-41$ & $-53$ & $-68$ & $-61$ & $-40$ \\
    V  & $-82$ & $-71$ & $-59$ & $-50$ & $-45$ & $-25$ & $-40$ & $-68$ & $-60$ & $-45$ \\
    Cr & $-72$ & $-54$ & $-50$ & $-44$ & $94$ & $-26$ & $-62$ & $-63$ & $-53$ & $-39$ \\
    Mn & $-54$ & $36$ & $-45$ & $94$ & $-46$ & $-56$ & $-62$ & $-59$ & $-49$ & $-35$ \\
    Fe & $-42$ & $-41$ & $-25$ & $-26$ & $-56$ & $5$ & $-63$ & $-60$ & $-50$ & $-33$ \\
    Co & $-39$ & $-53$ & $-40$ & $-62$ & $-62$ & $-63$ & $-66$ & $-61$ & $-47$ & $-30$ \\
    Ni & $-39$ & $-68$ & $-68$ & $-63$ & $-59$ & $-60$ & $-61$ & $-51$ & $-37$ & $-18$ \\
    Cu & $-32$ & $-61$ & $-60$ & $-53$ & $-49$ & $-50$ & $-47$ & $-37$ & $-20$ & $2$ \\
    Zn & $-13$ & $-40$ & $-45$ & $-39$ & $-35$ & $-33$ & $-30$ & $-18$ & $2$ & $24$ \\
    \hline
    Y  & $-33$ & $-47$ & $-48$ & $-39$ & $-25$ & $-17$ & $-12$ & $-6$ & $3$ & $20$ \\
    Zr & $-81$ & $-98$ & $-80$ & $-59$ & $-48$ & $-45$ & $-56$ & $-67$ & $-62$ & $-41$ \\
    Nb & $-91$ & $-88$ & $-68$ & $-55$ & $-53$ & $-65$ & $-82$ & $-86$ & $-79$ & $-62$ \\
    Mo & $-84$ & $-69$ & $-58$ & $-56$ & $-66$ & $-57$ & $-89$ & $-89$ & $-81$ & $-65$ \\
    Tc & $-64$ & $-55$ & $-55$ & $-66$ & $-13$ & $-27$ & $-88$ & $-86$ & $-76$ & $-58$ \\
    Ru & $-47$ & $-53$ & $-55$ & $-18$ & $-67$ & $-76$ & $-81$ & $-79$ & $-65$ & $-47$ \\
    Rh & $-41$ & $-63$ & $-69$ & $-65$ & $-70$ & $-70$ & $-74$ & $-66$ & $-51$ & $-34$ \\
    Pd & $-40$ & $-70$ & $-67$ & $-62$ & $-58$ & $-59$ & $-59$ & $-48$ & $-34$ & $-16$ \\
    Ag & $-24$ & $-51$ & $-49$ & $-42$ & $-38$ & $-38$ & $-36$ & $-26$ & $-9$ & $12$ \\
    Cd & $3$ & $-21$ & $-26$ & $-20$ & $-15$ & $-15$ & $-12$ & $-2$ & $18$ & $40$ \\
    \hline
    La & $41$ & $28$ & $31$ & $56$ & $79$ & $95$ & $95$ & $79$ & $78$ & $94$ \\
    Hf & $-88$ & $-109$ & $-104$ & $-88$ & $-77$ & $-67$ & $-69$ & $-75$ & $-70$ & $-47$ \\
    Ta & $-97$ & $-112$ & $-97$ & $-82$ & $-76$ & $-80$ & $-92$ & $-94$ & $-85$ & $-68$ \\
    W  & $-97$ & $-96$ & $-83$ & $-76$ & $-81$ & $-88$ & $-98$ & $-96$ & $-88$ & $-74$ \\
    Re & $-80$ & $-79$ & $-74$ & $-114$ & $-142$ & $-87$ & $-96$ & $-92$ & $-84$ & $-67$ \\
    Os & $-62$ & $-68$ & $-76$ & $-28$ & $-74$ & $-83$ & $-87$ & $-84$ & $-71$ & $-55$ \\
    Ir & $-47$ & $-67$ & $-74$ & $-68$ & $-71$ & $-74$ & $-79$ & $-70$ & $-57$ & $-40$ \\
    Pt & $-35$ & $-64$ & $-62$ & $-60$ & $-58$ & $-60$ & $-60$ & $-50$ & $-37$ & $-19$ \\
    Au & $-15$ & $-42$ & $-43$ & $-36$ & $-33$ & $-34$ & $-32$ & $-22$ & $-7$ & $14$ \\
    Hg & $25$ & $-1$ & $-5$ & $1$ & $4$ & $4$ & $7$ & $17$ & $36$ & $58$ \\
    \hline\hline
  \end{tabular}
  \label{tab:daa_seg_ag111_ls_3d}
\end{table}

\begin{table}[ht]
  \centering
  \renewcommand{\arraystretch}{1.2}
  \caption{\textbf{Segregation energies} for \textbf{dual-atom} sites in \textbf{Ag(111)} for all combinations of \textbf{4\textit{d}} transition metals (columns) and all other transition metals (rows), as obtained from initial \textbf{low} configurations. Energies are reported in kJ~mol$^{-1}$.}
  \vspace{8pt}
  \begin{tabular}{@{\hspace{8pt}}*{11}{c@{\hspace{8pt}}}}
    \hline\hline
    TM & Y & Zr & Nb & Mo & Tc & Ru & Rh & Pd & Ag & Cd \\
    \hline
    Sc & $-33$ & $-81$ & $-91$ & $-84$ & $-64$ & $-47$ & $-41$ & $-40$ & $-24$ & $3$ \\
    Ti & $-47$ & $-98$ & $-88$ & $-69$ & $-55$ & $-53$ & $-63$ & $-70$ & $-51$ & $-21$ \\
    V  & $-48$ & $-80$ & $-68$ & $-58$ & $-55$ & $-55$ & $-69$ & $-67$ & $-49$ & $-26$ \\
    Cr & $-39$ & $-59$ & $-55$ & $-56$ & $-66$ & $-18$ & $-65$ & $-62$ & $-42$ & $-20$ \\
    Mn & $-25$ & $-48$ & $-53$ & $-66$ & $-13$ & $-67$ & $-70$ & $-58$ & $-38$ & $-15$ \\
    Fe & $-17$ & $-45$ & $-65$ & $-57$ & $-27$ & $-76$ & $-70$ & $-59$ & $-38$ & $-15$ \\
    Co & $-12$ & $-56$ & $-82$ & $-89$ & $-88$ & $-81$ & $-74$ & $-59$ & $-36$ & $-12$ \\
    Ni & $-6$ & $-67$ & $-86$ & $-89$ & $-86$ & $-79$ & $-66$ & $-48$ & $-26$ & $-2$ \\
    Cu & $3$ & $-62$ & $-79$ & $-81$ & $-76$ & $-65$ & $-51$ & $-34$ & $-9$ & $18$ \\
    Zn & $20$ & $-41$ & $-62$ & $-65$ & $-58$ & $-47$ & $-34$ & $-16$ & $12$ & $40$ \\
    \hline
    Y  & $-9$ & $-48$ & $-57$ & $-53$ & $-34$ & $-18$ & $-10$ & $-6$ & $9$ & $34$ \\
    Zr & $-48$ & $-97$ & $-92$ & $-71$ & $-57$ & $-55$ & $-63$ & $-70$ & $-53$ & $-22$ \\
    Nb & $-57$ & $-92$ & $-80$ & $-68$ & $-66$ & $-80$ & $-90$ & $-87$ & $-67$ & $-42$ \\
    Mo & $-53$ & $-71$ & $-68$ & $-68$ & $-81$ & $-91$ & $-94$ & $-88$ & $-69$ & $-47$ \\
    Tc & $-34$ & $-57$ & $-66$ & $-81$ & $-90$ & $-95$ & $-93$ & $-86$ & $-64$ & $-41$ \\
    Ru & $-18$ & $-55$ & $-80$ & $-91$ & $-95$ & $-93$ & $-91$ & $-78$ & $-54$ & $-30$ \\
    Rh & $-10$ & $-63$ & $-90$ & $-94$ & $-93$ & $-91$ & $-83$ & $-64$ & $-41$ & $-18$ \\
    Pd & $-6$ & $-70$ & $-87$ & $-88$ & $-86$ & $-78$ & $-64$ & $-46$ & $-24$ & $-1$ \\
    Ag & $9$ & $-53$ & $-67$ & $-69$ & $-64$ & $-54$ & $-41$ & $-24$ & $0$ & $26$ \\
    Cd & $34$ & $-22$ & $-42$ & $-47$ & $-41$ & $-30$ & $-18$ & $-1$ & $26$ & $54$ \\
    \hline
    La & $58$ & $24$ & $21$ & $46$ & $75$ & $94$ & $93$ & $76$ & $84$ & $108$ \\
    Hf & $-53$ & $-105$ & $-112$ & $-98$ & $-82$ & $-74$ & $-73$ & $-75$ & $-60$ & $-28$ \\
    Ta & $-59$ & $-109$ & $-106$ & $-94$ & $-87$ & $-91$ & $-95$ & $-92$ & $-73$ & $-46$ \\
    W  & $-60$ & $-94$ & $-93$ & $-88$ & $-95$ & $-100$ & $-100$ & $-93$ & $-75$ & $-53$ \\
    Re & $-45$ & $-77$ & $-86$ & $-95$ & $-102$ & $-102$ & $-97$ & $-90$ & $-72$ & $-49$ \\
    Os & $-27$ & $-68$ & $-89$ & $-99$ & $-101$ & $-98$ & $-92$ & $-83$ & $-60$ & $-37$ \\
    Ir & $-13$ & $-65$ & $-92$ & $-96$ & $-95$ & $-93$ & $-86$ & $-69$ & $-47$ & $-23$ \\
    Pt & $0$ & $-63$ & $-82$ & $-84$ & $-84$ & $-78$ & $-65$ & $-49$ & $-27$ & $-3$ \\
    Au & $20$ & $-42$ & $-58$ & $-63$ & $-60$ & $-50$ & $-37$ & $-21$ & $2$ & $29$ \\
    Hg & $55$ & $0$ & $-21$ & $-27$ & $-21$ & $-11$ & $1$ & $18$ & $44$ & $72$ \\
    \hline\hline
  \end{tabular}
  \label{tab:daa_seg_ag111_ls_4d}
\end{table}

\begin{table}[ht]
  \centering
  \renewcommand{\arraystretch}{1.2}
  \caption{\textbf{Segregation energies} for \textbf{dual-atom} sites in \textbf{Ag(111)} for all combinations of \textbf{5\textit{d}} transition metals (columns) and all other transition metals (rows), as obtained from initial \textbf{low} configurations. Energies are reported in kJ~mol$^{-1}$.}
  \vspace{8pt}
  \begin{tabular}{@{\hspace{8pt}}*{11}{c@{\hspace{8pt}}}}
    \hline\hline
    TM & La & Hf & Ta & W & Re & Os & Ir & Pt & Au & Hg \\
    \hline
    Sc & $41$ & $-88$ & $-97$ & $-97$ & $-80$ & $-62$ & $-47$ & $-35$ & $-15$ & $25$ \\
    Ti & $28$ & $-109$ & $-112$ & $-96$ & $-79$ & $-68$ & $-67$ & $-64$ & $-42$ & $-1$ \\
    V  & $31$ & $-104$ & $-97$ & $-83$ & $-74$ & $-76$ & $-74$ & $-62$ & $-43$ & $-5$ \\
    Cr & $56$ & $-88$ & $-82$ & $-76$ & $-114$ & $-28$ & $-68$ & $-60$ & $-36$ & $1$ \\
    Mn & $79$ & $-77$ & $-76$ & $-81$ & $-142$ & $-74$ & $-71$ & $-58$ & $-33$ & $4$ \\
    Fe & $95$ & $-67$ & $-80$ & $-88$ & $-87$ & $-83$ & $-74$ & $-60$ & $-34$ & $4$ \\
    Co & $95$ & $-69$ & $-92$ & $-98$ & $-96$ & $-87$ & $-79$ & $-60$ & $-32$ & $7$ \\
    Ni & $79$ & $-75$ & $-94$ & $-96$ & $-92$ & $-84$ & $-70$ & $-50$ & $-22$ & $17$ \\
    Cu & $78$ & $-70$ & $-85$ & $-88$ & $-84$ & $-71$ & $-57$ & $-37$ & $-7$ & $36$ \\
    Zn & $94$ & $-47$ & $-68$ & $-74$ & $-67$ & $-55$ & $-40$ & $-19$ & $14$ & $58$ \\
    \hline
    Y  & $58$ & $-53$ & $-59$ & $-60$ & $-45$ & $-27$ & $-13$ & $0$ & $20$ & $55$ \\
    Zr & $24$ & $-105$ & $-109$ & $-94$ & $-77$ & $-68$ & $-65$ & $-63$ & $-42$ & $0$ \\
    Nb & $21$ & $-112$ & $-106$ & $-93$ & $-86$ & $-89$ & $-92$ & $-82$ & $-58$ & $-21$ \\
    Mo & $46$ & $-98$ & $-94$ & $-88$ & $-95$ & $-99$ & $-96$ & $-84$ & $-63$ & $-27$ \\
    Tc & $75$ & $-82$ & $-87$ & $-95$ & $-102$ & $-101$ & $-95$ & $-84$ & $-60$ & $-21$ \\
    Ru & $94$ & $-74$ & $-91$ & $-100$ & $-102$ & $-98$ & $-93$ & $-78$ & $-50$ & $-11$ \\
    Rh & $93$ & $-73$ & $-95$ & $-100$ & $-97$ & $-92$ & $-86$ & $-65$ & $-37$ & $1$ \\
    Pd & $76$ & $-75$ & $-92$ & $-93$ & $-90$ & $-83$ & $-69$ & $-49$ & $-21$ & $18$ \\
    Ag & $84$ & $-60$ & $-73$ & $-75$ & $-72$ & $-60$ & $-47$ & $-27$ & $2$ & $44$ \\
    Cd & $108$ & $-28$ & $-46$ & $-53$ & $-49$ & $-37$ & $-23$ & $-3$ & $29$ & $72$ \\
    \hline
    La & $111$ & $27$ & $20$ & $31$ & $60$ & $82$ & $90$ & $88$ & $97$ & $127$ \\
    Hf & $27$ & $-113$ & $-124$ & $-117$ & $-100$ & $-86$ & $-76$ & $-69$ & $-48$ & $-5$ \\
    Ta & $20$ & $-124$ & $-127$ & $-115$ & $-104$ & $-100$ & $-99$ & $-87$ & $-63$ & $-24$ \\
    W  & $31$ & $-117$ & $-115$ & $-107$ & $-107$ & $-109$ & $-103$ & $-88$ & $-68$ & $-33$ \\
    Re & $60$ & $-100$ & $-104$ & $-107$ & $-113$ & $-110$ & $-101$ & $-88$ & $-67$ & $-29$ \\
    Os & $82$ & $-86$ & $-100$ & $-109$ & $-110$ & $-104$ & $-96$ & $-83$ & $-56$ & $-17$ \\
    Ir & $90$ & $-76$ & $-99$ & $-103$ & $-101$ & $-96$ & $-89$ & $-69$ & $-43$ & $-3$ \\
    Pt & $88$ & $-69$ & $-87$ & $-88$ & $-88$ & $-83$ & $-69$ & $-50$ & $-24$ & $15$ \\
    Au & $97$ & $-48$ & $-63$ & $-68$ & $-67$ & $-56$ & $-43$ & $-24$ & $4$ & $47$ \\
    Hg & $127$ & $-5$ & $-24$ & $-33$ & $-29$ & $-17$ & $-3$ & $15$ & $47$ & $90$ \\
    \hline\hline
  \end{tabular}
  \label{tab:daa_seg_ag111_ls_5d}
\end{table}

\clearpage

\section{Hetero dimer candidates}

\begin{figure}[htbp]
    \includegraphics[width=12cm,height=\textheight,keepaspectratio]{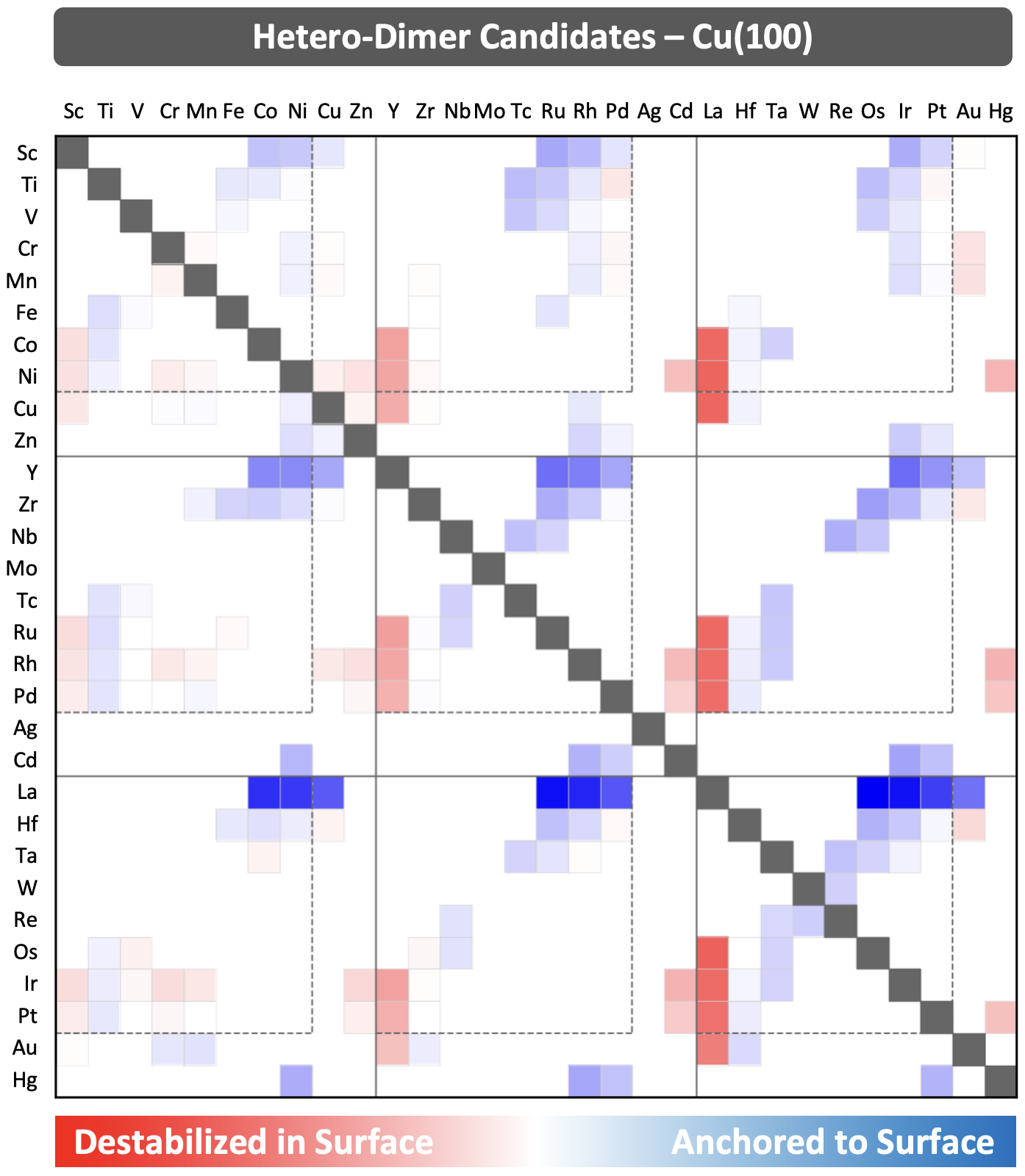}
    \caption{
        Anchoring energies ($E_\text{anchor}$), as also shown in Fig.~5 of the main text.
        Blue indicates increased stabilization in the Cu(100) surface layer and red indicates increased stabilization in the bulk, relative to the reference SAA defined by the column.
        Only combinations that preferentially form hetero-dimers over both SAAs and homo-dimers are shown. 
        }
    \label{fig:Cu100-candidates}
\end{figure}

\begin{figure}[htbp]
    \includegraphics[width=12cm,height=\textheight,keepaspectratio]{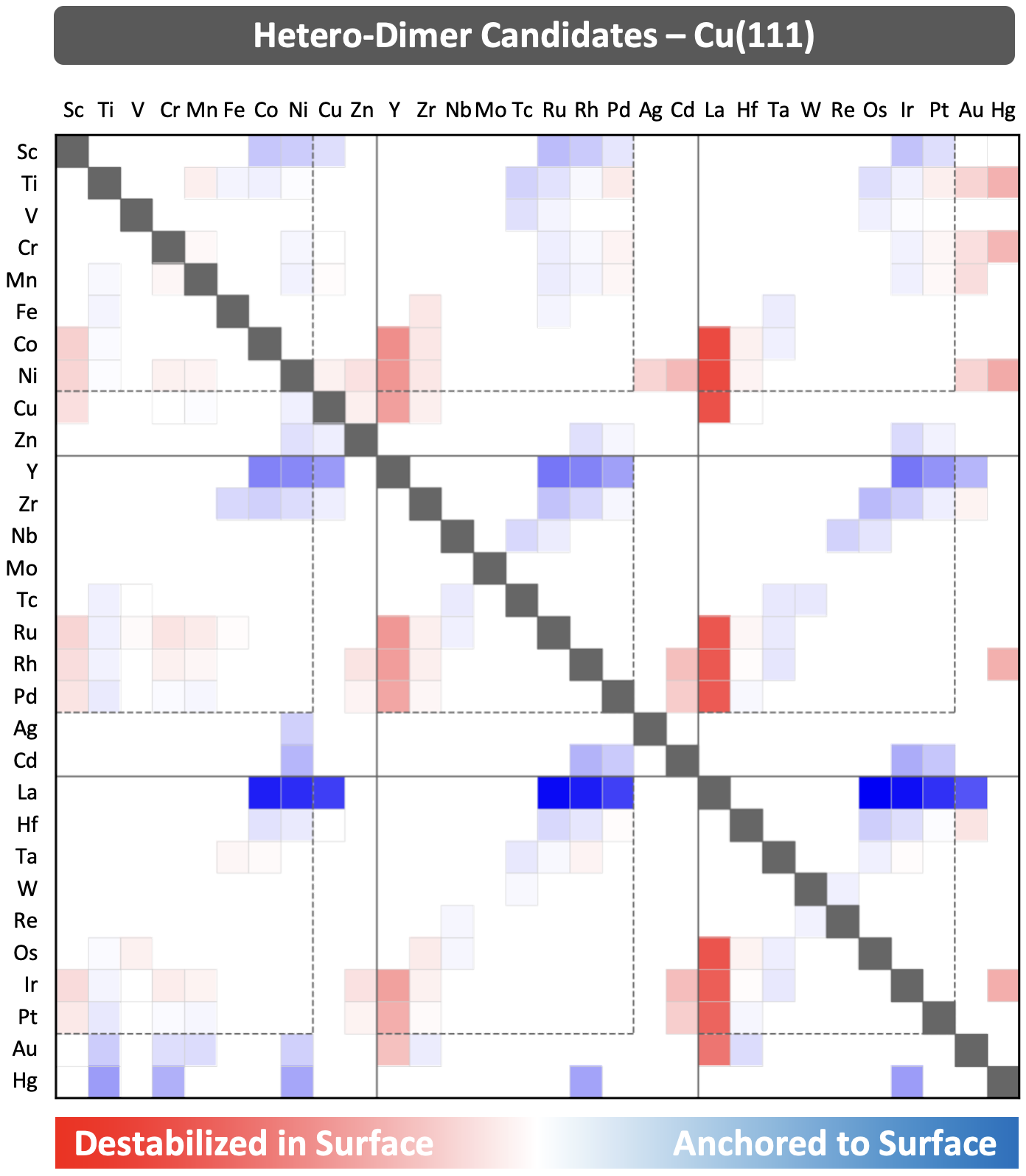}
    \caption{
        Anchoring energies ($E_\text{anchor}$), analogous to Fig.~5 in the main text.
        Blue indicates increased stabilization in the in the Cu(111) surface layer and red indicates increased stabilization in the bulk, relative to the reference SAA defined by the column.
        Only combinations that preferentially form hetero-dimers over both SAAs and homo-dimers are shown. 
        }
    \label{fig:Cu111-candidates}
\end{figure}

\begin{figure}[htbp]
    \includegraphics[width=12cm,height=\textheight,keepaspectratio]{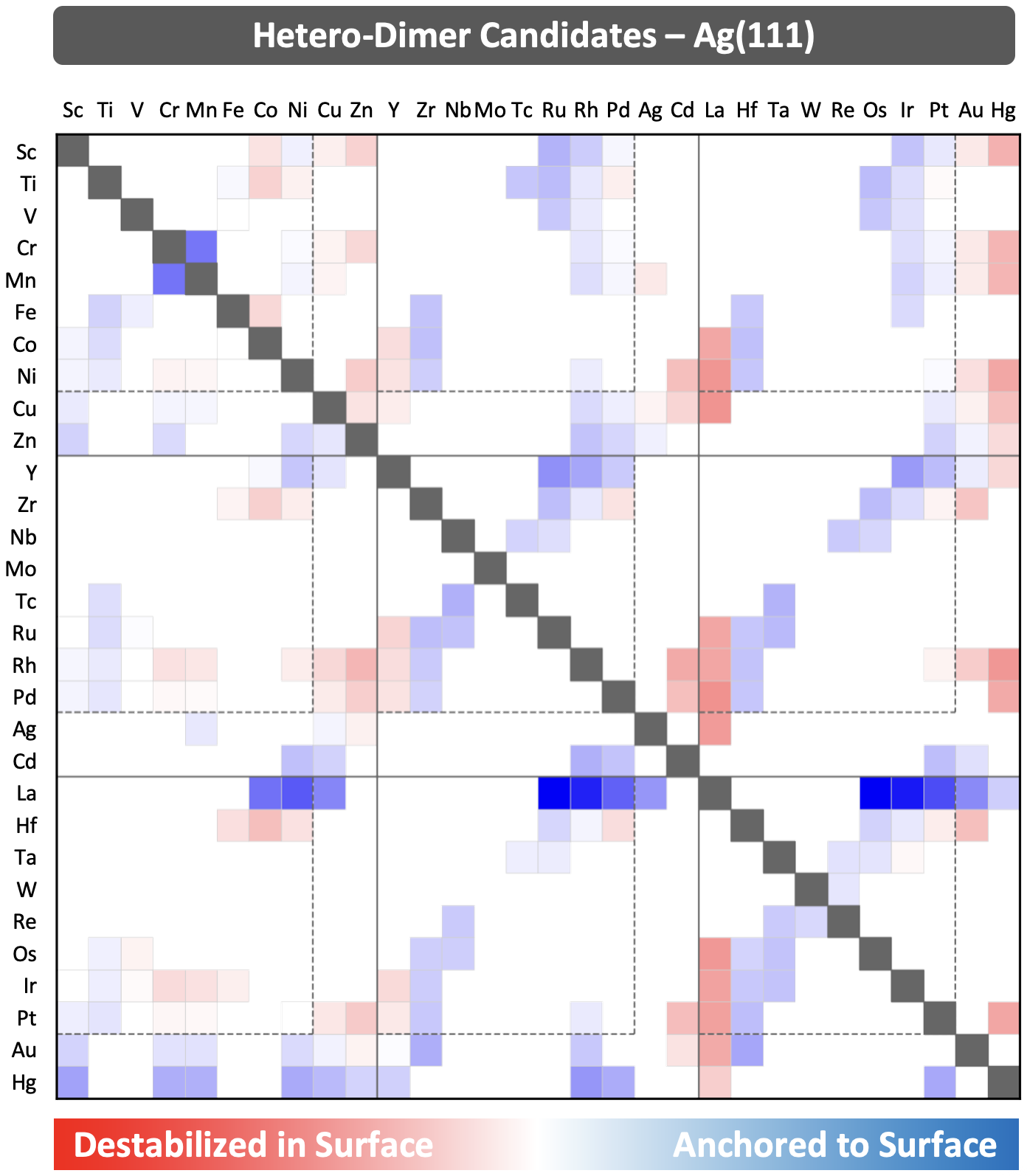}
    \caption{
        Anchoring energies ($E_\text{anchor}$), analogous to Fig.~5 in the main text.
        Blue indicates increased stabilization in the Ag(111) surface layer and red indicates increased stabilization in the bulk, relative to the reference SAA defined by the column.
        Only combinations that preferentially form hetero-dimers over both SAAs and homo-dimers are shown. 
        }
    \label{fig:Ag111-candidates}
\end{figure}

\clearpage

\clearpage
\bibliography{refs}